\documentclass[aip,amsmath,amssymb,reprint]{revtex4-1}
\usepackage{color}
\usepackage{graphicx}
\usepackage{dcolumn}
\usepackage{bm}
\DeclareUnicodeCharacter{2212}{-}

\usepackage[utf8]{inputenc}
\usepackage[T1]{fontenc}
\usepackage{mathptmx}
\usepackage{etoolbox}

\def\bra#1{\left\langle{#1}\right|}
\def\ket#1{\left|{#1}\right\rangle}
\def\braket#1#2{\left\langle{{#1}}\mathrel{\left|{\vphantom{{#1}{#2}}}\right.\kern-\nulldelimiterspace}{{#2}}\right\rangle}

\newcommand{\red}[1]{{\color{red}{#1}}}

\makeatletter
\def\@email#1#2{%
 \endgroup
 \patchcmd{\titleblock@produce}
  {\frontmatter@RRAPformat}
  {\frontmatter@RRAPformat{\produce@RRAP{*#1\href{mailto:#2}{#2}}}\frontmatter@RRAPformat}
  {}{}
}%
\makeatother
\begin{document}

\preprint{AIP/123-QED}

\title[Entanglement-enhanced quantum metrology: from standard quantum limit to Heisenberg limit]{Entanglement-enhanced quantum metrology: from standard quantum limit to Heisenberg limit}

\author{Jiahao Huang}
 \affiliation{Institute of Quantum Precision Measurement, State Key Laboratory of Radio Frequency Heterogeneous Integration, College of Physics and Optoelectronic Engineering, Shenzhen University, Shenzhen 518060, China}
 \affiliation{Laboratory of Quantum Engineering and Quantum Metrology, School of Physics and Astronomy, Sun Yat-Sen University (Zhuhai Campus), Zhuhai 519082, China}
 
\author{Min Zhuang}%
 \affiliation{Institute of Quantum Precision Measurement, State Key Laboratory of Radio Frequency Heterogeneous Integration, College of Physics and Optoelectronic Engineering, Shenzhen University, Shenzhen 518060, China}
 \affiliation{Quantum Science Center of Guangdong-Hong Kong-Macao Greater Bay Area (Guangdong), Shenzhen 518045, China}
 
\author{Chaohong Lee}
\email{chleecn@szu.edu.cn}
 \affiliation{Institute of Quantum Precision Measurement, State Key Laboratory of Radio Frequency Heterogeneous Integration, College of Physics and Optoelectronic Engineering, Shenzhen University, Shenzhen 518060, China}
 \affiliation{Quantum Science Center of Guangdong-Hong Kong-Macao Greater Bay Area (Guangdong), Shenzhen 518045, China}

\date{\today}

\begin{abstract}
Entanglement-enhanced quantum metrology explores the utilization of quantum entanglement to enhance measurement precision.
When particles in a probe are prepared into a suitable quantum entangled state, they may collectively accumulate information about the physical quantity to be measured, leading to an improvement in measurement precision beyond the standard quantum limit and approaching the Heisenberg limit.
The rapid advancement of techniques for quantum manipulation and detection has enabled the generation, manipulation, and detection of multi-particle entangled states in synthetic quantum systems such as cold atoms and trapped ions. 
This article aims to review and illustrate the fundamental principles and experimental progresses that demonstrate multi-particle entanglement for quantum metrology, as well as discuss the potential applications of entanglement-enhanced quantum sensors.
\end{abstract}

\maketitle

\tableofcontents

\section{\label{sec:1}Introduction}

Quantum metrology is a field that harnesses quantum principles for measurement purposes, with the objective of attaining high measurement precision using quantum resources. 
It has emerged as an interdisciplinary science that employs fundamental principles of quantum mechanics to measure physical quantities~\cite{PhysRevD.23.1693,PhysRevA.33.4033,PhysRevLett.71.1355,Giovannetti2004,Giovannetti2006,PhysRevLett.97.150402,PhysRevLett.101.040403,PhysRevLett104103602,Maccone2011,Giovannetti2011,Gross2012,PhysRevLett.113.250801,Sekaski2016Q,Degen2017,PhysRevX.8.021022,Pezze2018_RMP,PhysRevLett.124.120504,PhysRevLett.129.070502,Len2022N} and surpass the limitations of traditional measurement schemes~\cite{PhysRevD.23.1693,PhysRevA.33.4033,Giovannetti2004, Giovannetti2006, Giovannetti2011,Maccone2011, PhysRevLett.97.150402,Pezze2018_RMP}. 
The precise measurement plays a critical role in advancing physics~\cite{PhysRevLett.132.190001}, encompassing tasks such as testing fundamental physical principles~\cite{Schnabel2010, Rosi2017,Takamoto2020,Safronova2021,Lehnert2022,Bothwell2022,DeMille2024}, determining fundamental physical constants~\cite{Rosenband2008, Yu2019, Morel2020,DeMille2024}, detecting gravitational waves~\cite{PhysRevX.13.021019,Schnabel2010,PhysRevD.94.124043,PhysRevLett.110.171102,PhysRevLett.121.031101,PhysRevLett.123.231107,YijunJiang2021, Badurina2022}, and searching for dark matter~\cite{PhysRevLett.114.161301, PhysRevD.91.015015, Backes2021,PRXQuantum.4.020101,PRXQuantum.3.030333,Peik2021,Dixit2021,Lehnert2022,YijunJiang2021,Badurina2022}.
Moreover, it spurs innovation and advancements in various scientific disciplines and technologies~\cite{Bondarescu2012}, including inertial navigation~\cite{Adams2021,doi:10.1126/sciadv.aax0800,Dutta2023,Saywell2023,https://doi.org/10.1002/j.2161-4296.2005.tb01726.x,Hines2020}, groundwater monitoring~\cite{Stray2022}, mineral exploration~\cite{Stray2022}, and more~\cite{PhysRevLett.124.200503,PhysRevResearch.2.023414,PhysRevLett.92.160406,Giovannetti2001QuantumenhancedPA,Xiang2011}.

Quantum parameter estimation is a fundamental aspect of quantum metrology. 
It involves determining unknown parameters, such as transition frequency~\cite{PhysRevA.54.R4649,Huelga1997,PhysRevLett.1.105,Simons2021,Nichol2022,Rosenband2008,campbell2008absolute,Nichol:23,PhysRevX.12.021061,science7009,Schulte2020,PhysRevApplied.16.064056,Young2020,PhysRevLett.124.150502}, magnetic field~\cite{Meinel2021,science5532,science1094025,Jones2009,Ripka2002M,Marina2009S,Balogh2010S,PhysRevLett.130.203602,degen2008scanning,loretz2014nanoscale,mamin2013nanoscale,Meinel2021,science7009,Maze2008,science5532,PhysRevLett109253605,Mamin2007,science1094025,Ramsey1986,Ockeloen2013,PhysRevLett.117.138501,Zhuang2020}, and acceleration~\cite{https://doi.org/10.1002/j.2161-4296.2005.tb01726.x,peters1999measurement,Altin_2013,Braun2018, Saywell2023,Hines2020}, based upon the combination of estimation theory and quantum mechanics. 
Quantum interferometry~\cite{Abend2020,Lee2012,Hradil2005} is the most commonly employed method for implementing high-precision parameter estimation.
This approach typically involves initially preparing a quantum probe into a desired initial state, allowing the quantum probe to evolve to accumulate a phase relevant to the physical quantity to be measured, and ultimately extracting the quantity information through quantum interference and observable measurement.
Choosing quantum probes of remarkable sensitivity to external fields, we can construct high-precision quantum sensors for practical purposes.
By utilizing versatile quantum control techniques~\cite{Maze2008,science1192739,PhysRevA79062324,science1220513,PhysRevLett119183603,PhysRevA84060302,PhysRevA84042329,science5532,Kotler2011,Shaniv2017,science7009}, high-precision measurements of numerous physical quantities have been achieved in experiments~\cite{Degen2017,Bongs2023}.

Entanglement-enhanced quantum metrology is a specialized area within quantum metrology that explores how quantum entanglement can be utilized to further enhance measurement precision. 
Quantum entanglement is a phenomenon where two or more particles become correlated in such a way that the state of one particle cannot be described independently of the others~\cite{Pezze2018_RMP,PhysRevLett.123.073001,Colombo2022}.
This unique characteristic of quantum mechanics holds immense value as a resource that lacks classical equivalents.
By combining advanced quantum interferometry techniques with quantum entanglement, it becomes possible to significantly improve measurement precision.
In entanglement-enhanced quantum metrology, quantum entangled states are created and manipulated to extract more information about the parameter to be measured, thus enabling measurements with higher precision. 

In conventional measurement schemes involving an ensemble of individual particles or multiple independent measurements of a single particle, the measurement precision is limited by the standard quantum limit (SQL)~\cite{Giovannetti2004, Giovannetti2011,Giovannetti2006}.
In statistics, the fluctuations of a measured quantity can be reduced by conducting repeated measurements and averaging them, resulting in improved precision. 
Using a linear generator, according to the central limit theorem, the corresponding measurement precision scales as $\Delta \theta \propto N^{-1/2}$, where $\Delta \theta$ represents the standard deviation of an estimated quantity $\theta$ and $N$ denotes the times of repeated measurements or the number of individual particles (without entanglement) measured simultaneously as an ensemble.
This precision scaling is commonly known as the shot-noise limit, also referred to as the SQL.  

By employing entanglement among particles, all particles collectively accumulate the information about the physical quantity to be measured and the measurement precision may surpass the SQL~\cite{Giovannetti2004, Giovannetti2011,Giovannetti2006}. 
It has been demonstrated that the SQL can be surpassed by using quantum entangled states, such as 
spin squeezed states~\cite{Ruschhaupt_2012,PhysRevLett.131.063401,PhysRevLett.79.4782,PhysRevLett.118.083604,Vasilakis2015,science3397,Franke2023,Muessel2015,PhysRevLett.125.223401,Ma2011,Riedel2010}, twin-Fock states~\cite{Lucke2011,Strobel2014,Colombo2022,Li2020,Li2022}, {spin cat states~\cite{Jones2009,PhysRevLett.104.043601,Huang2015}, as input states. 
Notably, for an $N$-particle quantum probe, the utilization of a maximally entangled state~\cite{dowling2008quantum} [such as Greenberger-Horne-Zeilinger (GHZ) state~\cite{Huelga1997,Zhao2021,PhysRevLett.82.1345,Li2020}] as an input state may enhance the measurement precision to the Heisenberg limit (HL) with a scaling of $N^{-1}$ for a linear generator~\cite{Giovannetti2004, Giovannetti2011,Giovannetti2006}.
In the case of a linear generator, the HL surpasses the corresponding SQL by a factor of $\sqrt{N}$, leading to a $\sqrt{N}$-fold improvement in measurement precision.

To realize entanglement-enhanced quantum metrology, the preparation of multi-particle entangled states is a crucial step.
In recent decades, rapid progresses in preparing, manipulating and detecting multi-particle entangled states have propelled entanglement-enhanced quantum metrology into a distinct and rapidly expanding field in quantum technology~\cite{Gross2010,Bohnet2016}.
Numerous groundbreaking experimental breakthroughs have emerged, focusing on leveraging multi-particle entangled states to improve measurement precision from the SQL to the Heisenberg limit.
These advancements encompass diverse types of entangled states, ranging from spin squeezed states to non-Gaussian entangled states, and utilize common platforms such as cold atoms in cavities\cite{Greve2022,Li2022,PhysRevLett.104.073602,Zhang2015,Greve2022}, Bose condensed atoms\cite{Bohnet2016,Ma2011,Franke2023,Luo2017,Ockeloen2013,Zou2018,Pezze2016,RevModPhys.80.517,Riedel2010,Strobel2014} and ultracold trapped ions\cite{Strobel2014,Roos2006,Gilmore2021, Bohnet2016, PhysRevLett.106.130506}.
The successful generation of various multi-particle entangled states has laid a robust foundation for achieving entanglement-enhanced metrology.

To fully harness their potentials in quantum metrology, one has to choose appropriate methods for detecting entangled states.
Conventional measurement techniques may be inadequate for precisely extracting estimated information from highly entangled states, especially in the presence of noise~\cite{PhysRevLett.129.070502,PhysRevLett.113.250801,Polino:19,PhysRevLett.111.033608,PhysRevA.81.052330,PhysRevA.76.042127,Wan2020,PhysRevLett.109.233601,Jiao2023}.
Low detection efficiency for large particle numbers constitutes one of the primary limitations to realizing entanglement-enhanced metrology in practice, as it obscures signals and decreases measurement precision\cite{PhysRevA.83.063836,Huang2018_1}. 
By utilizing novel readout methods such as nonlinear detection~\cite{science3397,PhysRevLett104103602,Davis2016,PhysRevLett116090801,Linnemann2016,PhysRevLett118150401,PhysRevLett119193601,Liu2022,Mao2023,Colombo2022,Burd2019}, the signal can be amplified to enable precise measurements that are robust against disruption from detection noise. 
Owing to rapid advances in generating, manipulating, and measuring entanglement, this field is now progressing from initial proof-of-concept experiments towards developing practical entanglement-enhanced sensing technologies.

The applications of entanglement-enhanced quantum metrology are extensive and diverse. 
The ultimate goal of quantum metrology is to advance the technology of quantum sensing for practical use in various fields.
It aims to scrutinize the fundamental laws of physics with increased accuracy, reveal new phenomena in physics, determine physical quantities with enhanced precision, and construct practical quantum sensing devices.
The research in this field not only facilitates the manipulation and detection of quantum entanglement but also opens up new opportunities in applied physics, such as the development of next-generation entanglement-enhanced quantum sensors. 
These sensors offer high precision and sensitivity and can include atomic clocks~\cite{10.1063/5.0121372,Ludlow2015,PhysRevLett.123.123401,Nichol:23,Willette2018,Meiser_2008,PhysRevLett.92.230801,PhysRevLett.111.090802,PhysRevLett.111.090801,
Bondarescu2012,Bloom2014,Oelker2019,Bloom2014,Schioppo2017,Oelker2019}, atomic magnetometers~\cite{Barry2019,Budker2007,Davis2016,PhysRevLett104093602,PhysRevLett104133601,PhysRevLett104093602,PhysRevLett104133601,PhysRevLett113103004,Ockeloen2013,PhysRevX5031010}}, quantum gravimeters~\cite{peters1999measurement,Altin_2013,Saywell2023,Hines2020,PhysRevLett.117.203003,doi:10.1126/sciadv.aax0800}, quantum gyroscopes~\cite{PhysRevApplied.14.034065,PhysRevA.86.052116}, and other similar devices.
They can be utilized for tasks such as maintaining communication and energy network synchronization, real-time imaging of brain signals, collecting precise meteorological data for improved climate modeling, and monitoring underground water levels and volcanic eruptions~\cite{Bongs2023}. 
Furthermore, the integration of quantum control with quantum entanglement is expected to significantly advance the progress of entanglement-enhanced sensing technologies.

\begin{figure*}[!htp]
\includegraphics[width=2\columnwidth]{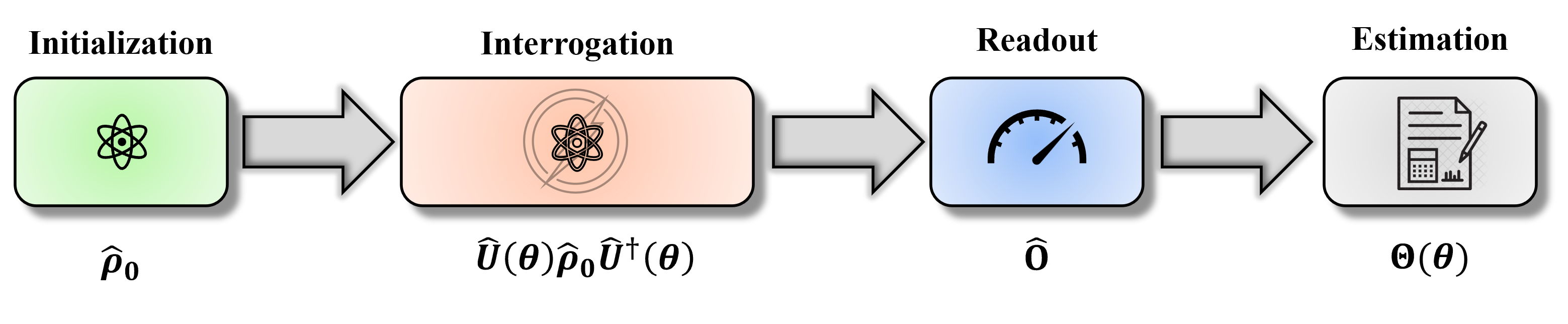}
\caption{\label{ProcessQPE}
  General procedure of quantum parameter estimation.
  (i) Initialization: a desired initial probe state is prepared.
  (ii) Interrogation: the probe undergoes a dynamical evolution dependent on the physical parameter $\theta$ to be measured.
  (iii) Readout: a recombination operation is performed and the final state is read out. 
  (iv) Estimation: the unknown parameter is finally estimated through an estimation procedure.
}
\end{figure*}

This review aims to provide a comprehensive understanding of entanglement-enhanced quantum metrology, covering its fundamental principles, experimental realization, and wide range of potential applications. 
In Sec.~\ref{sec:2}, we introduce the fundamental concepts necessary for grasping quantum parameter estimation and precision bounds.
This section is particularly valuable for readers who are new to quantum metrology and sensing.
In Sec.~\ref{sec:3}, we explore commonly employed entangled states that are beneficial for quantum metrology and discuss their preparation in various quantum systems, such as Bose-Einstein condensates (BECs), cold atoms in cavities, trapped ions, Rydberg tweezer arrays, and superconductors. 
The methods commonly used for entanglement generation include one-axis twisting, twist-and-turn dynamics, spin-mixing dynamics
, and crossing through quantum phase transitions (QPTs).
In Sec.~\ref{sec:4}, we delve into the techniques for effective and efficient quantum parameter estimation, focusing on interrogation and readout methods, with a special emphasis on nonlinear detection and interaction-based readout.
In Sec.~\ref{sec:5}, we discuss the potential applications of entanglement-enhanced quantum sensors, including atomic clocks, magnetometers, gravimeters, and gyroscopes.
Finally, in Sec.~\ref{sec:6}, we provide a summary and outlook for this rapidly advancing field.

\section{\label{sec:2}Fundamentals of quantum parameter estimation theory}

Quantum parameter estimation focuses on conducting measurements on a specific physical quantity and evaluating the resulting measurement precision~\cite{Giovannetti2004,Giovannetti2006,Maccone2011,Pezze2018_RMP}. 
Examples of quantum parameter estimation include magnetic field estimation~\cite{Jones2009,Ockeloen2013}, frequency estimation~\cite{PhysRevLett.117.143004,Pedrozo-Penafiel2020,Nichol2022}, weak force estimation~\cite{Shaniv2017}, and so on.
In a realistic measurement, the estimated parameter often relates to a phase shift, which can be measured using interferometric techniques~\cite{Hradil2005}.
This section primarily focuses on single parameter estimation scenarios. 
We provide an overview of fundamental concepts in quantum parameter estimation, such as the general procedure~\cite{PhysRevA.90.022117,PhysRevA.92.012312,liu2015quantum}, the Fisher information~\cite{Helstrom1969}, the Cram\'{e}r-Rao bound~\cite{Helstrom1969,PhysRevLett.72.3439}, and the precision bounds represented by the standard quantum limit~\cite{Giovannetti2004,PhysRevLett.102.100401,PhysRevLett.71.1355} and the Heisenberg limit~\cite{Giovannetti2004,PhysRevLett.102.100401,PhysRevLett.71.1355}.
Additionally, we illustrate different strategies of quantum metrology~\cite{Maccone2011}, emphasizing the importance of quantum entanglement in the preparation stage.

\subsection{\label{sec:2-1}General procedure of quantum parameter estimation}

A quantum parameter estimation process aims to measure an unknown physical parameter (denoted as $\theta$ in our review), which could represent a field, a frequency, a force, or any other physical quantity.
By leveraging the interaction between a probe and the system, it becomes possible to encode parameter-dependent information into the probe state.
The fundamental procedure of quantum parameter estimation includes the following four steps (as shown in Fig.~\ref{ProcessQPE}): 
(i) Preparation of a probe state $\rho_{0}$, which is sensitive to variations in the unknown parameter $\theta$.
(ii) The probe undergoes a unitary dynamical evolution $\hat{U}_{\theta}$ that depends on $\theta$. After the dynamical evolution, the state of the probe becomes $\rho_{\theta}=\hat{U}_{\theta}\rho_{0}\hat{U}_{\theta}^{\dag}$, which contains information about the unknown parameter $\theta$.
(iii) The information about $\theta$ is extracted by performing a practical measurement $\hat{E}$ on the final state.
(iv) Finally, a parameter estimation $\theta_{\textrm{est}}$ is derived from the obtained measurement results.
Here, for simplicity, we concentrate our discussion on unitary evolution and unbiased estimators, although it can be extended to non-unitary evolution and biased estimators.
Therefore, as a result of unbiased estimators, the statistical average may precisely gives the true parameter value: $\theta_{\textrm{est}}=\theta$. 
To evaluate the performance of an unbiased estimation, one may analyze the measurement precision of the estimated parameter $\theta$, which is characterized by the standard deviation $\Delta \theta = \sqrt{\langle \theta^2 \rangle-\langle \theta \rangle^2}$.

Parameter estimation is commonly achieved by associating the physical parameter with a phase shift, which can be obtained through  interferometry~\cite{Hradil2005}.
Optical interferometry involves the coherent combination of two or more light beams, resulting in interference patterns that enable the determination of their relative phase.
Quantum interferometry surpasses traditional interferometry by utilizing the wave nature of particles to achieve more precise measurements, thus enhancing measurement precision compared to conventional interferometry~\cite{Lee2012}.
In this context, we provide an introduction to two common types of quantum interferometry: SU(2) quantum interferometry and SU(1,1) quantum interferometry~\cite{PhysRevA.94.023834,PhysRevA.33.4033,10.1063/1.3606549,PhysRevA.87.023825,10.1063/1.4774380,Hudelist2014,PhysRevLett.111.033608,Li_2014,Ou2020,PRXQuantum.3.010202}.

\begin{figure*}[!htp]
 \includegraphics[width=2\columnwidth]{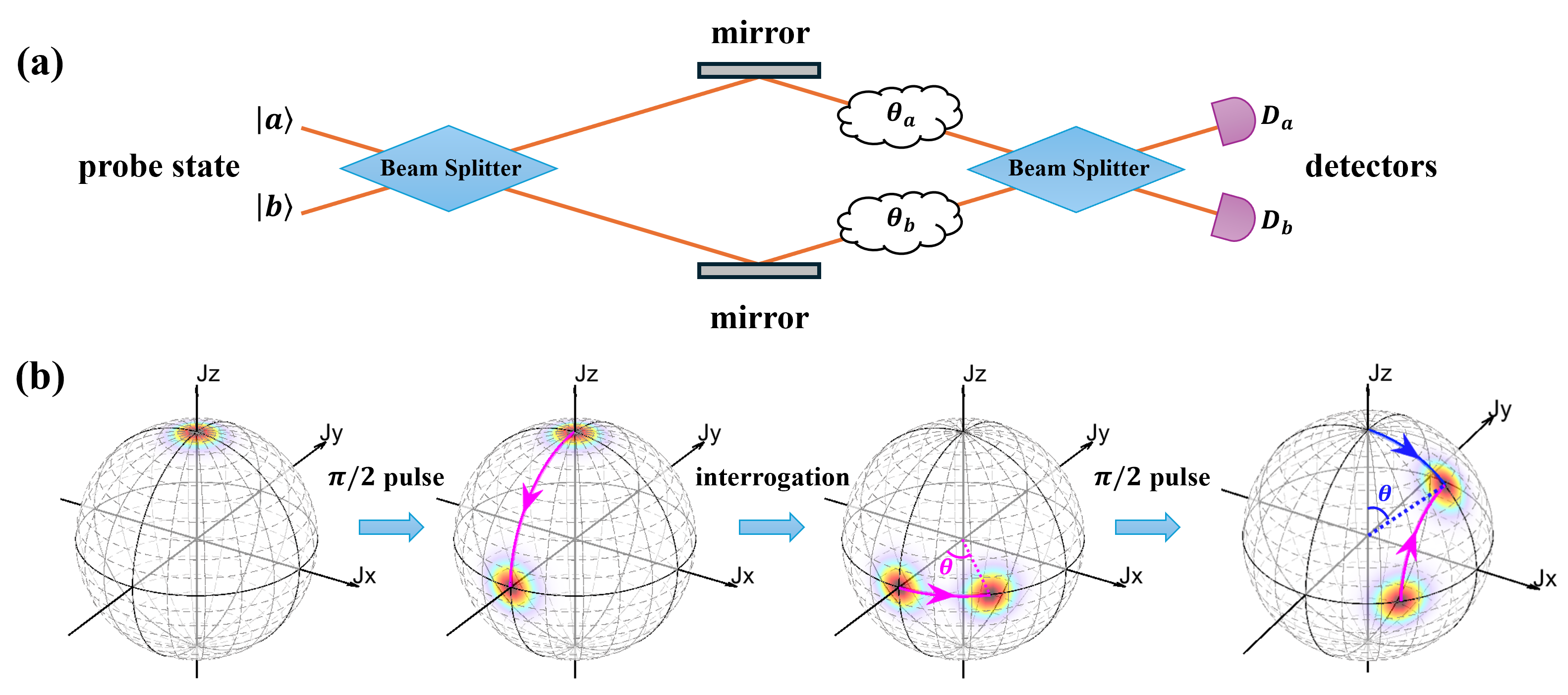}
  \caption{\label{Fig-Interferometry}
  Two-mode quantum interferometers.
(a) two modes $\ket{a}$ and $\ket{b}$ are combined by a balanced beam splitter, followed by a relative phase shift $\theta=\theta_a-\theta_b$ between the two arms, and finally recombined on a second balanced beam splitter.
 (b) Equivalent representation of two-mode quantum interferometer operations as rotations of the collective spin on the generalized Bloch sphere.
The initial state here is pointing toward the north pole. }
\end{figure*}

SU(2) quantum interferometry operates with quantum systems of SU(2) symmetry.
Consider a system of $N$ two-mode particles with the two modes labelled as $\ket{a}$ and $\ket{b}$, the system can be well described by a collective spin with length $J=\frac{N}{2}$.
The collective spin operators $\hat{J}_{\alpha}=\frac{1}{2}\sum_{l=1}^{N}\sigma_{\alpha}^{(l)}$ with $\alpha=\{x, y,z\}$ satisfy the angular momentum commutation relations $[\hat{J}_{\alpha}, \hat{J}_{\beta}]=i\epsilon_{\alpha\beta\gamma}\hat{J}_{\gamma}$, where $\alpha,\beta,\gamma=\{x,y,z\}$, $\sigma_{\alpha,\beta,\gamma}^{(l)}$ are the Pauli matrices for the $l$-th particle, and $\epsilon_{\alpha\beta\gamma}$ is the Levi-Civita symbol.
The conserved quantity of the associated SU(2) group is $\hat{J}^2=\hat{J}_{x}^2+\hat{J}_{y}^2+\hat{J}_{z}^2$, which is related to the total particle number $N$.
The angular momentum commutation relations satisfy the SU$(2)$ Lie algebra and so that this interferometry is called as SU$(2)$ quantum interferometry.
By using the Schwinger representation, the collective spin operators can be written as
$\hat{J}_{x}=\frac{1}{2}(\hat{a}^{\dag}\hat{b}+\hat{a}\hat{b}^{\dag})$,
$\hat{J}_{y}=\frac{1}{2i}(\hat{a}^{\dag}\hat{b}-\hat{a}\hat{b}^{\dag})$,
and $\hat{J}_{z}=\frac{1}{2}(\hat{a}^{\dag}\hat{a}-\hat{b}^{\dag}\hat{b})$ with the annihilation operators $\hat{a}$ and $\hat{b}$ for particles in $\ket{a}$ and $\ket{b}$.
Therefore the states can be represented by using the Dicke basis $\{|J,m\rangle\}$ obeying the eigen-equation $\hat J_z |J,m\rangle =m |J,m\rangle$ with the half population difference between the two modes $m = \{-J,-J + 1, ..., J-1, J\}$.

A common SU$(2)$ quantum interferometry employs two modes accumulating a relative phase encoded by the parameter to be estimated.
The two modes can be either two spatial paths, such as in a SU$(2)$ Mach-Zehnder interferometer ~\cite{VanFrank2014,PhysRevLett.100.073601,PhysRevLett.65.1348,RevModPhys.84.777}, or two hyperfine levels of an atom, such as in a Ramsey interferometer~\cite{Ramsey1980,Ramsey1986,Dalton2012}.
In a Mach-Zehnder interferometer, the input particle is divided into two parts by a linear balanced beam splitter and then the two parts pass through two different spatial paths to accumulate a relative phase between the two parts $\theta =\theta_b-\theta_a$, as shown in Fig.~\ref{Fig-Interferometry}~(a).
Finally, the two parts are recombined for interference via another linear balanced beam splitter to extract the relative phase from the interference fringe.
A common Ramsey interferometry consists of two $\pi/2$ pulses and a free evolution process, as shown in Fig.~\ref{Fig-Interferometry}~(b).
In comparison with a Mach-Zehnder interferometry, the two $\pi/2$ pulses act as the two beam splitters and the free evolution process accumulates the relative phase between the two involved levels.
For a single two-mode particle initially prepared in the probe state $\ket{a}$,
the particle is transformed to $(\ket{a}+\ket{b})/\sqrt{2}$ by the first balanced beam splitter of the Mach-Zehnder interferometer or the first resonant $\pi/2$ pulse in a Ramsey interferometer.
During the interrogation process, $\ket{a}$ and $\ket{b}$ respectively acquire the phases $\theta_a=-\omega T/2$ and $\theta_b=\omega T/2$ determined by $T$ the evolution time and $\omega$ the transition frequency between the two modes.
Thus the evolved state becomes $(e^{-i\theta_a}\ket{a}+e^{-i\theta_b }\ket{b})/\sqrt{2}$ with a relative phase $\theta=\theta_b-\theta_a=\omega T$.
Then, a second beam splitter or the second resonant $\pi/2$ pulse is applied and the final state reads as $\cos(\theta/2)\ket{a}+\sin(\theta/2)\ket{b}$.
Finally, through performing the half population difference measurement $\hat{J}_{z}=\frac{1}{2}(\hat{a}^{\dag}\hat{a}-\hat{b}^{\dag}\hat{b})$ on the final state, the relative phase $\theta$ can be estimated via $\theta$-dependent expectation values $\langle\hat{J}_{z}\rangle=\frac{1}{2}\cos{\theta}$.
Thus one can obtain the unknown parameter $\theta$ by an estimator function $\theta_{\textrm{est}}=\arccos[2\langle\hat{J}_{z}\rangle]$.
Generally, the evolution time $T$ is well-controlled and so that one can infer the transition frequency $\omega=\theta_{\textrm{est}}/T$ from the estimated relative phase $\theta_{\textrm{est}}$. 

In the case of multiple two-mode particles, using the picture of a collective spin, the process of parameter estimation via quantum interferometry is similar to the single-particle case. 
The whole process can be represented on a generalized Bloch sphere with a radius of the collective spin length $J = \frac{N}{2}$, as shown in Fig.~\ref{Fig-Interferometry}~(b).
Firstly, the initial state $\ket{\Psi_{\text{in}}}$ of $N$ two-mode particles is transformed by a balanced beam splitter, which implements a rotation $e^{i\frac{\pi}{2} \hat{J}_{x}}$.
Then, the system undergoes a dynamical evolution leading to a relative phase, which can be denoted by $e^{-i\theta \hat{J}_{z}}$.
Finally, the second balanced beam splitter $e^{-i\frac{\pi}{2} \hat{J}_{x}}$ is applied and the accumulated relative phase can be extracted from the half-population difference $\hat J_z$.
Combining the above three transformations, the final state is given as $\ket{\Psi_{\text{f}}(\theta)}=\hat{U}\ket{\Psi_{\text{in}}}$ with $\hat{U}=e^{-i\frac{\pi}{2} \hat{J}_{x}}e^{-i\theta \hat{J}_{z}}e^{i\frac{\pi}{2} \hat{J}_{x}}$.
Therefore, the information of an unknown relative phase $\theta$ is encoded into the final state $\ket{\Psi_{\text{f}}(\theta)}$.
The relative phase $\theta$ can be estimated by measuring the half population difference $\langle \hat{J}_{z}\rangle =\sum_{m=-J}^JmP(m|\theta)$ with $P(m|\theta)=|\langle{\Psi_{\text{f}}(\theta)}|{J,m}\rangle|^2$ denoting the conditional probability with outcome $m$ given the relative phase $\theta$.

\begin{figure}[!htp]
 \includegraphics[width=\columnwidth]{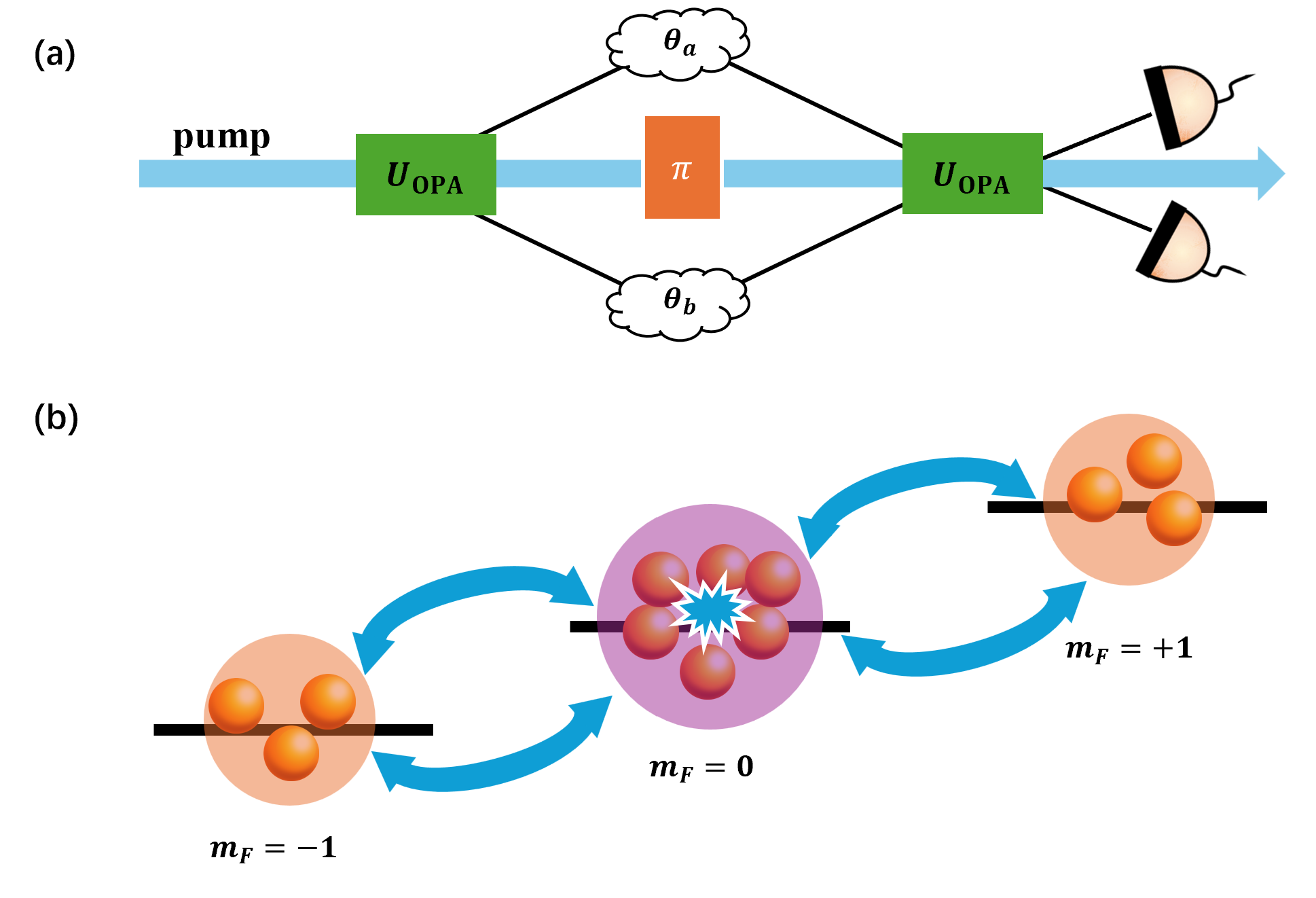}
  \caption{\label{Fig-Interferometry-SU11}
  Schematic of SU$(1,1)$ quantum interferometers. (a) An optical SU$(1,1)$ interferometer. The basic idea of this interferometric scheme is to replace the linear beam splitters of a standard interferometric scheme with OPAs that create or annihilate pairs of particles. (b) Spin-changing collisions of atoms can be regarded as an analogue of OPA in atom optics. For example, BEC in $m_F = 0$ acts as the pump beam. The nonlinear spin-changing collisions populate the modes $m_F = \pm 1$ which are the analogue to the signal and idler beams. }
\end{figure}

SU$(1,1)$ quantum interferometry~\cite{PhysRevA.33.4033} works with quantum systems of SU$(1,1)$ symmetry.
In contrast to SU$(2)$ quantum interferometry, whose beam splitters are implemented by using linear operations, the beam splitters of SU$(1,1)$ interferometry are achieved by using nonlinear wave mixing of parametric amplifiers.
Similar to a SU$(2)$ quantum interferometry with linear splitters obeying SU$(2)$ commutation relation, a SU$(1,1)$ quantum interferometry contains two nonlinear beam splitters obeying SU$(1,1)$ commutation relation, see Fig.~\ref{Fig-Interferometry-SU11}.

In a SU$(1,1)$ quantum interferometry~\cite{PhysRevA.33.4033}, the procedure of parameter estimation can be describe by three operators: $\hat{K}_{x}=\frac{1}{2}(\hat{a}^{\dag}\hat{b}^{\dag}+\hat{a}\hat{b})$, $\hat{K}_{y}=\frac{1}{2i}(\hat{a}^{\dag}\hat{b}^{\dag}-\hat{a}\hat{b})$, and $
\hat{K}_{z}=\frac{1}{2}(\hat{a}^{\dag}\hat{a}+\hat{b}^{\dag}\hat{b})$.
These operators satisfy the commutation relations: $[\hat{K}_{x},\hat{K}_{y}]=-i\hat{K}_{z}$, $[\hat{K}_{y},\hat{K}_{z}]=i\hat{K}_{x}$, and $[\hat{K}_{z},\hat{K}_{x}]=i\hat{K}_{y}$, for a SU$(1,1)$ group.
Besides, it is easy to find that $\hat{K}_{\textrm{tot}}^2=\hat J_z(\hat J_z+1)$ and $\hat J_z$ commute with all $\hat K_{x,y,z}$~\cite{PhysRevA.33.4033}.
Different from a SU$(2)$ group, the conserved quantity (Casimir invariant) of this SU$(1,1)$ group is given by $\hat{K}_{\textrm{tot}}^2=\hat{K}_{z}^2-\hat{K}_{x}^2-\hat{K}_{y}^2$, which is related to the ever fixed atom number imbalance.
In a SU$(1,1)$ interferometer, beam splitting and recombination are implemented by using optical parametric amplification that may generate correlated photon pairs.
The optical parametric amplification (OPA) can be described by $\hat{U}_{OPA}=e^{i \beta \hat{K}_{y}}$ with $\beta$ dependent upon the effective nonlinear coupling strength.
Such a beam splitter is inherently related to the parametric down conversion of a strong pump beam going through a nonlinear crystal. 
In parametric down conversion~\cite{PhysRevLett.59.2044}, the pump beam undergoes a nonlinear process, resulting in the generation of two new beams: the signal beam and the idler beam.  
Due to energy conservation, we have $\nu_p=\nu_s+\nu_v$ with $\nu_p$, $\nu_s$ and $\nu_v$ respectively denoting the frequencies of the pump beam, the signal beam and the idler beam. 
In the degenerate case, we have $\nu_s=\nu_v=\nu_p/2$. 

In Fig.~\ref{Fig-Interferometry-SU11}~(a), we show an optical SU$(1,1)$ interferometer. Starting from the vacuum state $\ket{0,0}$, the first OPA generates correlated photon pairs via $\hat{U}_{OPA}$.
Then the two modes accumulate phases via $\hat{U}_{ph}=e^{-i(\theta_a \hat{a}^{\dag}\hat{a}+\theta_b\hat{b}^{\dag}\hat{b})}=e^{-i \theta \hat K_z +\bar \theta \hat J_z}$ with $\theta=\theta_a+\theta_b$ and $\bar \theta=\theta_a-\theta_b$.
Finally, the second OPA is applied and the accumulated total phase can be extracted from the population $\hat{K}_{z}$.
Therefore the whole SU$(1,1)$ interferometer sequence can be expressed as $\hat{U}=e^{-i\beta\hat{K}_{y}} e^{-i \theta \hat K_z} e^{-i\beta \hat{K}_{y}}\ket{0,0}$ and the final expectation value of $\hat{K}_{z}$ is given as $\langle\hat{K}_{z}\rangle=(1-\cos\theta)\sinh^2(\beta)$.

The phase sensitivity of SU$(1,1)$ interferometer can be improved to the Heisenberg scaling~\cite{PhysRevA.33.4033,Li_2014}.
The difference between SU$(1,1)$ and SU$(2)$ interferometers lies in the beam splitting and mixing processes.
The parametric amplifiers are active to generate quantum fields, while the linear beam splitters are passive and rely on injection of quantum states to achieve quantum enhancements~\cite{Ou2020}. 
On one hand, parametric amplifiers allow coherent superposition of waves and lead to amplified noises.  
On the other hand, parametric amplifiers may generate quantum entanglement, leading to correlated quantum noises, which can be canceled at destructive interference. 
This gives rise to higher signal amplification than noise amplification and thus improves the measurement precision~\cite{Ou2020}. 

In recent years, SU$(1,1)$ interferometers have been implemented experimentally with different physical systems~\cite{PhysRevLett.109.183901,PhysRevLett.115.043602,PhysRevLett.119.223604,Linnemann_2017,10.1063/1.3606549,PhysRevA.87.023825,10.1063/1.4774380,Hudelist2014,PhysRevLett.111.033608,Li_2014}.
Particularly, the SU$(1,1)$ interferometry has been demonstrated by using the spin-mixing dynamics in spin-1 systems, which will be briefly illustrated in Sec.~\ref{sec:3-3}.

\subsection{\label{sec:2-2}Fisher information and  Cram\'{e}r-Rao bound}

Fisher information is a mathematical concept in statistics that measures the amount of information  about an unknown parameter.
Measurement precision refers to the degree of uncertainty with which a physical quantity can be measured.
Fisher information provides a theoretical bound on the best achievable precision of an unbiased estimator for the parameter to be estimated.
In practice, higher Fisher information corresponds to better measurement precision. 

In a classical estimation problem with a set of measurement data from $\nu$ times of identical experiments using an unbiased estimator, the measurement precision is limited by the Cram\'{e}r-Rao bound (CRB), 
\begin{equation}\label{CRB}
    \Delta \theta \ge \frac{1}{\sqrt{\nu F(\theta)}},
\end{equation}
with the Fisher information
\begin{eqnarray}\label{FI_def}
    F(\theta) &=& \sum_{i} P(x_i|\theta) \left(\frac{\partial \ln [P(x_i|\theta)]}{\partial \theta}\right)^2 \nonumber \\
    &=& \sum_{i} \frac{1}{P(x_i|\theta)} \left(\frac{\partial P(x_i|\theta)}{\partial \theta}\right)^2,
\end{eqnarray}
where $P(x_i|\theta)$ is the conditional probability of measuring the experimental data $x_i$ given a specific value of $\theta$.
%

In quantum mechanics, the measurement is described by a positive operator-values measure (POVM)~\cite{Nielsen2010}.
A POVM is a set of Hermitian operators $\{\hat{E}(\mu)\}$ parameterized by $\mu$ that are positive, $E(\mu) \geq 0$.
$ P(\mu|\theta)=\textrm{Tr}[\hat{\rho}(\theta)\hat{E}(\mu) ] \geq 0$, to guarantee non-negative probabilities and satisfy $\sum_{\mu} \hat{E}(\mu)=\hat{I}$, to ensure normalization $\sum_{\mu} P(\mu|\theta)= 1$.
In the estimation scenario whose probe and measurement are both determined, the classical Fisher information (CFI) provide an achievable bound on measurement precision, which is a quantity able to catch the amount of information encoded in output probabilities of the estimation process~\cite{Fisher_1925}.
With the probability distribution $P(\mu|\theta)$, the CFI can be defined as
\begin{eqnarray}\label{Eq:CFI}
{F}_C(\theta;\hat{E})=\sum_{\mu}\frac{1}{P(\mu|\theta)}\left(\frac{\partial{P(\mu|\theta)}}{\partial{\theta}}\right)^2.
\end{eqnarray}
For an unbiased estimator, the measurement precision of the parameter $\theta$ is bounded by the classical CRB~\cite{Helstrom1976},
\begin{eqnarray}\label{Eq:CRB}
\Delta \theta \geq \frac{1}{\sqrt{\nu{F}_C(\theta;\hat{E})}}
\end{eqnarray}
where $\Delta \theta=\sqrt{\langle\theta^2\rangle-\langle\theta\rangle^2}$ is the standard deviation of the estimated phase $\theta$ and $\nu$ denotes the number of trials.
The CFI provides an asymptotic measure of the amount of information about the parameters of a system under specific measurements.

\subsection{\label{sec:2-3}Quantum Fisher information and quantum  Cram\'{e}r-Rao  bound}

In the above subsection, we illustrate the scenario that both probe and measurement are determined.
However, the selection of $\hat E(\mu)$ has a significant influence on $P(\mu|\theta)$, which in turn directly impacts the CFI ${F}_C(\theta)$.%
To achieve optimal measurement precision, the POVM measurement should be carefully selected. 
%
An upper bound to the CFI can be obtained by maximizing Eq.~\eqref{Eq:CFI} over all possible POVMs $\{\hat{E}(\mu)\}$ for the involved quantum system~\cite{PhysRevLett.72.3439}, leading to
\begin{eqnarray}\label{Eq:QFI}
{F}_Q(\theta)=\max_{\hat{E}}{F}_C(\theta;\hat{E}),
\end{eqnarray}
which is called quantum Fisher information (QFI).
Therefore the measurement precision of the parameter $\theta$ is bounded by the the quantum Cram\'{e}r-Rao bound (QCRB),
\begin{eqnarray}\label{Eq:CRB2}
\Delta \theta \geq \frac{1}{\sqrt{\nu{F}_Q(\theta)}}.
\end{eqnarray}
Since the property ${F}_Q(\theta)\geq {F}_C(\theta;\hat{E})$, the QCRB is the ultimate achievable limit of the CRB.

According to the QCRB~\eqref{Eq:CRB2}, for a given probe state and an interferometer transformation, the QFI and the QCRB allow one to calculate the ultimate achievable precision bound regardless of the measurements.
A general expression of the QFI can be expressed as~\cite{Care_1983,Helstrom1976,Helstrom1969} 
\begin{eqnarray}\label{Eq:FQ1}
{F}_Q(\theta)=(\Delta \hat{L}_{\theta})^2=\textrm{Tr}[\hat{\rho}_\theta \hat{L}_{\theta}^{2}],
\end{eqnarray}
which is the variance of the symmetric logarithmic derivative operator $\hat{L}_{\theta}$ defined by $\partial_{\theta}\hat{\rho}_\theta = (\hat{\rho}_\theta \hat{L}_{\theta}+\hat{L}_{\theta} \hat{\rho}_\theta)/2$.
Moreover, the QFI can be expressed in terms of the spectral decomposition of the output state $\hat{\rho}_\theta=\sum_{n} a_n |n_{\theta}\rangle\langle n_{\theta}|$ with $\theta$-dependent eigenvalues $a_n \geq 0$ and eigenvectors $|n_{\theta}\rangle$.
Thus, the QFI can be explicitly written as~\cite{BRAUNSTEIN1996135,PhysRevA.82.042103}
\begin{eqnarray}\label{Eq:FQ2}
{F}_Q(\theta)=\sum_{n}\frac{(\partial_\theta a_{n})^2}{a_{n}}+2\sum_{i\neq j}\frac{( a_{i}- a_{j})^2}{a_{i}+ a_{j}}|\langle n_{i}|\partial_\theta n_{j}\rangle|^2,
\end{eqnarray}
with $|\partial_\theta n_{j}\rangle\equiv \partial_\theta |n_{j}\rangle$.
In Eq.~\eqref{Eq:FQ2}, the first term quantifies the information about $\theta$ encoded in $|n_{\theta}\rangle$ and it corresponds to the Fisher information obtained via projecting on the eigenstates of $\hat{\rho}_\theta$, and the second term accounts for the change of eigenstates with $\theta$.

According  to Eq.~\eqref{Eq:FQ2}, the QFI for a pure state $\hat{\rho}_\theta=|\Psi_{\text{f}}(\theta)\rangle\langle\Psi_{\text{f}}(\theta)|$ can be written as~\cite{Fujiwara1995}
\begin{eqnarray}\label{Eq:QFI3}
{F}_Q(\theta)=4 \left[\langle\partial_{\theta}\Psi_{\text{f}}(\theta)|\partial_{\theta}\Psi_{\text{f}}(\theta)\rangle-|\langle\partial_{\theta}\Psi_{\text{f}}(\theta)|\Psi_{\text{f}}(\theta)\rangle|^2 \right]\nonumber \\
\end{eqnarray}
with $|\partial_{\theta}\Psi_{\text{f}}(\theta)\rangle=\partial|\Psi_{\text{f}}(\theta)\rangle/\partial \theta$.
Furthermore, if the final state is evolved from a pure initial states $\hat{\rho}_0=|\Psi_{\text{in}}\rangle\langle\Psi_{\text{in}}|$ under an unitary evolution $U_{\theta}=e^{-i\hat{H}_{\theta}}$, the QFI can be given as
\begin{eqnarray}\label{Eq:QFI4}
{F}_Q(\theta)=4(\Delta \hat{H}_{\theta})^2,
\end{eqnarray}
where $(\Delta \hat{H}_{\theta})^2$ is the variance of $\hat{H}_{\theta}$ for the initial state $|\Psi_{\text{in}}\rangle$.

However in realistic experiments, one needs to find a suitable observable to approach the above theoretical precision bounds.
For an observable $\hat{O}$ with eigenvalue $\mu$, we have the standard deviation of $\hat{O}$ on the final state $\hat{\rho}_\theta$ are
\begin{equation}\label{Eq:Deviation}
\Delta{\hat{O}}=\sqrt{\langle\hat{O}^2\rangle_{\text{f}}-\langle\hat{O}\rangle_{\text{f}}^2}=\sqrt{\textrm{Tr}[\hat{\rho}_\theta \hat{O}^2]-\textrm{Tr}[\hat{\rho}_\theta \hat{O}]^2}.
\end{equation}
For a pure final state, one can obtain
\begin{equation}\label{Eq:Expectation}
\langle\hat{O}\rangle_{\text{f}}=\sum_{\mu}\mu P(\mu|\theta)=\bra{\Psi_{\text{f}}(\theta)} \hat{O} \ket{\Psi_{\text{f}}(\theta)},
\end{equation}
and
\begin{equation}\label{Eq:Expectation2}
\langle\hat{O}^2\rangle_{\text{f}}=\sum_{\mu}\mu^2 P(\mu|\theta)=\bra{\Psi_{\text{f}}(\theta)} \hat{O}^2 \ket{\Psi_{\text{f}}(\theta)}.
\end{equation}%
From the general quantum estimation theory, the measurement precision of the estimated parameters can be given via the error-propagation formula~\cite{PhysRevLett.72.3439},
\begin{equation}\label{Eq:Parameter uncertainty}
\Delta \theta=\frac{\Delta{\hat{O}}}{|\partial{\langle\hat{O}\rangle_{\text{f}}/ \partial{\theta}|}}.
\end{equation}

\subsection{\label{sec:2-4}Standard quantum limit and Heisenberg limit}

In this subsection we will illustrate how to optimize QFI over the initial states and attain the ultimate precision limits allowed by quantum mechanics~\cite{Giovannetti2006,Giovannetti2011}.
%
For simplicity, we first consider a single-particle system with a Hamiltonian $\hat{h} {\theta}$ for parameter encoding.
According to Eq.~\eqref{Eq:QFI4}, we have
\begin{equation}
   4(\Delta \hat{h})^2\leq (\lambda_\textrm{{max}}-\lambda_\textrm{{min}})^2, 
\end{equation}
where $\lambda_\textrm{max}$ and $\lambda_\textrm{min}$ are the maximum and minimum eigenvalues of $\hat{h}$, corresponding to eigenvectors $\ket{\lambda_\textrm{max}}$ and $\ket{\lambda_\textrm{min}}$.
The variance of $\hat{h}$ is maximized when $\ket{\Psi_{\textrm{in}}}=\frac{1}{\sqrt{2}}\left(\ket{\lambda_\textrm{max}}+\ket{\lambda_\textrm{min}}\right)$, so the maximal Fisher information is
\begin{eqnarray}\label{Eq:QFI-m}
{F}_Q(\theta)=\left({\lambda_\textrm{max}}-{\lambda_\textrm{min}}\right)^2.
\end{eqnarray}
When there are $N$ identical systems, the whole Hamiltonian for phase encoding is given by 
\begin{equation}\label{H}
   \hat{H} {\theta}= \left[\hat{h}^{(1)} +\hat{h}^{(2)} ... + \hat{h}^{(N)}\right] {\theta}, 
\end{equation} 
where $\hat{h}^{(l)}$ is the generator for the $l$-th system.

For $N$ probes that are classically combined without entanglement, 
the whole state can be then written as $\rho=\rho^{(1)} \bigotimes \rho^{(2)} ...\bigotimes \rho^{(N)}$, with $\rho^{(l)}$ is the state of the $l$-th system.
Using the convexity and additivity properties of QFI~\cite{Pezze2018_RMP}, its value for state $\rho$ satisfies
\begin{eqnarray}\label{Eq:QFI-N}
{F}_Q(\theta)=\sum_{l=1}^{N}{F}_Q^{(l)}(\theta)  \leq N \left[{\lambda_\textrm{max}^{(l)}}-{\lambda_\textrm{min}^{(l)}}\right]^2.
\end{eqnarray}
The condition for the equality is that each probe is in the same state of $\ket{\Psi_{\textrm{in}}^{(l)}}=\left(\ket{\lambda_\textrm{max}^{(l)}}+\ket{\lambda_\textrm{min}^{(l)}}\right)/{\sqrt{2}}$.  
Since $\left[{\lambda_\textrm{max}^{(l)}}-{\lambda_\textrm{min}^{(l)}}\right]^2$ are the same, the measurement precision versus the number of the probes $N$ is 
\begin{eqnarray}\label{Eq:SQL1}
    \Delta\theta \propto 1/\sqrt{N}.
\end{eqnarray}
Thus for a system of $N$ particles (labeled as $l= 1, 2,..., N$) and each particle regarding as a qubit with $\hat{h}^{(l)}=\frac{1}{2} \sigma_z^{(l)}$,  $\lambda_{\textrm{max}}^{(l)}=1/2$, $\lambda_{\textrm{max}}^{(l)}=-1/2$, 
one can get
\begin{eqnarray}\label{Eq:SQL}
    \Delta \theta_{\textrm{SQL}} = 1/\sqrt{N},
\end{eqnarray}
which is the so-called standard quantum limit (SQL)~\cite{PhysRevA.55.2598}. 

By employing quantum entanglement, the SQL can be surpassed~\cite{PhysRevLett.87.270404}.
For $N$ entangled particles whose density matrix $\rho$ cannot be expressed in the product form,  the maximum and minimum eigenvalues of $\hat{H}$ in Eq.~\eqref{H} are $N\lambda_\textrm{max}$ and $N\lambda_\textrm{min}$, respectively.
Here, the $\hat{h}^{(l)}$ for each system is in the same form with maximal and minimal eigenvalues $\hat{h}^{(l)}\ket{\lambda_\textrm{max}^{(l)}}=\lambda_\textrm{max} \ket{\lambda_\textrm{max}^{(l)}}$ and $\hat{h}^{(l)}\ket{\lambda_\textrm{min}^{(l)}}=\lambda_\textrm{min}\ket{\lambda_\textrm{min}^{(l)}}$.  
%
%
According to Eq.~\eqref{Eq:QFI-N}, one can obtain
\begin{equation}
    F_{Q}(\theta)|_\textrm{max}=N^2 ({\lambda_\textrm{max}}-{\lambda_\textrm{min}})^2.
\end{equation}
Thus the corresponding measurement precision bound becomes
\begin{eqnarray}\label{Eq:HL}
    \Delta\theta \propto 1/N.
\end{eqnarray} 
Similarly, for a system with $N$ qubits with $\hat{h}^{(l)}=\frac{1}{2} \sigma_z^{(l)}$, one can obtain 
\begin{equation}
    \Delta\theta_\textrm{HL} = 1/N,
\end{equation}
which is the so-called Heisenberg limit (HL)~\cite{Giovannetti2004,PhysRevLett.102.100401,PhysRevLett.71.1355}.
The HL implies a $\sqrt{N}$ improvement of the precision scaling over the SQL.
A primary goal in quantum metrology is to attain this ultimate precision limit by using optimal quantum parameter estimation process~\cite{PhysRevLett.131.150802}. 

\subsection{\label{sec:2-5}Strategies of quantum metrology}
\begin{figure}[!htp]
 \includegraphics[width=\columnwidth]{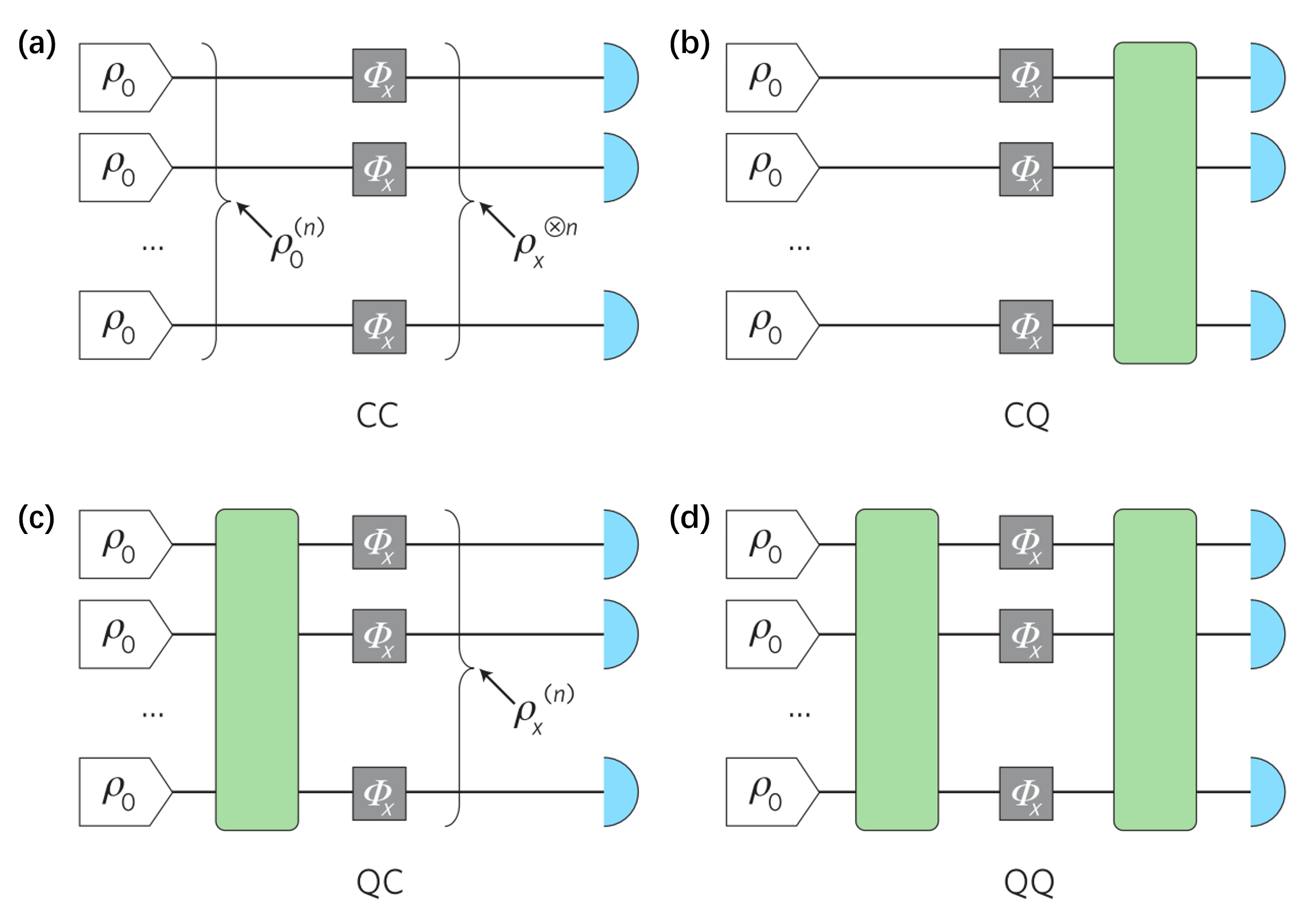}
  \caption{\label{Fig-Strategies}
  Four strategies of quantum metrology~\cite{Giovannetti2011}. (a) The CC strategy. The probes are prepared in a separable state $\rho_0^{(n)}$ and local measurements are made at the output independently. (b) The CQ strategy.  The probes are prepared in a separable state $\rho_0^{(n)}$ but entanglement among the probes is generated after phase encoding and before the detection. (c) The QC strategy. Entanglement among the probes is generated before phase encoding, but no entanglement resource is employed at the detection stage. (d) The QQ strategy. The entanglement can be used both at the probe preparation and detection stages. The QQ strategy provides the best performance, and the CC strategy provides the worst. Reproduced with permission from Giovannetti \textit{et al.}, Nat. Photon. \textbf{5}, 222 (2011). Copyright 2011 Springer Nature.}
\end{figure}

Based on the general procedure of quantum parameter estimation, below we will introduce different strategies for implementing quantum metrology.
As illustrated previously, the procedures for the parameter estimation generally contains four stages: initialization, interrogation, readout and estimation.
In analogy to quantum communication, four different scenarios are possible to be implemented~\cite{Giovannetti2006, Maccone2011}:
(i) classical-classical (CC) strategy, which do not employ quantum effects in both stages, as shown in Fig.~\ref{Fig-Strategies}~(a);
(ii) classical-quantum (CQ) strategy, which employs quantum effects only in the measurement stage stage, as shown in Fig.~\ref{Fig-Strategies}~(b);
(iii) quantum-classical (QC) strategy, which employs quantum effects only in the interrogation stage, as shown in Fig.~\ref{Fig-Strategies}~(c);
(iv) quantum-quantum (QQ) strategy, which employs quantum effects only in both stages, as shown in Fig.~\ref{Fig-Strategies}~(d).

It is proved that the QC and QQ strategies that using entanglement in the preparation stage can help to improve the measurement precision while for CQ strategy where entanglement operations only in readout cannot. 
In the following, we give brief derivations for the ultimate precision bounds in the above four different scenarios and show their performances.
We assume the interrogation stage for each probe is through a unitary operator $\hat U_{\theta}=e^{-i{\theta }\hat{h}^{(l)}}$, where $\theta$ is the parameter to be estimated and the generator for the $l$-th probe $\hat{h}^{(l)}$ is a known Hermitian operator.
Taking $\rho_0^{(N)}$ as the state of the $N$ probes, it will be transformed into $\hat U_{\theta}^{\bigotimes N} \rho_0^{(N)} \hat U_{\theta}^{\dagger \bigotimes N}$, where $\hat U_{\theta}^{\bigotimes N}$ is the unitary transformation generated by $\hat{H}= \sum_{l=1}^{N} \hat{h}^{(l)}$.
According to Eq.~\eqref{Eq:QFI4}, we have
\begin{eqnarray}\label{Eq1}
    \Delta\theta \geq \frac{1}{2\Delta \hat{H}}.
\end{eqnarray}

For the CC and CQ strategies with separable state, $\Delta \hat{H}=\left[\sum_{l=1}^{N} \Delta^2 \hat{h}^{(l)}\right]^{1/2}$.
As illustrated in Sec.~\ref{sec:2-4}, the maximum $\Delta \hat{H}= \sqrt{N}(\lambda_\textrm{max}-\lambda_\textrm{min})/2$ that can be achieved via preparing each probe in the equally weighted superpositions of the eigenvectors $\ket{\lambda_\textrm{{max}}}$ and $\ket{\lambda_\textrm{{max}}}$ of $\hat{h}^{(l)}$.
%
%
%
Thus, measurement precision of parameter $\theta$ for the optimal CC and CQ strategies is
\begin{eqnarray}\label{Eq:CC&CQ}
\Delta\theta \geq \frac{1}{\sqrt{N}(\lambda_\textrm{max}-\lambda_\textrm{min})}.
\end{eqnarray}
This result indicates that the CC strategy is as precise as the CQ strategy. 
Thus for non-entangled probes, the subsequent entangled measurements have no enhancement on the precision in Eq.~\eqref{Eq:CC&CQ}, which is just the SQL.

In contrast, for the QC and QQ strategies with entangled probes, e.g., in the maximally entangled state $\rho_0^{(N)} =\ket{\Psi} \bra{\Psi}$ that maximizing $\Delta \hat{H}$
\begin{equation}\label{state_QQ}
  \ket{\Psi}=(\ket{\lambda_\textrm{{max}}}^{\otimes N}+\ket{\lambda_\textrm{{min}}}^{\otimes N})/\sqrt{2},  
\end{equation}
which is the equally weighted superposition of the eigenvectors relative to the maximum and minimum eigenvalues of the global generator $\hat{H}$, the measurement precision of parameter $\theta$ for the optimal QC and QQ strategies can be enhanced to the Heisenberg limit
\begin{eqnarray}\label{Eq:QC&QQ}
    \Delta\theta \geq \frac{1}{{N}(\lambda_\textrm{max}-\lambda_\textrm{min})},
\end{eqnarray}
with a $\sqrt{N}$ improvement over Eq.~\eqref{Eq:CC&CQ}.
The bound of Eq.~\eqref{Eq:QC&QQ} is obtained by using the fact that the maximum and minimum eigenvalues both grow linearly with the number of systems $N$, the maximum and minimum eigenvalues of the global generator $\hat{H}$ are $N\lambda_\textrm{max}$ and  $N\lambda_\textrm{min}$, respectively.

According to Eqs.~\eqref{Eq:CC&CQ} and~\eqref{Eq:QC&QQ}, one can find that the ultimate measurement precision for the CC and CQ strategies just can approach the SQL, i.e., $\propto 1/\sqrt{N}$, while the
ultimate measurement precision for the QC and QQ strategies can approach to the Heisenberg limit, i.e., $\propto 1/N$.
This implies that quantum entanglement at the preparation stage is useful to increase the measurement precision, while the entanglement operations at the measurement stage may not.
There are many different types of multi-particle entangled states that have been used to improve the precision.
In the next section, we will introduce some typical multi-particle entangled states in practical quantum metrology.

\section{\label{sec:3}Metrologically useful entangled states and their preparation}

As discussed in the previous section, entanglement plays a crucial role in enhancing the measurement precision of quantum sensors. 
Entanglement is a unique phenomenon in quantum mechanics that holds significant value as a resource without classical equivalents~\cite{RevModPhys.80.517}.
It refers to a phenomenon characterized by fascinating correlations that are exclusively observable at the minuscule scale of the quantum realm and do not manifest in our macroscopic world.
In 1935, Einstein, Podolsky, and Rosen (EPR) originated the famous ``EPR paradox''~\cite{Einstein1935, Reid2009}, which states that two spatially separated particles can have perfectly correlated positions and momenta.
This seemingly contradicts the Heisenberg uncertainty principle, where the position and momentum of a particle cannot be simultaneously determined. 
The EPR argument pinpoints a contradiction between local realism and the completeness of quantum mechanics. 
In their attempt to support local realism, they aimed to demonstrate that the lack of precise predictions of measurement results is due to the incompleteness of quantum mechanics.
Schrodinger was the first to recognize that the wavefunction of the two particles is entangled~\cite{Schrodinger1935}. 
This implies that the global states of a composite system cannot be expressed as a product of the states of individual subsystems.
This phenomenon, known as entanglement, serves as the foundation for quantum technology to break through bottlenecks of conventional technologies. 

\subsection{\label{sec:3-1} Metrologically useful entanglement}

Formally, quantum entanglement is a basic phenomenon in which the quantum states of two or more particles get coupled in a manner where the state of one particle is intrinsically dependent on the state of the others.
In a system of $N$ particles,
a pure quantum state $|\Psi\rangle_{sep}$ is separable if it can be expressed as a product state~\cite{RevModPhys.81.865}, 
\begin{equation}
    \label{sep}
    |\Psi\rangle_{sep}=|\psi^{(1)}\rangle \otimes |\psi^{(2)}\rangle \otimes \cdot \cdot \cdot \otimes |\psi^{(N)}\rangle,
\end{equation}
with $|\psi^{(k)}\rangle$ being the $k$-th particle's state and the state of each individual particle is independent on the others. 
Such a state can be written as a density matrix, i.e.,
\begin{equation}\label{rho}
    \hat \rho = \hat \rho^{(1)} \otimes  \hat \rho^{(2)} \otimes \cdot \cdot \cdot \otimes \hat \rho^{(N)},
\end{equation}
where $\hat \rho^{(i)} = |\psi^{(i)}\rangle \langle \psi^{(i)} |$ describing the state of the $i$-th subsystem.
For a mixed state, it is separable if it can be expressed as a mixture of product states.  
More generally, a separable state of $N$ particles, also known as a non-entangled state, is defined as a linear combination of density matrices multiplied by positive weights $p_l$,
\begin{equation}\label{rho_nonentangled}
    \hat \rho = \sum_l p_l\hat \rho^{(1)}_l \otimes  \hat \rho^{(2)}_l \otimes \cdot \cdot \cdot \otimes \hat \rho^{(N)}_l,
\end{equation}
where $\hat \rho^{(i)}_l$ is the density matrix of the $i$-th particle in the $l$-th term of the weighted sum.
Then, quantum states that can not be expressed as product states in forms of Eq.~\eqref{sep} or Eq.~\eqref{rho_nonentangled} are entangled states~\cite{Pezze2018_RMP}. 
Similarly, if we cannot write the state in the form of Eq.~\eqref{rho_nonentangled}, there must be more than classical correlations and it can be considered as an entangled state.
This section will provide some typical multi-particle entangled states and demonstrate the methods for generating these states in different many-body quantum systems. 

%
For a pure state $\hat \rho$, we always have $\text{Tr} (\hat \rho^2)=1$. 
If one divide  the whole system into two sub systems $A$ and $B$, it follows that $\hat \rho$ is entangled if and only if the von Neumann entropy measure  
\begin{equation}
    S[\rho]=-\text{Tr} \hat \rho \text{ln} \hat \rho
\end{equation}
of either reduced density matrix $\hat \rho^A = \text{Tr}_B \hat \rho$ or $\hat \rho^B = \text{Tr}_A \hat \rho$ is positive~\cite{Reid2009}. 
In the case of two particles, any quantum state is either separable or entangled.
For the system containing massive number of particles, it might exhibit multi-particle entanglement and not only pairwise entanglement, which needs to be classified. 
Such multi-particle entanglement is best characterized by the entanglement depth~\cite{PhysRevLett.86.4431} which is defined as the number of particles in the largest non-separable subset of a state.
A quantitative measure of entanglement in a multi-particle system is the number of elements that must at least have gone together in entangled states. 
We define a $k$-particle entangled state to be a state of $N$ particles which cannot be decomposed into a convex sum of products of density matrices with all density matrices involving less than $k$ particles: at least one of the terms is a $k$ particle entangled density matrix.

Similar to Eq.~\eqref{sep}, if the system state can be written as
\begin{equation}
    \label{k-sep}
    |\Psi_{k,sep}\rangle=|\psi_{N_1}\rangle \otimes |\psi_{N_2}\rangle \otimes \cdot \cdot \cdot \otimes |\psi_{N_M}\rangle,
\end{equation}
where $\ket{\psi_{N_l}}$ is a state of $N_l \le k$ particles with $\sum_{l=1}^M = N$, the state is $k$-particle entangled.
The state with Eq.~\eqref{k-sep} implies that there are at least one state of $k$ particles cannot be factorized. 
This is also referred to an entanglement depth larger than $k-1$.
Similarly, a mixed state is $k$-particle entangled if it can be written in terms of a mixture of $k$-particle entangled pure states, i.e.,
\begin{equation}
    \label{k-sep-rho}
    \rho_{k,sep} = \sum_{l} p_l |\Psi_{k,sep}\rangle_{l} {}_{l}\langle \Psi_{k,sep}|.
\end{equation}
Thus, if each particle is entangled with each other completely, the system state is maximally entangled with $k=N$. 

Theoretically, one can use QFI~\eqref{Eq:FQ2} or \eqref{Eq:QFI3} to qualify the metrological usefulness of a quantum state since QFI is only related to the properties of the state.  
According to the QCRB of inequality~\eqref{Eq:CRB2}, if the QFI of an entangled state is larger than the particle number $N$, 
\begin{equation}\label{QFI-SQL}
    F_Q>N,
\end{equation}
the state can be metrologically useful since it has the potential to attain a measurement precision better than the SQL. 
An entangled state with large QFI necessitates a substantial level of entanglement depth. 
For a state in the form of Eq.~\eqref{k-sep}, the QFI should satisfy~\cite{PhysRevA.85.022321, PhysRevA.85.022322}
\begin{equation}\label{ent-depth}
    F_Q \le s k^2 +r^2,
\end{equation}
where $s$ equals the integer part of $N/k$ and $r=N-s k$. 
If the bound in Eq.~\eqref{ent-depth} is violated, the state should contain $(k+1)$-particle entanglement. 
When $k=N$, the QFI has an upper bound, i.e., $F_Q \le N^2$, which corresponds to the Heisenberg limit. The state such as GHZ state that can have the maximal QFI $F_Q=N^2$ is considered to be ideal for metrological use. 
It should be mentioned that the condition of Eq.~\eqref{QFI-SQL} is for the linear generator.  
While for nonlinear generator, the dependence of QFI on total particle number $N$ may be different.  

The relation between entanglement and metrological usefulness have been theoretically studied~\cite{PhysRevA.82.012337, PhysRevLett.120.020506, PhysRevResearch.3.023101, Vitagliano2017,Ozaydin2014,Pezze2016,Li2020}.
For example, the quantum states with positive partial transpose can be useful for quantum metrology~\cite{PhysRevLett.120.020506, PhysRevResearch.3.023101}. 
However, it should be noted that, not every entangled state is equally useful for quantum metrology applications~\cite{PhysRevA.82.012337}. 
While several entangled states can provide benefits for quantum metrology, their level of usefulness may be different. 
For instance, despite being an $N$-particle entangled state, the $W$ state has a QFI of only $3N-2$, which can only attain the SQL precision scaling when $N$ is large~\cite{Ozaydin2014,Pezze2016,Li2020}. 
Interestingly, some of the metrologically useless states may be activated to be useful in multicopy scenarios~\cite{PhysRevLett.125.020402,Trenyi2024}.  

In practise, multi-particle entangled states that typically exhibit high QFI and are thus well-suited for quantum metrology include spin squeezed states, twin-Fock states, spin cat states, and etc.
These types of entangled states have the potential to surpass conventional precision limits and enable entanglement-enhanced quantum metrology~\cite{Geza2014J, Pezze2018_RMP}.
Besides QFI, squeezing parameters can be used to assess the metrological usefulness of an entangled state. 
There are several different squeezing parameters for characterizing spin squeezing. 
For Gaussian entangled states,  there are two commonly used squeezing parameters. 
The first squeezing parameter, proposed by Kitagawa and Ueda~\cite{Kitagawa1993}, is inspired by the concept of photon squeezing. The second squeezing parameter, introduced by Wineland et al.~\cite{Wineland1992}, is based on standard Ramsey spectroscopy. 
However, these two well-known metrological linear squeezing parameters can only quantify the sensitivity of Gaussian states, and they are insufficient to characterize the much wider class of highly sensitive non-Gaussian states~\cite{PRXQuantum.2.030204}.
For non-Gaussian entangled states,
%
a class of metrological nonlinear squeezing parameters have been introduced via optimizing measurement observables among a given set of accessible (possibly nonlinear) operators~\cite{Gessner2019}. 
These nonlinear squeezing parameters allow for the metrological characterization of non-Gaussian entangled states of both discrete and continuous variables. 

The quantum features of large systems become fragile against environmental noises, leading to significant decoherence of entangled states involving many particles~\cite{PhysRevLett.100.210401,Li2009EPJB, PhysRevA.81.052330,PhysRevA.76.042127,Wan2020,PhysRevLett.109.233601}. 
Even a weak coupling with the environment can disrupt the quantum characteristics of such states, causing them to exhibit classical behavior.
Consequently, creating large-scale entanglement is an extremely challenging task.
In 1999, experimental efforts to achieve entanglement between more than two particles began with the successful creation of a GHZ state involving three entangled photons~\cite{PhysRevLett.82.1345,Huang2011}. 
In 2000, the entanglement of four ions was demonstrated ~\cite{Sackett2000}.
Then multi-particle entanglement has been achieved with up to $14$ ions~\cite{Haffner2005,PhysRevLett.106.130506} and $8$ photons~\cite{Yao2012}. 
Other systems, such as superconducting qubits~\cite{DiCarlo2010,Neeley2010} and nitrogen-vacancy defect centers in diamond~\cite{Neumann2008}, have also been utilized to create tripartite entanglement.
Up to now, various multi-particle entangled states, including spin squeezed states~\cite{Ruschhaupt_2012,PhysRevLett.131.063401,PhysRevLett.79.4782,PhysRevLett.118.083604,Vasilakis2015,science3397,Colombo2022,Franke2023,Muessel2015,PhysRevLett.125.223401,Bao2020}, twin-Fock states~\cite{Lucke2011,Strobel2014,Colombo2022}, and spin cat states~\cite{Jones2009,PhysRevLett.104.043601}, have been generated in different many-body quantum systems~\cite{PhysRevA.68.052315,PhysRevLett.110.181101,PhysRevLett.59.2153,Xie2021,Zhong2021}, demonstrating entanglement-enhanced measurement precisions.
In the following sections, we will introduce the experimental and theoretical advancements in this field. 

\subsection{\label{sec:3-2}Spin squeezed states}

Spin squeezing plays a crucial role in understanding quantum entanglement and serves as a significant quantum resource for achieving high-precision measurements. 
In this subsection, we will first introduce bosonic squeezing, a concept extensively investigated in the realm of quantum optics.  Subsequently, we will present several definitions of spin squeezing in collective spin systems. 
Meanwhile, we will showcase diverse approaches employed to generate spin-squeezed states across a range of quantum many-body systems.\\ 

\noindent \textit{(1). Bosonic squeezing}\\

First, let us go over some fundamental concepts of bosonic squeezing in quantum optics.
Squeezed states exist in various systems of bosonic particles such as photons, phonons and atoms. 
Consider a simple harmonic oscillator obeying the Hamiltonian
\begin{equation}\label{H_osc}
    H=\frac{\hat p^2}{2m} + \frac{1}{2} m \omega^2 \hat x^2,
\end{equation}
where the scaled Planck constant $\hbar=h/(2\pi)=1$, and the position operator $\hat x$ and the momentum operator $\hat p$ satisfy the commutation relation $[\hat x, \hat p]=i$.
According to the Heisenberg uncertainty relation, 
\begin{equation}
   \Delta \hat x \Delta \hat p \ge \frac{1}{2},
\end{equation}
where $\Delta \hat x$ and $\Delta \hat p$ are respectively the standard deviations of $\hat x$ and $\hat p$. 

To connect with number states, one may introduce the creation and annihilation operators $\hat a^{\dagger}\equiv \sqrt{\frac{m\omega}{2}}\hat x -i \sqrt{\frac{1}{2m\omega}}\hat p$ and $\hat a\equiv \sqrt{\frac{m\omega}{2}}\hat x +i \sqrt{\frac{1}{2m\omega}}\hat p$ for the phonons in the system.
Thus the Hamiltonian~\eqref{H_osc} reads
\begin{equation}\label{H_osc_diag}
H=\omega\left(\hat a^{\dagger} \hat a + \frac{1}{2}\right).
\end{equation}
The eigenstate $|n\rangle$ is a number state, that is, eigenstate of the number operator $\hat a^{\dagger} \hat {a} |n\rangle =n|n\rangle$.
For convienence, we consider the following dimensionless quadrature operators 
\begin{equation}\label{quadrature1}
    \hat X=\sqrt{2m\omega} x = \hat a +\hat a^{\dagger},
\end{equation}
and 
\begin{equation}\label{quadrature2}
    \hat P=\sqrt{\frac{2}{m\omega}} p = \frac{\hat a -\hat a^{\dagger}}{i}.
\end{equation}
One can find that $[\hat X,\hat P]=2i$ and the corresponding Heisenberg uncertainty relation is given as
\begin{equation}\label{Heisenberg_uncertainty}
   \Delta \hat X \Delta \hat P \ge 1.
\end{equation}
If the quadratures of a state satisfy the minimum uncertainty
\begin{equation}
    \Delta \hat X = \Delta \hat P=1,
\end{equation}
this state is a coherent state. 
In general, a coherent state is defined as the eigenstate of the annihilation operator $\hat{a}\ket{\alpha}=\alpha\ket{\alpha}$.
Moreover, an arbitrary coherent state $\ket{\alpha}$  with the  complex number $\alpha =|\alpha| e^{i \phi}$ can be generated from the vacuum state $\ket{0}$, that is,
\begin{equation}\label{coherent_state}
    |\alpha\rangle=\hat D(\alpha)|0\rangle
\end{equation}
by using the displacement operator $\hat D(\alpha)=\exp(\alpha \hat a^{\dagger} -\alpha^{*}\hat a)$. 

The Heisenberg uncertainty relation remains inviolable, but it is possible to choose a condition in which the uncertainty in either $\Delta \hat X$ or $\Delta \hat P$ is less than 1, at the expense of having the uncertainty in the other variable exceed 1.
The state of such an uncertainty less than 1 is referred to as a squeezed state. 
An example of squeezed states is the squeezed coherent state, 
\begin{equation}\label{squeezed_coherent_state}
    |\alpha,\eta\rangle = \hat S(\eta) |\alpha\rangle
\end{equation}
which can be generated by the nonlinear Hamiltonian
\begin{equation}
    H=i(g\hat a^{\dagger2}-g^{*} \hat a^2).
\end{equation}
Here, the squeezing operator is given as $\hat S(\eta)=\exp(-i H t)=e^{\frac{1}{2}\eta^{*}\hat a^2 - \frac{1}{2}\eta \hat a^{\dagger 2}}$ with $r=-2|g|t$ and the complex squeezing parameter $\eta = r e^{i \theta}$.

To illustrate the squeezing, one can rotate the quadrature operators by an angle $\theta/2$, thus the rotated quadrature operators can be expressed as 
\begin{equation}
     {\tilde  X}=\hat a^{\dagger} e^{i \frac{\theta}{2}} +\hat a e^{-i \frac{\theta}{2}},
\end{equation}
and 
\begin{equation}
     {\tilde P}=i \left(\hat a^{\dagger} e^{i \frac{\theta}{2}} -\hat a e^{-i \frac{\theta}{2}}\right).
\end{equation}
Consequently, the standard deviation for the rotated quadratures are  
\begin{equation}
    \Delta \tilde X = e^{-r} , \quad \Delta \tilde P = e^{r}.
\end{equation}
Obviously, the standard deviation of $\tilde X$ is decreased by a factor of $e^{-r}$, yet the product of the two standard deviations $\Delta \tilde X \Delta \tilde P = 1$ does not violate the Heisenberg uncertainty principle. 

The use of squeezed coherent states in interferometers has shown promising potential in improving measurement precision. 
Initially proposed by Caves~\cite{PhysRevD.23.1693}, the sensitivity of optical interferometers can be improved by introducing quadrature-squeezed states. 
By replacing the vacuum state with squeezed states, the injection of these states into interferometers effectively reduces vacuum quantum noise~\cite{PhysRevD.23.1693}.  
This injection of squeezed states results in a reduction of detection noise below the shot-noise level, thereby enhancing phase measurement sensitivity. 
Notably, significant progress has been made in generating and applying these quantum states to optical interferometry systems, as demonstrated by various experimental efforts~\cite{PhysRevLett.68.3020,PhysRevLett.59.278,PhysRevLett.59.2153,Vaartjes2024}.
Recently, this technique has been employed in large-scale interferometers spanning kilometers, with the aim of improving sensitivity for gravitational wave detection~\cite{NP2011-LIGO,PhysRevLett.110.181101,Schnabel2017}.
Besides, by combining unconventional nonlinear interferometers~\cite{PhysRevA.33.4033} and stimulated emission of squeezed light, a scalable scheme for achieving unconditional and robust quantum metrological advantage beyond NOON states has been proposed and demonstrated with photons~\cite{PhysRevLett.130.070801}, making it possible for practical quantum metrology at the low photon flux regime.\\

\noindent \textit{(2). Spin squeezing}\\

Below we concentrate on spin squeezing, which refers to the phenomenon of reducing uncertainty in spin measurements. 
We will primarily illustrate spin squeezing using an ensemble of $N$ indistinguishable spin-1/2 particles. 
The system can be described by the following collective spin operators
\begin{equation}
    \hat J_{k} = \frac{1}{2} \sum_{l=1}^N \sigma_{k}^{(l)} ,
\end{equation}
where $\sigma_{k}^{(l)}$ is the Pauli matrix for the $l$-th particle and $k=\{x, y, z\}$. 
Similar to a coherent state, a spin coherent state consists of $N$ identical single spin states aligned along the same direction and it can be expressed as~\cite{PhysRevA.6.2211}
\begin{equation}\label{spin_coherent_state}
    |\theta,\varphi\rangle = \otimes_{l=1}^N \left[\cos \frac{\theta}{2} |\uparrow\rangle^{(l)} + e^{i \varphi} \sin \frac{\theta}{2} |\downarrow\rangle^{(l)}\right],
\end{equation}
where $|\uparrow\rangle^{(l)}$ and $|\downarrow\rangle^{(l)}$ are the eigenstates of $\sigma_{z}^{(l)}$ with eigenvalues $+1$ and $-1$, respectively. 
The mean spin direction is given by the vector $\vec J =\{\sin \theta \cos \varphi , \sin \theta \sin \varphi, \cos \theta\}$.
A spin coherent state $|\theta,\varphi\rangle$ can be generated from the state  $\otimes_{l=1}^N \ket{\downarrow}_l$ of all particles in spin-down state $\ket{\downarrow}$~\cite{PhysRevA.6.2211}, that is,  
\begin{eqnarray}\label{spin_coherent_state2}
    |\theta,\varphi\rangle &=& \otimes_{l=1}^N R_J (\theta,\varphi)|\downarrow\rangle_l \nonumber \\
    &=& \otimes_{l=1}^N \exp(\xi \sigma_{+}^{(l)} - \xi^{*} \sigma_{-}^{(l)}) |\downarrow\rangle_l,
\end{eqnarray}
where $\sigma_{\pm}^{(l)}=\sigma_{x}^{(l)} \pm \sigma_{y}^{(l)}$, and $\xi=\frac{\theta}{2} e^{-i\varphi}$. 
In the Dicke basis $\{|J,m\rangle\}$, which is defined by $\hat J_z|J,m\rangle=m|J,m\rangle$, the spin coherent state~\eqref{spin_coherent_state2} can be written as
\begin{eqnarray}\label{spin_coherent_state3}
    |\theta,\varphi\rangle = R(\theta,\varphi)|J,-J\rangle= \exp(\xi \hat J_{+} - \xi^{*} \hat J_{-}) |J,-J\rangle,
\end{eqnarray}
where $\hat J_{\pm}=\hat J_x \pm \hat J_y$ and $|J,-J\rangle \equiv \otimes_{l=1}^N |\downarrow\rangle_l$ is the eigenstate of $\hat J_z$ with eigenvalue $J=N/2$. 
After some algebra calculations, the spin coherent state can be given as 
\begin{eqnarray}
    \label{SCS_Dicke}
     |\theta,\varphi\rangle=\left[\frac{(2J)!}{(J+m)!(J-m)!}\right]^{1/2} \cos^{J-m}(\frac{\theta}{2}) \nonumber \\
     \sin^{J+m}(\frac{\theta}{2}) e^{i(J+m)\varphi  }|J,m\rangle. 
\end{eqnarray}
When measuring the spin component that is orthogonal to the mean collective spin $\vec J$, the variance $\Delta^2 \hat J_{\perp}$ is determined by the summation of variances from individual spin-1/2 particles, resulting in $\Delta^2 \hat J_{\perp}=\frac{N}{4}$. 
For the spin coherent state $|\pi/2,0\rangle=\otimes_{l=1}^{N} \left[ \frac{1}{\sqrt 2} (|\uparrow\rangle^{(l)} +  |\downarrow\rangle^{(l)})  \right]$ along $x$-axis, its variances along two orthogonal directions are isotropic, that is,
\begin{equation}
    \Delta^2 \hat J_{y}=\Delta^2 \hat J_{z}=\frac{N}{4}.
\end{equation}

Spin squeezing occurs when the uncertainty in measuring the spins of a collection of particles is reduced below the limit imposed by spin coherent states. 
This process involves manipulating the quantum states of spinor particles to enhance their correlation and decrease the fluctuations in their collective spin.
Spin squeezed states, akin to squeezed coherent states~\eqref{squeezed_coherent_state}, represent a kind of collective spin states that minimize the variance of spin along a specific direction while increasing the variance of the anti-squeezed spin along a perpendicular direction~\cite{Wineland1992,Wineland1994-1,Ma2011}.
Spin squeezing is currently regarded as one of the most effective means of achieving significant quantum entanglement and demonstrating measurement precision that surpasses the SQL~\cite{Braverman2019}.

The degrees of spin squeezing can be characterized by squeezing parameter. 
According to the Heisenberg uncertainty relations for collective spin operators, a collective spin state with $\xi^2_H =\frac{2(\Delta \hat J_{\alpha})^2}{|\langle \hat J_{\gamma}\rangle|}<1$ [where $\alpha \in (x,y,z)$, $\gamma \in (x,y,z)$ and $\alpha \neq \gamma$] can be regarded as a spin squeezed state. 
However, the squeezing parameter $\xi_H$ cannot indicate the optimal spin squeezing.

Inspired by the concept of photon squeezing, Kitagawa and Ueda introduced a squeezing parameter~\cite{Kitagawa1993} 
\begin{equation}
    \label{squeezing1}
    \xi_S^2=\frac{4 \text{min} (\Delta \hat J_{\vec n_{\perp}})^2}{N}
\end{equation}
determined by the minimum variance along the direction perpendicular to the mean collective spin direction.
Therefore a state is squeezed if $\xi_S^2<1$.
By minimizing $\xi_S$ over all feasible perpendicular directions $\vec n_{\perp}$, the optimal squeezing direction can be determined. 
This knowledge is crucial for implementing quantum-enhanced measurement. 

\begin{figure}[!htp]
 \includegraphics[width=\columnwidth]{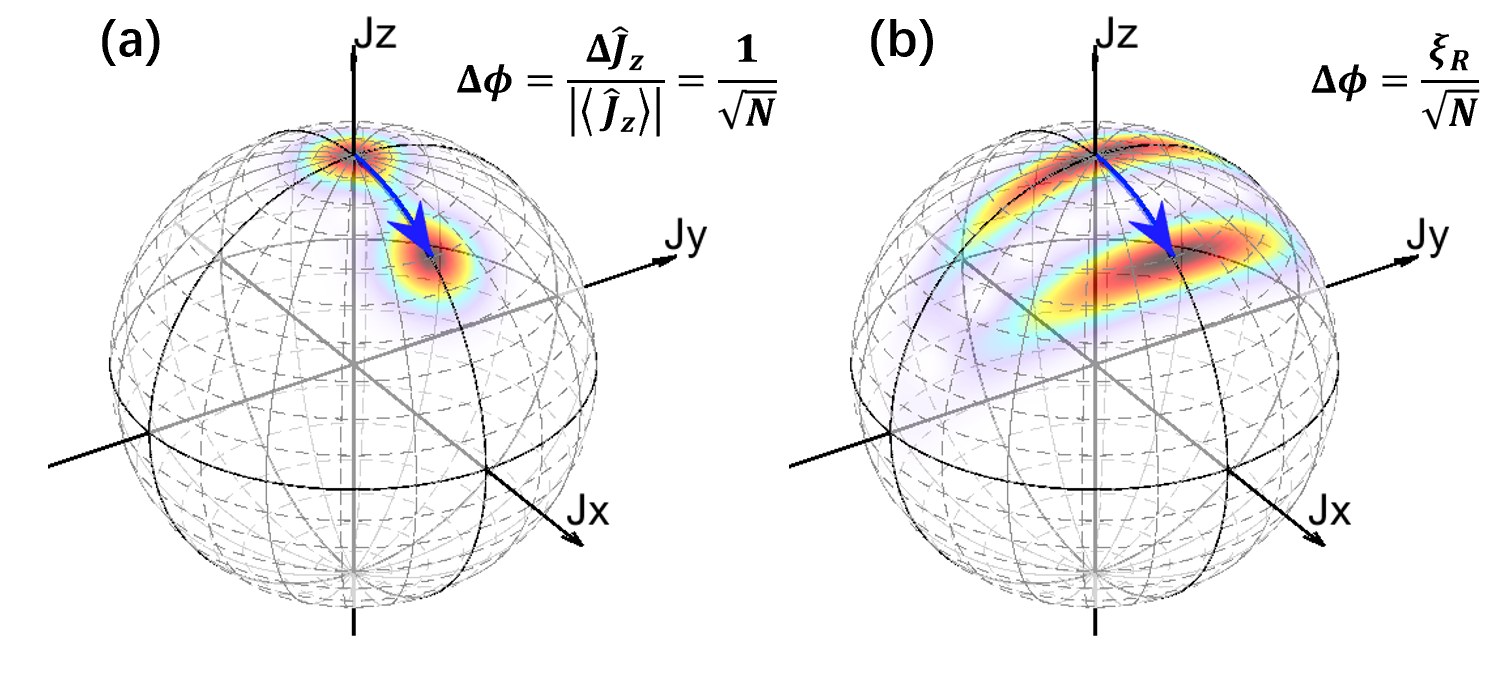}
  \caption{\label{Fig-Sec3-spin-squeezed}  Spin coherent states and spin squeezed states~\cite{Ma2011}. The Husimi distribution on the generalized Bloch sphere for (a) spin coherent state and (b) spin squeezed state. Implementing spin coherent state and spin squeezed state for Ramsey interferometry, the phase measurement precision can be achieved $\Delta \phi =\frac{1}{\sqrt N}$ and $\Delta \phi =\frac{\xi_R}{\sqrt N}$, respectively. Reproduced with permission from Ma \textit{et al.}, Phys. Rep. \textbf{509}, 89 (2011). Copyright 2011 Elsevier.}
\end{figure}

Alternatively, in the context of Ramsey interferometry, Wineland et al. introduced a squeezing parameter
\begin{equation}
    \label{squeezing2}
    \xi_R^2=\frac{\Delta \phi}{\Delta \phi_{SCS}}=\frac{N (\Delta \hat J_{\vec n_{\perp}})^2}{|\langle \hat J \rangle|^2},
\end{equation}
which is the ratio of the phase measurement precision obtained via the spin squeezed state and the spin coherent state~\cite{Wineland1992,Wineland1994-1}.
Here, $(\Delta \hat J_{\vec n_{\perp}})^2$ can be obtained by measuring the population difference between two sensor levels and $|\langle \hat J \rangle|$ can be inferred from the Ramsey fringes contrast. 
Consequently a squeezed state corresponds to $\xi_R^2 <1$.
As shown in Fig.~\ref{Fig-Sec3-spin-squeezed}, implementing spin squeezed states for Ramsey interferometry, the phase measurement precision can be improved to $\Delta \phi \propto \xi_R/\sqrt N$. 
Notably, if $\xi_R = 1/\sqrt{N}$, the measurement precision attains the Heisenberg limit: $\Delta \phi \propto 1/N$.

From Eq.~\eqref{squeezing1} and Eq.~\eqref{squeezing2}, one can easily find the relation between the above two spin squeezing parameters, 
\begin{equation}
    \xi_R^2 = \left(\frac{N}{2|\langle \hat J \rangle|}\right)^2 \xi_S^2 \ge \xi_S^2.
\end{equation}
This means that the state has $\xi_R^2<1$ always implying $\xi_S^2<1$.

Although bosonic squeezing and spin squeezing looks very different, they may connected through the Holstein-Primakoff transformation.
In the limit of large particle number and small excitations, the spin squeezing can be reduced to the bosonic squeezing.
Assume the total particle number $N\gg 1$ and the number of excited-state particles $\langle \hat b^{\dagger} \hat b\rangle \ll N$, one can perform the Holstein-Primakoff transformation~\cite{Holstein1940}
\begin{equation}
    \hat J_+ = \hat b \sqrt{N-\hat b^{\dagger} \hat b},~~\hat J_- = \hat b^{\dagger} \sqrt{N-\hat b^{\dagger} \hat b},~~\hat J_z = \frac{N}{2}-\hat b^{\dagger} \hat b,
\end{equation}
which is equivalent to making the mean-field approximation. 
As $\langle \hat b^{\dagger} \hat b\rangle \ll N$, $\sqrt{N-\hat b^{\dagger} \hat b}$ can be approximated as $\sqrt{N}$, therefore we have
\begin{equation}
    \hat J_+/\sqrt{N} \rightarrow \hat b, \quad \hat J_-/\sqrt{N} \rightarrow \hat b^{\dagger}, \quad 2\hat J_z/N \rightarrow 1.
\end{equation}
The scaled spin operators can be respectively mapped onto the position and momentum quadrature operators,
\begin{equation}
    \sqrt{2/N} \hat J_x \rightarrow \hat X=(\hat b +\hat b^{\dagger})/\sqrt{2},
\end{equation}
and
\begin{equation}
    \sqrt{2/N} \hat J_y \rightarrow \hat P=(\hat b -\hat b^{\dagger})/(i\sqrt{2}).
\end{equation}
Thus we have
\begin{equation}
    \xi^2_R = 2(\Delta Q)^2
\end{equation}
with $\hat Q = (\hat b e^{-i\phi} + \hat b^{\dagger} e^{i\phi})/\sqrt{2}=\hat X \cos \phi +\hat P \sin \phi$ and $0\le \phi<2\pi$. 
This indicates that the appearance of spin squeezing $\xi^2_R<1$ corresponds to the variance of the quadrature below the vacuum limit $(\Delta \hat Q)^2 < 1/2$.\\ 

\begin{figure}[!htp]
 \includegraphics[width=\columnwidth]{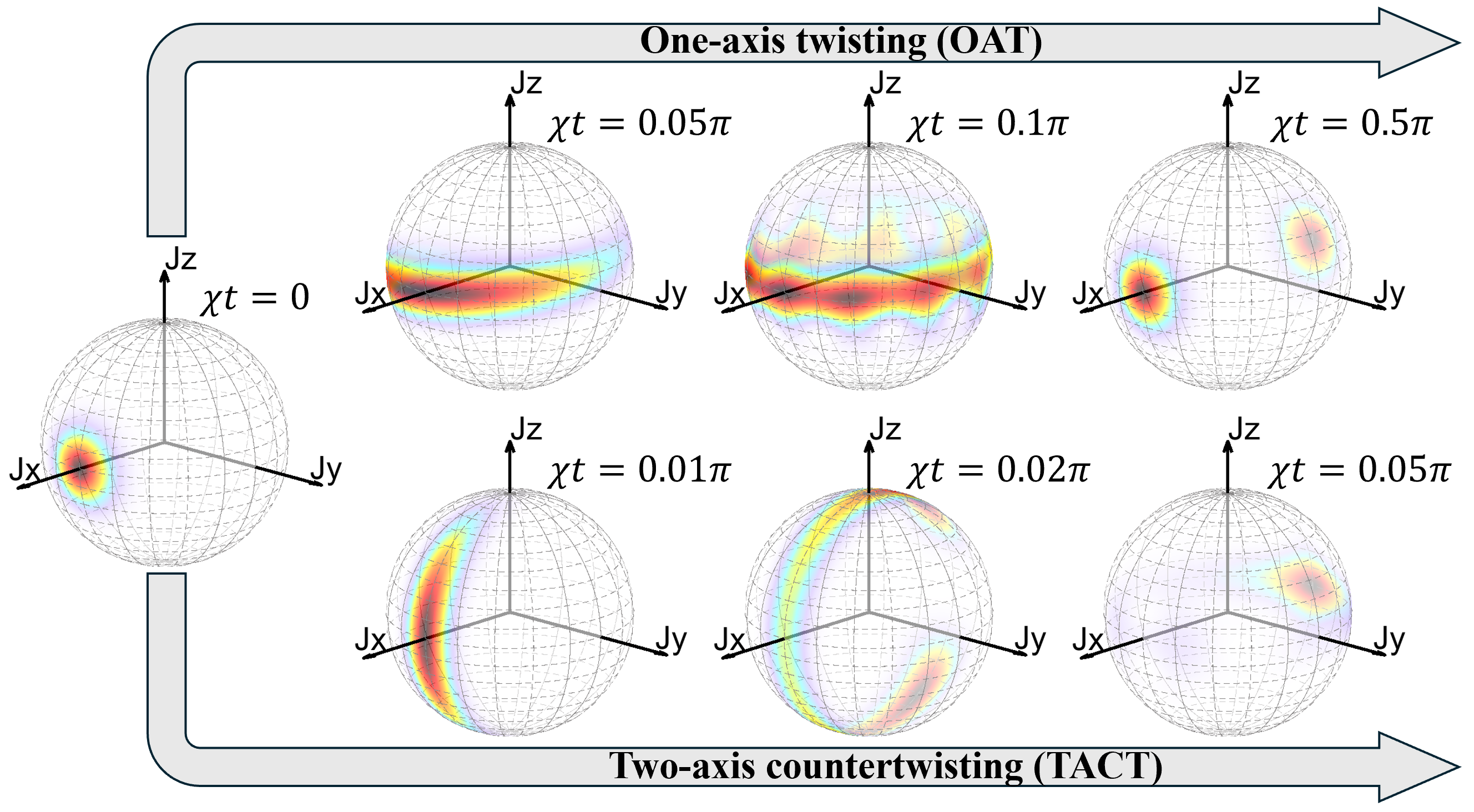}
  \caption{\label{Fig-Sec3-OAT-TACT}  State evolution via OAT and TACT dynamics~\cite{Kitagawa1993}. The Husimi distribution on the generalized Bloch sphere for (a) one-axis twisting (OAT) dynamics and (b) two-axis countertwisting (TACT) dynamics. Here, the Hamiltonian for TACT is chosen as $\tilde H_{TACT}=-\chi (\hat J_y \hat J_z+\hat J_z \hat J_y)$, which is obtained via a unitary transformation $\tilde H_{TACT}=e^{-i\frac{\pi}{2}\hat J_y} H_{TACT} e^{i\frac{\pi}{2}\hat J_y}$ on $H_{TACT}$~\cite{Kajtoch2015}. Reproduced with permission from Kitagawa \textit{et al.}, Phys. Rev. A \textbf{47}(6), 5138 (1993). Copyright 1993 American Physical Society.}
\end{figure}

\noindent \textit{(3). One-axis twisting}\\

One-axis twisting (OAT) is a widely recognized approach for producing entangled states that already have vast practical applications in quantum metrology~\cite{Jin_2009}. 
In principle, through the OAT dynamics from a spin coherent state, spin squeezed states, over-squeezed states, and spin cat states can be prepared. 
Generally, the OAT Hamiltonian reads 
\begin{equation}
    \label{OAT}
    H_{OAT}=\chi \hat J_z^2,
 \end{equation}
where $\chi$ denotes the twisting strength. This Hamiltonian can be realized in various synthetic quantum systems such as BECs~\cite{Gross2010,Riedel2010,Gross2012,Strobel2014}, trapped ions~\cite{Gilmore2021}, cold atoms in optical cavity~\cite{Zhang2015,Greve2022,Li2022}, and etc. 

Starting from the spin coherent state $|\pi/2,0\rangle$, the state after OAT reads
\begin{equation}
    |\Psi(t)\rangle_{OAT}=e^{-i \chi t \hat J_z^2}|\pi/2,0\rangle.
\end{equation}
During the OAT dynamics, the uncertainties are redistributed among certain orthogonal components in the $(y,z)$-plane based on the squeezing angle, see Fig.~\ref{Fig-Sec3-OAT-TACT}. 
The minimum and maximum values correspond to the spin components that are squeezed and anti-squeezed, respectively.
Analytically, one can obtain the squeezing angle
\begin{equation}
    \alpha = \frac{1}{2} \arctan (B/A)
\end{equation}
with $A=1-\cos^{N-2}(2\chi t)$ and $B=4\sin(\chi t) \cos^{N-2}(\chi t)$.
Then the squeezed and anti-squeezed spin operators can be explicitly expressed as
\begin{equation}
    \hat J_s =\hat J_z \cos(\alpha) -\hat J_y \sin (\alpha),
\end{equation}
and
\begin{equation}
    \hat J_a =\hat J_y \cos(\alpha) +\hat J_z \sin (\alpha).
\end{equation}
The expectation of the $x$ component evolves according to
\begin{equation}
    \langle \hat J_x \rangle = \frac{N}{2} \cos^{N-1}(\chi t),
\end{equation}
where the contrast gradually diminishes. 
The variances of squeezed (denoted by $s$) and anti-squeezed (denoted by $a$) spin components are expressed as 
\begin{equation}
    \Delta^2 \hat J_{s,a}=\frac{N}{4}\left[1+\frac{1}{4}(N-1)A \mp \frac{1}{4}(N-1)\sqrt{A^2+B^2}\right].
\end{equation}
Therefore the squeezing parameter can be given as
\begin{equation}
    \xi_S^2 =\frac{N(\Delta \hat J_s)^2}{\langle \hat J_x \rangle^2} = \frac{1+(N-1)(A-\sqrt{A^2+B^2})/4}{\cos ^{2N-2} (\chi t)}.
\end{equation}
In the limit of $N \gg 1$ and $\chi t \ll 1$, the variances can be approximated as
\begin{equation}
    \Delta^2 \hat J_{s} \approx \frac{N^2 \chi t}{2},
\end{equation}
and
\begin{equation}
    \Delta^2 \hat J_{a} \approx \frac{1}{4N\chi^2 t^2}+\frac{N^3\chi^4 t^4}{24}.
\end{equation}
When $\chi t \le 1/\sqrt{N}$, the state becomes spin squeezed, that is, the squeezing parameter satisfies $\xi_S^2 <1$. 
Particularly at the point of $\chi t=3^{1/6} N^{-2/3}$, $\Delta^2 \hat J_{s, opt} \approx (\frac{3^{2/3}}{8}) N^{1/3}$, the spin squeezing becomes optimal~\cite{Kitagawa1993}, 
\begin{equation}
    \xi^2_{S, opt} \approx (\frac{3^{2/3}}{8}) N^{-2/3}.
\end{equation}
Later for $\chi t > 1/\sqrt{N}$, the state stretches into a over-squeezed state, whose Husimi distribution turns to be non-Gaussian. 
Specifically, at the points of $\chi t =\pi/M$ ($2 \le M < \pi \sqrt N/2$), the state evolves into a superposition state of $M$ spin coherent states distributed evenly on the equator of the generalized Bloch sphere, which can be used for generating spin cat states, see Sec.~\ref{sec:3-4} below. 
%

Implementing Ramsey interferometry with optimal spin squeezed state, the best measurement precision obeys the scaling, 
\begin{equation}
    \Delta \phi \propto N^{-5/6},
\end{equation}
which beats the SQL but does not attain the Heisenberg limit.\\

\noindent \textit{(4). Two-axis countertwisting}\\

The degree of spin squeezing achieved through OAT depends on the number of particles $N$ and the duration of the evolution $T$, as the optimal squeezing angle varies accordingly. 
Furthermore, even with optimal spin squeezing, it is not possible to reach the Heisenberg limit precisely.
In theory, this limitation could be overcome by simultaneously twisting clockwise and counterclockwise around two perpendicular axes within the plane perpendicular to the mean spin direction.
The two orthogonal axes can be selected in the directions with $\theta=\pi/2$ and $\varphi=\pm \pi/4$.
The spin operators corresponding to these two orientations are
\begin{equation}
    \hat J_{\pi/2,\pm \pi/4}=\frac{1}{\sqrt{2}}(\hat J_x \pm \hat J_y).
\end{equation}
This kind of twisting is known as two-axis countertwisting (TACT), and it obeys the following Hamiltonian,
\begin{eqnarray}\label{H_TACT}
    H_{TACT} &=& \chi (\hat J^2_{\pi/2,\pi/4}-\hat J^2_{\pi/2,-\pi/4}) \nonumber \\
    &=& \chi (\hat J_x \hat J_y+\hat J_y \hat J_x)=\frac{\chi}{2i} (\hat J_+^2 -\hat J_-^2).
\end{eqnarray}

Nevertheless, the TACT model is not amenable to analytical solutions when the value of $N$ is large.  
Starting from spin coherent state $|\pi,0\rangle$, the mean spin direction is along the $z$-axis.
When the quasi-probability Husimi distribution covers almost half of the generalized Bloch sphere, the squeezed component achieves a minimal variance of $1/2$, while the anti-squeezed component reaches $N^2/8$.
The optimal squeezing angle is invariant during the TACT evolution.
If $\chi t $ exceeds the optimal value, the quasi-probability distribution will bifurcate into two distinct components~\cite{Kitagawa1993}. 
Compared with OAT, the degree of spin squeezing is higher. 
The TACT can also be achieved from any two orthogonal axes, which can be seen in Fig.~\ref{Fig-Sec3-OAT-TACT}.\\

\noindent \textit{(5). Preparing spin squeezed states via OAT}\\

Significant progress has been made in generating and applying spin squeezed states in various physical systems. 
Various methods, including atom-atom collision~\cite{PhysRevLett.79.337, Gross2010, Riedel2010}, quantum non-demolition measurement~\cite{Kuzmich1998,Meiser_2008,Greve2022}, quantum feedback control~\cite{Berni2015,PRXQuantum.3.020310,PhysRevLett.130.240803}, and Rydberg dressing~\cite{PhysRevLett.131.063401,Bornet2023}, have been employed to achieve spin squeezing. 
Experiments involving ultracold atoms, trapped ions, and Rydberg atoms have successfully created spin squeezed states, pushing the boundaries of what can be achieved in quantum control.
Below we will explore the experimental progress made in preparing spin squeezed states via OAT in different quantum many-body systems. 

\begin{figure}[!htp]
 \includegraphics[width=\columnwidth]{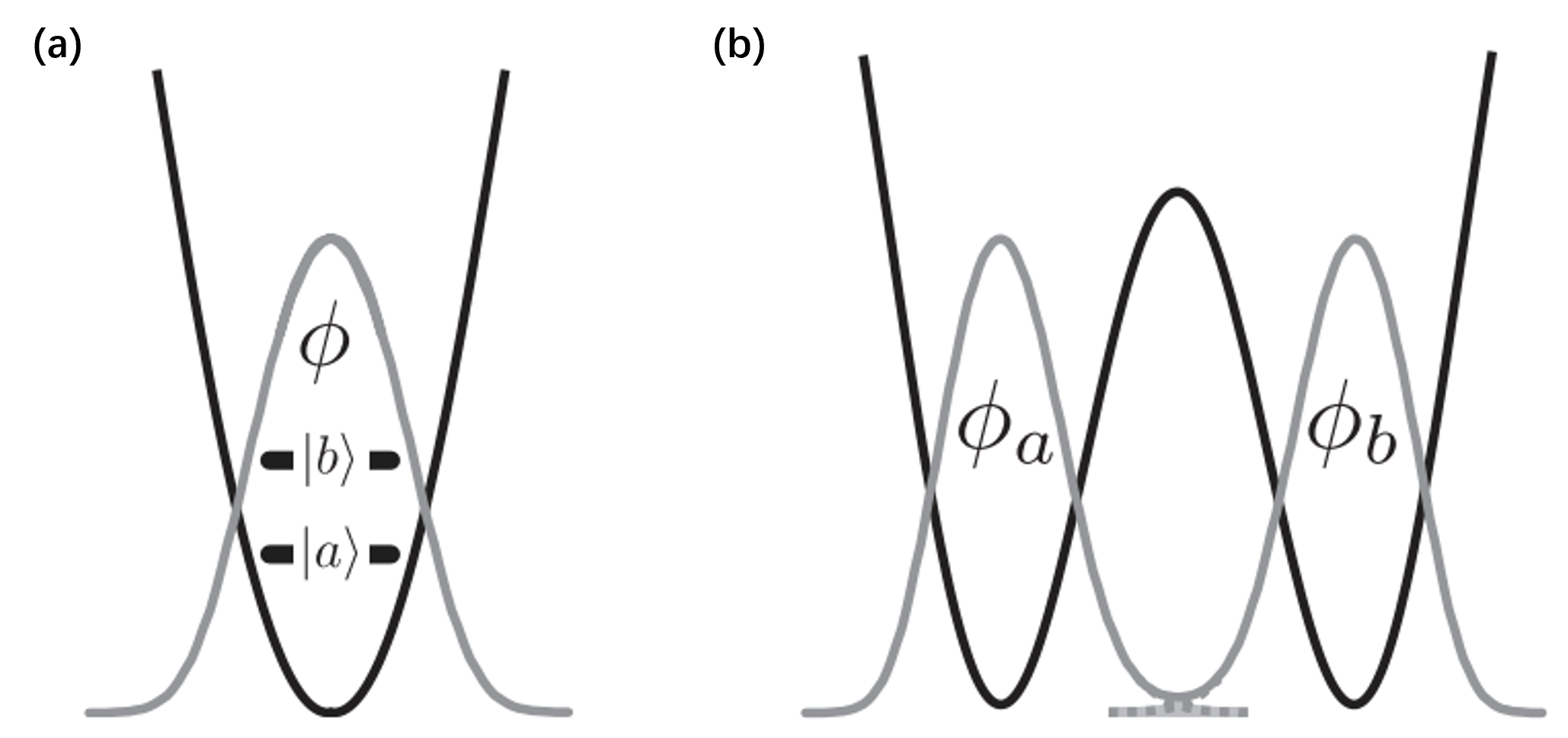}
  \caption{\label{Fig-Sec3-BJJ} Schematics of Bose-Josephson junctions~\cite{Gross2012}. (a) Internal Bose-Josephson junction realized with with a Bose-Einstein condensate in a single-well potential occupying on two different hyperfine levels coupled by laser fields. (b) External Bose-Josephson junction realized with a BECs confined in a double-well potential. Reproduced with permission from Gross \textit{et al.}, J. Phys. B: At. Mol. Opt. Phys. \textbf{45}, 103001 (2012). Copyright 2012 IOP Publishing.}
\end{figure}

Tunable atom-atom interactions are naturally occur in BECs and offer a practical means for manipulating twisting dynamics.
Considerable endeavors have been dedicated to the creation of spin squeezed states in atomic systems.
Typically, spin squeezing can be achieved using two main approaches. 
One method involves utilizing atomic collisions inside BECs~\cite{PhysRevLett.79.337}, while the other method involves exploiting atom-photon interactions within atomic ensembles. 
We first discuss how to generate the OAT Hamiltonian from a two-component BEC, which can be regarded as an atomic BEC occupying two internal states, or alternatively, as a BEC confined inside a double-well potential.
The former and the latter ones can be well described by the internal and external Bose-Josephson junction (BJJ) respectively, as shown in Fig.~\ref{Fig-Sec3-BJJ}. 
Denoted the two modes by $a$ and $b$, in the Swinger representation, the unified BJJ Hamiltonian can be written as 
\begin{equation}\label{Ham_BJJ}
    H_{BJJ}=-\Omega \hat J_x + \chi \hat J_z^2 +\delta \hat J_z,
\end{equation}
where $\hat J_x = \frac{\hat a^{\dagger} \hat b + \hat a \hat b^{\dagger}}{2}$,  $\hat J_y = \frac{\hat a^{\dagger} \hat b - \hat a \hat b^{\dagger}}{2i}$, and $\hat J_x = \frac{\hat a^{\dagger} \hat a - \hat b \hat b^{\dagger}}{2}$.

For an internal BJJ, the two hyperfine states $|a\rangle$ and $|b\rangle$ are assumed in the spatial single-mode wavefunctions $\phi_a(\vec r)$ and $\phi_b(\vec r)$ coupled by a driving field with Rabi frequency $\Omega_R$ and detuning $\Delta$ to the resonant transition between $|a\rangle$ and $|b\rangle$.
The coefficients in the many-body Hamiltonian~\eqref{Ham_BJJ} are given as $\Omega = \Omega_R \int d\vec r \left[\phi^{*}_a(\vec r) \phi_b (\vec r)\right]$, $\delta = \epsilon_b-\epsilon_a + N(U_{bb}-U_{aa})/4$ and $\chi = U_{aa} + U_{bb} -2U_{ab}$,
with $\epsilon_{k}=\int d\vec r \left[\phi^{*}_{k}(\vec r) \left(-\frac{\nabla^2}{2m}+V(\vec r) + \Delta_k\right) \phi_{k} (\vec r)\right]$, $\Delta_{a,b}=\mp \Delta/2$, $U_{kl}=\frac{g_{kl}}{2} \int d\vec r |\phi_k (\vec r) |^2 |\phi_l (\vec r) |^2$.
Here $g_{kl}=4\pi a_s^{kl}/m$ with $a_s^{aa}$, $a_s^{bb}$, and $a_s^{ab}$ denoting the intraspecies and interspecies s-wave scattering lengths, and $m$ being the atomic mass. 
For $g_{aa}=g_{bb}$ and $\phi_a=\phi_b=\phi$ [as shown in Fig.~\ref{Fig-Sec3-BJJ}~(a)], $\delta =\Delta$, $\Omega=\Omega_R$, we have $\chi \propto a_s^{aa}+a_s^{bb}-2a_s^{ab}$. 
In particular for $\Omega=0$ and $\delta=0$, the Hamiltonian of Eq.~\eqref{Ham_BJJ} is equivalent to the OAT Hamiltonian of Eq.~\eqref{OAT}. 
Thus one can tune the nonlinear interaction $\chi$ and use it for dynamical generation of spin squeezed states.

\begin{figure*}[!htp]
 \includegraphics[width=2\columnwidth]{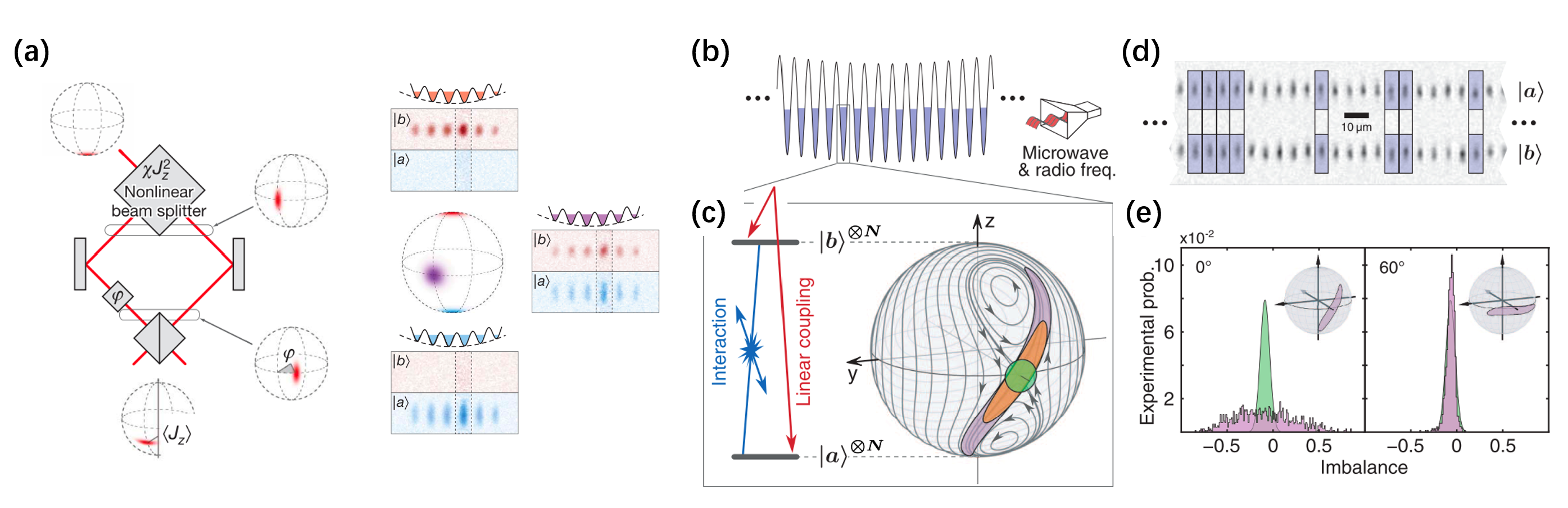}
  \caption{\label{Fig-Sec3-Feshbach} Generation of spin squeezed~\cite{Gross2010} and non-Gaussian entangled states~\cite{Strobel2014} with BECs trapped in optical lattices. (a) The many-body Ramsey interferometry with spin squeezed state. Six independent $^{87}$Rb BECs occupying two hyperfine states in a one-dimensional optical lattice are prepared. The quantum states on the generalized Bloch sphere in one of the wells are shown. Reproduced with permission from Gross \textit{et al.}, Nature \textbf{464}, 1165 (2010). Copyright 2010 Springer Nature. (b) A group of BECs is manipulated inside an optical lattice potential using microwave and radio frequency fields. (c) The combination of nonlinear interaction (blue) and weak Rabi coupling (red) between the internal states leads to an unstable fixed point in the classical phase space. The gray lines represent the paths followed by the mean-field equations of motion. The initial spin coherent state (represented by the color green), undergoes an ideal transformation into a spin squeezed state (represented by the color orange). As time progresses, the spin state further evolves into non-Gaussian states (represented by the color violet). The boundaries of the shaded regions represent the outlines of the Husimi distribution at its peak. (d) An experimental absorption image is shown, which displays the site- and state-resolved optical lattice after a Stern-Gerlach separation. (e) Displayed are histograms illustrating the disparity following the nonlinear progression of 25 milliseconds and final rotation (as shown by the angles in the graphs), in comparison to the ideal spin coherent state of same nature (represented by the green Gaussian). Reproduced with permission from Strobel \textit{et al.}, Science \textbf{345}, 424 (2014). Copyright 2014 The American Association for the Advancement of Science.}
\end{figure*}

By loading Bose condensed atoms into optical lattices, the entanglement of about $170$ $^{87}$Rb atoms has been successfully prepared~\cite{Gross2010}. 
As shown in Fig.~\ref{Fig-Sec3-Feshbach}~(a), a BEC of $^{87}$Rb atoms occupying the hyperfine state $|F=1,m_{F}=-1\rangle$ was firstly prepared in an optical dipole trap. 
Then, through supposing a one-dimensional optical lattice potential, the dipole trap is split into six, which allows to perform $6$ independent experiments in parallel. 
Before applying the first $\frac{\pi}{2}$ pulse, the atoms are swept from the state $|F=1,m_{F}=-1\rangle$ to the state $|a\rangle=|F=1,m_{F}=1\rangle$. 
Only two hyperfine states $|a\rangle=|F=1,m_{F}=1\rangle$ and $|b\rangle=|F=2,m_{F}=-1\rangle$ are involved. 
The effective nonlinear interaction $\chi \varpropto a_s^{aa}+a_s^{bb}-2a_s^{ab}$ was tuned by changing inter-species interaction with the technique of Feshbach resonance.
By carefully tuning the magnetic field at $B=9.10$~G, $\chi=2\pi\times0.063$~Hz can be achieved.  
The Rabi frequency $\Omega$ can be switched rapidly from $0$ to $2\pi\times600$~Hz, changing the system from Rabi regime to Fock regime.
When the Rabi frequency is switched off, the system stays in the Fock regime and its state evolves under the nonlinear term.
Thus individual systems localized in each lattice site can be described by the OAT Hamiltonian with Eq.~\eqref{OAT}.
The OAT evolution induces a squeezing angle $\alpha_{0}$ with respect to $z$-direction. A rotation of the uncertainty ellipse around its center by $\alpha=\alpha_{0}+\pi/2$ is followed. 
After spin squeezing preparation, the modes $|a\rangle$ and $|b\rangle$ experience a $\tau=2\mu s$ phase accumulation period and recombine via another $\frac{\pi}{2}$ pulse before the readout of population imbalance.
In comparison to the ideal phase sensitivity obtained by spin coherent states, their experimental data show that the phase sensitivity is enhanced by $15\%$.  
With spin noise tomography, the inferred spin squeezing parameter can be up to $\xi_R^2=-8.2$ dB.
If introducing the linear coupling between the two states, the spin squeezing can be generated faster, which will be brief discussed at the end of this section. 
The experimental demonstration is shown in Fig.~\ref{Fig-Sec3-Feshbach}~(b)-(e).

\begin{figure}[!htp]
 \includegraphics[width=\columnwidth]{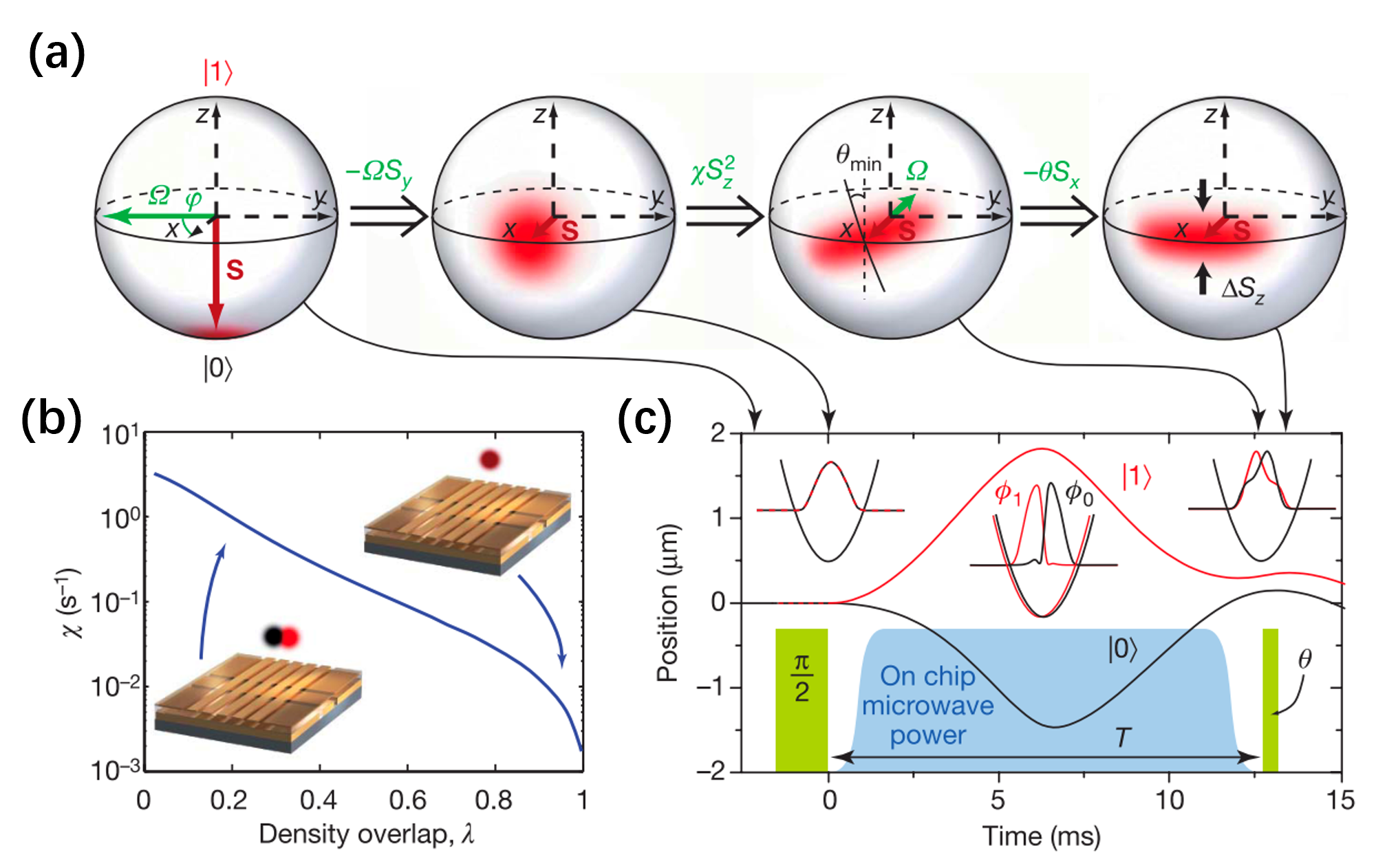}
  \caption{\label{Fig-Sec3-atomchip} Atomic spin squeezing with BEC on an atom chip~\cite{Riedel2010}. (a) The BEC internal state undergoes evolution on the generalized Bloch sphere. By first setting all atoms in the state $\ket{0}$, a $\pi/2$-pulse is applied to create a spin coherent state. After undergoing subsequent nonlinear development, the noise circle is transformed into an ellipse, resulting in a spin-squeezed state that exhibits decreased noise at an angle $\theta_{\textrm{min}}$. A second pulse induces rotation of the state around the $x$ axis by an angle $\theta$, which is variable. Subsequently, the detection of half population difference occurs.  (b) Manipulation of the nonlinearity $\chi$ on the atom chip. The value of $\chi$ is determined by the disparity between atomic interactions within a state and those between states. The relationship between the two BEC components is demonstrated by analyzing the normalized density overlap $\lambda$. This value is determined by evaluating the stationary mode functions in potentials with increasing spacing. (c) Sequence of experiments. During the intervals for internal-state manipulation (green), a microwave potential that depends on the state of the system is activated (blue indicates the lengths of the pulses). During the time $T$, it dynamically divides and merges the two BEC components, resulting in a positive value of $\chi$. Reproduced with permission from Riedel \textit{et al.}, Nature \textbf{464}, 1170 (2010). Copyright 2010 Springer Nature.}
\end{figure}

The nonlinear interaction $\chi$ can also be tuned by controlling the spatial overlap between two spin components. 
Using this technique, spin squeezed states can be generated on atomic chips. 
By loading Bose condensed atoms into an atomic chip and generating spin squeezed states via OAT dynamics, the measurement precision can also be improved beyond the SQL~\cite{Riedel2010}. In the experiment, two spin states $\left|F=1,m_{F}=-1\right\rangle$ and $\left|F=2,m_{F}=1\right\rangle$ of $^{87}$Rb atoms are involved and the system obeys the OAT Hamiltonian of Eq.~\eqref{OAT}. Different from using Fechbach resonance, the effective interaction is controlled by adjusting the spatial overlap between two spin components.
Except for controlling the nonlinear interaction, the procedures for spin squeezing generation are similar, as shown in Fig.~\ref{Fig-Sec3-atomchip}. First, a spin coherent state is prepared by a resonant $\frac{\pi}{2}$ pulse for 120 $\mu s$. During the pulse, the coupling term dominates, $\Omega\gg\chi N$, so that the atom-atom interaction can be neglected. 
The state-dependent microwave potential is turned on within 50 $\mu s$ to cause a sudden separation of trap minima for the two hyperfine states. The two components begin to oscillate oppositely, the overlap of the modes wavefunction reduces, which leads to the decreasing of the inter-component interaction and the increasing of effective nonlinearity $\chi$. The nonlinearity can attain $\chi=1.5$ $s^{-1}$ at the maximum separation. The two components overlap again after 12.7 $ms$ and the nonlinear interaction squeezing dynamics stops.
The results give a spin squeezing parameter $\xi_R^2 = -1.2$ dB and with detection noise subtracted $\xi_R^2 = -7.0$ dB can be achieved in recent experiments.

The external BJJ can be realized with BEC confined in a double-well potential $V_{dw}(\vec r)$.
For sufficiently high barrier and weak interaction, the system can be simplified by two-mode approximation. 
The two modes can be considered the spatial modes localized in the left and right wells constructed by the lowest two quasi-degenerate symmetric and antisymmetric eigenstates ($|\phi_g\rangle$ and  $|\phi_e\rangle$) for the corresponding Gross-Pitaevskii equation, i.e., $|a\rangle =(|\phi_g\rangle+|\phi_e\rangle)/\sqrt{2}$ and $|b\rangle =(|\phi_g\rangle-|\phi_e\rangle)/\sqrt{2}$, as shown in Fig.~\ref{Fig-Sec3-BJJ}~(b). The corresponding wavefunctions can be written as $\phi_a = \frac{1}{\sqrt{2}}(\phi_g+\phi_e)$ and $\phi_b = \frac{1}{\sqrt{2}}(\phi_g-\phi_e)$.
For a double-well with sufficiently high barrier, we have the integrals $\int d\vec r |\phi_g|^2 |\phi_e|^2 \approx \int d\vec r |\phi_g|^4 \approx \int d\vec r |\phi_e|^4$ and the coefficients $\Omega = \mu_e-\mu_g$ and $\chi = 2g \int d\vec r |\phi_g(\vec r)|^2 |\phi_e (\vec r)|^2$
with $\mu_{g,e}= \int d\vec r \left\{\left[\phi_{g,e}^{*} \left(-\frac{\nabla^2}{2m}+V_{dw}(\vec r)\right)\phi_{g,e}\right]+gN|\phi_{g,e}|^4\right\}$ and $g=4\pi a_s/m$. 

In principle, the external BJJ can be realized by loading BEC in a double-well potential.
However, the main challenge lies in generating a stable double-well potential. 
One approach utilizes the adiabatic dressed state potential, which has been used to achieve coherent splitting of a BEC and matter-wave interferometry in experiments ~\cite{Schumm2005}.
Another method involves employing a radio-frequency dressed state potential on an atom chip, which enables the creation of double-well potentials for neutral atoms~\cite{Schumm2006}.
Additionally, a double-well potential can be realized using all-optical potentials by combining a 3D harmonic trapping potential with a periodic lattice potential ~\cite{PhysRevLett.96.130404}. 
The resulting potential, which combines a dipole trap and an optical lattice, exhibits a symmetric double-well shape at the center with a separation of approximately $4.4 \mu$m between the two wells.  
Another demonstrated approach for establishing weak coupling between two spatially separated BECs is through the use of Bragg beams~\cite{PhysRevLett.95.170402}. 
In this setup, the atoms in the left and right wells are coupled via two Bragg beams, allowing for coherent tunneling. 
In an external BJJ, two important properties are the number and coherence fluctuations. 
As the coherence fluctuation increases with $\chi$, it is accompanied by number squeezing, which can be considered a form of spin squeezing.

Besides atom-atom collision, spin squeezing can also be generated through atom-light interactions. 
Numerous proposals focus on transferring squeezing from light to atoms as a means of generating spin squeezing.
The key parameters to be controlled in this process are the atom-light coupling and the detuning between the light and atoms.
In the regime of large detuning, an effective Hamiltonian can be derived that includes a dispersive interaction and a nonlinear interaction term. 
The magnitudes of these terms can be adjusted accordingly. 
The dispersive interaction between light and atoms is a type of quantum non-demolition (QND) measurement.
This measurement allows for the determination of certain properties in a quantum system without disturbing or altering the system itself.
The observable being measured commutes with the system's Hamiltonian, which means that the observable and the Hamiltonian share a common set of eigenstates.  
Through a QND measurement, an atomic ensemble and a light beam are coupled in such a way that direct measurements on the light can provide indirect information about the atomic system.  
QND measurements offer a means of preparing entangled and spin squeezed states for atomic ensembles.
By performing a QND measurement on the light, the atomic ensemble can be conditioned to specific measurement outcomes, leading to squeezed states.
The effective interaction between atoms is in form of an OAT interaction, which can be used to directly generate spin squeezing. 

\begin{figure}[!htp]
 \includegraphics[width=\columnwidth]{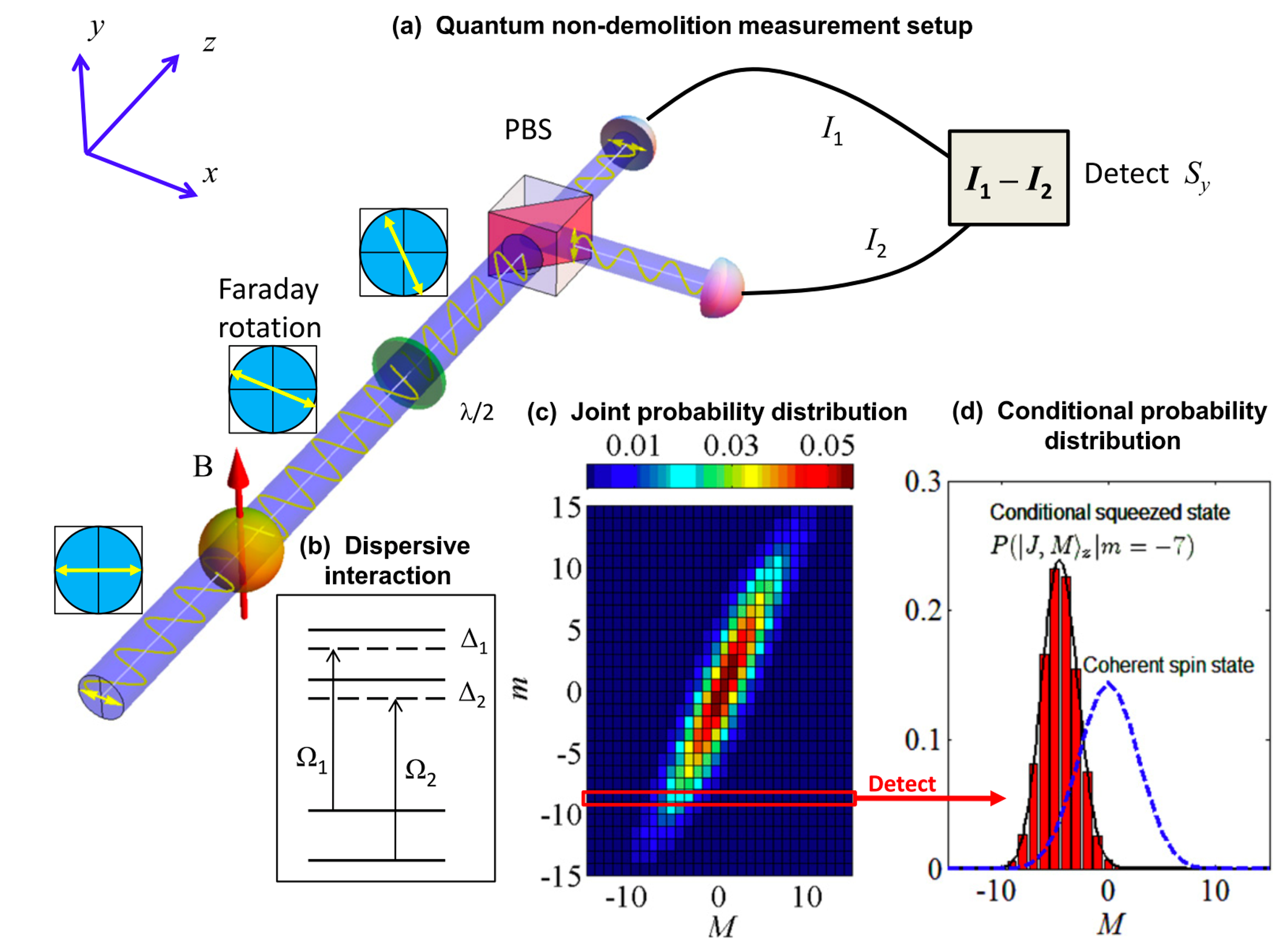}
  \caption{\label{Fig-Sec3-QND} Spin squeezing generation via QND measurement. (a) Diagram illustrating the schematic for the QND measurement. The input probe beam is polarized in the $x$-direction, and the atoms are initially aligned along the $x$-axis. The probe beam's two circularly-polarized components interact with atoms in a dispersive manner, with detuning $\Delta_1$ and $\Delta_2$~\cite{PhysRevLett.79.4782}. Reproduced with permission from Kuzmich \textit{et al.}, Phys. Rev. Lett. \textbf{79}(24), 4782 (1997). Copyright 1997 American Physical Society. (b) The terms $\Omega_1$ and $\Omega_2$ refer to the couplings between atoms and electromagnetic fields. Dispersion interactions cause a rotation of the polarization plane, known as Faraday rotation. The $\lambda/2$ wave plate, along with the subsequent PBS and two detectors, function as a QND measurement of $\hat J_z$ by detecting $\hat S_y$, which is then measured by the difference in photocurrent. (c) Joint probability distributions with $N = n = 30$ and $\chi = 0.1$. (d) When the measurement of the observable $\hat S_y$ yields a value of $m = -7$, the squeezed state under this condition is graphed and juxtaposed with the spin coherent state. Reproduced with permission from Ma \textit{et al.}, Phys. Rep. \textbf{509}, 89 (2011). Copyright 2011 Elsevier.}
\end{figure}

The far-off resonant dispersive interaction between an atomic ensemble and two-mode light beams can be used for implementing the QND Hamiltonian.
Suppose two probe light beams coupling to different hyperfine atomic levels respectively.
In large detuning condition, the effective Hamiltonian reads~\cite{Kuzmich1998}
\begin{eqnarray}\label{QND}
    H_{QND}&=&\frac{g_1^2 \Delta_1}{\gamma_1^2/4+\Delta_1^2} \hat c_1^{\dagger} \hat c_1 \hat a^{\dagger} \hat a + \frac{g_2^2 \Delta_1}{\gamma_2^2/4+\Delta_2^2} \hat c_2^{\dagger} \hat c_2  \hat b^{\dagger} \hat b\nonumber \\
    &=& \Omega (\hat S_z \hat J_z +\frac{1}{4} \hat n \hat N),
\end{eqnarray}
where $\hat c_{1,2}$ are the annihilation operators for two probe light beams and $\Delta_{1,2}$ are the corresponding detunings. $g_{1,2}$ and $\gamma_{1,2}$ are the atom-light coupling constants and the spontaneous decay rates of the excited levels. $\hat S_z=(\hat c^{\dagger}_1 \hat c_1 -\hat c^{\dagger}_2 \hat c_2)/2$ denotes the Stokes vector operator, $\hat J_z=(\hat a^{\dagger} \hat a -\hat b^{\dagger} \hat b)/2$ stands for the collective spin operator, $\hat n = \hat c_1^{\dagger} \hat c_1 + \hat c_2^{\dagger} \hat c_2$ and $\hat N =\hat a^{\dagger} \hat a+\hat b^{\dagger} \hat b$ are the operators for total photon number and atom number. 
The final form of Hamiltonian of Eq.~\eqref{QND} is obtained by choosing suitable detunings satisfying the following way 
\begin{equation}
    \frac{g_1^2 \Delta_1}{\gamma_1^2/4+\Delta_1^2}=\frac{g_2^2 \Delta_1}{\gamma_2^2/4+\Delta_2^2}\equiv \frac{\Omega}{2}.
\end{equation}
The Heisenberg equations can be given as
\begin{eqnarray}
    \hat J_x(t) &=& \cos(\beta \hat S_z) \hat J_x(0) -\sin(\beta \hat S_z) \hat J_y(0), \\
    \hat J_y(t) &=& \sin(\beta \hat S_z) \hat J_x(0) +\cos(\beta \hat S_z) \hat J_y(0), \\
    \hat J_z(t) &=& \hat J_z(0),
\end{eqnarray}
and
\begin{eqnarray}
    \hat S_x(t) &=& \cos(\beta \hat J_z) \hat S_x(0) -\sin(\beta \hat J_z) \hat S_y(0), \\
    \hat S_y(t) &=& \sin(\beta \hat J_z) \hat S_x(0) +\cos(\beta \hat J_z) \hat S_y(0), \\
    \hat S_z(t) &=& \hat S_z(0),
\end{eqnarray}
where $\beta=\Omega t$.
It is shown that the Hamiltonian acts as a Faraday rotation to the light polarization and as an artificial magnetic field to the collective spin of atoms. 
 
When the light and atoms are both in their own spin coherent states along the $x$ direction initially, i.e.,
\begin{equation}
    |\psi(0)\rangle =|S,S\rangle_x |J,J\rangle_x,
\end{equation}
the expectation
\begin{equation}
    \langle \hat S_y(t)\rangle =\frac{n}{2}\langle \sin(\beta \hat J_z)\rangle \approx \langle \hat S_y(0)\rangle +\frac{n\beta}{2}\langle \hat J_z(0) \rangle
\end{equation}
is related to the expectation of $\hat J_z$. Since $\langle \hat S_y(0)\rangle = 0$ and $\langle \hat J_z(0)\rangle = 0$, $\langle \hat S_y(t)\rangle \approx 0$ while the variance $\Delta^2 \hat S_y(t)\approx \langle \hat S_y^2(0)\rangle + \beta^2 \langle \hat S_x^2(0)\rangle \langle \hat J_z^2(0)\rangle = \frac{n}{4}+\beta^2\frac{n^2}{4}\frac{N}{4}=n(1+\kappa^2)/4$ with $\kappa^2=nN\beta^2/4$.
The evolution of the system state reads
\begin{eqnarray}
     |\psi(t)\rangle &=& \exp(-i \beta \hat S_z \hat J_z) |\psi(0)\rangle \nonumber \\
     &=&\sum_{M=-J}^J \exp(-i\beta M \hat S_z)\sqrt{P_J(M)} |S,S\rangle_x |J,J\rangle_x.
\end{eqnarray}
As shown in Fig.~\ref{Fig-Sec3-QND}, performing the measurement of $\hat S_y$ on $|\psi(t)\rangle$ with result $m$ would cause the photon state collapsing into the eigenstate $|S,m\rangle_y$. Meanwhile, the atomic state would change according to the result $m$.
When $N \gg 1$, the binomial distribution can be well approximated by a Gaussian distribution, i.e., $P_J(M) \approx \frac{1}{\sqrt{\pi N/2}} \exp (-\frac{2 M^2}{N})$.
For small rotated angle $\beta M$, one can obtain the conditional probability distribution of the atomic state given by the measurement result $m$ of $\hat S_y$ as
\begin{equation}
    P(M|m)=\frac{1}{\sqrt{\pi \xi_R^2 N/2}} \exp \left[-\frac{(M-\beta m \xi_R^2 N/2)^2}{\xi_R^2 N/2}\right],
\end{equation}
where $\xi_R^2 =\frac{1}{1+\kappa^2}$ is the spin squeezing parameter and $\xi_R^2 <1$ when $t>0$.
Thus, after measuring $\hat S_y$ the atomic state becomes spin squeezed and the expectation $\langle \hat J_z\rangle$ is shifted to $\beta m \xi_R^2 N/2$.

In free space, the experimental demonstration of spin squeezing via QND measurements was realized~\cite{Appel2009} in 2009.
In this experiment, the two hyperfine states $\left|\uparrow\right\rangle \equiv |F=4, m_F=0\rangle$ and $\left|\downarrow\right\rangle \equiv |F=3, m_F=0\rangle$ of Cs atoms are referred to clock levels with total atom number over $10^5$.
Initially, the Cs atoms are prepared in $\left|\downarrow\right\rangle$ by using optical pumping. A resonant $\frac{\pi}{2}$ microwave pulse at the clock frequency is applied to prepare a spin coherent state. 
Then, successive QND measurements of the population difference $N_{\uparrow}-N_{\downarrow}$ are performed by measuring the state dependent phase shift of the off-resonant probe light in a balanced homodyne configuration. 
After the QND measurement, all atoms are pumped into the $F=4$ level to determine the total atom number $N=N_{\uparrow}+N_{\downarrow}$. Two identical linear polarized beams $P_{\uparrow}$ and $P_{\downarrow}$ off-resonantly probe the transitions $|F=3\rangle$ to $|F'=4\rangle$ and $|F=4\rangle$ to $|F'=5\rangle$, respectively. Each beam gains a phase shift proportional to the number of atoms in the corresponding clock states,
\begin{eqnarray}
\phi_{\uparrow}=k_{\uparrow}N_{\uparrow},
\phi_{\downarrow}=k_{\downarrow}N_{\downarrow},\nonumber
\end{eqnarray}
where $k_{\uparrow({\downarrow})}$ are the coupling constants and the detuning $\Delta_{\uparrow(\downarrow)}$ are tuned to make $k_{\uparrow}=k_{\downarrow}=\beta$. The phase difference between the two arms of the Mach-Zehnder interferometer is related to the measurement of $J_z$ and the shot noise of the photons,
\begin{eqnarray}
\phi=\frac{\delta n}{n} + \beta (N_{\uparrow}-N_{\downarrow})=\frac{\delta n}{n} + 2\beta J_z.
\end{eqnarray}
The variance of the phase difference,
\begin{eqnarray}
\textrm{Var}(\phi)=\frac{1}{n} + \beta^2 \textrm{Var}(N_{\uparrow}-N_{\downarrow})=\frac{1}{n} + \beta^2 N.
\end{eqnarray}
%
%
The spin squeezing can be verified by estimating the correlations between two successive QND measurements.
First use $n_1$ photons to measure $\hat J_z$ and obtain a measurement result of $\phi_1$, then use $n_2$ photons to measure $\hat J_z$ and obtain a measurement result of $\phi_2$ on the same atomic ensemble. 
The best estimator for $\phi_2$ is $\epsilon \phi_1$, which results in a conditionally reduced variance,
\begin{equation}
    \textrm{Var}(\phi_2-\epsilon \phi_1)=\frac{1}{n_2} + \frac{\beta^2}{1+\kappa^2} N
\end{equation}
with $\kappa^2 = n_1 \beta^2 N$ and $\epsilon=\textrm{Cov}(\phi_1,\phi_2)/\textrm{Var}(\phi_1)=\frac{\kappa^2}{1+\kappa^2}$.
For $\kappa^2=3.2$ with $N=1.2\times 10^5$ atoms, the projection noise can be reduced to $-(5.3\pm0.6)$ dB and metrologically relevant spin squeezing of $\xi_R^2=-(3.4\pm0.7)$ dB on the Cs microwave clock transition has been realized.
With extremely large number of atoms around $10^{11}$ in the vapor cell at high temperature, spin squeezing generated via stroboscopic QND measurements~\cite{Bao2020} can further offer benefits for entanglement-enhanced magnetic field sensing. 

\begin{figure}[!htp]
 \includegraphics[width=\columnwidth]{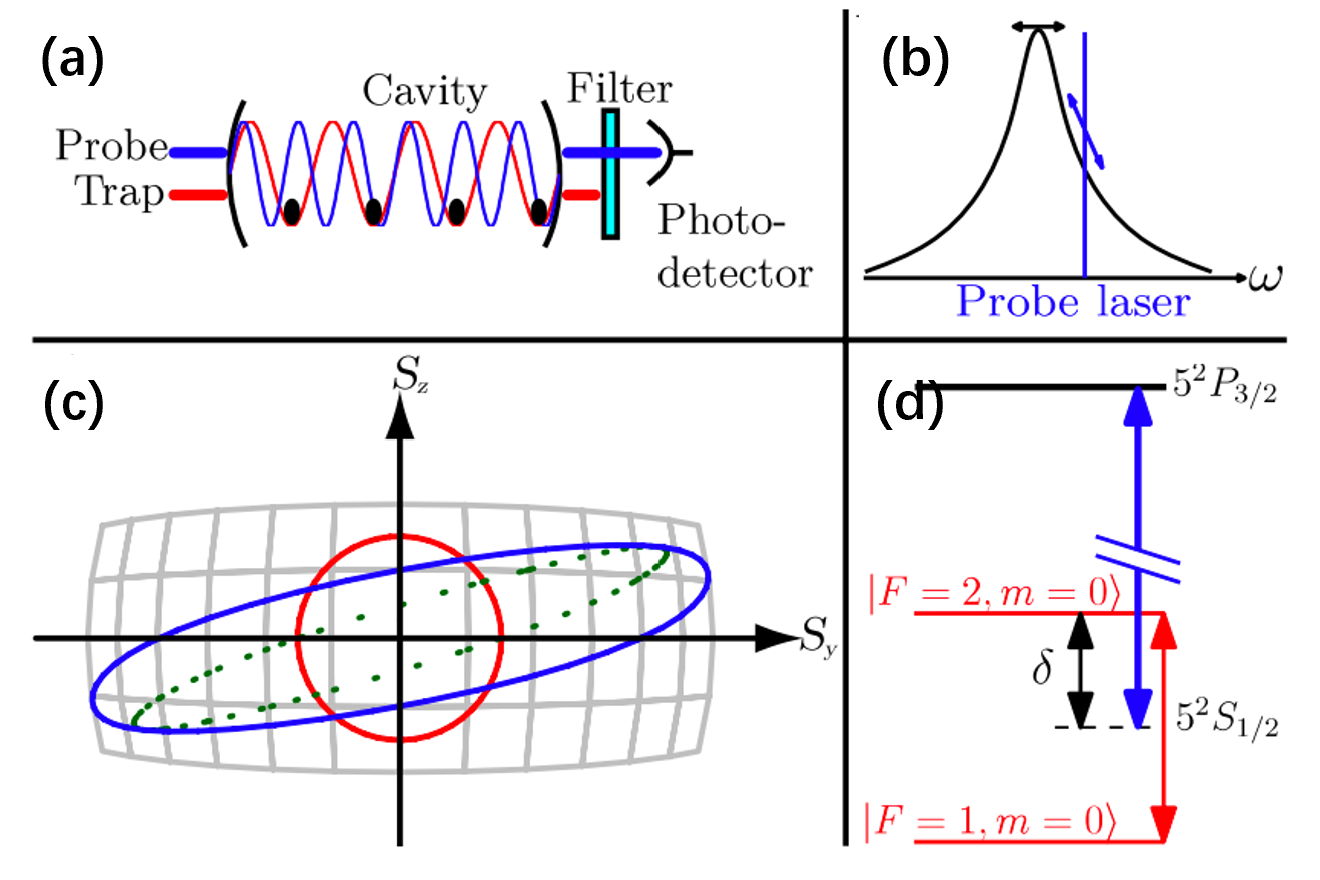}
  \caption{\label{Fig-Sec3-Cavity} Squeezing via cavity feedback~\cite{PhysRevLett.104.073602}. (a) The atoms are confined within a dipole trap formed by a standing-wave pattern, which is located inside an optical resonator. (b) The probe laser is adjusted to be off resonance from the cavity by half of its linewidth. This allows for changes in the cavity frequency caused by atoms to affect the transmitted power in proportion to $\hat J_z$. (c) The light shift that depends on $J_z$ causes the circular uncertainty region of the initial coherent spin state (red circle) to become an elliptical shape (dotted ellipse).  The presence of photon shot noise leads to the broadening of the phase, resulting in an increase in the area of the solid ellipse. (d) The cavity is tuned halfway between the optical transition frequencies for the two clock states, where $\delta$ is the effective detuning. Reproduced with permission from Leroux \textit{et al.}, Phys. Rev. Lett. \textbf{104}(7), 073602 (2010). Copyright 2010 American Physical Society.}
\end{figure}


The interaction strength between light and atoms can be enhanced by confining atomic ensemble into an optical cavity. 
The optical cavity allows a photon coupling with the confined atoms multiple times on successive round trips, which significantly enhances the atom-light interaction.
Using high finesse cavity, the effect of spin squeezing can be dramatically increased. 
The generation of spin squeezed states of trapped $^{87}$Rb atoms by cavity-aided QND measurement with a far-detuned light field was demonstrated~\cite{PhysRevLett.104.073604}.
Besides, the application of cavity has advantages for practical applications such as atomic clocks~\cite{Bloom2014,PhysRevLett.121.070403,Oelker2019}, which will be discussed in Sec.~\ref{sec:5-1}. 

It has been demonstrated that the strength of atom-light entanglement, regardless of light detuning or coupling strength, can be determined by multiplying the single-atom cooperativity, the normalized cavity transmission, and the total number of photons scattered by atoms into free space~\cite{Li2022}. 
The entanglement between atoms and photons is primarily generated through the lowest order of atom-cavity interaction.  
When the laser frequency closely matches the cavity resonance and the transmission is high, the dominance lies in the atom-photon entanglement. 
In such cases, the measurement of the light field can generate atom-atom interaction.
By transferring the atomic spin state information onto the light field, one can conditionally project the state of atoms onto an entangled state, known as the measurement-based squeezing approach. 
On the other hand, higher-order atom-cavity interaction can result in effective atom-atom and photon-photon interactions. 
When the laser detuning significantly deviates from the cavity resonance (exceeding the cavity linewidth) and the transmission is low, the atom-light entanglement weakens, and the higher-order terms become crucial in generating light-mediated atom-atom interaction.
In this scenario, atomic entanglement can be unconditionally created without the need for any measurements, referred to as the cavity feedback squeezing approach.

\begin{figure*}[!htp]
 \includegraphics[width=2\columnwidth]{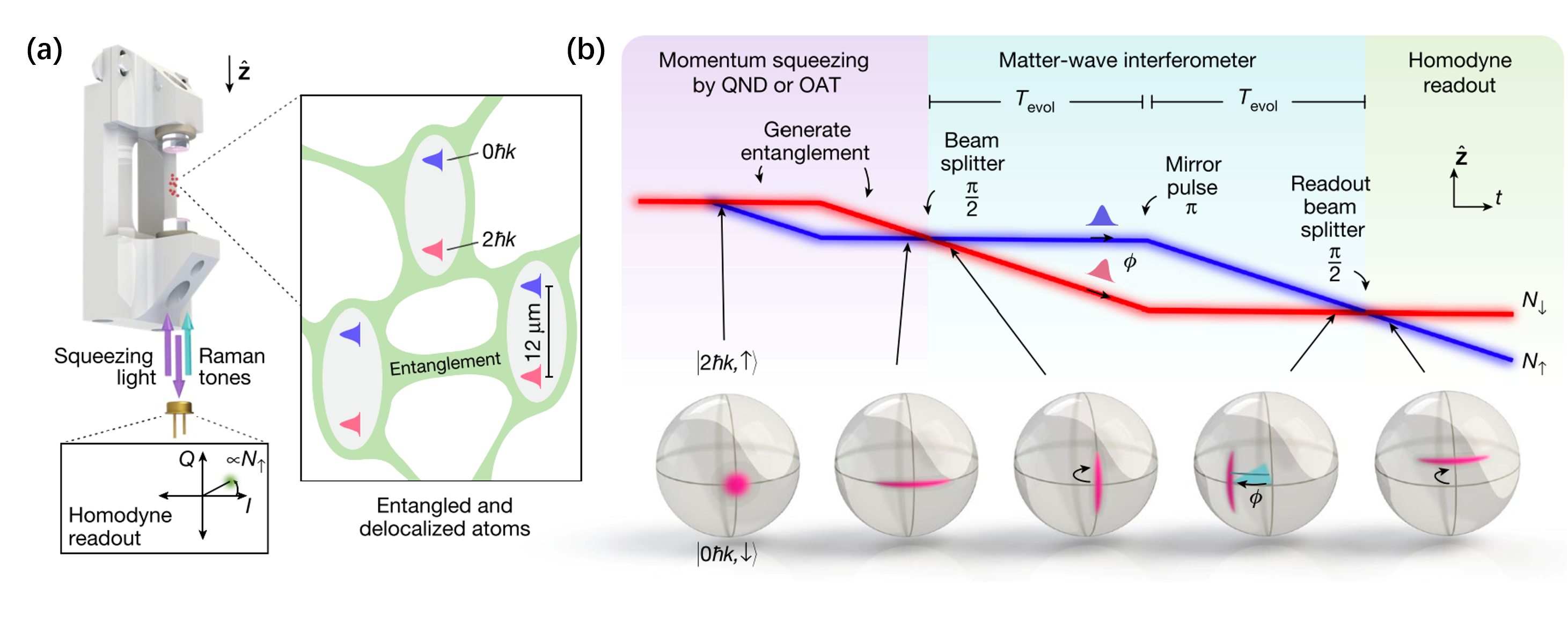}
  \caption{\label{Fig-Sec3-Cavity-momentum} Matter-wave interferometry with spin squeezing in a high-finesse cavity~\cite{Greve2022}. (a) Ultracold atoms experience controlled downward motion within a vertical chamber that has a high level of reflectivity. The atomic wave packets undergo division and subsequent fusion by the use of two-photon Raman transitions, resulting in discrete momentum impulses being imparted to the atoms. Entanglement between atoms is generated by the use of intracavity atomic probe light. This may be achieved either using OAT dynamics or QND measurements. The entanglement is created by measuring the quadrature of the reflected atomic probe field using a homodyne detector. The entanglement between atoms is observable even when the wave packets are separated by distances greater than 12 $\mu$m. (b) The entanglement is generated and injected into a Mach-Zehnder matter-wave interferometer, and this process is depicted using space-time and Bloch sphere representations. Squeezing is initially produced inside the population, and then, a Raman beam splitter pulse aligns the squeezing to improve the sensitivity of the interferometer phase. The two routes, shown by the red and blue colors, accumulate a relative phase $\phi$ as time progresses. The mirror pulse is used to bring the wave packets back together, while the readout beam splitter pulse generates interference that can be measured as a population difference with sub-SQL sensitivity. The Bloch sphere displays noise distributions for different sites in the interferometer. Reproduced with permission from Greve \textit{et al.}, Nature \textbf{610}, 472 (2022). Copyright 2022 licensed under a Creative Commons Attribution (CC BY) license.}
\end{figure*}

For example, we consider the systems of three-level atoms in a cavity.
Each atom has two hyperfine levels labelled by $|a\rangle$ and $|b\rangle$ with transition frequency $\omega$, and an excited state $|e\rangle$ possessing a linewidth $\Gamma$.
The system can be described the following Hamiltonian,
\begin{equation}\label{Ham_c}
    H_{c}= \omega_c \hat c^{\dagger} \hat c +\frac{2g^2}{\Delta} \hat c^{\dagger} \hat c \hat J_z +\omega \hat J_z,
\end{equation}
where $\hat c$ is the annihilation operator for photons in the cavity,  $\omega_c$ is the resonance frequency of the cavity, $\Delta=\pm \omega/2$ is the detuning between the cavity and the transition frequency of $|a\rangle \leftrightarrow |e\rangle$ and $|b\rangle \leftrightarrow |e\rangle$, $g$ is the atom-photon effective intracavity Rabi frequency, and $\hat J_z$ is the collective spin operator for the atoms. 
The above Hamiltonian is derived in the assumption of homogeneous interaction and large detuning with low intracavity photon number, where $\langle \hat c^{\dagger} \hat c \rangle \ll \Delta^2/g^2$ and $\Delta \gg \kappa_c, \Gamma, \sqrt{N}g$ with $\kappa_c$ the linewidth of the cavity and $N$ the atom number~\cite{PhysRevA.81.021804, Reiter2012}. 
By grouping the last two terms in Hamiltonian of Eq.~\eqref{Ham_c}, one can see that the light induces an ac Stark shift proportional to cavity photon number onto the atomic transition frequency. 
This shift can be detected by injecting a laser into the cavity and performing a QND measurement of $\hat J_z$. 
However, the performance of entanglement generation using QND measurement is generally limited by the efficiency of photon collection. 

To relax the need for high-efficiency photon detection, the cavity-feedback squeezing technique has been proposed and successfully demonstrated. 
This method enables the unconditional squeezing of the atomic spin state and the deterministic generation of spin squeezed states by utilizing light-mediated interactions between distant atoms within the cavity. 
The approach can be conceptualized as a quantum feedback process, where the atomic ensemble imprints quantum fluctuations on the light field, which in turn acts back on the atomic spin state, reducing its quantum fluctuations.  
The origin of spin squeezing lies in the atom-light interaction term $\frac{2g^2}{\Delta} \hat c^{\dagger} \hat c \hat J_z$, where the intracavity photon number $\hat c^{\dagger}\hat c$ is proportional to $\hat J_z$. 
This leads to the OAT Hamiltonian $H_{OAT}=\chi \hat J_z^2$, with an effective twisting strength $\chi$. 
In 2010, the spin squeezing via cavity-based QND measurement was experimentally conducted using an ensemble of $5\times 10^4$ $^{87}$Rb atoms ~\cite{PhysRevLett.104.073602}, see Fig.~\ref{Fig-Sec3-Cavity}. 
By preparing the atomic state in a superposition of two hyperfine clock levels, the experiment successfully generated an atomic spin squeezed state, surpassing the SQL by achieving a $3.0\pm0.8$ dB improvement in precision.
Later, a subsequent experiment using a cavity with higher cooperativity and uniform coupling between the probe light and all the atoms demonstrated spin squeezing of $-20.1\pm0.3$ dB via an optical-cavity-based measurement involving a million $^{87}$Rb atoms in their clock states~\cite{Hosten2016}.

In addition to hyperfine states, spin squeezing can also be achieved between momentum states~\cite{Shankar_2019,PhysRevA.106.043711}.
This development offers a promising avenue for combining particle delocalization and entanglement in various applications such as inertial sensors, searches for new physics and particles, future precision measurements, and exploring beyond mean-field quantum many-body physics. 
However, the experimental realization of an entanglement-enhanced matter-wave interferometer remained elusive until the groundbreaking demonstration in 2022~\cite{Greve2022}, see Fig.~\ref{Fig-Sec3-Cavity-momentum}.
In that experiment, the entanglement of external degrees of freedom was successfully realized to construct a matter-wave interferometer involving 700 atoms.
Remarkably, each individual atom simultaneously traversed two distinct paths while entangled with the other atoms. 
The experimental process involved preparing the atoms in superpositions of two momentum states, each associated with different hyperfine spin labels.  
Subsequently, both QND measurements and cavity-mediated interactions were employed to generate squeezing between momentum states. 
The resulting entangled state was then injected into a Mach-Zehnder light-pulse interferometer, yielding a directly observed metrological enhancement of $1.7^{+0.5}_{-0.5}$ dB, marking a significant milestone in this field.

In entanglement-enhanced atomic interferometers with momentum states, the atomic momentum states are tagged to the atomic spin states and thus the entanglement of spin states may map onto the entanglement of momentum states.
Recently, significant progress has been made in realizing momentum-exchange interactions, wherein atoms exchange their momentum states through the collective emission and absorption of photons from a common cavity mode~\cite{Luo2023}.
This process, akin to a spin-exchange interaction, was experimentally demonstrated by pairs of atoms exchanging their momentum states via the collective emission and absorption of cavity photons~\cite{PhysRevResearch.5.L032039}. 
The momentum-exchange interaction manifests as an all-to-all Ising-like interaction, which proves to be instrumental in generating entanglement. 
This tunable momentum-exchange interaction introduces a new capability for entanglement-enhanced matter-wave interferometry, opening up avenues for exploring exotic phenomena such as simulating superconductors and dynamical gauge fields.

The van der Waals interaction between Rydberg atoms provides another resource for generating spin squeezing.
The interaction between atoms in clock states is very small. 
In contrast, the van der Waals interaction between Rydberg atoms is strong. 
For atoms in an lattice clock of $\lambda = 813$ nm, the nearest neighbor interaction $C_6/r^6$ between two Rydberg atoms can exceed $10$ GHz~\cite{Vaillant2012}.
This provides an excellent resource for creating entangled states.
A promising approach for generating spin squeezing in optical lattice clocks  has been theoretically proposed via optically coupling one clock state to a highly excited Rydberg state~\cite{Gil2014}, which can be described by an Ising-type Hamiltonian $H_{Ryd} = \Omega \hat J_x + \sum_{i<j}^N V_{ij} \hat \sigma_z^{(i)} \hat \sigma_z^{(j)} + \sum_i \delta_i \hat \sigma_z^{(i)}$.
Here $\Omega$ is the transverse field strength, $\delta_i$ is the inhomogeneous longitudinal field strength, $V_{ij}=V_0 \frac{R_c^6}{|\vec r_i-\vec r_j|^6+R_c^6}$ is the effective two-body interaction, and $R_c=|C_6/2\Delta|^{1/6}$  is the critical distance. 
The transverse and longitudinal terms can be switched on and off independently via controlling the intensities of the two laser fields. 
This kind of switchable atom-atom interaction provides great flexibility for the creation of entangled many-body states.

\begin{figure}[!htp]
 \includegraphics[width=\columnwidth]{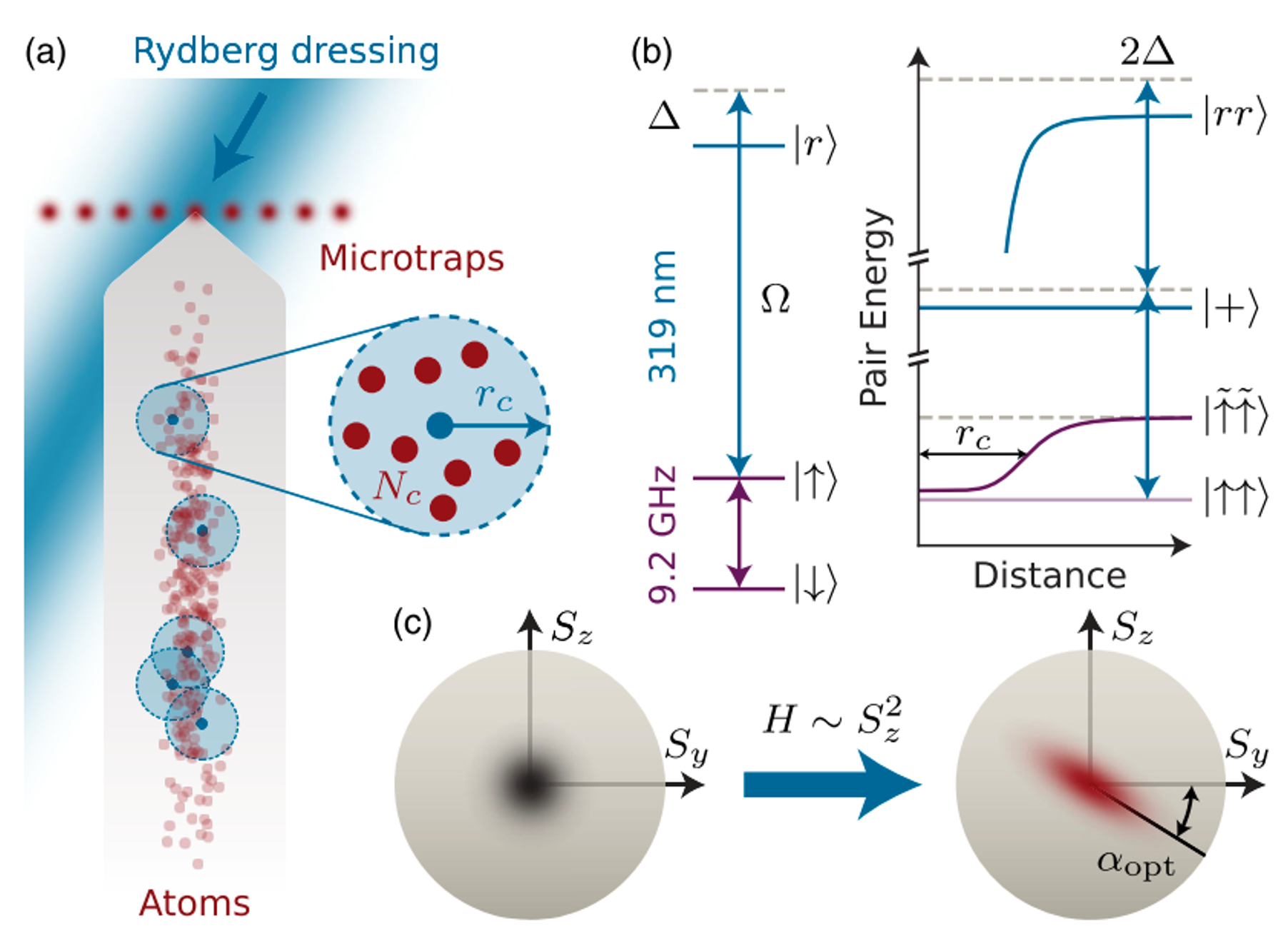}
  \caption{\label{Fig-Sec3-Rydberg1} Experimental setup of spin squeezing via Rydberg dressing~\cite{PhysRevLett.131.063401}. (a) An array of atomic ensembles is locally illuminated with 319 nm Rydberg dressing light, inducing interactions of characteristic range $r_c$. (b) Level diagrams for one atom (left) and a pair of atoms (right), where $\ket{+}=(\ket{r\uparrow}+\ket{\uparrow r})/\sqrt{2}$ and $\ket{\tilde \uparrow}$ denotes the Rydberg-dressed state. (c) Inter-collective spin $S$ that shears a spin coherent state (left) into a spin squeezed state (right). Reproduced with permission from Hines \textit{et al.}, Phys. Rev. Lett. \textbf{131}(6), 063401 (2023). Copyright 2023 American Physical Society.}
\end{figure}

Rydberg dressing are appealing for generating entanglement in optical tweezers.
Optical tweezers provide a new platform that atoms are individually trapped, and arranged in one- or two-dimensional arrays, offering high tunability and control of inter-atomic distance. 
By combining Rydberg dressing and optical tweezers, it becomes possible to finely tune atom-atom interactions, thereby enabling the realization of a wide range of interactions. 
This approach allows for the transformation of the interactions from an effective nearest-neighbor interaction to a nearly homogeneous interaction throughout the entire ensemble.
Recently, an array of spin-squeezed ensembles of cesium atoms has been prepared via Rydberg dressing~\cite{PhysRevLett.131.063401}, see Fig.~\ref{Fig-Sec3-Rydberg1}.
The interactions induced by Rydberg dressing can be understood as a superposition of the ac Stark shift that the dressing light imparts to each atom due to the influence of nearby atoms. 
This effect becomes particularly significant when considering an ensemble of $N$ atoms confined within a critical length scale $R_c$, where the van der Waals interaction $V_R=C_6/r^6$ surpasses the pair-state detuning $2\Delta$. 
In this limit, the system obeys the Hamiltonian $H\approx U_0 \hat J_z -\frac{\chi}{N} \hat J_z^2$, where $U_0\approx \Omega^2/(4\Delta)$ is an overall ac Stark shift and $\chi\approx N\Omega^4/(16\Delta^3)$ is the mean-field interaction. 
Using spin echo, the influence of $U_0$ can be removed and so that one can use OAT dynamics to create spin squeezing.
In the experiment, a squeezed state of $N=200$ atoms has been successfully prepared, exhibiting a squeezing parameter of $\xi_R^2=0.77(9)$. 

Moreover, short-range interactions have been proposed~\cite{PhysRevLett.125.223401,PhysRevLett.129.150503,PhysRevLett.129.113201} and demonstrated~\cite{Bornet2023} for creating scalable spin squeezing.
The utilization of a dipolar Rydberg quantum simulator, comprising up to $N=100$ atoms, has demonstrated the generation of spin squeezing through quench dynamics originating from a polarized initial state. 
The degree of spin squeezing increases with the system size, reaching a maximum of $-3.5\pm0.3$ dB (prior to correcting detection errors, or roughly $-5\pm0.3$ dB after correction)~\cite{Bornet2023}. 
In addition, two independent refinements was introduced~\cite{Bornet2023}. 
Firstly, a multistep spin-squeezing protocol was implemented, resulting in an additional enhancement of approximately 1 dB. 
Secondly, through the utilization of Floquet engineering to realize Heisenberg interactions, the lifetime of the squeezed state can be extended by freezing its dynamics.

\begin{figure}[!htp]
 \includegraphics[width=\columnwidth]{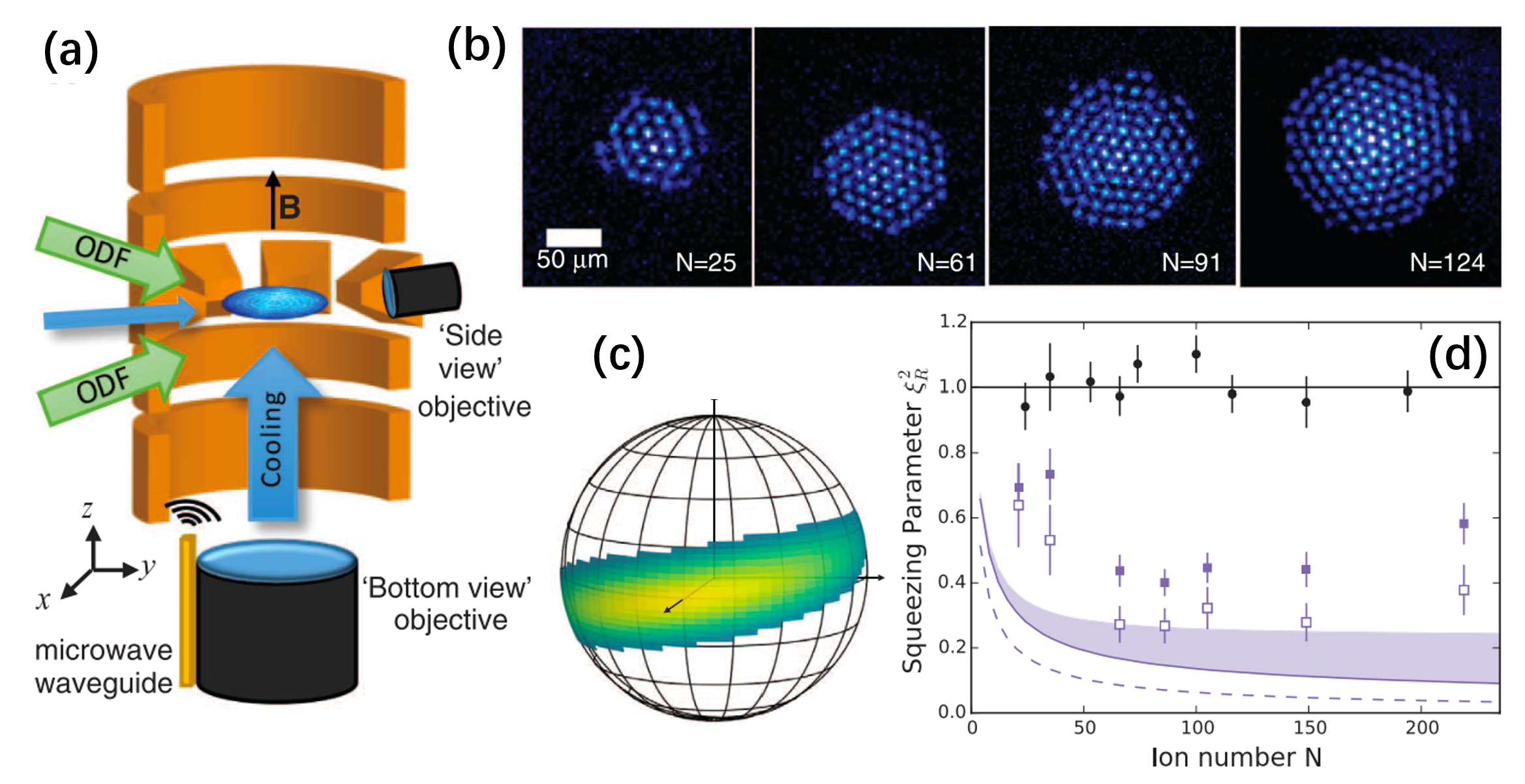}
  \caption{\label{Fig-Sec3-Ion-OAT} Spin squeezed states generated in a Penning trap with multiple ions~\cite{Bohnet2016}. The ions are restricted to the $x-y$ plane by a powerful uniform magnetic field and axially constrained by a quadrupole electric field. The green ODF arrows represent bichromatic light fields employed for the creation of entanglement. The inset displays a depiction of a two-dimensional crystal composed of 91 ions. Reproduced with permission from Bohnet \textit{et al.}, Science \textbf{352}, 1297 (2016). Copyright 2016 The American Association for the Advancement of Science.}
\end{figure}

\begin{figure*}[!htp]
 \includegraphics[width=1.9\columnwidth]{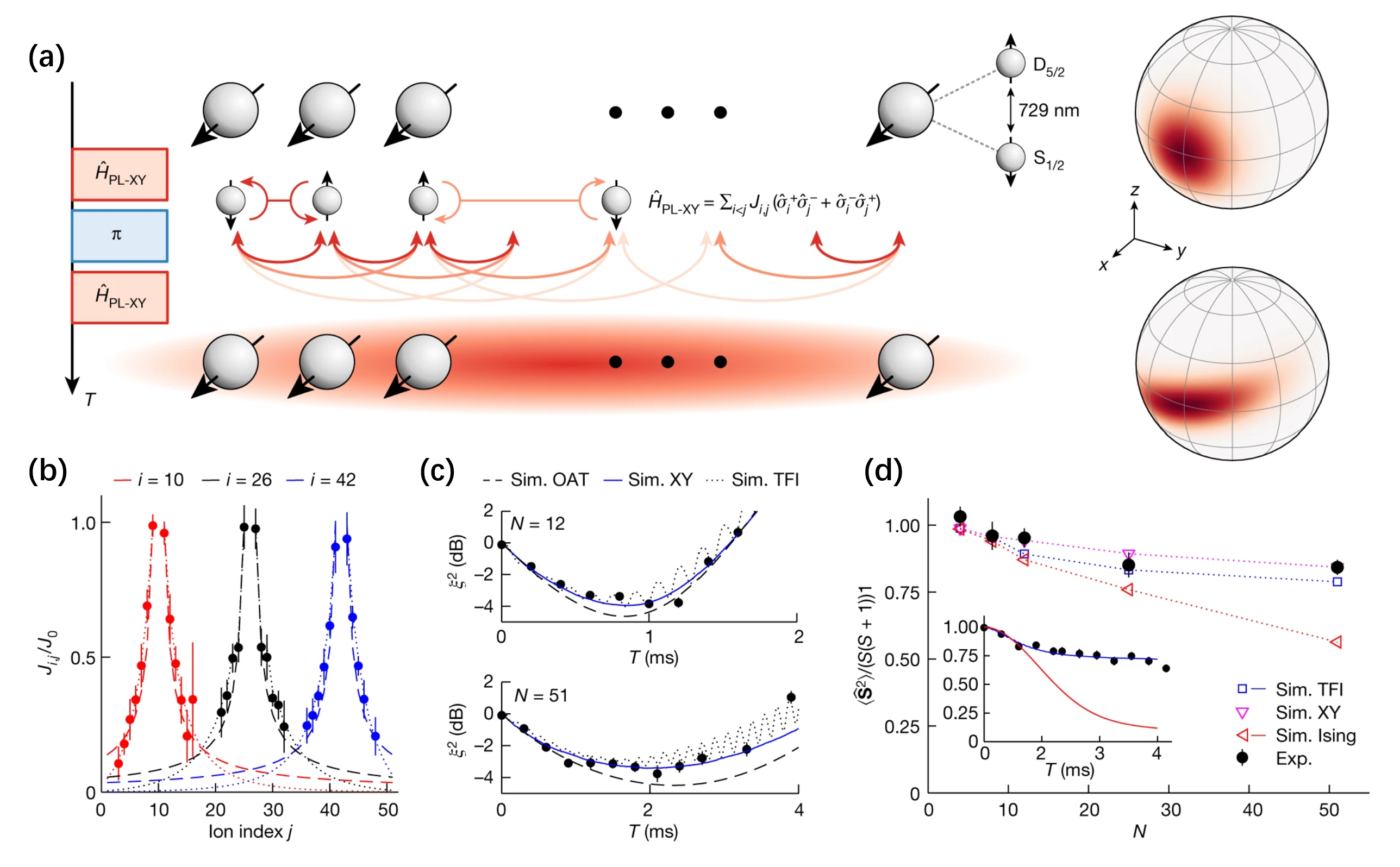}
  \caption{\label{Fig-Sec3-Ising} Generation of spin squeezing with trapped-ion quantum simulator~\cite{Franke2023}.  (a) Experimental sequence detailing the process of squeezing the total spin vector by exchange-interaction.  These interact by power-law XY interactions, where the strength of the connection diminishes as the distance between them increases. The sequence employed involves the integration of interaction pulses (red) and pulses from a global addressing beam (blue) to transform an initial spin coherent state into a spin squeezed state, as seen on the generalized Bloch sphere. (b) The spin–spin couplings $J_{i,j}$ were measured in a 51-ion chain. The couplings were fitted using a power-law interaction with an approximate value of $\alpha \approx 0.9$. The anticipated coupling for the given mode structure and laser detuning is represented by the dashed line. (c) The squeezing measurement is shown versus the interaction time for two different scenarios. In the first scenario, the value of $\alpha$ is about 1, $J_0$ is equal to $560$ rad s$^{-1}$, and the chain consists of $12$ ions. In the second scenario, the value of $\alpha$ is around $0.9$, $J_0$ is equal to $216$ rad s$^{-1}$, and the chain consists of $51$ ions. Numerical calculations are conducted to compare the dynamics of the XY model (represented by a solid blue line) with the dynamics in the presence of a finite transverse field (represented by a dotted black line). Additionally, theoretical results for the OAT model (represented by dashed black lines) is considered, including the decoherence. (d) The total spin, which is measured at the point of maximum spin-squeezing in the XY model and then normalized by its maximum value. The collective spin of $51$ ions is displayed as a function of interaction time $T$. Theoretical findings for the Ising models, both without (red line) and with (blue line) a transverse field, are also shown for comparison.  Reproduced with permission from Franke \textit{et al.}, Nature \textbf{621}, 740 (2023). Copyright 2023 Springer Nature.}
\end{figure*}

Due to their clean environment and well-developed manipulation techniques, entanglement between ultracold trapped ions can be created through the Coulomb interaction. 
Penning traps, in particular, have emerged as a viable option for performing quantum tasks (including quantum metrology) with hundreds of ions. 
Experimental demonstrations of metrological spin squeezing have been achieved by using over 200 trapped $^9$Be$^+$ ions~\cite{Bohnet2016}.
In Fig.~\ref{Fig-Sec3-Ion-OAT}, the quantum spin dynamics was observed in an engineered, homogeneous Ising interaction in a two-dimensional array in a Penning trap.
Through OAT, the spin squeezed states with multiple ions are generated. 
The spin squeezing parameter $\xi_R^2$ was measured with different numbers of ions ($N$) ranging from 21 to 219, and a minimum value of $\xi_R^2=-4.0\pm0.9$ dB was observed for $N=84$. 
Recently, the metrological utility of spin squeezed states in a Ramsey-type interferometer was also demonstrated using $51$ ions held in a macroscopic linear Paul trap~\cite{Franke2023}. 
The measurement uncertainty was reduced by $-3.2\pm0.5$ dB below the SQL.
In this 51-ion quantum simulator, a pseudo-spin is encoded in two electronic states of $^{40}$Ca$^{+}$: $\left|\downarrow \right\rangle =\left|{{\rm{S}}}_{1/2},m=+1/2\right\rangle$ and $\left|\uparrow \right\rangle =\left|{{\rm{D}}}_{5/2},m=+5/2\right\rangle$. 
These states are collectively coupled by a global laser beam, and spin-spin interactions between the ions are engineered using a two-tone laser that couples the ions' internal electronic states to their ground-state-cooled transverse motional modes.
The dynamics, under the influence of a strong driving transverse to the interaction axis, are described by the power-law transverse-field Ising model, ${{H}}_{{\rm{PL}}-{\rm{TFI}}}={\sum }_{i < j}{J}_{ij}{\hat{\sigma }}_{i}^{x}{\hat{\sigma }}_{j}^{x}+B{\sum }_{i}{\hat{\sigma }}_{i}^{z}$. 
In the rotating frame of the driving, this system obeys a power-law XY model ${{H}}_{{\rm{PL}}-{\rm{XY}}}=\sum _{i < j}{J}_{i,j}\left({\hat{\sigma }}_{i}^{+}{\hat{\sigma }}_{j}^{-}+{\hat{\sigma }}_{i}^{-}{\hat{\sigma }}_{j}^{+}\right)$,
with the interaction strengths ${J}_{i,j}=J_0|i-j|^{-\alpha}$ between sites $i$ and $j$. 
The interaction strengths are parameterized in terms of the nearest-neighbour strength $J_0$ and a tunable exponent $0<\alpha<3$ describing the interaction range.
Using the OAT sketched in Fig.~\ref{Fig-Sec3-Ising}~(a), spin squeezed states can be prepared. 
To begin, prepare a spin coherent state polarized along the $+x$ direction. 
Subsequently, evolve this state under the XY interaction for a variable time $T$. 
In the experiment, the interaction period is split by an echo pulse that cancels site-dependent Stark shifts along $z$ direction and increases the coherence time of the system.\\

\noindent \textit{(6). Preparing spin squeezed states beyond OAT}\\

Besides OAT, various schemes for generating spin squeezing have been proposed and demonstrated. 
To speed up the entanglement generation, one can apply an additional rotation along with the OAT, which is the so-called twist-and-turn (TNT) dynamics~\cite{PhysRevA.63.055601, Muessel2015, PhysRevA.99.022329} obeying the Hamiltonian
\begin{equation}
    \label{TNT}
    H_{TNT}=\chi \hat J_z^2 + \Omega \hat J_x,
 \end{equation}
where the turning term $\Omega \hat J_x$ rotates along the $x$-axis simultaneously with the OAT interaction.
The TNT dynamics is experimentally studied from the starting point of a spin coherent state pointing in the $+x$ or $-x$
direction. The interaction is suddenly switched to a finite value
of $\Lambda=N\chi/\Omega$ in presence of linear coupling.
The TNT dynamics can be described as the evolution of an effective relative-number wave packet~\cite{PhysRevA.86.023615,Pezze2018_RMP}. 
Using two internal hyperfine levels of $^{87}$Rb atoms, Fig.~\ref{Fig-Sec3-Feshbach}~(b)-(e) shows the TNT dynamics in an experiment with BEC. 
The nonlinearity of the system is adjusted through Feshbach resonance, while the linear coupling between the two hyperfine states is achieved using radio frequency and microwave fields.  
With about $400$ atoms, the spin squeezing of $\xi^2_R=-4.5\pm0.2$ dB was achieved. 
Later, spin squeezing with an inferred $\xi^2=-7.1$ dB (taking into account the reduced mean spin length) with a single BEC and $\xi^2=-2.8$ dB with up to $10^4$ atoms were reported~\cite{Muessel2015}.

The generation of extreme spin squeezed states has attracted a lot of attention since they can approach the Heisenberg limit.
However, to generate extreme spin squeezing, more complex nonlinear interactions such as TACT interactions~\eqref{H_TACT} are required~\cite{Kitagawa1993,Kajtoch2015}.
It is worth noting that the realization of TACT interactions in known quantum systems still has not been achieved.
An alternative way is using the TNT dynamics augmenting the OAT interaction with a well-designed sequence of transverse coherent field.
Promising analytic proposals based on this idea include transforming OAT to TACT by using a periodic train of $\pm \pi/2$ spin rotation pulses~\cite{PhysRevLett.107.013601,PhysRevA.90.013604}, or a periodically modulated transverse field~\cite{PhysRevA.91.043642}. 
However, these scheme demand stringent experimental conditions, either requiring a large number of high-precision pulses~\cite{PhysRevLett.107.013601} or a high modulation frequency~\cite{PhysRevA.92.063610}. 
One can also makes use of only a few optimized pulses along the mean spin direction, but the required evolution time is greatly prolonged~\cite{PhysRevA.87.051801}.
To efficiently generate an extreme spin squeezed state, one can even seek help from machine learning such as deep reinforcement learning~\cite{Chen2019} and stochastic gradient decent~\cite{PhysRevApplied.17.064050}, which can achieve an optimized policy and providing perspectives and solutions that is difficult to find with analytic approach.

In cavity-QED systems, it was shown that effective two-axis twisting Hamiltonians can be realized by applying bichromatic lasers~\cite{Borregaard_2017}. 
Spin squeezing of the atoms in the absence of dissipation can reach the ideal Heisenberg limit. However, the attainable spin squeezing is limited by collective decay and this scheme may be suited for Rydberg blockade system without strong collective decay. By using cavity-assisted Raman transitions, tunable two-axis spin model and spin squeezing may be realized in two cavities with all parameters can be tuned independently~\cite{Yu_2016}. The two-axis-twisting spin squeezing may also be generated by multipass quantum erasure, which may be extendable to an optical ring cavity system~\cite{PhysRevA.96.013823}.

Except for OAT and TACT interactions, the more general form of the quadratic collective-spin interaction can be described by the Lipkin-Meshkov-Glick (LMG) model~\cite{Lipkin1965,Meshkov1965,Glick1965,Lee2012,PhysRevLett.102.070401}. 
A way to implement general LMG model with independent multi-parameter  control as well as direct spin squeezing by both TACT and OAT is proposed, with a BEC in a ring trap with periodic angular modulation of the external potential and of the inter-particle interactions~\cite{PhysRevA.91.053612}.
It is also found that the squeezing properties are determined by the initial states and the anisotropic parameters. 
To achieve the Heisenberg-limited spin squeezing, a pulse rotation scheme is proposed to transform the model into a two-axis twisting model~\cite{Hu_2023}. 
While in generic coupled spin systems with collective spin-spin interactions, a universal scheme by continuous drivings is also proposed~\cite{PhysRevA.107.042613}.  

In a spin-1 BEC, the spin squeezed ground states is recently demonstrated using a double-quench method with a final Hamiltonian tuned close to the quantum-critical point~\cite{PhysRevLett.131.133402}. 
It is related to a kind of shortcut to adiabaticity~\cite{RevModPhys.91.045001} for fast production of highly entangled states. 
In contrast to typical nonequilibrium methods by quenching through a QPT, these squeezed ground states are time stationary. 
A squeezed ground state with $6\mbox{-}8$ dB of squeezing with a constant squeezing angle is demonstrated. 
The spin squeezing can be maintained for over $2$ s at a constant orientation angle and gradually decreases after that due to atom loss. 
This double-quench method can also be easily adapted to (pseudo) spin$-1/2$ systems such as bosonic Josephson junctions mentioned before. 

\subsection{\label{sec:3-3}Twin-Fock state}
Twin-Fock state is another kind of metrological useful entangled state~\cite{PhysRevLett.71.1355,Dunningham2002,Campos2003}, which belongs to a kind of Dicke states~\cite{PhysRevLett.103.020504, PhysRevLett.112.155304}.
A Dicke state can be described by a specific Dicke basis $|J,m\rangle=|J+m\rangle_a |J-m\rangle_b$, which is a two-mode Fock state with $J+m$ particles in mode $a$ and $J-m$ particles in mode $b$, respectively. Thus, a Dicke state has precise relative particle number between the two modes while the relative phase is completely undefined~\cite{PhysRevLett.71.1355}. 
It was proposed that the depth of entanglement can be detected by using the variance and second moments of the collective spin components, which may be possible to define spin squeezing parameters for Dicke states~\cite{Vitagliano2017}.
However, it was pointed out that a Dicke state is not spin squeezed~\cite{Wang2002}. 
The Dicke states have the potential to improve the measurement precision. 
One can first calculate the QFI, and find that
\begin{equation}\label{F_Q_Dicke}
    F_Q = \frac{N^2}{2}-2m^2 +N.
\end{equation}
Clearly, $m=\pm N/2$ is indeed a spin coherent state with $\theta=0$ or $\pi$ and the QFI $F_Q=N$ reduces to the case without entanglement. While for $-N/2 <m< N/2$, the QFI $F_Q >N$ indicates the metrological usefulness.  
%

In particular, when $m=0$ with $N$ an even number, it corresponds to a twin-Fock state 
\begin{equation}
    \label{TWIN}
    |\Psi\rangle_{TWIN} =|N/2,0\rangle=|N/2\rangle_{a} |N/2\rangle_{b},
\end{equation}
which is a two-mode Fock state with definite equal particle number in both modes. 
According to Eq.~\eqref{F_Q_Dicke}, the QFI for a twin-Fock state 
becomes $F_Q = N^2/2+N$. When the particle number is sufficiently large $N \gg 1$, $F_Q \approx N^2/2$, the ultimate phase measurement precision bound can be 
\begin{equation}
    \Delta \phi \ge 1/\sqrt{F_Q} \sim \frac{\sqrt{2}}{N}
\end{equation}
approaching the Heisenberg scaling with a factor of $\sqrt{2}$ above the exact Heisenberg limit.

For a twin-Fock state, applying a $\pi/2$ pulse can convert it to a state with well-defined relative phase, 
\begin{equation}
    |\psi_{AS}\rangle = \sum_{k=0}^{N/2} (-1)^{N/2-k} \sqrt{\frac{1}{2^{N}} C_{k}^{2k} C_{N/2-k}^{N-k}} |N/2,-N/2+2k\rangle
\end{equation}
which is called the arcsine state~\cite{Campos2003}  and can be visualized as a ring around the equator of the generalized Bloch sphere. 
Then accumulating phase $\phi$ between two modes and applying another $\pi/2$ pulse can finalize a whole Ramsey process. 
In practise, one cannot directly exploit the population difference measurement to extract the phase information since $\langle \hat J_z\rangle$ always equals $0$.
An effective way to extract the phase is using the square of the population difference $\hat J_z^2$.  
When $\phi \approx 0$, one can find that $\hat J_z^2$ can attain the ultimate phase measurement precision bound of the twin-Fock state~\cite{Lucke2011}.
Theoretically, it was proposed that the parity measurement $\hat \Pi = (-1)^{N-\hat J_z}$, which distinguishes the population number in one mode either even or odd is also capable to estimate $\phi$~\cite{Campos2003}.
However, in systems with large particle number, parity measurement is not feasible to be realized in experiments.

\begin{figure}[!htp]
 \includegraphics[width=\columnwidth]{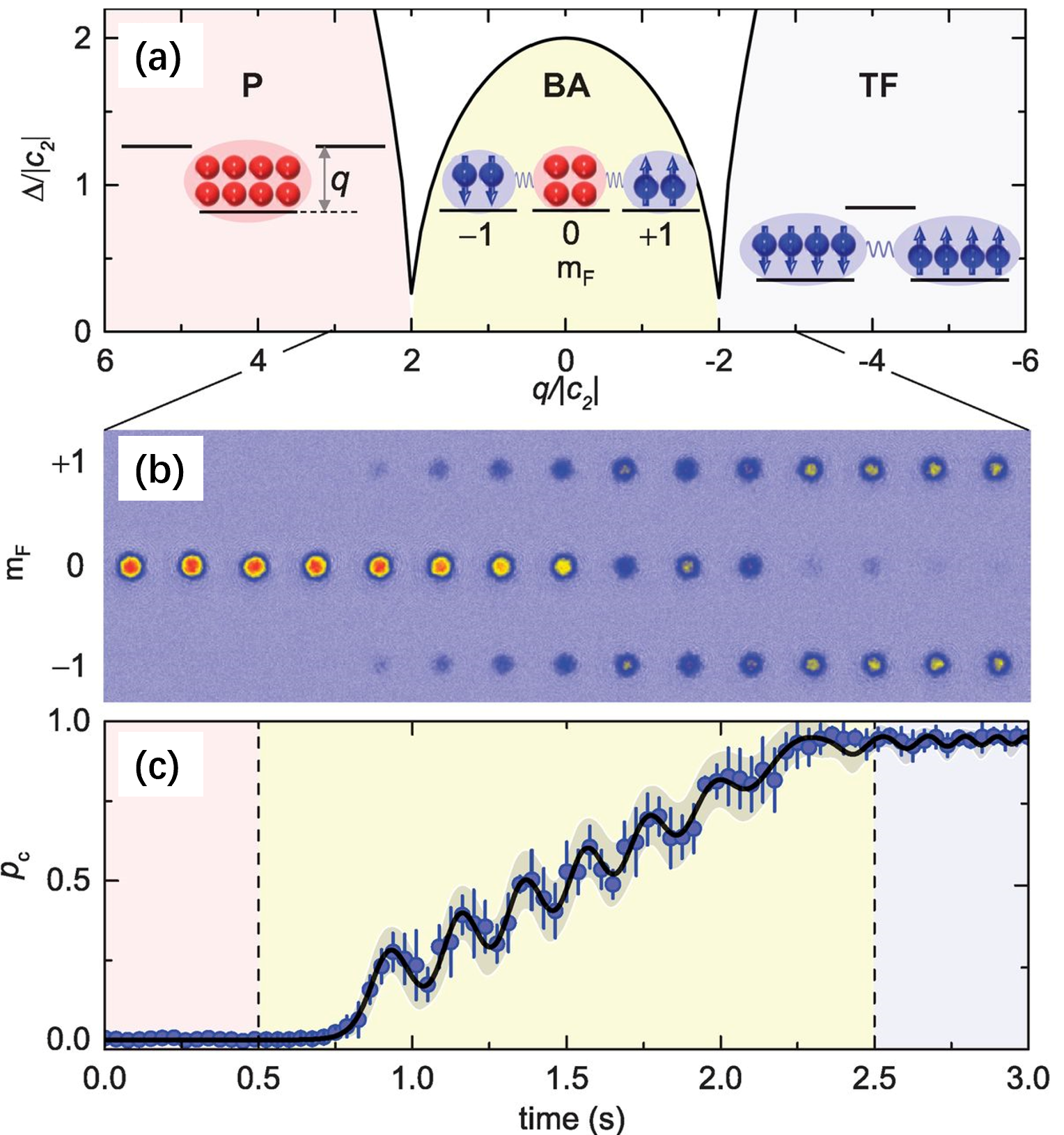}
  \caption{\label{Fig-Sec3-twin-QPT} Generation of twin-Fock state via adiabatic sweeping~\cite{Luo2017}. (a) The thick black solid line denotes the gap $\Delta$ between the first excited and the ground state of Hamiltonian. The two minima at $q=\pm 2|c_2| $ defines three quantum phases, illustrated by their atom distributions in the three spin components. Here $c_2$ is the same as the spin-dependent collision energy $c$. The first-order Zeeman shifts are not shown because they are inconsequential for a system with zero magnetization. (b) Absorption images of atoms in the three spin components after Stern-Gerlach separation, showing efficient conversion of a condensate from a polar state into a twin-Fock state by adiabatic sweeping. (c) Conversion efficiency as a function of time. Reproduced with permission from Luo \textit{et al.}, Science \textbf{355}, 620 (2017). Copyright 2017 The American Association for the Advancement of Science.}
\end{figure}

In spinor BECs, colliding atoms can exchange spin.
The spin exchange collisions can generate entangled atom pairs, which is suitable for generating twin-Fock state~\cite{Lucke2011,Luo2017}.
The many-body Hamiltonian for a spin-1 BEC under the spin-exchanging interaction can be expressed as
\begin{eqnarray}\label{H_spin1}
    H_{SM} &=& 2 c (\hat a_0^{\dagger} \hat a_0^{\dagger} \hat a_{1} \hat a_{-1} +\hat a_0 \hat a_0 \hat a_{1} ^{\dagger}\hat a_{-1}^{\dagger}) \nonumber \\
    &+& [q+c(2\hat N_0-1)](\hat N_{+1} + \hat N_{-1})
\end{eqnarray}
where $\hat a_{m_F}$ ($\hat a_{m_F}^{\dagger}$) denotes annihilation (creation) operator and $\hat N_{m_F}=\hat a_{m_F}^{\dagger} \hat a_{m_F}$ is the atom number in $m_F$ component with the total atom number $N=\sum_{m_F} N_{m_F}$ with $m_F=0, \pm 1$. 
The Hamiltonian is obtained under the assumption of all the three spin components being in the same spatial wavefuntion $\phi (\vec r)$. 
$c=\frac{2\pi(a_2-a_0)}{3m} \int d\vec r |\phi(\vec r)|^4$ denotes spin-dependent collision energy where atoms in $m_F=0$ can be transferred to $m_F=\pm 1$ in pairs and vice versa. 
Here, $m$ is the atomic mass, $a_0$ and $a_2$ respectively denote the s-wave scattering length for total spin $G=0$ and $G=2$ allowed by two-particle collisions~\cite{Lucke2014-thesis}. 
$c<0$ ($c>0$) corresponds to ferromagnetic  (antiferromagnetic) interaction respectively~\cite{PhysRevA.93.033608}. 
$q$ is an effective quadratic Zeeman shift, which can be tuned by the external magnetic field (proportional to the square of magnetic field strength) or near-resonant microwave dressing field~\cite{Luo2017}.
In the language of angular momentum with conserved magnetization $\hat N_{+1}-\hat N_{-1}=0$, the Hamiltonian of Eq.~\eqref{H_spin1} can be rewritten as
\begin{equation}\label{H_spin1_L}
     H_{SM}=c \hat  L^2 - q \hat N_0,
\end{equation}
where $\hat L=\{ \hat L_x, \hat L_y, \hat L_z\}$,  $\hat L_{\mu} = \sum_{m,n} \hat a^{\dagger}_m [f_{\mu}]_{mn} \hat a_n$ ($\mu = x,y,z $ and $m,n = 0, \pm 1$) with $[f_{\mu}]_{mn}$ the matrix element for spin-1 angular momentum. 
The ground state of Hamiltonian with Eq.~\eqref{H_spin1} is determined by the competition between $c$ and $q$. 
It is convenient to express the system state in the three-mode Fock basis $|N-2k\rangle_{0} |k\rangle_{-1}  |k\rangle_{+1}$ with $k=0,1,...,N/2$.

\begin{figure}[!htp]
 \includegraphics[width=\columnwidth]{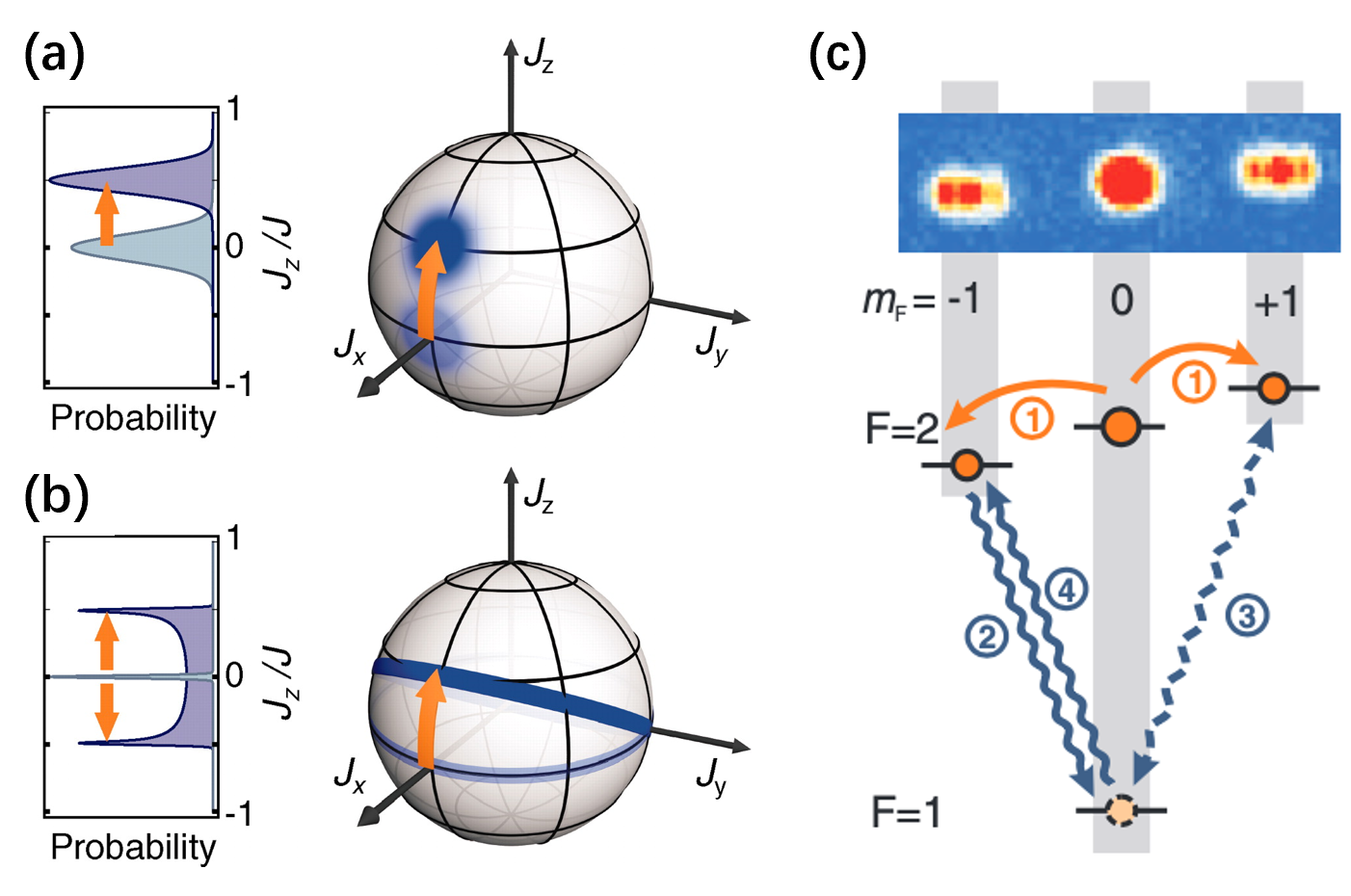}
  \caption{\label{Fig-Sec3-twinFock} Matter-wave interferometry with twin-Fock state generated through spin mixing dynamics~\cite{Lucke2011}. (a) An unentangled spin coherent state and the corresponding state after rotation. (b) An input twin-Fock state and the corresponding state after rotation. (c) Schematic of the sequence of the realization of the beam splitter. Three microwave pulses are sequentially applied to achieve the coupling of the two hyperfine states $|F=2,m_F=\pm1 \rangle$. The total effect is equivalent to a rotation around the $x$-axis by an angle. Reproduced with permission from Lucke \textit{et al.}, Science \textbf{334}, 773 (2011). Copyright 2011 The American Association for the Advancement of Science. }
\end{figure}

For a ferromagnetic BEC with $c<0$, the ground state can be divided into three distinct phases.
For $q\gg 2 |c|$, the ground state $|N\rangle_{0} |0\rangle_{-1}  |0\rangle_{+1}$ is in the polar phase with all atoms condensed in the $m_F=0$ component.
For $q\ll -2 |c|$, the ground state $|0\rangle_{0} |N/2\rangle_{-1}  |N/2\rangle_{+1}$ becomes a twin-Fock state with the atoms equally partitioned into the $m_F=\pm 1$ components. 
In the middle,  two critical points $q=\pm 2|c|$ correspond to the broken-axisymmetry phase whose ground state acquires a transverse magnetization, spontaneously breaking the SO(2) symmetry of the system. 

To generate a twin-Fock state, one can start in the polar phase and ramp down $q(t)$ adiabatically from $q\gg 2 |c|$ to $q\ll -2 |c|$ crossing the two critical points $q=\pm 2|c|$~\cite{PhysRevLett.111.180401}. 
At the QPT points, the energy gap $\Delta E \sim 1/N^{1/3}$. 
The energy gap decreases slowly with the total particle number $N$, which suggests the possibility to maintain the adiabaticity even for $N\sim 10^5$. 
It had been demonstrated that near-deterministic generation of a twin-Fock state with more than 10000 atoms can be achieved with $^{87}$Rb BEC~\cite{Luo2017}.  
Later, in a subsequent experiment~\cite{Zou2018}, an interferometric precision of $2.42$ dB beyond the three-mode SQL, using balanced spin-1 (three-mode) Dicke states containing thousands of entangled atoms are demonstrated. 
The input quantum states are also deterministically generated by controlled QPT and exhibit close to ideal quality. 
This work also demonstrated an interferometric precision of $8.44$ dB beyond the two-mode SQL. 

For an antiferromagnetic BEC with $c>0$, the system exhibits two phases of ground state. 
In the limit of $N \rightarrow \infty$, the ground state is a polar state $|N\rangle_{0} |0\rangle_{-1}  |0\rangle_{+1}$ when $q>0$ and a twin-Fock state $|0\rangle_{0} |N/2\rangle_{-1}  |N/2\rangle_{+1}$ when $q<0$.
For a finite total particle number $N$, the minimal energy gap $\Delta E \sim 1/N$ occurs at the critical point $q/c = 3.7688/N^2$, which is close to $q=0$ if $N$ is large~\cite{PhysRevLett.123.073001}.  
Ideally, by adiabatically tuning the effective quadratic Zeeman energy $q$ from positive infinity to negative infinity, the twin-Fock state can be generated. 
However, since the energy gap  decreases inversely proportional to $N$, which drops much faster than the one in a ferromagnetic spin-1 BEC, the adiabatic evolution is difficult to realize in experiment.
It was suggested that utilizing the adiabatic and multilevel-oscillation method~\cite{PhysRevLett.123.073001}, the high-fidelity twin-Fock state can be efficiently generated.  
By partially replacing the adiabatic evolution near the QPT point with multilevel oscillations, this method accelerates the state generation process and relaxes the requirement on the control accuracy of quadratic Zeeman splitting  from microgauss to milligauss.  

In this many-body system with Eq.~\eqref{H_spin1_L}, excited-state quantum phase transitions (ESQPTs) can also be appealing, which refers to quantum criticality aroused in ground state. 
The existence of QPTs and ESQPTs in an antiferromagnetic spin-1 condensate are shown and their correspondence with dynamical phase transition is demonstrated, which can be characterized by the QFI~\cite{PhysRevResearch.5.013087}. 
The dynamical phase transition with the condensate initially prepared in a spin coherent state can be used to probe the quantum criticality in excited states, which gives rise to a peak value of the QFI. 
It can also be used to implement sub-SQL estimation on the effective quadratic Zeeman energy $q$. 
It is interesting to note that the ground-state phase transitions from symmetry-broken states to the symmetry-restored spin-singlet state can also be indicated by the dynamical phase transition. 


In the approximation that $N_0\approx N \gg 1$, the initial condensate of $m_F=0$ can serve as an unlimited pump source for parametric amplification of the output modes $m_F=\pm 1$. 
In this case, one can make use of the parametric approximation $\hat a_0 \approx \sqrt{N_0}$ and the Hamiltonian~\eqref{H_spin1} becomes 
\begin{eqnarray}\label{H_spin1_largeN0}
    H_{SM} \approx A (\hat a_{1} \hat a_{-1} +\hat a_{1} ^{\dagger}\hat a_{-1}^{\dagger})+ B (\hat a_{1}^{\dagger} \hat a_{1} +\hat a_{-1} ^{\dagger}\hat a_{-1}),
\end{eqnarray}
where $A=2c N_0$ and $B=q+c(2N_0-1)$.
Evolving from $|0\rangle_{+1}|0\rangle_{-1}$ under Hamiltonian in the form of Eq.~\eqref{H_spin1_largeN0}, a two-mode squeezed vacuum state (a superposition state of multiple twin-Fock state with different total particle number) can be obtained~\cite{PhysRevLett.117.143004}. 

In particular for the resonance case with $B=0$, the spin-mixing Hamiltonian becomes 
\begin{equation}
    H_{SM}=2 c N_0 (\hat a_{1} \hat a_{-1} +\hat a_{1} ^{\dagger}\hat a_{-1}^{\dagger}),
\end{equation}
and the evolved state can be calculated as 
\begin{equation}
    |\psi_{SM}(t)\rangle =\sum_{n=0}^{\infty} \frac{(-i \tanh r)^n}{\cosh r} |n\rangle_{+1}|n\rangle_{-1},
\end{equation}
where $r=2c N_0 t$.
Projecting onto a certain total particle number $2n=N$, one can immediately get the twin-Fock state $|N/2\rangle_{+1}|N/2\rangle_{-1}$.

The quantum interferometry with twin-Fock state had been experimentally demonstrated with spin-1 BEC via spin mixing~\cite{Lucke2011}.
In the experiment, spin dynamics in BECs was used to create large ensembles of pair-correlated atoms with an interferometric sensitivity beyond SQL.  
The experiment starts with creating a $^{87}$Rb condensate of $2.8\times10^4$ atoms in the hyperfine state $|F=2,m_F=0\rangle$ in an optical dipole trap. Then the spin-exchange collision gradually produces correlated pairs of atoms with spins up and down. The total number of the correlated pairs of atoms in $|F=2,m_F=\pm1\rangle$ increase exponentially with time due to spin mixing. Afterwards, the trap is switched off and the three hyperfine states are split by a strong magnetic field gradient, and all three hyperfine states are recorded by absorption imaging. As the hyperfine states $|F=2,m_F=\pm1\rangle$ are generated in pairs, twin-Fock state in these two modes is prepared.
The twin-Fock state was effectively obtained by post-selection of a certain particle number.

Spin-mixing dynamics in cold atoms can also be used to realize a SU$(1,1)$ interferometer~\cite{PhysRevA.86.023844, PhysRevLett.115.163002}, which is analogous to parametric down conversion in optics. 
As shown in Fig.~\ref{Fig-Interferometry-SU11}~(b), for spin-1 BEC, the spin-exchanging collision plays the role of Kerr nonlinearity, and the condensate atoms in $m_F=0$ can be regarded as the optical pump field. 
The atoms that transferred to $m_F = \pm 1$ correspond to the signal and idler beams. 
Similar to the optical parametric down conversion, the amplification can be triggered by the fluctuations of spin orientation. 
Let us indicate with $N$ the total average number of atoms transferred in pairs from a condensate prepared in $m_F=0$ to the initially empty $m_F=\pm1$ modes. After spin mixing, the system acquires a relative
phase $\theta = 2\theta_0 - (\theta_{+1} + \theta_{-1})$ between $m_F=0$ and $m_F=\pm1$
modes. A second spin-mixing dynamics ends up the interferometer. The final populations in $m_F=0, \pm1$ depend on $\theta$.
A nonlinear SU$(1,1)$ interferometry with $F=2$ manifold of $^{87}$Rb atoms was experimentally demonstrated~\cite{Linnemann2016}. The phase imprinting is achieved via adding a second-order Zeeman shift and read out by counting the mean atom number in $m_F=\pm1$. 
In addition, if one apply a linear coupling among the $m_F=0,\pm 1$ before and after phase imprinting, realizing the so-called "pumped-up" SU$(1,1)$ interferometer~\cite{PhysRevLett118150401}, the measurement precision may be further improved.

\subsection{\label{sec:3-4}Spin cat states}
Spin cat states are promising candidates for achieving Heisenberg-limited quantum metrology~\cite{Huang2015,Huang2018_1}.
A spin cat state is a typical kind of macroscopic superposition of spin coherent states (MSSCS) ~\cite{PhysRevLett.113.090401,PhysRevA.95.043642, Nolan2018}.
Generally, an MSSCS is a superposition of multiple spin coherent states~\cite{Micheli2003, Ferrini2010, Spehner2014, Nolan2018}, which can be written in the form of
\begin{eqnarray}
    |\Psi(\theta, \varphi)\rangle_{\textrm{M}}=\mathcal{N}_{C}(|\theta,\varphi\rangle + |\pi-\theta,\varphi\rangle)
\end{eqnarray}
where $\mathcal{N}_{C}$ is the normalization factor and $\left|\theta,\varphi\right\rangle$ denotes a spin coherent state. 
Here, without loss of generality, we assume $\varphi=0$ and expressing in terms of Dicke basis,
\begin{equation}
    |\Psi(\theta)\rangle_{\textrm{M}}=\mathcal{N}_{C} \left[\sum^{J}_{m=-J} c_m(\theta)\left(\left|J,m\right\rangle+\left|J,-m\right\rangle\right)\right].
\end{equation}
Since $c_m(\theta)=c_{-m}(\pi-\theta)$, the coefficients are symmetric about $m=0$.

\begin{figure}[!htp]
 \includegraphics[width=\columnwidth]{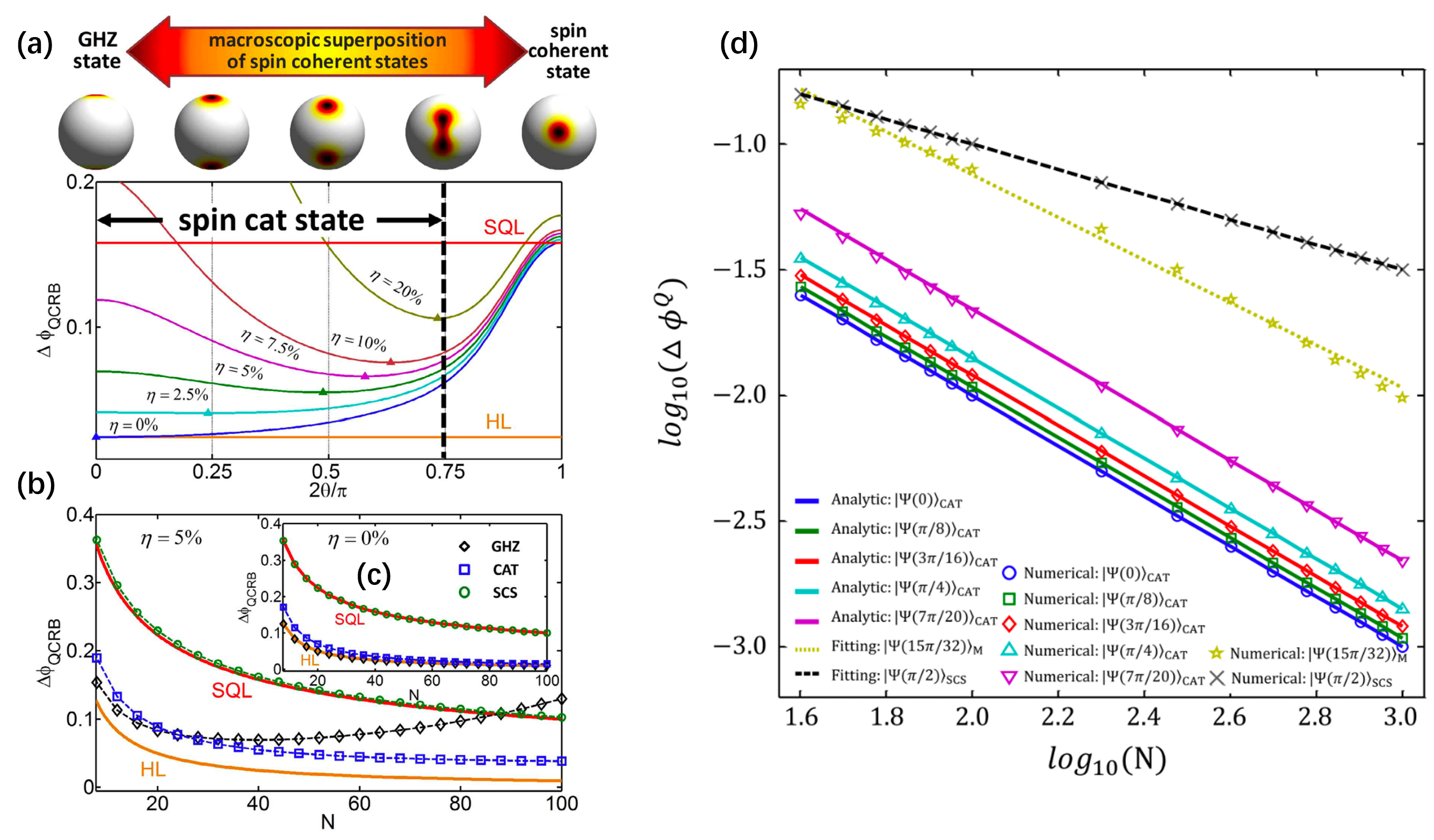}
  \caption{\label{Fig-Sec3-spin-cat} Measurement precisions offered by spin cat states~\cite{Huang2015,Huang2018_1}. (a) The achievable precisions versus $\theta$ for different atom loss ratios. The triangles denote the best optimal precisions. The precision becomes worse when the atom loss ratio becomes larger. For nonzero loss ratio, instead of the GHZ state, the optimal state becomes a spin cat states with modest $\theta$. Here, the initial total atomic number $N=40$. And the left side of the thick black dashed line indicates the region of spin cat states. (b) The achievable precisions versus the initial total atomic number $N$ for three typical input states (GHZ state, modest spin cat state and the spin coherent state under loss ratio $\eta=5\%$.  (c) The measurement precision versus $N$ for the three input states in the absence of loss (i.e. $\eta=0\%$). Reproduced with permission from Huang \textit{et al.}, Sci. Rep. \textbf{5}, 17894 (2015). Copyright 2015licensed under a Creative Commons Attribution (CC BY) license. (d) The ultimate measurement precision of spin cat states. Reproduced with permission from Huang \textit{et al.}, Phys. Rev. A \textbf{98}(1), 012129 (2018). Copyright 2018 American Physical Society.}
\end{figure}

The properties of the MSSCS depend on $\theta$.
When $\theta=\pi/2$, it corresponds to a spin coherent state $\left|\pi/2,0\right\rangle$.
As $\theta$ decreases, the two superposition spin coherent states become separated. When $\theta \lesssim \theta_c\equiv\sin^{-1}\left\{2\left[\frac{\left((J-1)!\right)^2}{2 (2J)!}\right]^{1/{2J}}\right\}$ is sufficiently small~\cite{Huang2015, Huang2018_1}, the two spin coherent states become quasi-orthogonal (or orthogonal), the MSSCS can be regarded as a spin cat state.  
In this case, we abbreviate the spin cat states as $\left|\Psi(\theta)\right\rangle_{\textrm{CAT}}$, and it can be approximated as
\begin{eqnarray}\label{CAT}
    \left|\Psi(\theta)\right\rangle_{\textrm{CAT}}&\approx& \frac{1}{\sqrt{2}}\left(|\theta,\varphi\rangle + |\pi-\theta,\varphi\rangle\right)\\\nonumber
    &=&\frac{1}{\sqrt{2}} \left[\sum^{J}_{m=-J} c_m(\theta)\left(\left|J,m\right\rangle+\left|J,-m\right\rangle\right)\right].
\end{eqnarray}
Note that spin cat states can be understood as a superposition of GHZ states with different spin length.
Particularly when $\theta=0$, it reduces to a GHZ state, i.e., $\left|\Psi(0)\right\rangle_{\textrm{CAT}}=\frac{1}{\sqrt 2}\left(|J,-J\rangle + |J,J\rangle \right)$.
The expectation of $\hat J_z$ for a spin cat state $_\textrm{CAT}\langle\Psi(\theta) |\hat J_z| \Psi(\theta)\rangle_\textrm{CAT}=\sum_{m} m c_m^{*}(\theta) c_m(\theta)=0$.
Hence, the variance of a spin cat state becomes $\Delta^2 \hat J_z=\langle\Psi(\theta) |\hat J_z^2| \Psi(\theta)\rangle_\textrm{CAT} = \sum_{m} m^2 c_m^{*}(\theta) c_m(\theta)$.
The variance can be analytically obtained $\Delta^2 \hat J_z \approx \frac{1}{4} N^2 \cos^2\theta$, which only depends on $\theta$ and $N$.
If spin cat state is input for quantum interferometry, the QFI for the output state can be computed as 
\begin{equation}\label{FQ_CAT}
    F_{Q}^{CAT}=4\Delta^2 \hat J_z \approx N^2 \cos^2 \theta,
\end{equation}
and the corresponding measurement precision can achieve the Heisenberg-limited scaling $\propto \frac{1}{N \cos\theta}$. 

OAT is one of the well-known strategies for generating entanglement in many-body quantum systems. 
Under the time-evolution of OAT interaction [Eq.~\eqref{OAT}], an initial spin coherent state (without entanglement) can evolve
to spin squeezed states, over-squeezed states and other kinds of nonclassical states~\cite{Pezze2018_RMP}. 
For evolution times beyond the optimal spin squeezing time, the Husimi distribution of the evolved state in the OAT model starts to develop into the non-Gaussian regime. 
Especially when $\chi t =\pi/2$, it evolves to the GHZ state~\cite{You2003}.
Recently in a trapped-ion quantum simulator, the multi-headed cat states are successfully prepared with $N=12$ ions~\cite{Franke2023}. 
Yet, the generation of spin cat state with this method~\cite{Micheli2003,PhysRevLett.102.100401,PhysRevA.78.051601,PhysRevA.95.063609} requires long evolution times and the generated states are not deterministically prepared. 
So far, experiments have demonstrated the generation of spin-squeezed~\cite{Gross2010,Riedel2010} and slightly non-Gaussian states~\cite{Strobel2014} in ultracold atomic ensembles.
The generation of spin cat states in an engineered Dicke model may also be achieved~\cite{PhysRevA.104.053721}. 
%
%
%
In an appropriate limit of the Dicke model, one can obtain OAT dynamics that leads to the generation of spin cat states. 
Additionally, the quantum jump trajectories that generate "kitten" states in a random fashion is investigated.
Besides,  the effects of decoherence are also explored on spin cat state generation and the experimental feasibility in trapped ions and atom ensembles with cavity-mediated Raman transitions is discussed.  

An effective and deterministic way to generate spin cat states is the adiabatic evolution~\cite{PhysRevLett.97.150402, Huang2015, Huang2018_2}.  
As shown in Fig.~\ref{Fig-Sec3-QPT}, adiabatic evolution crossing QPTs can be used for generating non-Gaussian entangled states.
The process can be described by a time-dependent Hamiltonian~\cite{Huang2018_2},
\begin{equation}\label{Ham_QPT}
    H_{BS1}(t) = R_1(t) H_1 + R_2(t) H_2.
\end{equation}
Here, the Hamiltonians $H_1$ and $H_2$ are interpolated with the time-varying parameters $R_1(t)$ and $R_2(t)$.
One can choose proper Hamiltonians such that the groundstate of $R_1(0) H_1$ is non-degenerate while the groundstate of $R_2(T) H_2$ at time $T$ is multi-fold degenerate.
Defining the ratio $\lambda(t)=R_1(t)/R_2(t)$, there exists a QPT at the critical point $\lambda_c$, that is, $R_1(t) H_1$ dominates the system when $|\lambda(t)|>|\lambda_c|$ and $R_2(t) H_2$ dominates the system when $0\le |\lambda(t)|<|\lambda_c|$.
Thus, starting from the non-degenerate groundstate of $R_1(0) H_1 + R_2(0) H_2$, an entangled groundstate can be obtained with high fidelity if $\lambda$ is adiabatically swept through $\lambda_c$.
For instance, preparing the ground state of a BEC trapped in a double-well potential with attractive interaction~\cite{Trenkwalder2016}, or in an ion crystal~\cite{PhysRevLett.121.040503}.
By adiabatic sweeping the control parameter across the spontaneous symmetry-breaking transition, spin cat states with different degree of entanglement can be prepared. 
However, due to the degeneracy of the spectrum, this method may be  fragile to symmetry-breaking perturbations and finite temperature~\cite{PhysRevLett.119.090401,Gabbrielli2018}.
It should be mentioned that, by driving the parameter inversely across the critical point again, one can realize recombination as a kind of nonlinear detection, which will be illustrated in Sec.~\ref{sec:4-3}.  

For the Hamiltonian of Eq.~\eqref{Ham_BJJ} with $\delta=0$, when $| \Omega/N\chi| \gg 1$, the system ground state is an SU(2) spin coherent state.
The sign of $\chi$ determines the properties of the ground state when $\Omega$ is not large enough.
When $\chi>0$ and $|\Omega/N\chi| \ll 1$, the ground state is a spin squeezed state.
While for $\chi<0$, the ground state becomes a spin cat state when $|\Omega/N\chi| \ll 1$~\cite{PhysRevLett.102.070401}.
To generate a spin cat state, it is natural to prepare the ground state of Hamiltonian~\eqref{Ham_BJJ} in the limit of $|\Omega/N \chi| \ll 1$ with $\chi<0$.
For atomic BEC system, the negative twisting strength can be achieved by tuning the interspecies s-wave scattering length via Feshbach resonance~\cite{Chin2010}.

%
Under the transformation $\hat a (\hat b) \rightarrow \hat b (\hat a)$, $\hat J_x \rightarrow \hat J_x$, $\hat J_z \rightarrow -\hat J_z$, the Hamiltonian of Eq.~\eqref{Ham_BJJ} with $\delta=0$ remains unchanged.
Thus, this Hamiltonian possesses a parity symmetry and it guarantees the symmetry-protected adiabatic evolution~\cite{PhysRevA.93.043615, Zhuang2020-2}.
Adiabatic evolution occurs due to the presence of a finite minimum energy gap between instantaneous eigenstates of the same parity.
With negative $\chi$, one can generate spin cat states via the adiabatic evolution.
Initially, we set $\Omega(0)$ to be sufficiently large, the ground state is nearly a spin coherent state along $x$ axis with even parity.
By sweeping the Rabi frequency across the critical point ${\Omega}/{N|\chi|}=1$, the two lowest eigenstates change from non-degenerate to degenerate.
Through adiabatically sweeping $\Omega(t)$ to zero, the evolved state will stay in the instant ground state and spin cat states can be prepared.

\red{\begin{figure}[!htp]
 \includegraphics[width=\columnwidth]{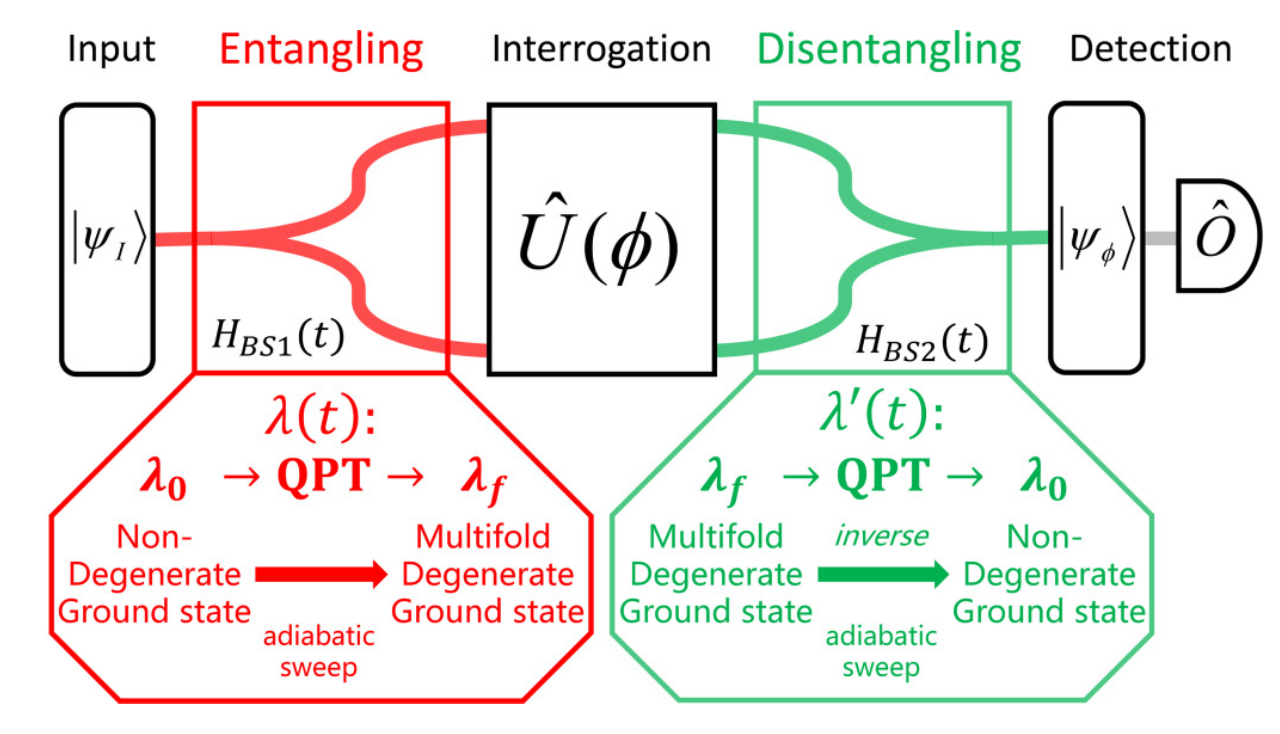}
  \caption{\label{Fig-Sec3-QPT} Schematic of quantum phase estimation via driving through quantum phase transitions. The two beam splitters are achieved by driving the parameter forwardly and inversely across the critical point~\cite{Huang2018_2}. In the first beam splitter, an entangled state can be prepared while in the second beam splitter, the state is gradually disentangled. Reproduced with permission from Huang \textit{et al.}, Phys. Rev. A \textbf{97}(3), 032116 (2018). Copyright 2018 American Physical Society.}
\end{figure}}

For the sweeping, one can linearly change $\Omega(t)=\Omega(0)+\upsilon t$ from the non-degenerate regime across to the degenerate regime with the fixed sweeping rate $\upsilon$. If $\upsilon$ is sufficiently small, the adiabatic evolution of the ground state can still be achieved with high fidelity.
However, this linear sweeping scheme is not time-saving.
In order to accelerate ground state adiabatic evolution, the sweeping rate can be modified temporally in accordance with the instantaneous energy gaps separating the ground state and the second excited state (both possessing the same parity), while maintaining the adiabatic parameter $\epsilon$.
%
It is one of the nonlinear sweeping schemes. Since $\epsilon$ is fixed, it is called adiabatic-parameter-fixed sweeping~\cite{Zhuang2020-2}.
It is shown that adiabatic-parameter-fixed sweeping can generate spin cat states with high fidelity~\cite{PhysRevA.105.062456}.

In this scheme, the time-varying Rabi frequency 
\begin{equation}
    \Omega(t)=\Omega(0)+\int_{0}^{t} \upsilon(t') dt', 
\end{equation}
where $\upsilon(t)=\dot{\Omega}(t)$ is the instant sweeping rate of the Rabi frequency.
Since the parity symmetry, for adiabatic-parameter-fixed sweeping, 
\begin{equation}
    \upsilon(t) = \frac{\epsilon{\left[{E_{1}}(t)-E_{3}(t)\right]}^{2}} {\left|\langle{\phi_{1}(t)}|\hat J_x| {\phi_{3}(t)}\rangle\right|}.
\end{equation}
Here, $E_1(t)$ and $E_3(t)$ respectively represent the energy of instant ground state $\phi_{1}(t)$ and the second excited state $\phi_{3}(t)$ of Hamiltonian with Eq.~\eqref{Ham_BJJ}.
Based on the adiabatic-parameter-fixed sweeping scheme, the total time for adiabatic evolution can be reduced compared with the naive linear sweeping.
The fidelities can exceed 0.9 with $\epsilon=0.05$ and total atom number $N=100$.

However, the adiabatic sweeping still demands the control parameter vary slowly to let the evolved state follow the instant ground state of the system.
Due to the limited coherence time in experiments, adiabatic process may be time-consuming and not easy to realize. 
Shortcut to adiabaticity is a well-known approach for quickly achieving the target state~\cite{Ruschhaupt_2012,RevModPhys.91.045001,PhysRevA.99.043621}. 
However for many-body quantum state generation, this method is not simple and may necessitate some approximations (such as in the semiclassical limit) in order to get the analytical best solution.  
To accelerate state preparation process, TNT dynamics is proposed in which an additional rotation is introduced along with OAT interaction as shown in Sec.~\ref{sec:3-2}. 
One may also use TNT dynamics with time-dependent rotation sequence for the preparation of spin cat states. 
Moreover, to obtain the better performance, one can apply piecewise time-modulation of rotations designed via machine optimization~\cite{PhysRevA.105.062456}. 
When comparing adiabatic evolution, the fidelity to the spin cat states can be increased while also significantly reducing the needed evolution time. 
It does not require large modification to the existing experimental setups and can be realized with state-of-the-art techniques in a BEC system or an optical cavity system with light-mediated interactions. 
%

\begin{figure}[!htp]
 \includegraphics[width=\columnwidth]{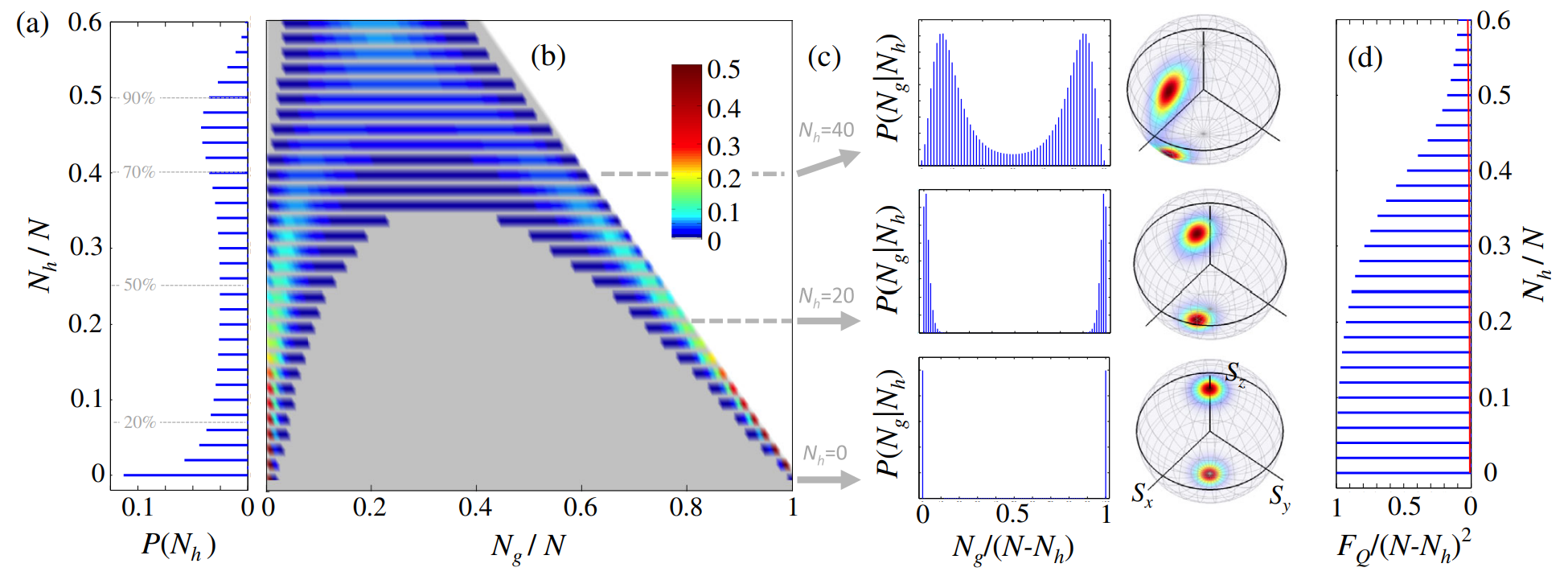}
  \caption{\label{Fig-Sec3-Herald} Stochastic generation of spin cat states under ideal conditions~\cite{PhysRevLett.123.260403}. (a) The probability of measuring $N_h$ particles in the ground state is represented by bars. The dashed lines represent cumulative probability thresholds. (b) Conditional probability to find $N_g$ particles in the $g$ mode given a heralded measurement of $N_h$ particles in the $h$ mode. The $g$ and $h$ modes are defined as the symmetric and antisymmetric combinations of $\ket{m_F=+1}$ and $\ket{m_F=-1}$. (c) Examples of probability (left) and Husimi distributions on the Bloch sphere (right) of rotated heralded states for different values of $N_h$. (d) QFI of the heralded states as a function of $N_h$. In all panels $N=100$. Reproduced with permission from Pezze \textit{et al.}, Phys. Rev. Lett. \textbf{123}(26), 260403 (2019). Copyright 2019 American Physical Society.}
\end{figure}

In a spin-1 BEC, which is described by Hamiltonian of Eq.~\eqref{H_spin1}, a proposal for generating macroscopic superposition states of a large number of atoms is presented~\cite{PhysRevLett.123.260403}.
As shown in Fig.~\ref{Fig-Sec3-Herald}, measuring the number of particles in one mode prepares with large probability highly entangled macroscopic superposition states in the two remaining modes. 
The macroscopic superposition states are heralded by the measurement outcome. This protocol is robust under realistic conditions in current experiments, including finite adiabaticity, particle loss, and measurement uncertainty.
It is demonstrated that MSSCS of a large number of atoms, $N > 100$, can be heralded with large success probability—in a ferromagnetic spin-1 BEC. 
First, one can consider the preparation of the ground state of the three-mode condensate by the quasiadiabatic crossing over a QPT point, as  demonstrated experimentally~\cite{Hoang2016,Luo2017}. 
Once the ground state is prepared, MSSCS can be generated stochastically, with up to 90$\%$ probability, by measuring the number of atoms in one of the modes. 
The state preparation is nondestructive - the spin cat states can be manipulated after its generation.
Remarkably, a numerical simulation of the proposed procedure reveals that the structure of the MSSCS survives under experimentally realistic conditions. 
Nonadiabatic crossing of the QPT and particle loss reduce the Fisher information to about 50$\%$ of its ideal value, which still implies significant entanglement and metrological sensitivity above the classical limit.

\red{\begin{figure}[!htp]
 \includegraphics[width=\columnwidth]{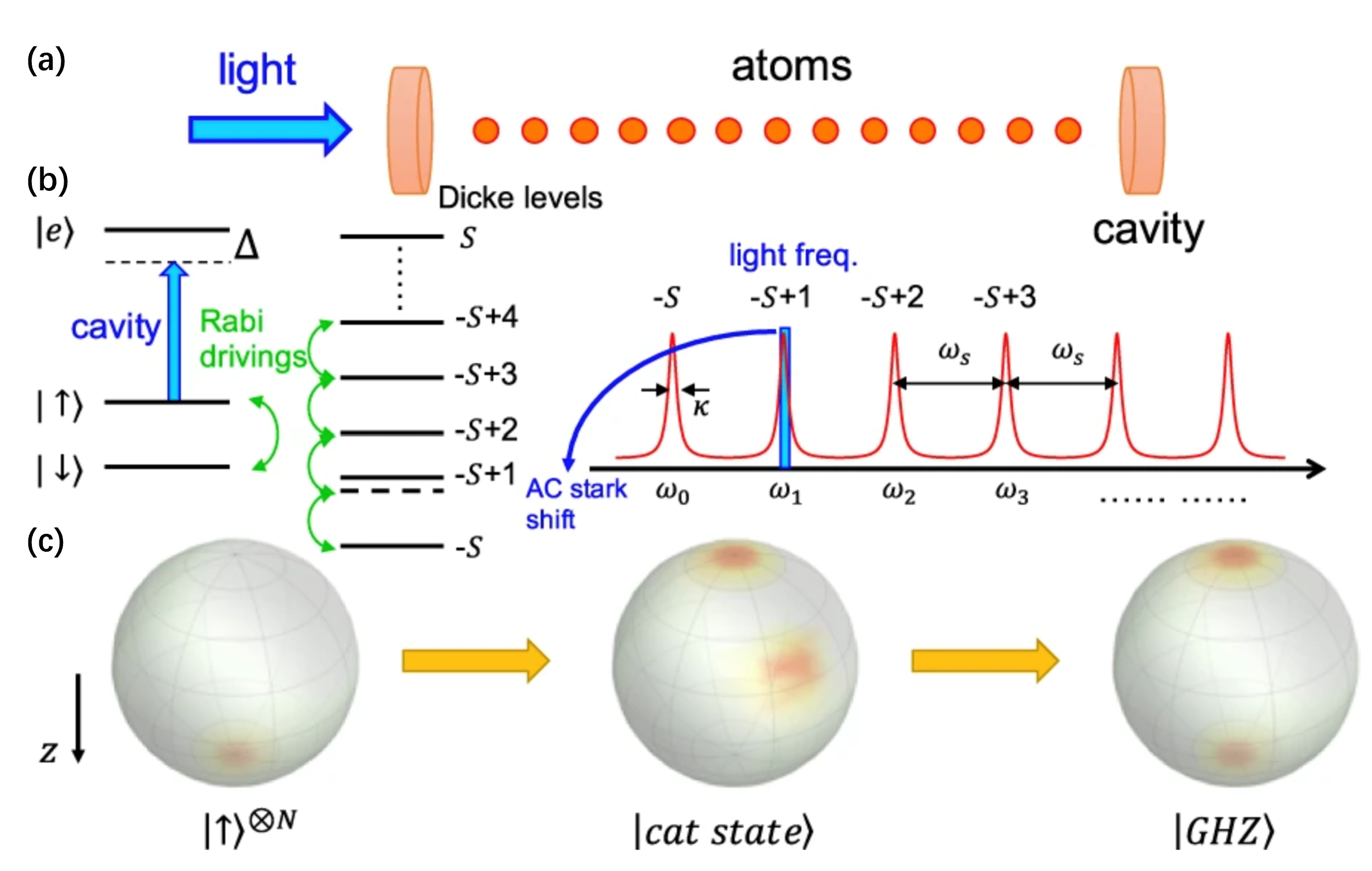}
  \caption{\label{Fig-Sec3-GHZ-cavity} Entanglement amplification for GHZ state generation with atomic ensemble trapped in an optical cavity~\cite{Zhao2021}. (a) A collection of $N$ atoms is connected to a high-finesse optical cavity. The cavity is exposed to a single-frequency light that can be switched on or off. (b) The atomic levels and the spectra of cavity transmission. The Rabi drivings pair the states $|\uparrow\rangle$ and $|\downarrow \rangle$. The cavity mode exhibits coupling between the state $(\left|\uparrow \right\rangle)$ and $|\text{e}\rangle$, characterized by a detuning $\Delta$. Since the intense connection between the atom and the cavity, each atom in the state $|\uparrow \rangle$ causes a shift in the resonance of the cavity. This shift is equal to $\omega_s=g^2/\Delta$, which is more than $\kappa$. Therefore, the incoming light with a frequency of $\omega_1 = \omega_c + \omega_s$ only alters the energy of the Dicke state $|-S+1\rangle$ and establishes a boundary in the Hilbert space at this state with $S$ the total spin length. (c) The Husimi-Q function is observed on the  generalized Bloch sphere before, during, and after the process of entanglement amplification. Reproduced with permission from Zhao \textit{et al.}, npj Quantum Inf. \textbf{7}, 24 (2021). Copyright 2021 licensed under a Creative Commons Attribution (CC BY) license.}
\end{figure}}

%
%

In an optical cavity with multi-atom ensemble, an entanglement-creation scheme named entanglement amplification is proposed, converting unentangled states into entangled states and amplifying less-entangled ones to maximally entangled GHZ states whose fidelity is logarithmically dependent on the atom number and robust against common experimental noises~\cite{Zhao2021}. 
As shown in Fig.~\ref{Fig-Sec3-GHZ-cavity}, the scheme starts with a multi-atom ensemble initialized in a spin coherent state. 
By shifting the energy of a particular Dicke state, the Hilbert space of the ensemble is broken into two isolated subspaces to tear the spin coherent state into two components so that entanglement is introduced. 
After that, the isolated subspaces are utilized to further enhance the entanglement by coherently separating the two components. 
By single-particle Rabi drivings on atoms in a high-finesse optical cavity illuminated by a single-frequency light~\cite{PhysRevLett.115.250502}, $2000$-atom GHZ states can be created with a fidelity above 80$\%$ in an experimentally achievable system.
This may make resources of ensembles at Heisenberg limit practically available for quantum metrology.

In experimental trapped-ion systems, GHZ state can also be generated and used for precision measurement. 
In 2001, an experiment demonstrated that the sensitivity of rotation angle estimation with a Ramsey spectroscopy can be improved by using entangled trapped ions~\cite{PhysRevLett.86.5870}. The experiment used two $^{9}$Be$^{+}$ ions that are confined in a linear radio-frequency trap. Two hyperfine states $\left|F=1,m_F=-1\right\rangle \equiv \left|\uparrow \right\rangle$ and $\left|F=2,m_F=-2\right\rangle \equiv \left|\downarrow \right\rangle$ form the basis of an effective spin-1/2 system. $\left|\uparrow \right\rangle$ and $\left|\downarrow \right\rangle$ are coupled by two-photon Raman transitions. The detection of the ions in states $\left|\uparrow \right\rangle$ and $\left|\downarrow \right\rangle$ is done by state-sensitive fluorescence. The use of entangled states for parity measurement and Ramsey spectroscopy has been demonstrated with $\left|\Psi_P \right\rangle =\left(e^{i\phi}\left|\uparrow \uparrow \right\rangle +\left|\downarrow \downarrow \right\rangle\right)/\sqrt{2}$ and $\left|\Psi_R \right\rangle =\left(\left|\uparrow \downarrow \right\rangle +\left|\downarrow \uparrow \right\rangle\right)/\sqrt{2}$, respectively. The experimental data show the measurement sensitivity is improved beyond the SQL and close to the Heisenberg limit.
Later the Ramsey spectroscopy with three entangled $^{9}$Be$^{+}$ ions in the GHZ state was demonstrated. The experimental data show that the spectroscopic sensitivity is 1.45(2) times as high as that of a perfect experiment with three independent ions, which approaches the Heisenberg limit~\cite{Leibfried2004}.
Then the precision spectroscopy of a pair of trapped Ca$^+$ ions in a decoherence-free subspace of specifically designed entangled states~\cite{Roos2006} was demonstrated. In addition to the enhancement of signal-to-noise ratio in frequency measurements, a suitably designed pair of ions enable atomic clock measurements in the presence of magnetic field noise.
Later, there have been many other experiments~\cite{PhysRevX.8.021012, PRXQuantum.2.020343, Franke2023} about the preparation of GHZ state with trapped atomic ions for quantum metrology.

Multi-particle GHZ states have also been experimentally generated in other quantum systems. 
The deterministic generation of an 18-qubit GHZ state and multi-component atomic spin cat states of up to 20 qubits have been demonstrated with a superconducting quantum processor interconnected by a bus resonator~\cite{Song2019}.  
By engineering an OAT Hamiltonian, the system of qubits, once initialized, coherently evolves to multi-component atomic spin cat states, which are superpositions of atomic coherent states including the GHZ state at specific time intervals as expected.
Moreover, a highly scalable method for producing multipartite cat states, also known as flying cat states, for itinerant microwave photons has also been presented~\cite{doi:10.1126/sciadv.abn1778}. 
By reflecting coherent-state photon pulses from a microwave cavity containing a superconducting qubit, flying cat states of microwave photons containing up to four photonic modes are successfully prepared.
In addition, the GHZ states of $20$ Rydberg atoms have been demonstrated by exploiting optimal control methods to optimize the preparation time~\cite{Omran2019}. 
The entanglement manipulation by using GHZ states to distribute entanglement to distant sites in the array has been demonstrated, establishing important ingredients for quantum metrology.

\section{\label{sec:4}Interrogation and readout}

Apart from state preparation, interrogation and readout are two important processes in quantum metrology.
The interrogation process is provided by a dynamical evolution of the probe and it determines the QCRB for the parameter $\theta$ with a probe state $\rho_{0}$. 
On the other hand, the readout process determines the CRB and the measurement precision through practical observable measurements.
In this section, we will discuss the advancements made in investigating the interrogation and readout processes.
Regarding the interrogation process, the signal can either be time-invariant or time-varying.
Therefore, we will review both the interrogation process with time-independent Hamiltonians and the one with time-dependent Hamiltonians. 
Regarding the readout process, we will cover both conventional readout methods and the recent developments in interaction-based readout.

\subsection{\label{sec:4-1}Interrogation}

The parameters to be estimated are typically embedded within the Hamiltonian or within the dynamics of a physical system.
During the interrogation process, these parameters are encoded into quantum states by evolving the probe state $\rho_{0}$ under the influence of the corresponding Hamiltonians or physical dynamics, represented by $\hat{U}_{\theta}$ (as shown in Fig.~\ref{ProcessQPE}).
Here, we illustrate the interrogation process with time-independent Hamiltonians~\cite{PhysRevLett.98.090401,PhysRevA.90.022117} and time-dependent Hamiltonians~\cite{ncomms14695,PhysRevLett.115.110401,PhysRevLett.117.160801,PhysRevLett.123.040501}, respectively.

A commonly studied and straightforward example in quantum metrology involves estimating a multiplicative parameter within a Hamiltonian.
For example, estimating the parameter $\theta$ in a unitary evolution $\hat{U}_{\theta}=e^{-i \theta \hat{H} }$, where the generator $\hat{H}$ is time-independent.
In this case, if a quantum probe with a pure state $\rho_{0}=|\Psi\rangle\langle\Psi|$ undergoing a unitary evolution $\hat{U}_{\theta}=e^{-i \theta \hat{H} }$ with $\theta=\omega T$, the QFI that determines the ultimate measurement precision of $\omega$ can be given as $F_{\omega}^{Q}=T^2 (\Delta \hat{H})^2$ and its maximum is $F_{\omega}^{Q}|_\textrm{{max}}=T^{2}(\lambda_\textrm{{max}}-\lambda_\textrm{{min}})^2$
with $\lambda_\textrm{{max}}$ and $\lambda_\textrm{{min}}$ respectively denoting the maximal and minimal eigenvalues of $\hat{H}$.
For a many-body SU$(2)$ quantum interferometry involving $N$ two-mode particles, one can choose a linear or nonlinear coupling of collective spin operators as the generator $\hat{H}$.

In most cases, one may choose a linear operator such as $\hat{H}=\hat{J}_{z}$ as the generator, which is commonly used in quantum metrology~\cite{Pezze2018_RMP} including magnetic field measurement~\cite{Maze2008,Kotler2011,science5532}, frequency measurement~\cite{science7009,Huelga1997,PhysRevLett.125.210503}, and more.
The maximal and minimal eigenvalues are $\lambda_\textrm{{max}}=N/2$ and $\lambda_\textrm{{min}}=-N/2$, respectively.
Thus, the maximum quantum Fisher information is $F_{\omega}^{Q}|_\textrm{{max}}=T^2(\lambda_\textrm{{max}}-\lambda_\textrm{{min}})^2=T^2 N^2$, which corresponds to the Heisenberg limit (HL).
Achieving the HL with the linear generator $H=\hat J_{z}$ requires the probe to be initialized in a maximally entangled state.

In the case of a nonlinear generator $\hat{H}=\left({\hat{J}}_{z}\right)^{k}$ (where the exponent $k>1$), the measurement precision can reach a scaling of $N^{-k}$.
When $k$ is an even number, the maximal and minimal eigenvalues of $\hat{H}$ are $\lambda_\textrm{{max}}=(\frac{N}{2})^{k}$ and $\lambda_\textrm{{min}}=0$, respectively.
The corresponding maximum of QFI and QCRB of $\omega$ are $F_{\omega}^{Q}|_\textrm{{max}}=T^2(\frac{N}{2})^{2k}$ and $\omega_\textrm{QCRB}=\frac{2^k}{T}\frac{1}{N^{k}}$, respectively. 
When $k$ is an odd number, the maximal and minimal eigenvalues of $\hat{H}$ are $\lambda_\textrm{{max}}=(\frac{N}{2})^{k}$ and $\lambda_\textrm{{min}}=-(\frac{N}{2})^{k}$, respectively.
The corresponding maximum of QFI and QCRB of $\omega$ are $F_{\omega}^{Q}=4T^{2}(\frac{N}{2})^{2k}$ and $\omega_\textrm{QCRB}=\frac{2^{k-1}}{T}\frac{1}{N^{k}}$, respectively.
Mathematically, when $k > 1$, the scaling of $N^{-k}$ is better than the Heisenberg scaling $N^{-1}$ for the case of linear generators~\cite{Nie2018, PhysRevLett.98.090401,PhysRevLett.100.220501,PhysRevX.2.041006}.
However, this does not mean that the measurement precision breaks the Heisenberg uncertainty relation.
The measurement precision offered by a nonlinear generator still obeys the Heisenberg uncertainty relation.
In other words, when we unify the definitions of HL for both linear and nonlinear generators as the ultimate precision limit dictated by the Heisenberg uncertainty relation, the scaling of HL depends on the form of the associated generator.
With this unified definition, it should be noted that the HL remains unbreakable regardless of the protocol employed.
A realistic possibility for taking advantage of nonlinear generator is the case of $k=2$, such as a quadratic OAT Hamiltonian $\hat{H}=\chi \hat{J}_{z}^{2}$.

Extensive studies have been dedicated to quantum metrology utilizing time-independent Hamiltonians.
However, the investigation of time-dependent Hamiltonians is crucial in order to broaden the applicability of quantum metrology to more complex scenarios.
Below we consider the quantum metrology with time-dependent Hamiltonians $\hat{H}_{\theta}(t)$~\cite{ncomms14695}.
A system is initialized in a state $\ket{\Psi}$, and then evolves under the time-dependent Hamiltonian $\hat{H}_{\theta}(t)$ for a time duration of $T$, its final state reads
\begin{eqnarray}\label{Eq:Finalstate}
\ket{\Psi_{\theta}(T)}=\hat{U}_{\theta}(T)\ket{\Psi},
\end{eqnarray}
where $\hat{U}_{\theta}(T)=e^{-i \int_0^T \mathcal{T} \hat H_{\theta}(t) dt}$ is the unitary evolution under the Hamiltonian $\hat{H}_{\theta}(t)$ with the time-ordering operator $\mathcal{T}$.
According to Eq.~\eqref{Eq:QFI3}, the QFI of $\ket{\Psi_{\theta}(T)}$ for estimating $\theta$ can be written as
\begin{eqnarray}\label{Eq:FQ-psi}
    F_{\theta}^{Q}=4 \left[\langle\Psi|\hat{h}^2_{\theta}(T)|\Psi\rangle-\langle\Psi|\hat{h}_{\theta}(T)|\Psi\rangle^2 \right],
\end{eqnarray}
where $\hat{h}_{\theta}(T)=i\hat{U}_{\theta}(T)^{\dag}\left[\partial_{\theta} \hat{U}_{\theta}(T)\right]$.
Therefore, the maximum $F_{\theta}^{Q}$ is the squared gap between the maximum and minimum eigenvalues of $\hat{h}_{\theta}(T)$, similar to the case of a static Hamiltonian.
This means that the key to give the optimal estimation precision is finding the maximum and minimum eigenvalues of $\hat{h}_{\theta}(T)$.
To obtain the maximum QFI, one can take an alternative approach that breaks the unitary evolution $\hat{U}_{\theta}(T)$ into products of small time intervals $dt$ and takes the
limit $dt\rightarrow0$.
That is,
\begin{eqnarray}\label{Eq:FQ-general}
\hat{h}_{\theta}(T)=\int_{0}^{T}\hat{U}_{\theta}(t)^{\dag} \partial_{\theta}\hat{H}_{\theta} \hat{U}_{\theta}(t)dt.
\end{eqnarray}
Thus one can decompose the global optimization of the eigenvalues of $\hat{h}_{\theta}(T)$ into the local optimizations at each time instant $t$.
As $\hat{U}_{\theta}(t)$ is unitary, $\hat{U}_{\theta}(t)|\psi\rangle$ is a normalized state and the upper bound of $\langle\Psi| \hat{U}_{\theta}(t)^{\dag} \partial_{\theta}\hat{H}_{\theta}(t) \hat{U}_{\theta}(t)|\Psi\rangle$ must be the maximum eigenvalue $\lambda_\textrm{{max}}(t)$ of $\partial_{\theta}\hat{H}_{\theta}(t)$ at time $t$.
Consequently, the maximum eigenvalue and the minimum eigenvalue of $\hat{h}_{\theta}(T)$ are bounded by $\int_{0}^{T}\lambda_\textrm{{max}}(t)dt$ and $\int_{0}^{T}\lambda_\textrm{{min}}(t)dt$, respectively.
Therefore the upper bound of the QFI can be given as
\begin{eqnarray}\label{Eq:FQ}
F_{\theta}^Q(T)\leq\left[\int_{0}^{T}(\lambda_\textrm{max}(t)-\lambda_\textrm{min}(t)) dt \right]^2.
\end{eqnarray}
This indicates that the upper bound of $F_{\theta}^Q(T)$ is determined by the integral of the gap between the maximum and minimum eigenvalues of $\partial_{\theta}\hat{H}_{\theta}(t)$ over the time interval from $0$ to $T$.

\begin{figure}[!htp]
 \includegraphics[width=\columnwidth]{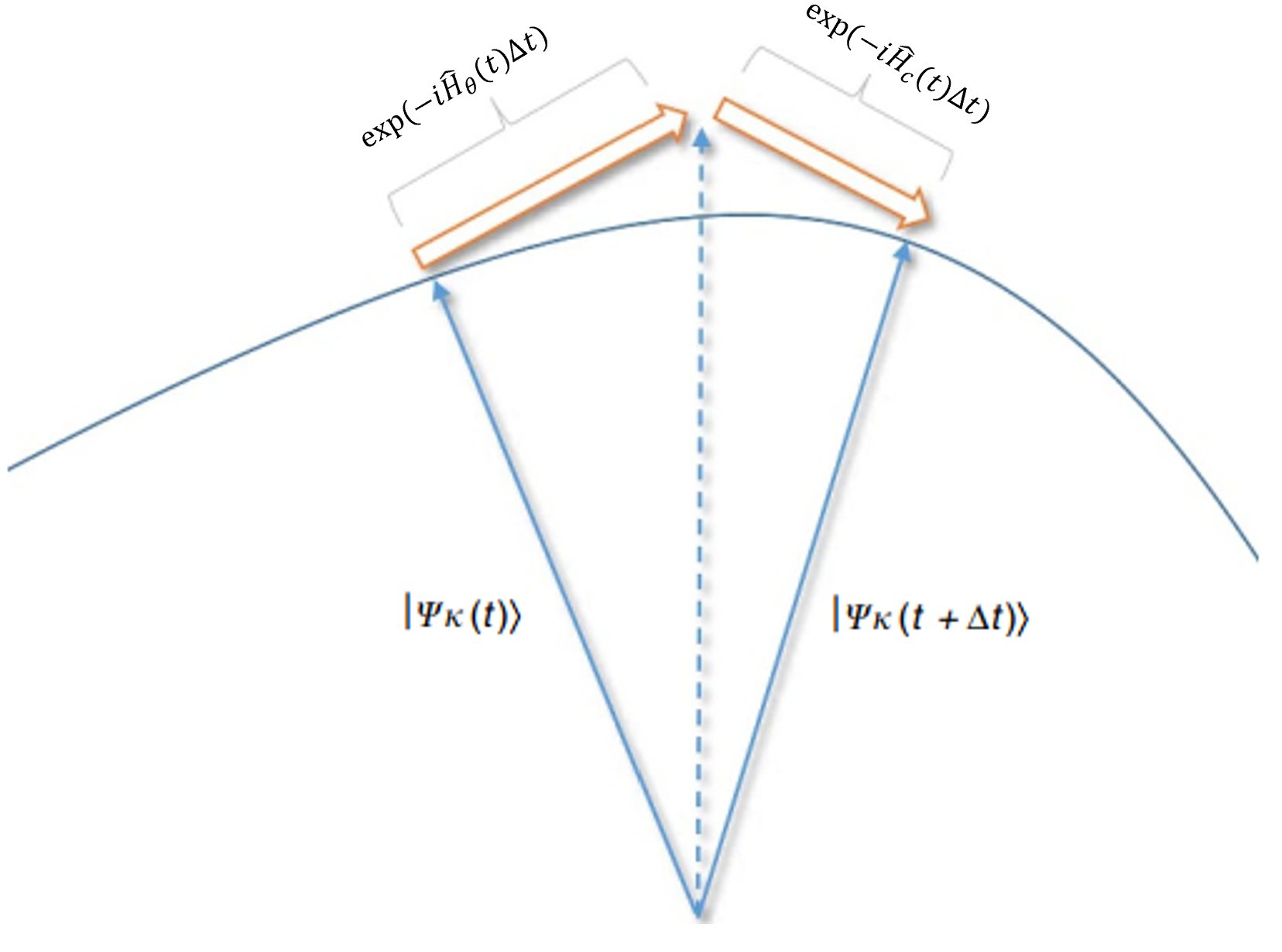}
  \caption{\label{Fig4.1.1}
  Optimal Hamiltonian control scheme~\cite{ncomms14695}. To achieve the maximal Fisher information, the optimal control Hamiltonian needs to keep the state evolution under the total Hamiltonian $\hat{H}_{\text{tot}}$ evolving along the tracks of the eigenstates of $\partial_{\theta} \hat H_{\theta}(t)$ at all times $t$. Reproduced with permission from Pang \textit{et al.}, Nat. Commun. \textbf{8}, 041006 (2017). Copyright 2017 licensed under a Creative Commons Attribution (CC BY) license.}
\end{figure}

From Eq.~\eqref{Eq:FQ}, in the case of a time-independent Hamiltonian $\hat{H}_{\theta}=\theta \hat{H}$, it reduces to $F_{\theta}^Q(T)|_\textrm{{max}}=T^2(\lambda_\textrm{{max}}-\lambda_\textrm{{min}})^2$.
For a time-dependent Hamiltonian $\hat{H}_{\theta}(t)$, a convenient way to achieve the upper bound of QFI is to introduce a control Hamiltonian $\hat{H}_c(t)$, which is independent of the parameter $\theta$,to the
original Hamiltonian.
That is, the system will evolve under the total Hamiltonian is $\hat{H}_{\text{tot}}=\hat{H}_{\theta}(t)+\hat{H}_c(t)$.
By applying the control Hamiltonian $\hat{H}_c(t)$, the state evolution under the total
Hamiltonian will always stay in the instantaneous eigenstates of $\partial_{\theta}\hat{H}_{\theta}(t)$.
To realize the above target, the optimal control Hamiltonian $\hat{H}_c(t)$ can be chosen as
\begin{eqnarray}\label{Hc5}
\hat{H}_c(t)=\sum_{k}f_{k}(t)|\Psi_{k}(t)\rangle\langle\Psi_{k}|-\hat{H}_{\theta}(t)+i\sum_{k}|\partial_{t}\Psi_{k}(t)\rangle\langle\Psi_{k}|,\nonumber \\
\end{eqnarray}
where $f_{k}(t)$ are arbitrary real functions, and $|\Psi_{k}\rangle$ is the $k$-th eigenstate of $\partial_{\theta}\hat{H}_{\theta}(t)$.
The aim of applying this control Hamiltonian is to guide the state evolution governed by the total Hamiltonian $\hat{H}_{\text{tot}}$, following the paths traced by the instantaneous eigenstates of $\partial_{\theta} \hat H_{\theta}(t)$. 
This approach allows for acquiring the maximum amount of information about $\theta$.
Fig.~\ref{Fig4.1.1} illustrates a schematic diagram for the state evolution under the optimal control Hamiltonian $\hat{H}_c(t)$.
This control approach can also be extended to multi-particle systems. However, the optimal control Hamiltonian may become more complex and challenging to implement.

By employing the above method, the precision of measuring the field amplitude and frequency of a rotating magnetic field can be significantly improved.
Similar ideas of steering quantum state evolution along specific trajectories have also been used in other studies~\cite{D.A.Garanin_2002,Berry_2009,Ruschhaupt_2012,PhysRevA.88.013818}.
A simple case is that the Hamiltonian is still time-independent but the parameter to estimate is not multiplicative, such as a spin-$1/2$ particle in a DC vector magnetic field~\cite{PhysRevLett.117.160801}.
In this case, the Hamiltonian can be represented as $\hat H_{\theta}(t)=\hat H_{\theta}$, and one have $|\partial_{t}\Psi_{k}(t)\rangle=0$.
Thus the optimal control Hamiltonian can be chosen as $\hat{H}_c(t)=\sum_{k}f_{k}(t)|\Psi_{k}(t)\rangle\langle\Psi_{k}|-\hat{H}_{\theta}(t)$.

%

In realistic entanglement-enhanced metrology, the time required to prepare entangled states is typically not short~\cite{PhysRevA.94.052320,Hayes2018}. 
Especially for highly entangled states, the preparation time is comparable with the interrogation time.
Hence, the simultaneous state preparation and signal interrogation becomes an effective approach to optimize time utilization in quantum metrology~\cite{Hayes2018}. 
In this approach, it is more advantageous to allocate the entire duration for the entanglement generation and signal interrogation, rather than dividing the time between creating an entangled state and separately using the prepared state for signal interrogation.
Intuitively, this approach can enhance sensitivity by accumulating a larger phase shift due to the extended time duration involved.
However, it generally depends on the trade-off between entanglement generation and signal interrogation~\cite{Hayes2018}. 
For instance, in the scenario of magnetic field sensing with twisted states, it is more advantageous to allocate the entire duration to the sensing process if the twisting strength is relatively weak or the total time available is limited.
However,  if the twisting strength is strong, it allows the twisting dynamics and the magnetic field sensing operate concurrently will be more effective. 
Moreover, chaotic dynamics can be utilized to achieve this purpose and develop the quantum chaotic sensors~\cite{Fiderer2018, PhysRevA.103.023309}. 
Meanwhile, the machine optimized concurrent state preparation and signal interrogation are developed~\cite{PhysRevLett.124.060402,Schuff2020,Huo_2022}, which can provide better measurement precision in a shorter time compared to traditional schemes.

\subsection{\label{sec:4-2}Conventional readout}

In addition to the preparation of multiparticle entangled states, the readout of such states poses a significant challenge in achieving entanglement-enhanced quantum metrology.
In a SU(2) quantum interferometry, it is usually to measure the population difference $n\equiv n_{\uparrow}-n_{\downarrow}$ between the two involved modes $\ket{\uparrow}$ and $\ket{\downarrow}$.
The population uncertainty $\Delta n$ limits the attainable measurement precision $\Delta \theta = \Delta n /(\partial \langle n \rangle/\partial \theta)$.
In situations involving non-entangled states, population measurement works well and can achieve the optimal measurement precision provided by the input state. 
However, when dealing with entangled states, approaching the optimal measurement precision (especially the Heisenberg limit), requires single-particle-resolved state detection. 
This task becomes progressively more challenging as the number of atoms increases.

For Gaussian entangled states, population measurement can still make the measurement precision beyond the SQL.
As an example, inputting a spin squeezed state and measuring the population difference measurement $\hat{J}_{z}$, the measurement precision reads
\begin{eqnarray}\label{Hc4}
\Delta \theta=\frac{\Delta{{\hat{J}_{z}}}}{|\partial{\langle\hat{J}_{z}\rangle/ \partial{\theta}|}}=\frac{\xi_{R}}{|\sin\theta|\sqrt{N}},
\end{eqnarray}
with $\xi_{R}$ being the squeezing parameter for the input state.
If $\theta=\pi/2$, the measurement precision $\Delta \theta$ reaches its minimum $\Delta \theta=\frac{\xi_{R}}{\sqrt{N}}$.
This means that spin squeezed states may be used to beat the SQL if $\xi_{R}<1$.
Especially, if the input state is squeezed to $\xi_{R}\propto 1/\sqrt{N}$, the measurement precision can reach the Heisenberg scaling: $\Delta \theta\propto 1/N$.

For non-Gaussian entangled states, such as twin Fock states, spin cat states, and over-squeezed states,  the population difference measurement cannot fully exploit their nonclassical features and achieve the optimal measurement precisions.
To resolve this issue, one can implement the parity measurement $\hat{\Pi}=e^{i\pi \hat b^{\dag} \hat b}$.
According to the error propagation formula, the measurement precision with parity measurement can be given as~\cite{PhysRevA.54.R4649,PhysRevA.82.013831,Campos2003}
\begin{eqnarray}\label{Hc3}
\Delta \theta=\frac{\Delta{\hat{\Pi}}}{|\partial{\langle\hat{\Pi}\rangle/ \partial{\theta}|}}.
\end{eqnarray}
For a twin Fock state~\cite{Campos2003}, the phase uncertainty $\Delta \theta$ versus the particle number $N$ approaches the Heisenberg limit $\Delta \theta=1/(2N)$ when $\theta\rightarrow 0$.
For certain values of $\theta$, the measurement precision $\Delta \theta$ may significantly increase for specific total particle numbers, while for other total particle numbers, it continues to approach the Heisenberg limit.
For a GHZ state, the measurement precision with parity measurement is always $\Delta \theta=1/N$, which is independent of the values of $\theta$~\cite{PhysRevA.54.R4649}.
However, in order to achieve entanglement-enhanced quantum metrology with parity measurement, the requirement of single-particle resolved detection poses a challenge, especially at larger atom numbers, and it also serves as a bottleneck in practical experiments.
Moreover, the fragility of quantum entangled states under inevitable sources of noises~\cite{Ferrini2010,Huang2015}, such as phase or detection noises, presents a major obstacle that hinders the enhancement of measurement precision.

\subsection{\label{sec:4-3}Interaction-based readout}

In addition to conventional readout protocols, which involve directly measuring the states output from an interferometer, interaction-based readout protocols have been introduced.
In interaction-based readout protocols~\cite{PhysRevA94010102,PhysRevA97043813,science3397,PhysRevA98030303,PhysRevA97053618,PhysRevA107052613,PhysRevLett104103602,Davis2016,PhysRevLett116090801,Linnemann2016,PhysRevLett118150401,PhysRevLett119193601,Huang2018_1,Burd2019, Liu2022,Colombo2022, Mao2023}, the states output from an interferometer undergo a unitary operation through specific nonlinear interactions before measurement takes place. 
The introduction of nonlinear operation before implementing measurement efficiently eliminates the need for single-particle resolved detection.

\begin{figure}[!htp]
 \includegraphics[width=0.75\columnwidth]{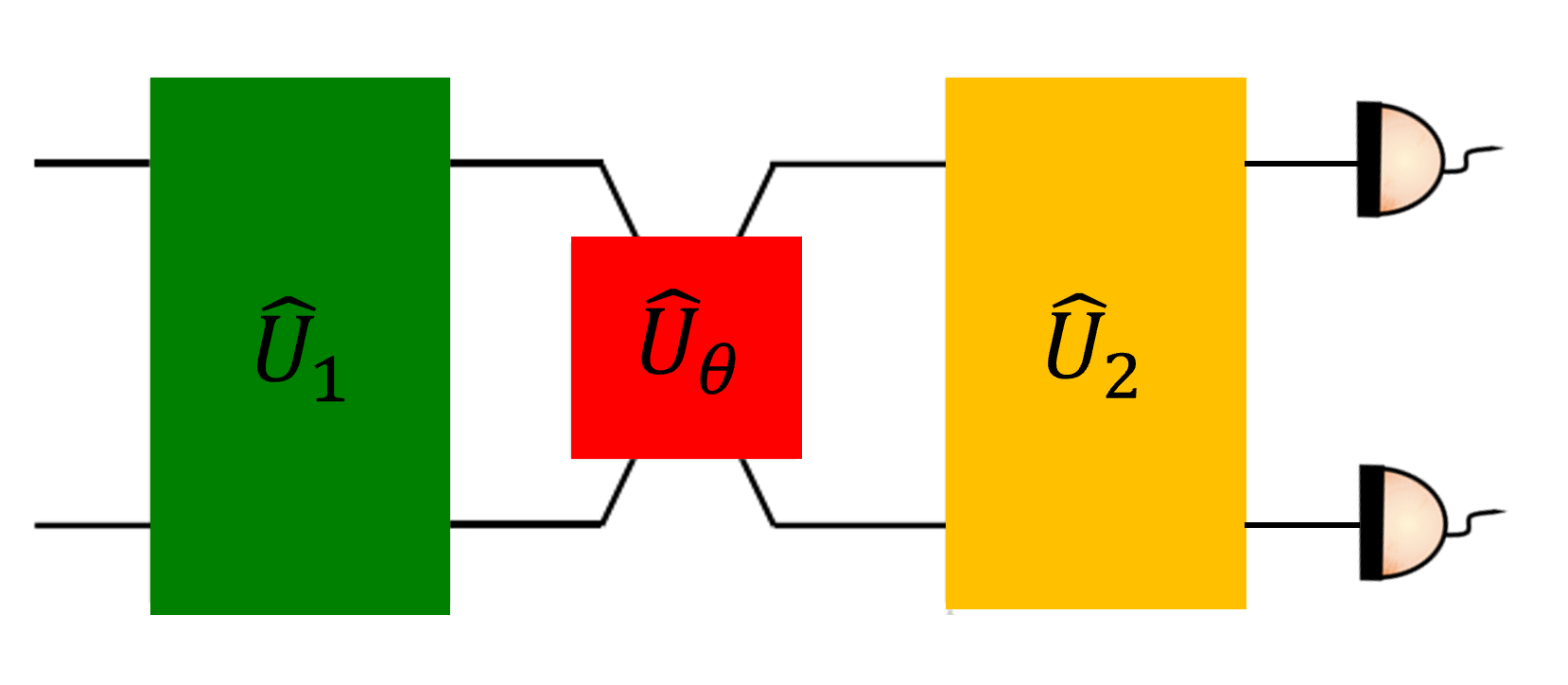}
  \caption{\label{Fig4.1.2}
  Schematic of interaction-based readout.
  Firstly, an unentangled initial state undergoes a unitary evolution $\hat{U}_{1}$ into a nonclassical state.
  Then, the estimated physical parameter $\theta$ is encoded onto the state in the interrogation stage via the unitary operator $\hat{U}_\theta$.
  Finally, another unitary operator $\hat{U}_{2}$ is applied before the final measurement.
  }
\end{figure}

In Fig.~\ref{Fig4.1.2}, we show how interaction-based readout works.
In the preparation stage, an unentangled initial state undergoes a unitary evolution $\hat{U}_{1}$ into a nonclassical state.
The physical parameter to be estimated, $\theta$, is then encoded onto the state via the interrogation stage, which is described by the unitary operator $\hat{U}_\theta$.
The interaction-based readout is then implemented by applying another unitary operator $\hat{U}_{2}$ before the final measurement.
Therefore the final state reads
\begin{eqnarray}\label{Hc1}
\ket{\Psi_{\text{final}}(\theta)}=\hat{U}_{2}\hat{U}_\theta \hat{U}_{1}\ket{\Psi_{\text{in}}}.
\end{eqnarray}
The QFI is just dependent on the unitary operator $\hat{U}_1$ and the initial state.
Although $\hat{U}_2$ does not alter the QFI, it may affect the CFI when a measurement is made in a particular basis.
Specifically, several studies have demonstrated that the protocol with perfectly time reverse the initial unitary operator, i.e., $\hat{U}_2 = \hat{U}_1^{\dag}$, the projection measurement onto the initial state is optimal which can saturate the QCRB~\cite{PhysRevLett119193601,PhysRevLett116090801,Linnemann2016,PhysRevLett118150401,PhysRevLett119193601,PhysRevA98030303}.
The corresponding interferometry also called time-reversed nonlinear
interferometry.
Meanwhile, the protocols with $\hat{U}_2\neq\hat{U}_1^{\dag}$ can also improve robustness against detection noise~\cite{Liu2022,PhysRevA97053618,PhysRevA107052613,PhysRevA.108.062611}.
Below we introduce these two interaction-based readout protocols.

\begin{figure}[!htp]
 \includegraphics[width=\columnwidth]{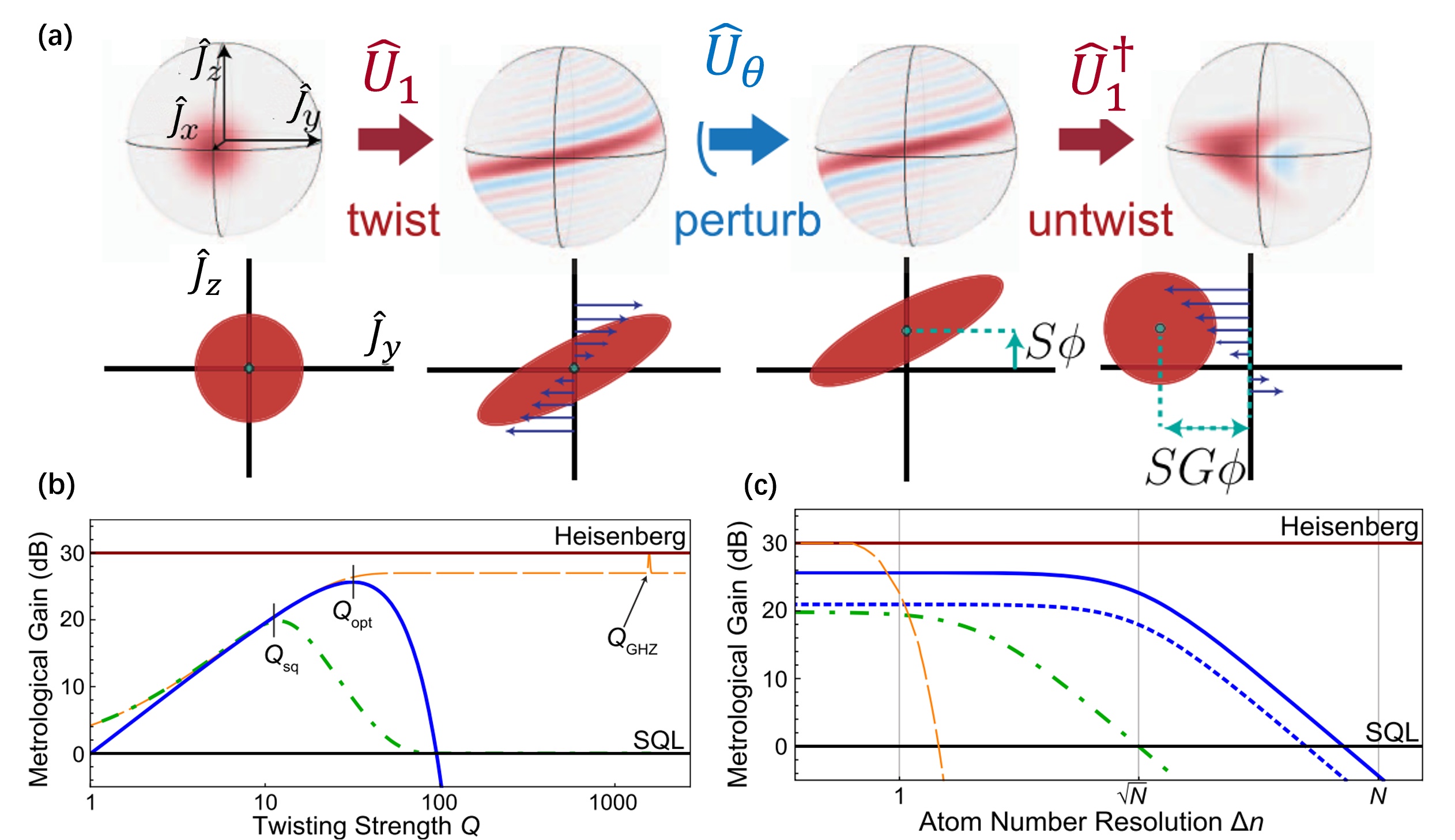}
  \caption{\label{Fig4.1.3}
  Interaction-based readout via one-axis twisting~\cite{Davis2016}.
  (a) A spin coherent state evolves under $\hat{H}_\textrm{OAT}(\chi)$ into an over-squeezed state $\ket{\Psi_e}$.
  Then, rotating $\ket{\Psi_e}$ about $\hat{\textbf{x}}$ by a small angle $\theta$.
  Finally, amplifying the perturbation into a large displacement by applying $\hat{H}_\textrm{OAT}(-\chi)$.
  (b) Metrological gain $1/[N(\Delta \theta)^2]$ versus twisting strength $Q$ for the unitary twisting echo (solid blue), compared to spin squeezing (dot-dashed green) and the quantum Cram\'{e}r-Rao bound (QCRB) on phase sensitivity (dashed orange).
  (c) Metrological gain $1/[N(\Delta \theta)^2]$ versus measurement uncertainty $\Delta n$ for echo with twisting strength $Q_\textrm{{opt}}$ (solid blue) or $Q_\textrm{{sq}}$ (dotted blue), compared to direct detection of the squeezed state at $Q_\textrm{{sq}}$ (dot-dashed green) and projective measurements with the GHZ state. Reproduced with permission from Davis \textit{et al.}, Phys. Rev. Lett. \textbf{116}(5), 053601 (2016). Copyright 2016 American Physical Society.}
\end{figure}

Firstly, we introduce the interaction-based readout protocols of $\hat{U}_2 = \hat{U}_1^{\dag}$ via OAT~\cite{Davis2016}.
The dynamics under a OAT Hamiltonian $\hat{H}_\textrm{OAT} (\chi)=\chi \hat{J}_{z}^2$ can generate different entangled states.
For spins initially polarized along $x$ direction, the lowest-order effect of $\hat{H}_\textrm{OAT}$ can generate spin squeezed state.
For a short twisting time, the generated state is a Gaussian state.
However, for a long twisting time, the generated state becomes a non-Gaussian state, such as spin cat state and over-squeezed state.
In particular, a GHZ state appears at $\chi t=\pi/2$.
Then the generated state undergoes a dynamical evolution to encode the information of the parameter $\theta$ to be estimated.
Before implementing measurement, a unitary operation via another OAT Hamiltonian $\hat{H}_\textrm{OAT} (-\chi)=-\chi \hat{J}_{z}^2$ is applied.

In Fig.~\ref{Fig4.1.3}~(a), we show the schematic of interaction-based readout with OAT.
From a spin coherent state along the x-direction $\ket{\hat{\textbf{x}}}$, applying $\hat{H}_\textrm{OAT}(\chi)$ for a duration $t$ yields a entangled state $\ket{\Psi_e}=e^{-i\hat{H}_\textrm{OAT}t}\ket{\hat{\textbf{x}}}$.
Then the parameter $\theta$ is encoded into the entangled state via a unitary operation $\hat{U}_\theta=e^{-i \theta \hat{J}_{y}}$.
Subsequently, one can use the interaction-based readout to undo the twisting by applying $\hat{H}_\textrm{OAT}(-\chi)$.
When $\theta=0$, the final state is identical to the original spin coherent state.
When $\theta\neq 0$, the $ \hat{J}_{z}$-dependent spin precession can produce a large final value of $\langle \hat{J}_{y}\rangle$.
Through measuring $ \hat{J}_{y}$, the measurement precision of $\theta$ is given by
\begin{eqnarray}\label{eqSec431}
    \Delta \theta=\frac{\Delta{{\hat{J}_{y}}}}{|\partial{\langle\hat{J}_{y}\rangle/ \partial{\theta}|}}.
\end{eqnarray}
The resulting metrological gain depends on the ``shearing strength'' $Q \equiv 2J\chi t$, see Fig.~\ref{Fig4.1.3}~(b).
At the optimal shearing strength $Q_\textrm{{opt}}=2J \textrm{arccot}\left(\sqrt{2J-1}\right)$ for $N\gg1$, the protocol yields the optimal measurement precision, 
\begin{eqnarray}\label{eqSec433}
\Delta{\theta}_\textrm{{min}}=\frac{\sqrt{e}}{N},
\end{eqnarray}
which is very close to the Heisenberg limit.
The scheme is highly robust against detection noise, as the ``untwisting" amplifies the spin rotation signal.
This implies that the measurement precision may achieve the Heisenberg limit, even when the detection resolution of atom number is approximately $\Delta n \approx \sqrt{N}$, see Fig.~\ref{Fig4.1.3}~(c).

Interaction-based readout with OAT interaction is also suitable for effective detection of non-Gaussian entangled states~\cite{PhysRevA98030303,Huang2018_1}. 
It is shown that the interaction-based readout enables spin cat states to saturate their ultimate precision bounds of Eq.~\eqref{FQ_CAT}. Compared with the twisting echo scheme on spin squeezed states, the scheme with with spin cat states via interaction-based readout is more robust against detection noise~\cite{Huang2018_1}. 

\begin{figure}[!htp]
 \includegraphics[width=\columnwidth]{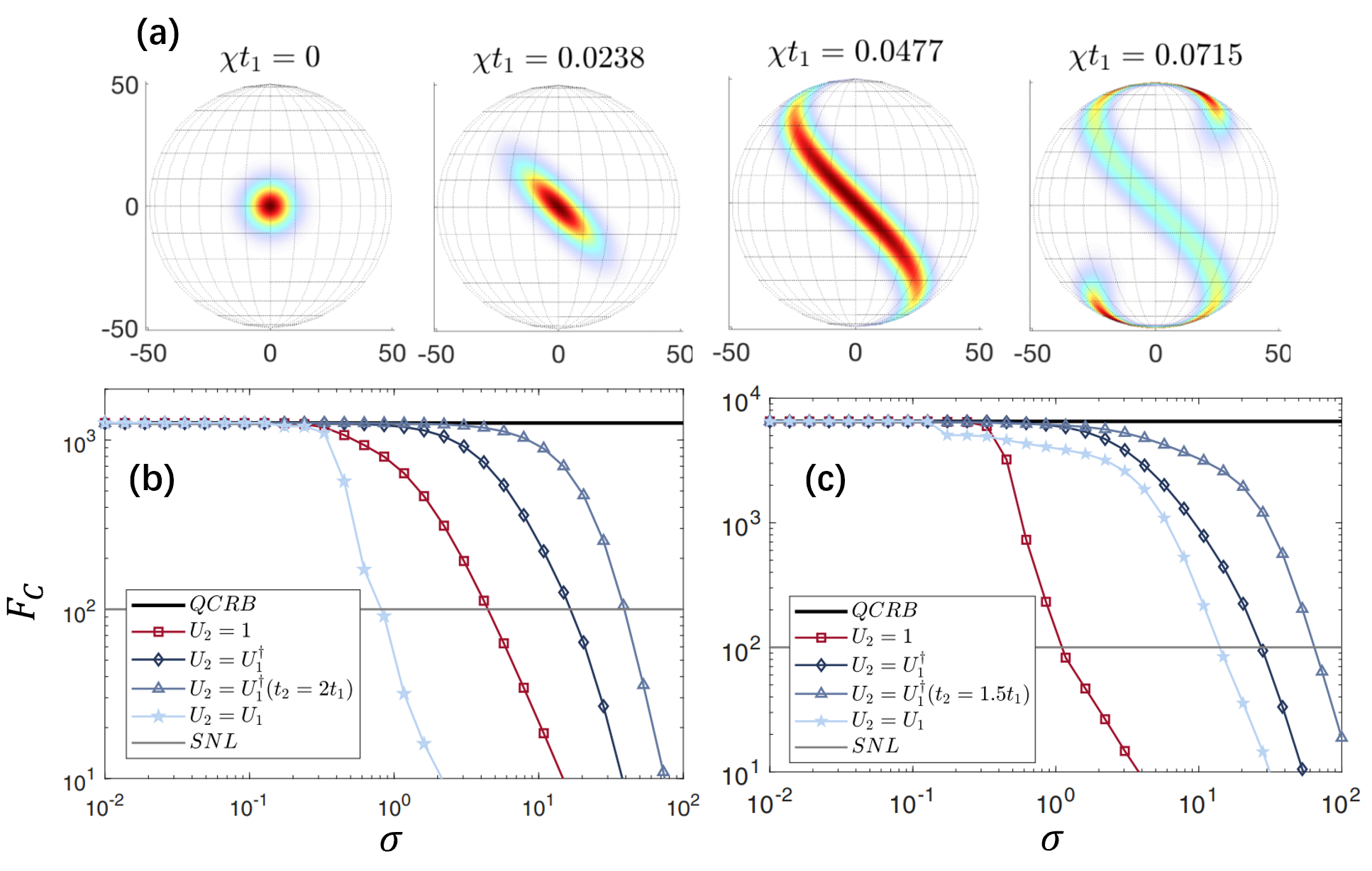}
  \caption{\label{Fig4.1.4}
  Twist-and-turn entanglement with interaction-based readout~\cite{PhysRevA97053618}. 
  (a) Husimi $Q$ function for the state $\ket{\Psi_\textrm{{in}}}$ under TNT, where $\ket{\Psi_\textrm{{in}}}$ is chosen as the eigenstate of $\hat{J}_x$ with minimum eigenvalue.
  Time evolution of the CFI for a many-body entangled state produced by the TNT Hamiltonian in presence of detection noise $\sigma$ for (b) $\chi  t_{1}=0.027$  and (c) $\chi t_{1}=0.072$ with $\hat{U}_2=1$ (red squares), a time-reversal echo $\hat{U}_2=\hat{U}_1^{\dag}$ (dark blue diamonds), an asymmetric echo $\hat{U}_2\neq\hat{U}_1^{\dag}$ (light blue triangles), and a pseudo echo $\hat{U}_2=\hat{U}_1$ (light turquoise pentagrams).
  Reproduced with permission from Mirkhalaf \textit{et al.}, Phys. Rev. A \textbf{97}(10), 053618 (2018). Copyright 2018 American Physical Society.}
\end{figure}

In addition to OAT, one may employ TNT to achieve interaction-based readout.
As mentioned earlier in Sec.~\ref{sec:3-2}, the TNT interaction exhibits a remarkable ability to generate entanglement at a faster rate compared to the OAT interaction. 
Additionally, it enables the rapid production of both spin squeezed states and entangled non-Gaussian spin states.
Fig.~\ref{Fig4.1.4} shows an interaction-based readout protocol with TNT Hamiltonian~\cite{PhysRevA97053618}.
The TNT Hamiltonian is
\begin{eqnarray}\label{eqSec434}
\hat{H}_\textrm{{TNT}}= \chi\hat{J}_z^2-\Omega \hat{J}_x
\end{eqnarray}
such that $\hat{U}_1=e^{-i \hat{H}_\textrm{{TNT}} t_1} $ with the adjustable parameter $t_1$ is the state preparation time.
To align the interferometer along the optimal Fisher information direction, one choose $\hat U_{\theta}=e^{-i \theta \hat{J}_\textbf{{n}}}$ with $\hat{J}_\textbf{{n}}=\alpha \hat{J}_{y}+\beta \hat{J}_z$ that normal to the mean spin direction $\hat{J}_{x}$.
%
%
Thus, the final state after the interaction-based readout is
\begin{eqnarray}\label{eqSec435}
\ket{\Psi_\textrm{{final}}}= \hat{U}_{2} e^{-i \theta \hat{J}_\textbf{{n}}} \hat{U}_1\ket{\Psi_\textrm{{in}}}
\end{eqnarray}
In general, the choices of $\hat{U}_{2}$ may affect the CFI for a measurement is made with a particular basis.
For a measurement that projects into the $\hat{J}_x$ eigenbasis (which can saturate the QCRB), we have
$F_{C}(\theta)=\sum_{m_x} \frac{1}{P_{m_x}(\theta)}\left(\frac{\partial P_{m_x}(\theta)}{\partial {\theta}}\right)^2$
with $P_{m_x}(\theta)=|\langle m_x|{\Psi_\textrm{{final}}} \rangle|^2$ being the ideal conditional probability via projecting into the eigenbasis $\ket{m_x}$ for a given parameter $\theta$.
Further, the robustness to detection noise with different $\hat{U}_{2}$ is shown in Fig.~\ref{Fig4.1.4}~(b, c).
To model detection noise, one can take the convolution of the probability distribution with a Gaussian distribution under detection noise $\sigma$~\cite{PhysRevLett116090801,PhysRevLett118150401,PhysRevLett119193601,PhysRevA98030303} and it is 
$\tilde{P}_{m_x}(\theta)=\sum_{m_{x}^{'}} C_{m_{x}^{'}}{e^{-(m_{x}-m_{x}^{'})^2/2{\sigma}^2}}{P_{m_x}(\theta)}$
with $C_{m_{x}^{'}}$ being a normalization factor, and the conditional probability depends on the detection noise $\sigma$.
$F_C$ as a function of detection noise for different choices of $\hat{U}_2$ are shown in Fig.~\ref{Fig4.1.4}~(b, c).
Interestingly, the optimal interaction-based readout may not be the case of perfect time-reversal echo.
In particular for weakly entangled initial states using an asymmetric echo provides better robustness than the symmetric case.

\begin{figure}[!htp]
 \includegraphics[width=\columnwidth]{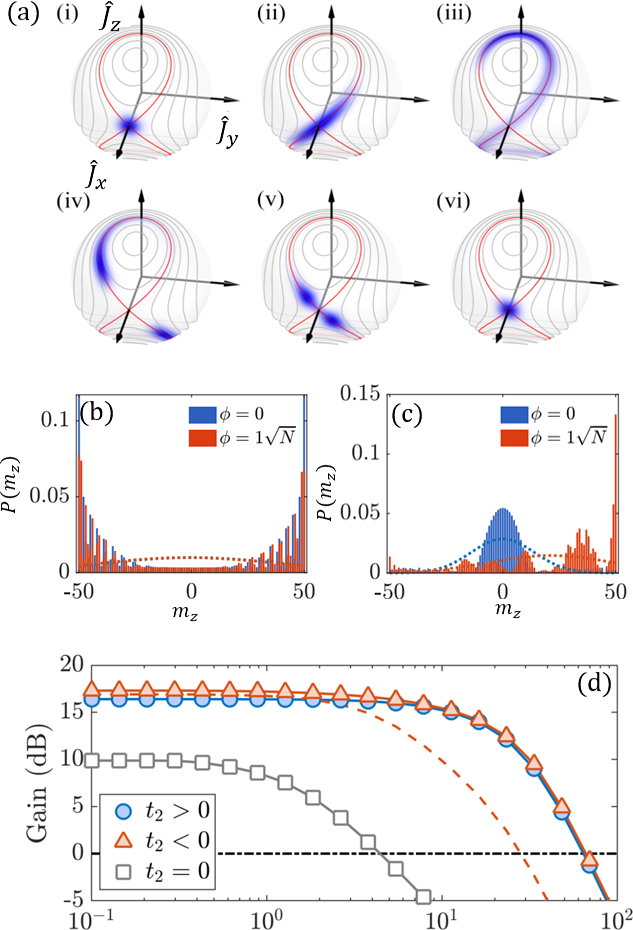}
  \caption{\label{Fig4.1.8}
  Quantum metrology with cyclic nonlinear dynamics.
  %
  %
  (a) Husimi representations of states with quasi-cyclic dynamics in the TNT model~\cite{PhysRevA107052613}.
  Spin coherent state (i) evolves into spin squeezed state (ii), non-Gaussian entangled states (iii)–(v), and back to the vicinity of the initial point (vi) in succession. 
  (b) Probability distributions of the probe states with the rotation $\theta$ in the $\hat{J}_{z}$ eigenbasis. Dashed lines denote the Gaussian fitting curves.
  (c) Probability distributions of the output states with the rotation $\theta$ via nonlinear detection with cyclic dynamics.
  %
  %
  (d) Robustness to detection noise for the TNT Hamiltonian.
  The nonlinear cyclic readout protocol (blue circles), the linear readout scheme (gray squares) and exact time-reversal scheme with $t_2 = -t_1$ (red dashed line). Reproduced with permission from Liu \textit{et al.}, Phys. Rev. A \textbf{107}(5), 052613 (2023). Copyright 2023 American Physical Society.}
\end{figure}

Interaction-based readout can also be realized via driving through quantum phase transitions (QPTs). 
As shown in Fig.~\ref{Fig-Sec3-QPT}, a nonlinear beam splitting process, in which an initial nondegenerate ground state evolves into a highly entangled state, can be achieved by adiabatically driving the system through a QPT~\cite{PhysRevLett.97.150402,PhysRevLett.102.070401}. 
Inversely, by driving the system from the entangled state back across the QPT again to the nondegenerate regime, the state becomes gradually disentangled~\cite{Huang2018_2}.
The reversed driving through QPTs can be regarded as a kind of nonlinear detection. 
Finally, the phase shift accumulated in the interrogation process can be easily inferred via population measurement without single-particle resolution~\cite{Huang2018_2,PhysRevApplied.16.064056}.

Moreover one may also use cyclic nonlinear dynamics to achieve the interaction-based readout.
Different from the time-reversed dynamics which drives a system back to its starting point, the cyclic nonlinear interferometer accomplish the same by forcing the system to travel along a `closed loop' instead of explicitly tracing back its antecedent path. 
For a TNT Hamiltonian of Eq.~\eqref{eqSec434}, using a spin coherent state as the input, the first stage of the nonlinear dynamics $\hat{U}_{1}=e^{-i\hat{H}_\textrm{{TNT}} t_{1}}$ serves as nonlinear splitting leading to the non-Gaussian entangled state. After sensing a spin rotation, the system undergoes a second stage of TNT dynamics $\hat{U}_{2}=e^{-i\hat{H}_\textrm{{TNT}} t_{2}}$ for nonlinear recombination.
The non-Gaussian entangled input state evolves directly back to the vicinity of the initial state. 
Compared with a direct linear readout of the non-Gaussian entangled state, the quasi-cyclic nonlinear readout can magnify the encoded signal and refocus the associated quantum noise, leading to a better signal-to-noise ratio (SNR), see Fig.~\ref{Fig4.1.8}.
It is found that the optimal metrological gain with population measurement nearly saturates the QCRB and follows the Heisenberg-limited scaling~\cite{PhysRevA107052613}.
Meanwhile, the interaction-based readout with cyclic nonlinear dynamics of TACT Hamiltonian is also investigated.

At the same time of generating multiparticle entangled states in experiments, numerous interaction-based readout protocols have been demonstrated for achieving entanglement-enhanced quantum metrology with Bose condensed atoms~\cite{Linnemann2016,Liu2022}, cavity-QED systems~\cite{Colombo2022}, and ultracold trapped ions~\cite{Burd2019}.

\begin{figure}[!htp]
 \includegraphics[width=\columnwidth]{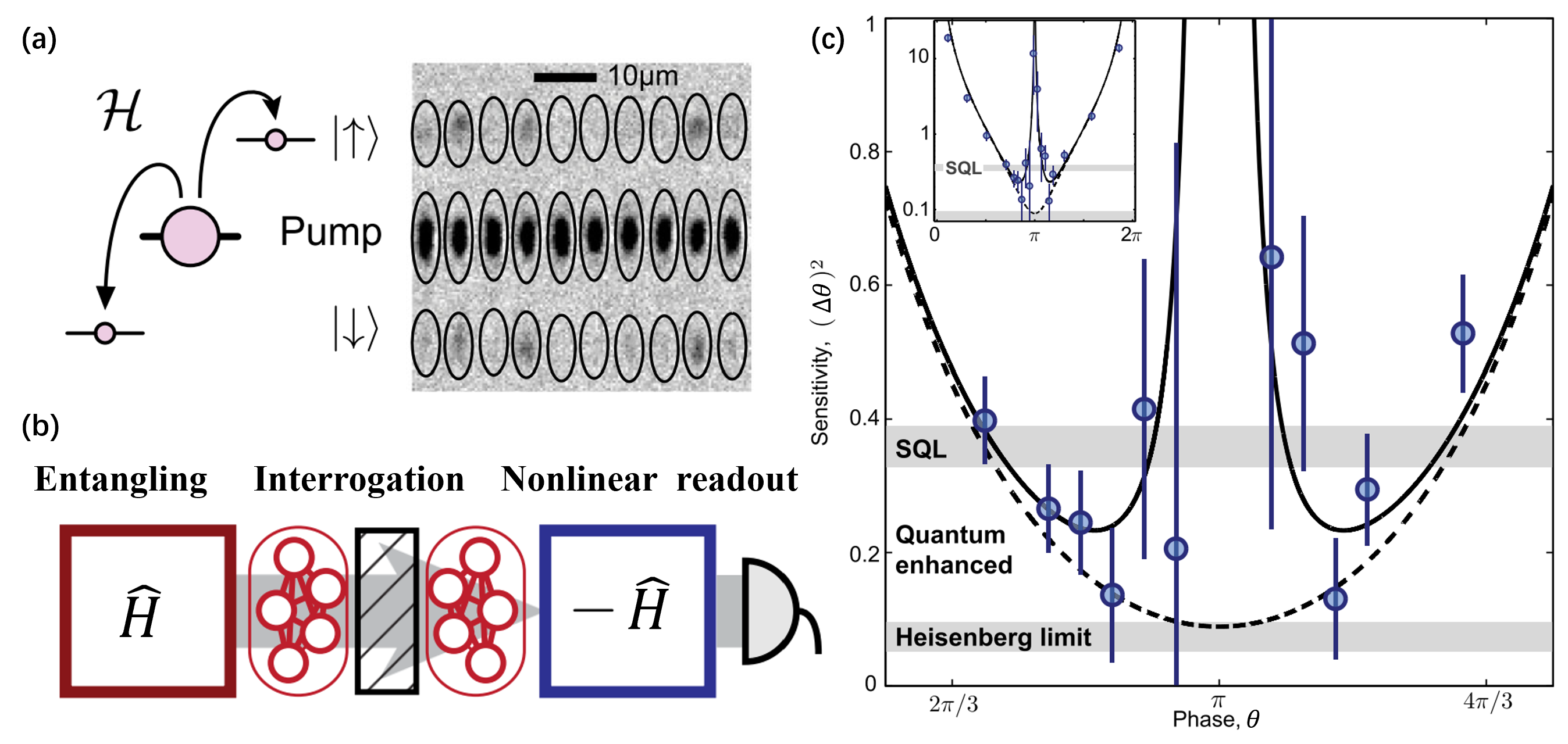}
  \caption{\label{Fig4.1.7}
  Entanglement-enhanced sensing based on time reversal of nonlinear dynamics in a spin-1 BEC~\cite{Linnemann2016}.
  (a) Spin-changing collisions in a BEC are used as the nonlinear process.
  Atom numbers are detected by high resolution absorption imaging after Stern-Gerlach separation. 
  (b) The interaction-based readout scheme exploits a time reversal sequence.
  The Hamiltonian $\hat{H}$ used for entangled state generation is inverted and reapplied for readout.
  %
  %
  (c) The phase sensitivity is experimentally extracted by Gaussian error propagation.
  The SQL (gray horizontal bar) is surpassed in close vicinity of phase $\theta=\pi$.
  At phase $\pi$ the sensitivity diverges due to the vanishing slope of the signal. Reproduced with permission from Linnemann \textit{et al.}, Phys. Rev. Lett. \textbf{117}(1), 013001 (2016). Copyright 2016 American Physical Society.}
\end{figure}

\begin{figure*}[ht]
 \includegraphics[width=1.9\columnwidth]{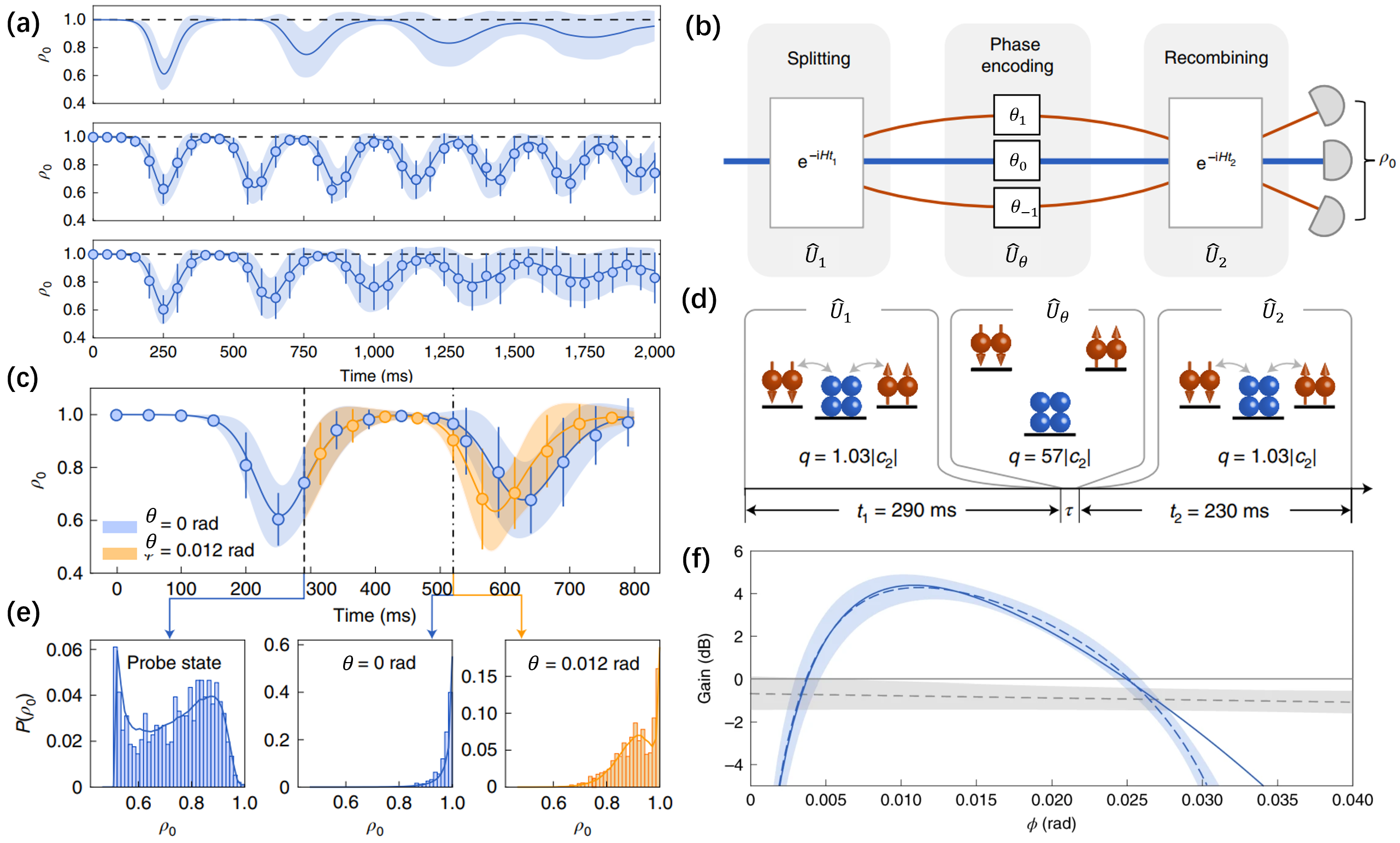}
  \caption{\label{Fig4.1.10}
  Quantum metrology with quasi-cyclic dynamics in a spin-1 BEC~\cite{Liu2022}.
  (a) The quasi-cyclic dynamics.
  Temporal evolution of $\rho_0$ starting from the polar state with all atoms in the $\ket{1,0}$ component.
  The upper panel denotes the ideal case without atom loss or technical noises.
  The middle and lower panels present experimentally measured data.
  (b) Schematic of the three-mode cyclic dynamics nonlinear interferometry based on spin mixing dynamics.
  (c) Temporal evolution of $\rho_0$ for different spinor phase $\theta$ encoded at the instant marked by the vertical dashed black line.
  (d) The specific sequence to implement the experiments with `path' splitting $(\hat{U}_1)$ and recombining $(\hat{U}_2)$ both from time-forward spin mixing dynamics for $t_1=290$ ms and $t_2=230$ ms, respectively.
  (e) Probability distributions of $\rho_0$ for the probe state and for the final states
  (f) Metrological gain $20 \log_{10}\left[\Delta \phi/\Delta \phi_\textrm{{SQL}}\right]$ ($\Delta \phi_\textrm{{SQL}} = 2/\sqrt{N}$) obtained from error propagation formula. Reproduced with permission from Liu \textit{et al.}, Nat. Phys. \textbf{18}(2), 167 (2022). Copyright 2022 Springer Nature.}
\end{figure*}

With a spin-$1$ BEC, its spin-exchange collision has been utilized to realize the time-reversal interaction-based readout~\cite{Linnemann2016}.
The spin-exchange collision is analog to the parametric
amplification, in which both the sign and the strength of the nonlinear coupling can be experimentally adjustable, see Fig.~\ref{Fig4.1.7}~(a).
As shown in Fig.~\ref{Fig4.1.7}~(b), the nonlinear detection is experimentally demonstrated by exploiting time-reversal dynamics to disentangle continuous variable entangled states for feasible readout.
As both preparation and readout of entangled states consisting of parametric amplification, an active atom SU$(1,1)$ interferometer is constructed, which is capable of utilizing the quantum entanglement by only detecting the mean atom numbers.
The phase sensitivity $(\Delta \theta)^2$ can surpass the SQL in the vicinity of $\theta=\pi$, see Fig.~\ref{Fig4.1.7}~(c).
Moreover, the time-reversal readout can also be achieved by a state flip operation.
Similar to the spin echo with a $\pi$-pulse state flip that reverses linear dynamics, this process switches the squeezing and anti-squeezing axes of the probe state, leading to effectively time-reversed nonlinear dynamics to realize the echo-based spin-nematic interferometry~\cite{Mao2023}.
A sensitivity of $15.6\pm0.5$ dB beyond the SQL for detecting small-angle Rabi rotation, as well as $16.6\pm1.1$ dB for phase sensing in a nonlinear Ramsey-like interferometry are observed.

\begin{figure}[!htp]
 \includegraphics[width=\columnwidth]{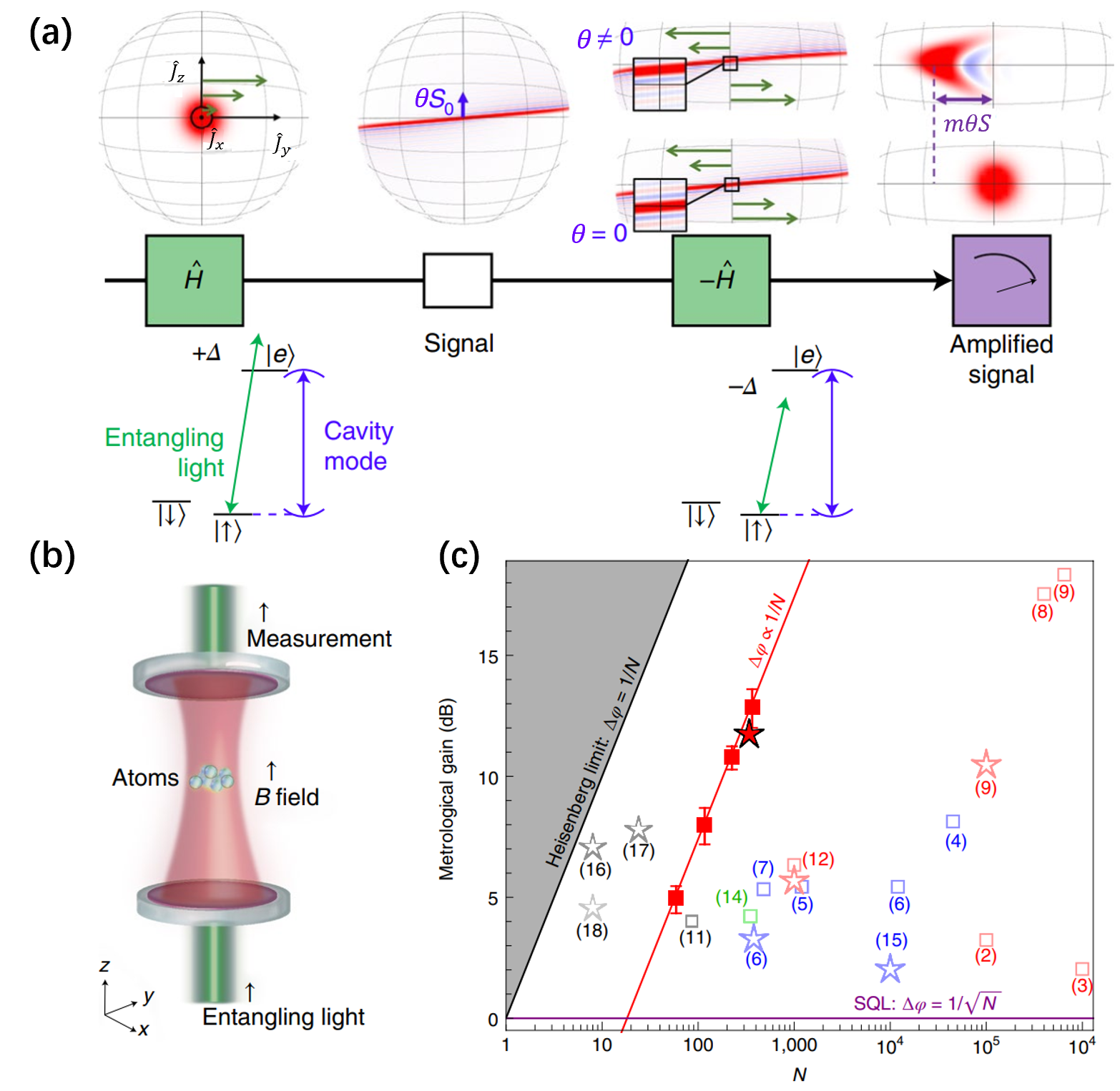}
  \caption{\label{Fig4.1.6}
  Time-reversal-based quantum metrology with entangled atoms in cavity~\cite{Colombo2022}.
  (a) The sequence of the interaction-based readout protocol.
  The protocol sequence with the quantum state evolution (top) and relevant energy levels and cavity mode (bottom).
  The Wigner quasi-probability distribution functions describing the collective quantum states.
  (b) $^{171}$Yb atoms are trapped inside an optical cavity in an optical lattice.
  %
  %
  (c) The scaling of the sensitivity with the atom number and a comparison with previous results.
  A comparison with previous results including BECs experiments (blue), thermal atoms (red), ions (dark grey) and Rydberg atoms (light grey).
  The green point shows the squeezing generated in an optical lattice clock.
  Squares show expected metrological gains obtained by quantum state characterization.
  Stars indicate directly measured phase sensitivity gains.
  Solid symbols are obtained in this work. Reproduced with permission from Colombo \textit{et al.}, Nat. Phys. \textbf{18}(8), 925 (2022). Copyright 2022 Springer Nature.}
\end{figure}

Besides the time-reversal dynamics, interaction-based readout can be achieved by the quasi-cyclic dynamics resulted from spin-exchange collision. 
It has demonstrated that a three-mode nonlinear interferometer with quasi-cyclic dynamics can attain a metrological gain of $5.01$ dB over the classical limit for a total of $26,500$ atoms can be achieved~\cite{Liu2022}.
As shown in Fig.~\ref{Fig4.1.10}, utilizing the spin mixing dynamics $\hat{U}_1$ at the beginning, paired atoms are created in $\ket{1,\pm1}$ components from the $\ket{1,0}$.
Then the dressing microwave is switches on and encodes a spinor phase $\theta=\theta_{1}+\theta_{-1}-2\theta_{0}$ via $\hat U_{\theta}=e^{i\theta \hat{N}_{0}/2}$.
At last, the microwave field is turned off and the spin mixing dynamics $\hat{U}_2$ is used again to annihilates paired atoms in $\ket{1,\pm1}$ components into the $\ket{1,0}$.
The encoded spinor phase $\theta$ can be extracted from the measured fractional population $\rho_0$ in the end.
The time evolution of the measured $\rho_{0}$ for different phase $\theta$ is shown in Fig.~\ref{Fig4.1.10}~(c).
Without an encoded phase, the system nearly returns to its initial state at the end of the interferometry.
However, a nonzero phase shift of $2/\sqrt{N}\approx 0.012$ rad causes an evident decrease in the mean value of $\rho_{0}$ as well as broadening of its distribution, as shown in Fig.~\ref{Fig4.1.10}~(e).
The metrological gain for different phase $\theta$ is shown in Fig.~\ref{Fig4.1.10}~(f).
Especially, a maximal gain of $4.28^{+0.56}_{0.58}$ dB beyond the SQL at $=0.011$ rad is observed.

With a cavity-QED system of $^{171}$Yb atoms, it has been experimentally demonstrated that the phase sensitivity can achieve a Heisenberg scaling of $(b/N)$ at a fixed distance $b=12.6$ dB from the Heisenberg limit by employing interaction-based readout with the OAT interaction~\cite{Colombo2022}. 
Especially, when used in a Ramsey sequence via an atomic interferometer, the highest metrological gain over the SQL with $11.8\pm0.5 $dB can be achieved.
%
%
%
As shown in Fig.~\ref{Fig4.1.6}, an entangling light pulse is passed through the cavity to generate the nonlinear OAT Hamiltonian $\hat{H}=\chi \hat{J}_{z}^{2}$ which shears the initial spin coherent state.
A rotation $\theta$ about the ${\hat{J}_y}$ axis displaces the state such that ${\langle\hat{J}_z\rangle=\theta S_{0}}$ with $S_{0}={N/2}$.
%
%
Subsequently, a (dis)entangling light pulse is sent through the cavity, which generates the negative OAT Hamiltonian $-\hat{H}$  causing the quantum state to evolve effectively ‘backwards in time’.
For a small angle ($\theta\neq0$) the final state is displaced by an angle $m\theta$ from the original spin coherent state, where $m$ is the signal amplification.
The scaling of the sensitivity with the atom number $N$ is shown in Fig.~\ref{Fig4.1.6}~(c), when the particle number $N$ varies between $50$ and $370$, a gain over the SQL is measured, which can achieve the Heisenberg scaling.
More Recently, a time-reversal protocol is used to observe a simultaneous exponential growth of both the metrological gain and the out-of-time-order correlation, thereby experimentally verifying the relation between quantum metrology and quantum information scrambling~\cite{Li2023-sci}. 
%
%

\begin{figure}[!htp]
 \includegraphics[width=\columnwidth]{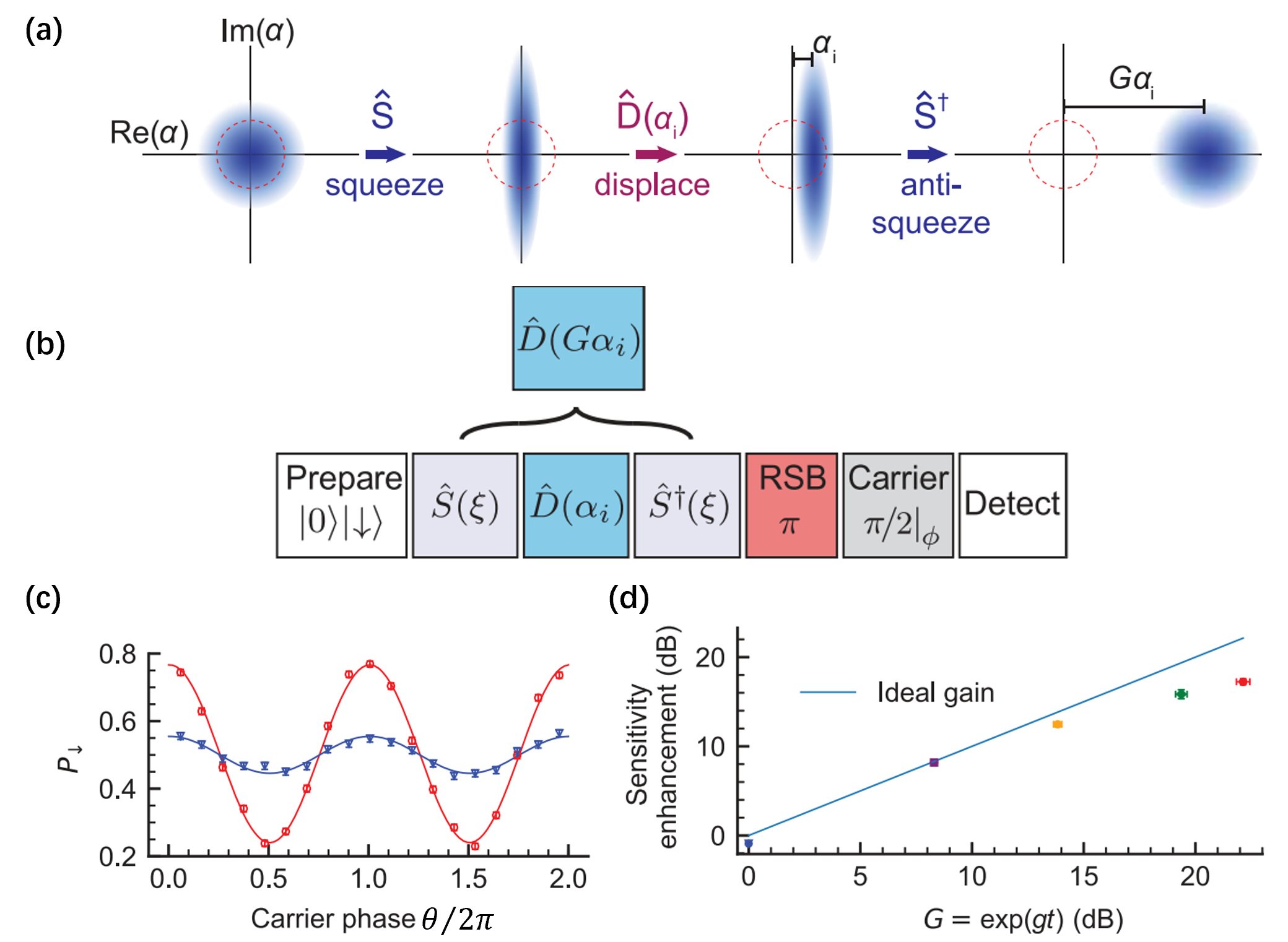}
  \caption{\label{Fig4.1.5}
  Quantum amplification in a trapped-ion mechanical oscillator~\cite{Burd2019}. 
  (a) Conceptual illustration of the amplification protocol.
  Each panel shows a Wigner function phase space distribution.
  A displacement $\alpha_i$ of an initially squeezed ground state is amplified by subsequent anti-squeezing, resulting in a final coherent state with amplitude $G \alpha_i$ with no added noise.
  (b) Pulse sequence for displacement sensing protocol with phase-sensitive red sideband (PSRSB) detection.
  (c) Population $P_{\downarrow}$ as a function of the carrier $\pi/2$ pulse phase with no squeezing (blue inverted triangles) and amplification (red circles).
  (d) Measurement sensitivity enhancement in the linear small-displacement regime as a function of the ideal gain $G=\exp(gt)$. Reproduced with permission from Burd \textit{et al.}, Science \textbf{364}, 1163 (2019). Copyright 2019 The American Association for the Advancement of Science.}
\end{figure}

With a trapped-ion mechanical oscillator, interaction-based readout has been employed to amplify and detect coherent motional displacements that are considerably smaller than the quantum zero-point fluctuations~\cite{Burd2019}.
As shown in Fig.~\ref{Fig4.1.5}~(a), by squeezing the motional state firstly, one can suppress the quantum fluctuations along a particular phase space quadrature.
Then, a small displacement $\alpha_{i}$ is applied along the squeezed axis.
Finally, by reversing the squeezing interaction, the oscillator returns to a minimum-uncertainty coherent state with a larger amplitude $\alpha_f = G\alpha_i$ with $G$ is the gain.
%
%
%
%
%
The maximum amplification is achieved when the displacement is along the squeezed axis. 
To demonstrate experimentally, a single trapped $^{25}$Mg$^{+}$ ion is used.
%
In each experiment, the ion is initialized in the electronic and motional ground state $\ket{\downarrow}\ket{0}$ via optical pumping, sideband laser cooling and microwave pulses.
The pulse sequence for displacement sensing protocol is shown in Fig.~\ref{Fig4.1.5}~(b).
%
The displacement $\alpha_f$ is obtained by measuring probability into state $\ket{\downarrow}$ with $P_{\downarrow}=\frac{1}{2}[1-C(|\alpha_f|)\cos \phi]$, where $C(|\alpha_f|)$ is the
signal contrast and $\phi$ is the phase of a carrier $\pi/2$ pulse.
Without amplification, $\alpha_{i}=\alpha_{f}$ and the contrast is $C(|\alpha_{i}|)$.
However, the initial displacement amplitude $\alpha_{i}$ is increased by utilizing the anti-squeezing and the contrast become $C(|G\alpha_{i}|)$.
This increase in contrast is shown in Fig.~\ref{Fig4.1.5}~(c).
Especially, if the displacement is along the squeezed axis, the measurement sensitivity enhancement as a function of the ideal gain $G=\exp(gt)$ in the linear small-displacement regime, as shown in Fig.~\ref{Fig4.1.5}~(d).
More recently, by modulating the trapping potential at close to twice of the center-of-mass mode frequency, a motional squeezing of $5.4 \pm 0.9$ dB below the ground-state motion has been achieved via a stroboscopic protocol~\cite{PhysRevA.107.032425}.

\section{\label{sec:5}Promising applications in quantum sensors}

Recent technological advancements have paved the way for the development of highly stable and precise instruments such as lasers, optical frequency combs, and microwave electronics.
These advancements have enabled the creation of quantum interferometers with unprecedented sensitivity and the world's most accurate and precise clocks.
In these devices, the stable quantum states of atoms play a crucial role. 
By harnessing controlled atomic entanglement, it is possible to achieve entanglement-enhanced atom interferometry, surpassing the precision limits of conventional methods.
Entanglement-enhanced atom interferometry allows for the precise measurement of various physical quantities, including frequency, magnetic field, gravity, acceleration, and rotation.
This breakthrough has significant implications for the development of atomic clocks, quantum magnetometers, quantum gravimeters, and quantum gyroscopes, offering unprecedented levels of precision in these applications.
Quantum sensors, empowered by the remarkable precision offered by quantum entanglement, have emerged as highly promising tools in various domains. 
They hold great potential in fields such as inertial navigation, mineral exploration, groundwater monitoring, dark matter detection, and space-based experiments aimed at testing theories of general relativity or quantum gravity.
Below we will introduce the exciting and promising applications of entanglement in the realm of quantum sensors.

\subsection{\label{sec:5-1}Atomic clocks}

\begin{figure}[!htp]
 \includegraphics[width=\columnwidth]{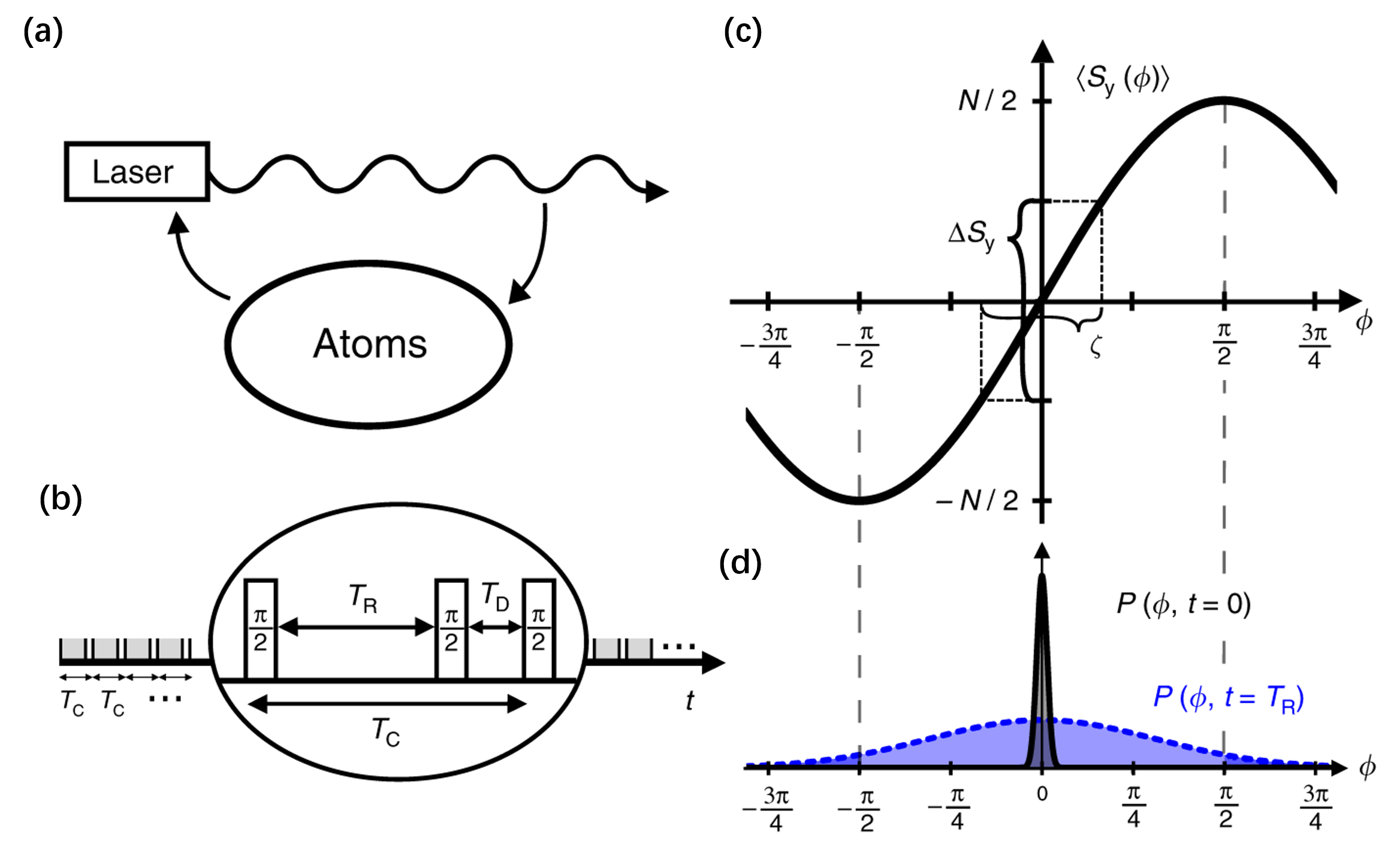}
  \caption{\label{Fig-Sec4-atomic-clock}
  Measurement noises of a squeezing-enhanced atomic clock~\cite{Schulte2020}. (a) Measurement and feedback loop to stabilize the laser frequency to an atomic transition. (b) Periodic measurements with Ramsey time $T_R$ and dead time $T_D$ in each cycle of total time $T_C$ lead to increased instability from the Dick effect. (c) Quantum projection noise for $N$ particles limits the clock stability for short interrogation times but can be decreased with squeezed states thus reducing the inferred phase uncertainty where $\zeta$ is the Wineland spin squeezing parameter of Eq.~\eqref{squeezing2}. (d) For longer $T_R$ the distribution of phases broadens substantially due the laser’s decoherence. Inefficient feedback for phases outside the interval gives the coherence time limit. Reproduced with permission from Schulte \textit{et al.}, Nat. Commun. \textbf{11}, 5955 (2020). Copyright 2020 licensed under a Creative Commons Attribution (CC BY) license.}
\end{figure}

Atomic clocks are the most accurate and precise clocks that have been developed to date.
Atomic clocks utilize the reliable transition frequency of stable long-lived energy levels found in atoms. 
They work by comparing the frequency of a local oscillator (LO) to the reference atomic transition frequency.
This comparison is used to ensure that the ticking period of the LO remains stable, as shown in Fig.~\ref{Fig-Sec4-atomic-clock}~(a).
For atomic clocks the selected states are often magnetically insensitive internal states. 
It makes them excellent frequency references as they do not experience the aging process observed in mechanical clocks. 
The atomic transition frequency is universally present and highly suitable for usage as a standard frequency.
For example, the current major frequency standard relies on the hyperfine transition in $^{133}$Cs, which is utilized to precisely define the SI second~\cite{PhysRevLett.1.105}.
The systematic error achieved by a solitary optical clock based on an aluminum ion~\cite{PhysRevLett.123.033201} can be $9.4\times 10^{-19}$.
Nevertheless, as systematic effects are fully understood and accounted for, atomic clocks that make use of a large number of atoms have the capacity to far exceed the performance of clocks that rely on a single ion.
Atomic clocks with a performance beating the SQL have been demonstrated in several experiments~\cite{Gross2010,Nichol2022,Schulte2020, Robinson2024}.

Entanglement-enhanced microwave-frequency clocks have been achieved using neutral atoms with spin-squeezed states as input~\cite{Gross2010, PhysRevLett.104.250801,Ockeloen2013,PhysRevLett.117.143004, Hosten2016}.
Although these proof-of-principle experiments serve as spectacular examples of entanglement-enhanced atomic clocks, they have not yet achieved the same level of frequency stability as state-of-the-art fountain clocks.
This is mostly due to their Ramsey interrogation time (limited by the noise of the microwave LO) is much shorter than the ones typically operating in fountain clocks using uncorrelated individual atoms~\cite{Weyers2018}.

The performance of atomic clocks could be greatly improved if the reference transition occurred in the optical range instead of the microwave range~\cite{PhysRevA.92.012106}. 
Typically, optical frequencies are approximately $10^4$ times higher than microwave frequencies, resulting in a corresponding improvement of the fractional stability by a factor of $10^4$.
With the advent of optical frequency combs spanning a wide range of octaves~\cite{PhysRevLett.40.847}, there has been notable progress in optical clock technology~\cite{Ludlow2015}. 
The lasers consists of a series of closely spaced spectral lines in frequency space. 
The separation between these lines is a few hundred MHz, and it is dictated by the repetition rate of the femtosecond laser. 
Therefore, by utilizing a frequency comb as an intermediate stage, two lasers operating at distinct wavelengths might effectively compete. 
The usage of optical frequency comb in an optical atomic clock enables an experimentally feasible way to measuring optical transition frequencies and comparing optical clocks. 

Optical transition-based atomic clocks have gained significant attention in recent years due to their remarkable precision and stability~\cite{Katori2011,10.1063/5.0121372}. 
Currently, optical atomic clocks are the most precise and accurate clocks. 
These clocks generally interrogate the optical transitions with groups of atoms that are confined in an optical lattice.
So far, the optical-transition atomic clocks can reach fractional stability below $10^{-18}$~\cite{Bloom2014,Schioppo2017,Oelker2019}, and approaching $10^{-19}$~\cite{Zheng2022,Li_2024}.
Technical noises from imperfect readout, intrinsic atom–atom interactions, or aliased frequency noise of the interrogating clock laser pose challenges for observing clock performance at the quantum projection noise (QPN) limit. 
Since the atomic collisions would disturb the transition frequency and the resulted collision shift would decrease the accuracy of the atomic clocks. 
Thus the collision shifts impose a limit on the total number of atoms used in atomic clocks. 
Typically, the atomic number is about $10^2 \sim 10^4$.
At such relatively small atom number, the SQL places a significant constraint on the clock frequency measurement.

\begin{figure*}[!htp]
 \includegraphics[width=2\columnwidth]{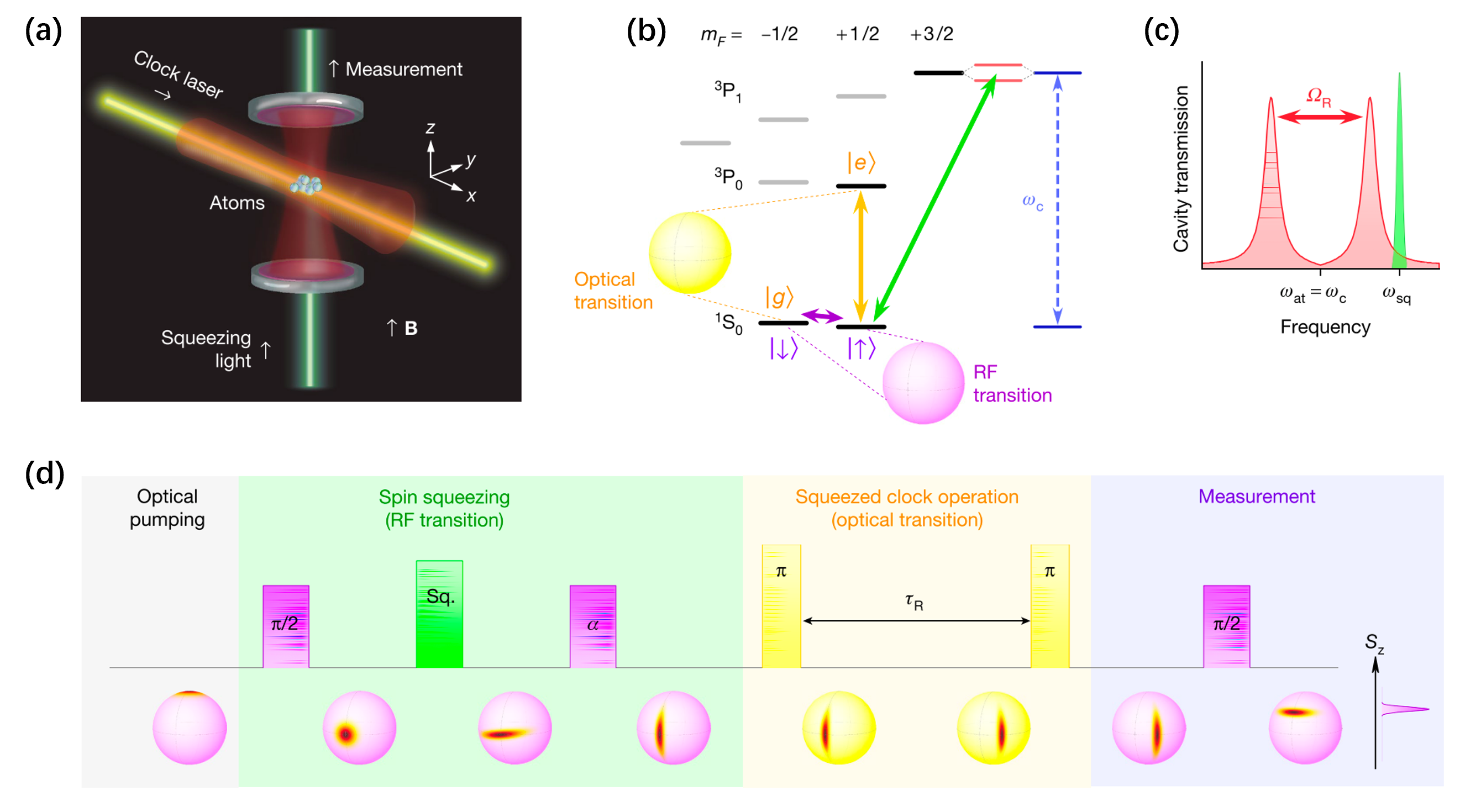}
  \caption{\label{Fig-Sec5-atomic-clock-cavity} Optical atomic clock operated with spin squeezing~\cite{Pedrozo-Penafiel2020}. (a) $^{171}$Yb atoms are confined within an optical cavity in a two-dimensional magic-wavelength optical lattice, namely in the $x$ and $z$ directions (shown by the color red). The optical pumping and spin squeezing light (green) is directed parallel to the cavity axis $z$, while the clock laser (yellow) travels parallel to the $x$-axis. (b) Levels of energy and changes between them. The ground-state RF transition $\ket{\downarrow} \rightarrow \ket{\uparrow}$ is shown by purple pulses, the squeezing transition at $\omega_{at}$ is represented by green pulses, and the optical-clock transition $\ket{\uparrow} \rightarrow \ket{e}$ is represented by yellow pulses. The system undergoes evolution inside either the ground-state manifold (the purple Bloch sphere) or the clock-state manifold (the yellow Bloch sphere). The resonance of the squeezing transition is achieved by adjusting the cavity frequency $\omega_c$. (c) The atoms' strong interaction with the cavity leads to the division of the cavity resonance into two distinct peaks, known as vacuum Rabi splitting (red peaks). A laser is used at a frequency $\omega_{sq}$ (green peak) that is detuned from the Rabi peak. This laser is employed to generate a spin squeezed state by cavity feedback. (d) Squeezed-clock sequence. A spin squeezed state is initially created in the lowest energy level and then rotated by an angle $\alpha$. It is then moved to the clock energy level and undergoes a Ramsey sequence for a duration of $\tau_R$. Finally, it is mapped back to the states {$\ket{\uparrow},\ket{\downarrow}$} and a measurement of the state is made. The quantum state's development is illustrated on the Bloch spheres, representing the RF (purple) and optical (yellow) transitions. Reproduced with permission from Pedrozo-Penafiel \textit{et al.}, Nature \textbf{588}, 414 (2020). Copyright 2020 Springer Nature.}
\end{figure*}

By using suitable entangled states such as spin squeezed states, the performance of optical atomic clocks can be improved beyond the SQL.
The spin squeezing has been used to improve an optical lattice clock of $^{171}$Yb atoms~\cite{Pedrozo-Penafiel2020}.
The strategy used is to generate a spin squeezed state in the ground state of $^{171}$Yb atoms and then transfer it to the ultra-narrow optical clock transition by applying an optical $\pi$ pulse, see Fig.~\ref{Fig-Sec5-atomic-clock-cavity}. 
Using the spin squeezed state on the optical transition as the input state, a Ramsey interferometry is perform in the optical domain with a five-order improved Q-factor compared to the microwave transitions. 
After the Ramsey interference, the spin squeezed state is mapped back onto the ground state and read out via the cavity. 
As a result, a measurement precision of 4.4 dB below the SQL is demonstrated, corresponding to a 2.8-fold reduction of the averaging time.
More recently, a protocol based on time-reversal dynamics called Signal Amplification through Time-Reversal Interaction was demonstrated for optical atomic clocks approaching the Heisenberg limit. 
As shown in Sec.~\ref{sec:4-2}, a non-Gaussian over-squeezed state is generated under a prolonged OAT dynamics and with a time-reversal process for readout, it enables one to reach near-Heisenberg sensitivity. 
%

The clock stability can be assessed by its Allan deviation, which depends on several factors such as the transition frequency, interrogation time, dead time, average time, and input state.   
For optical atomic clocks, the laser is stabilized by a control loop to an optical atomic transition of frequency $\nu_0$, as shown in Fig.~\ref{Fig-Sec4-atomic-clock}~(a). 
The Ramsey interrogation $T_R$ and the dead time $T_D$ comprise of the duration $T_C=T_R+T_D$ for each cycle. 
By measuring the spin component $\langle \hat S_y \rangle$, one can get the information of $\nu_0$ from the laser frequency.
The measurement result is converted into an error signal that is used to correct the laser frequency. 
Averaging over a long time $\tau \gg T_C$, the clock stability can be measured in terms of Allan deviation $\sigma_y(\tau)$. 
For an atomic clock with spin squeezing, the stability limited only by the QPN of the spin measurement results in the Allan deviation 
\begin{equation}
    \sigma_{QPN}(\tau)=\frac{1}{2\pi\nu_0 T_R} \sqrt{\frac{T_C}{\tau}}\frac{\xi_R}{\sqrt{N}},
\end{equation}
where $N$ is the number of atoms participating in the clock, $\xi_R$ is the Wineland spin squeezing parameter. 
It is important to note that under practical conditions, optical atomic clocks are not exclusively limited by QPN. 

\begin{figure*}[!htp]
 \includegraphics[width=2\columnwidth]{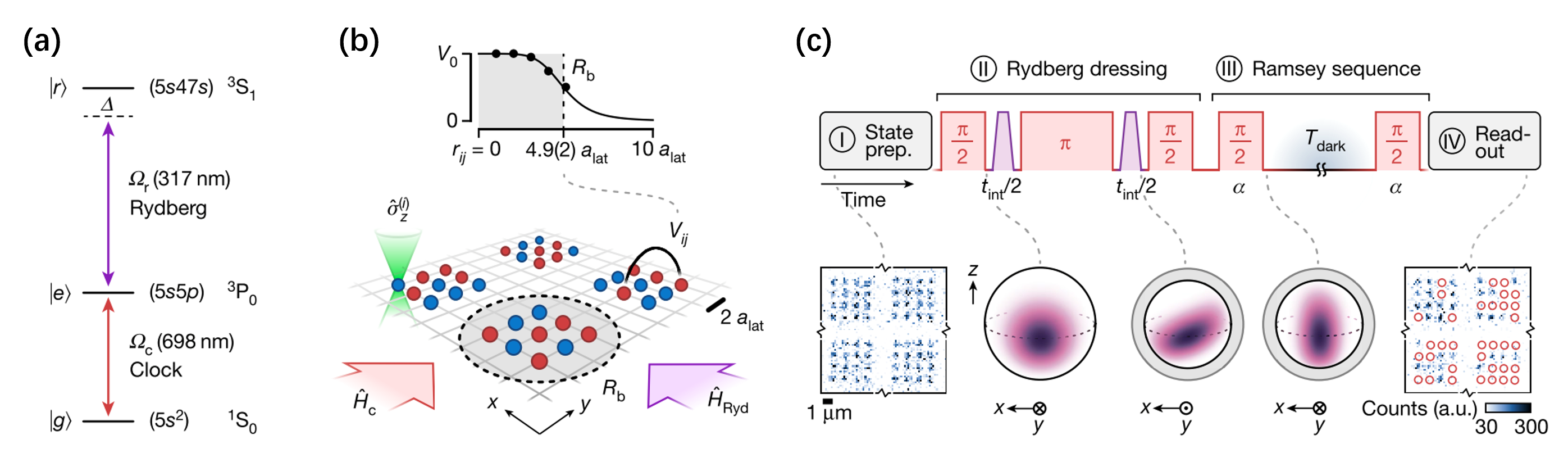}
  \caption{\label{Fig-Sec3-Rydberg2} Generation of spin squeezing in a Rydberg-dressed array of $^{88}$Sr atoms~\cite{eckner2023realizing}. (a) Atomic states (black lines) and transitions (colored arrows) of $^{88}$Sr relevant for clock interrogation and Rydberg dressing. (b) Diagram illustrating the configuration of the experimental arrangement. The spacing of $a_{\textrm{lat}}$ is approximately $575$ nm, and it is organized into several subarrays. The upper plot illustrates the relationship between atoms at a distance $r_{ij}$, which is defined by the potential energy $V_{ij}$. The black circles represent the data points. The solid black line represents a numerical fit. The laser beams that connect the states in (a) are shown by colored arrows and correspond to the Hamiltonians $H_c$ and $H_{\textrm{Ryd}}$. Optical tweezers, which have a green double cone shape and operate at a wavelength of $515$ nm, allow for the application of the operator $\hat \sigma_z^{(i)}$ to a specific atom $i$. (c) Diagram depicting the procedural method for generating spin squeezed states in an experimental setting. The pulse sequences for (II) and (III) display clock and Rydberg laser pulses in the colors red and purple, respectively. In this context, the symbol $\alpha$ represents the phase of the clock laser relative to a reference. Individual photographs are captured after the process of preparing the state (located at the bottom left) and for the purpose of determining the atomic populations (located at the bottom right). In this context, the red circles represent atoms in the state $\ket{g}$ that have been deliberately eliminated prior to the final imaging. The generalized Bloch spheres, located in the bottom center, depict the progression of the atomic state for a subarray consisting of 16 atoms, where $N = 4 \times 4$. Reproduced with permission from Eckner \textit{et al.}, Nature \textbf{621}, 734 (2023). Copyright 2023 Springer Nature.}
\end{figure*}

In practice, the overall stability of a clock is influenced by various sources of noise, including laser phase noise and dead time. 
Among these factors, the Dick effect and the laser coherence time limit play significant roles in compromising the stability. 
The Dick effect arises from the interrupted interrogation of the atomic system while the finite laser coherence time contributes to the stability with an additional diffusion process, which refers to the laser coherence time limit. 
Incorporating these effects, the optimal operating point of the control loop is determined by a trade-off between QPN, Dick effect, and coherence time limit.
This trade-off is achieved by minimizing the combined stability~\cite{Schulte2020}, 
\begin{equation}\label{total-stability}
    \sigma_y(\tau) =\sqrt{\sigma_{QPN}^2(\tau)+\sigma_{Dick}^2(\tau)+\sigma_{CTL}^2(\tau)}. 
\end{equation}
Considering all these critical factors of instability, it has been revealed that a typical single-ensemble Ramsey clock holds the potential for improving performance through spin squeezing, especially in systems with relatively low atom numbers~\cite{Schulte2020}. 
The primary limitations arise from the instability of the lasers in the system. 
However, these limitations can be addressed by employing an adaptive measurement scheme~\cite{PhysRevLett.111.090801}. 
This scheme involves performing a weak measurement of the collective spin state, followed by a feedback process that rotates the state by an amount determined by the measurement outcome. 
By doing so, the measurement backaction is minimized, resulting in a final state that remains unaffected by undesired noises.

Using two clock ensembles for comparison, the common-mode noises such as the LO decoherence can be cancelled and so that the Dick effect can be completely eliminated~\cite{Schioppo2017}.
Likewise, employing cascaded systems with multiple ensembles can effectively mitigate the impact of laser coherence limit~\cite{PhysRevLett.111.090802, PhysRevLett.112.190403}.
An experimental implementation of a "multiplexed" one-dimensional optical lattice clock has been realized, where spatially resolved ensembles of strontium atoms are trapped in the same optical lattice, allowing for simultaneous interrogation by a shared clock laser and parallel read-out processes~\cite{Zheng2022}.
The results of these experiments showcase that applications involving optical clock comparisons can overcome the limitations imposed by LO instability, making it a viable option for implementing a quantum network of clocks as well~\cite{Komar2014}.

To improve spin-squeezing-enhanced atomic clocks, it is crucial to consider the coupling of anti-squeezing, as it couple with the measurement projection, and then introduce extra noise and reduce the contrast. 
To overcome this problem, one may utilize adaptive measurement for a single clock ensemble  ~\cite{PhysRevLett.111.090801,PhysRevLett.110.210503,PhysRevLett.116.093602}, or combination of measurement and feed forward for multiple clock ensembles ~\cite{PhysRevLett.111.090802,PhysRevA.93.032138,PhysRevLett.125.210503}.
Recently, the protocol of multiple clock ensembles with feedback was analyzed~\cite{PhysRevLett.125.210503,PhysRevA.105.053116}, where one squeezed clock was considered together with another unentangled clock. 
It is shown that by feeding forward the measurement result of the unentangled ensemble to the squeezed ensemble, it is possible to sustain the entanglement enhancement for extended periods of time, even when there is LO noise present. 
The two methods can thereby increase the interrogation times that still allow an enhancement due to spin squeezing.

Through integrating an optical clock platform with collective strong-coupling cavity QED for QND measurements, the direct comparison of two spin-squeezed optical clock ensembles has been achieved~\cite{Robinson2024}. 
The QND measurements of the clock state~\cite{PhysRevX.10.041052} can be implemented for spin squeezing and clock readout. 
Optimizing the competition between spin measurement precision and loss of coherence, a metrological enhancement for a large ensemble of atoms beyond the initial spin coherent state can be measured. 
Using a movable lattice, the cavity can individually address two independent sub-ensembles, enabling one to spin squeeze two clock ensembles successively and compare their performance without the influence of clock laser noise.
Although the clock comparison remains above the effective SQL, the performance directly verifies $1.9(2)$ dB clock stability enhancement at the $10^{-17}$ level without subtracting any technical noise contributions.
Moreover, resonant electric dipole-dipole interactions in a cubic array of atoms in the many-excitation limit has been observed directly~\cite{Hutson2024}.
With the $^{87}$Sr atoms loaded in a cubic optical lattice, the effects of resonant dipole-dipole interactions is measured. 
The ensemble-averaged shifts can be suppressed below the level of evaluated systematic uncertainties for optical atomic clocks.
It is found that the interactions caused a tiny clock shift, the magnitude of which could be controlled by varying the relative orientation of the probe light and the atomic dipoles. 

In comparison to an optical lattice clock, the trapped ion clock offers the advantage of less dead time.
This is because that the same ion can be trapped and interrogated for extended durations due to the presence of a strong confinement potential.
Despite their lower signal-to-noise ratios when compared to optical lattice clocks, ion clocks remain among the most stable clocks now accessible.
The physics component of trapped ion clock systems is much more amenable to miniaturization, and ongoing attempts are being made to enable the transmission and reception of light from the ions through phonics integrated into the trap structure itself~\cite{Schulte2020}. 
In recent, several promising trapped ion-based clocks such as highly charged ion clocks~\cite{King2022} and nuclear clocks~\cite{Peik2021} are being pursued and they offer further improvement in the performance of these types of clocks.
In particular, the enhanced frequency comparisons using an elementary quantum network of two entangled trapped-ion atomic clocks has been demonstrated~\cite{Nichol2022}. 
The network's ability to generate entanglement with high fidelity and speed, resulting in a large signal and an efficient duty cycle, demonstrates the potential for practical enhancements in atomic clocks.
Furthermore, as shown in Fig.~\ref{Fig-Sec3-Ion-OAT}, a potential avenue for leveraging large-scale entanglement is demonstrated on an optical transition by utilizing one-dimensional chains of ions. 
The observation of emergent collective behaviors related to OAT with finite-range interactions represents a significant milestone in the integration of entanglement into high-performance ion clocks that operate with a large number of ions.

Another intriguing avenue for clock applications involves utilizing long-range Rydberg-dressing interactions to generate spin-squeezed and non-Gaussian entangled states.
With the versatile capabilities of optical tweezer arrays in manipulating atomic ensembles, optical clocks have been developed using current breakthroughs in optical tweezers~\cite{Young2020, Schine2022, eckner2023realizing}.
The tweezer array optical clocks combine the advantages of both single-ion-based optical clocks and optical lattice clocks. 
They incorporate the superb isolation and high duty cycles demonstrated by single-ion-based optical clocks, while also benefiting from the large ensembles and consequent low QPN characteristic of optical lattice clocks.
The demonstration of a half-minute-scale atomic coherence and high relative stability in a tweezer clock, as reported in a recent experiment~\cite{Young2020}, lays a solid foundation for the engineering of entanglement on an optical-clock transition. 
The utilization of expansive 2D arrays and close proximities in this context is crucial for utilizing limited-range Rydberg interactions. 
This approach enables for the study of larger samples with enhanced interconnections, more potent interactions, and therefore, increased entanglement. 
Despite the exponential decline in many-body entanglement due to single-particle decoherence, the coherence times demonstrated in the experiment show remarkable potential for achieving a highly precise entangled optical clock. 
This clock may operate with tens of atoms and interrogation durations lasting several seconds, making it suitable for metrological applications. 
Furthermore, by incorporating precise control and detection of individual particles along with a diverse set of tunable entangling interactions, the experimental generation of spin squeezing was achieved in an optical atomic clock. 
It has been demonstrated that such a clock can be built upon a programmable array of interacting optical qubits~\cite{eckner2023realizing}, see Fig.~\ref{Fig-Sec3-Rydberg2}.
In subarrays of $16$ atoms and $70$ atoms, spin squeezed states with $\xi_R^2=-3.8$ dB and $\xi_R^2=-3.4$ dB have been prepared, respectively. 
This enables the performance of a synchronous clock comparison with a stability that surpasses the SQL by up to $3.5$ dB, which paves an avenue to entanglement-enhanced optical clocks.

\subsection{\label{sec:5-2}Magnetometers}

The precise measurement of magnetic fields with high sensitivity is a critical requirement in various domains, including physics~\cite{PhysRevLett751879,PhysRevLett751879,Mamin2007,Marina2009S}, biology~\cite{Taylor2016,Thiel2016}, and medical applications~\cite{ROMALIS2011258,pnas1004037107,science1094025}, among others.
For instance, it is essential for detecting geomagnetic anomalies, studying magnetic fields in space, and mapping electric and magnetic fields generated in the brain for bio-magnetic measurements.
Quantum magnetometry uses the principles of quantum mechanics to measure magnetic fields with high precision and sensitivity. 
Based upon the interaction between quantum systems and external magnetic fields, various quantum systems including atoms, superconducting quantum interference devices, and nitrogen-vacancy centers have been employed to detect magnetic fields.
The atomic magnetometers mainly include optical Faraday magnetometers~\cite{Budker2007,Davis2016,PhysRevLett104093602,PhysRevLett104133601,PhysRevApplied.11.044034,PhysRevLett.131.133602}, coherent-population-trapping magnetometers~\cite{PhysRevA.105.L010601,PhysRevA.103.042607,PhysRevLett.95.123601,Hu:14,Andryushkov:22,Fang2021}, spin-exchange relaxation-free magnetometers~\cite{PhysRevA.80.013416,Savukov2017,PhysRevA.77.033408,Karaulanov_2016,PhysRevApplied.17.024004,Zhang:19,Hong2021ChipScaleUF}, and cold-atom magnetometers~\cite{PhysRevLett104093602,PhysRevLett104133601,PhysRevLett113103004,Ockeloen2013}.
Below, we will highlight the progresses in entanglement-enhanced magnetometry.

\begin{figure}[!htp]
 \includegraphics[width=\columnwidth]{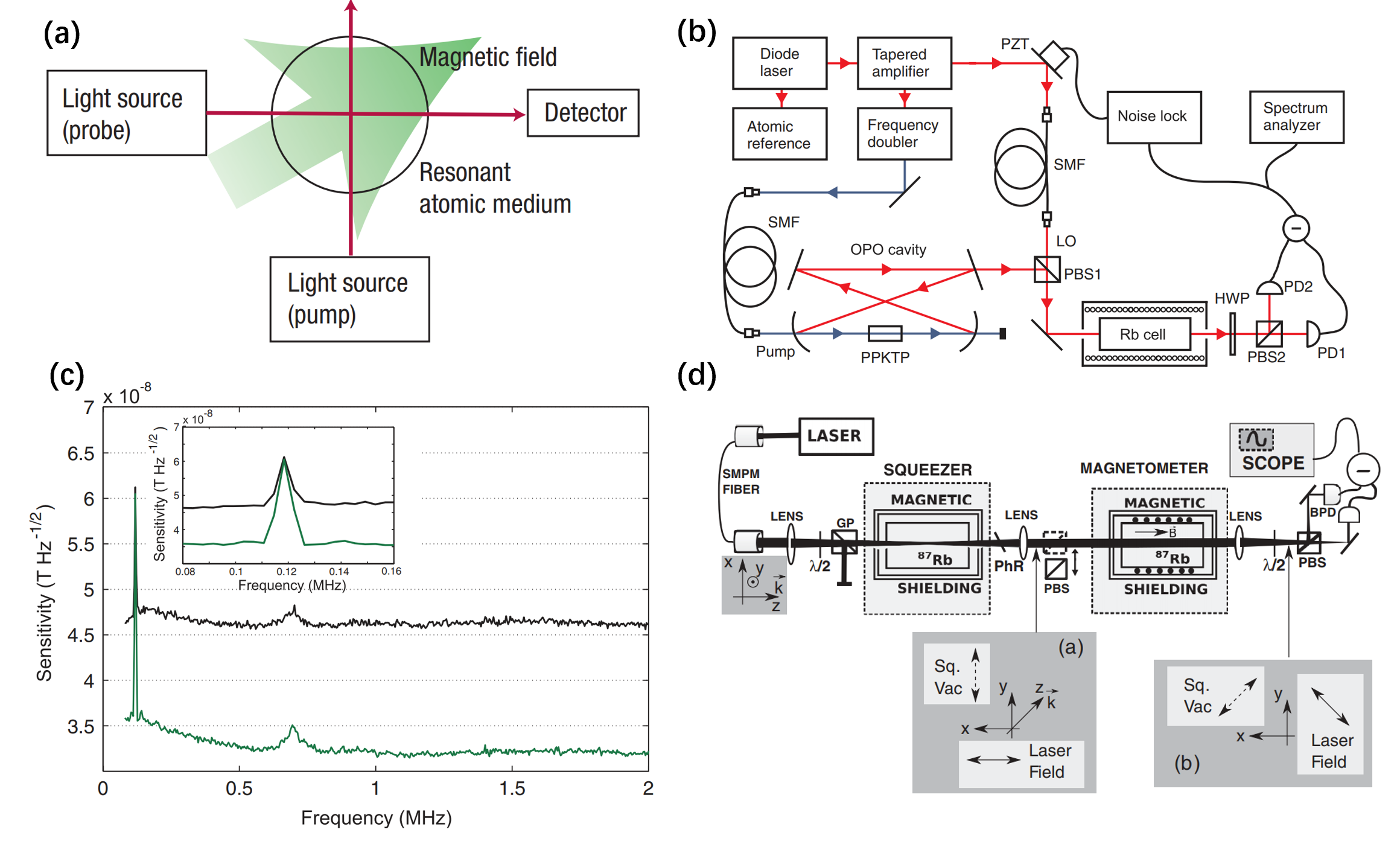}
  \caption{\label{Fig5.2.1}
  Optical Faraday magnetometers.
  (a) The schematic of an all-optical atomic magnetometer~\cite{Budker2007}. Pump light polarizes the atoms, atomic evolves in the magnetic field, and the resultant state of the atoms is detected by the probe light. Reproduced with permission from Budker \textit{et al.}, Nat. Phys. \textbf{3}(4), 227 (2007). Copyright 2007 Springer Nature.
  (b) Experimental setup of optical atomic magnetometer with optical parametric oscillator~\cite{PhysRevLett105053601}. Rb cell, rubidium vapor cell with magnetic coil and magnetic shielding; OPO, optical parametric oscillator; PPKTP, phase-matched nonlinear crystal; LO, local oscillator beam; PBS, polarizing beam splitter; HWP, half-wave plate; SMF, single-mode fiber; PD, photodiode. Reproduced with permission from Wolfgramm \textit{et al.}, Phys. Rev. Lett. \textbf{105}(5), 053601 (2010). Copyright 2010 American Physical Society.
  (c) Faraday rotation measurement.
  The  black curve shows the sensitivity with a polarized (but not squeezed) probe.
  The green line depicts the sensitivity with a polarization-squeezing probe.
  A zoomed view around the calibration peak at $120$ kHz is shown in the inset~\cite{PhysRevLett105053601}.
  Reproduced with permission from Wolfgramm \textit{et al.}, Phys. Rev. Lett. \textbf{105}(5), 053601 (2010). Copyright 2010 American Physical Society.
  (d) Experimental setup of optical atomic magnetometer with atomic squeezer~\cite{PhysRevA86023803}.
  The squeezer prepares an optical field with reduced noise properties which is used as a probe for the magnetometer.
  SMPM fiber: single-mode polarizationb maintaining fiber; $\lambda/2$: half-wave plate; PhR: phase-retarding wave plate; PBS: polarizing beam splitter; GP: Glan-laser polarizer;  BPD: balanced photodetector.
  Insets show the polarization of the squeezed-vacuum field and the laser field before the magnetometer cell (a) and right before the last PBS (b). Reproduced with permission from Horrom \textit{et al.}, Phys. Rev. A \textbf{86}(2), 023803 (2012). Copyright 2012 American Physical Society.
  }
\end{figure}

Optical Faraday magnetometers operate by measuring the polarization rotation of light as it traverses a group of atoms exposed to a magnetic field~\cite{Budker2007}.
Generally, the device consists of $N$ spin-$F$ atoms initially optically pumped into a fully polarized state (pointing along the $y$ axis, for instance), such that the collective spin of the ensemble has length $F_N$.
A weak magnetic field along the $x$ axis causes the precession of the collective spin in the $y-z$ plane at a rate $g\gamma B$, where $g$ is the
gyromagnetic ratio, $\gamma$ is the Bohr magneton, and $B$ is the
magnetic field strength.
Polarization of light propagating along $z$ will be rotated proportional to the collective atomic spin $J_z$ due to the Faraday effect.
Thus, the atomic spin precession is detected by measuring transmission or rotation of the polarized probe light, as show in Fig.~\ref{Fig5.2.1}~(a). 
%
%
%
The rotation of the polarized probe light are described via the Stokes parameters $S_x$, $S_y$, $S_z$, and the detected signal is~\cite{PhysRevLett105053601}
%
\begin{eqnarray}\label{Eq.5.21}
    S_y^{\text{out}}=S_y^{\text{in}}+S_x(\nu B_{z}+\alpha J_{z}) l
\end{eqnarray}
where $\nu$ is the Verdet constant of the vapor, $B$ is the magnetic field, $\alpha$ is proportional to the vector component of the atomic polarizability, and $l$ is the length of the medium.
Quantum mechanics sets fundamental limits on the best sensitivity that can be achieved in a magnetic-field measurement.
%
%
According to Eq.~\eqref{Eq.5.21}, the sensitivity of this magnetometers is fundamentally limited by quantum noise main in the form of photon shot noise presented in $S_y^{\text{in}}$ and atomic projection noise presented in $J_z$.
%
It is obvious that one can realize the entanglement-enhanced magnetic-field measurement via using a squeezed state in $S_{y}$ as an input.
%

\noindent

In experiments, the improvement in magnetic field sensing via using polarization-squeezed probe lights has been demonstrated with Rb atoms~\cite{Davis2016,PhysRevA86023803}.
By using off-resonant, polarization-squeezed probe light, the sensitivity of the magnetometer is improved by $3.2$ dB.
The magnetometer consists of a source of polarization squeezed light, a rubidium vapor cell at room temperature, and a shot-noise-limited polarimeter.
%
%
The experimental setup is shown schematically in Fig~\ref{Fig5.2.1}~(b).
Firstly, the squeezed vacuum state is produced by pumping a sub-threshold optical parametric oscillator (OPO).
The nonlinear medium in the OPO is used to polarize the squeezed vacuum.
Then, the polarization-squeezed light is sent through a atomic cell at room temperature.
At last, the optical rotation is detected by a shot-noise-limited polarimeter.
The sensitivity of the Faraday rotation measurement is shown in Fig.~\ref{Fig5.2.1}~(c).
%
Moreover, one can also use an atomic squeezer to generate the squeezed probe light, which is based on the polarization self-rotation (PSR) effect~\cite{PhysRevA86023803}.
%
The setup is depicted in Fig.~\ref{Fig5.2.1}~(d), the atomic squeezer can generate about $2$ dB of noise reduction, and the magnetometer with sensitivities close to $1 \text{pT}/\sqrt{\text{Hz}}$.
More recently, by using squeezed probe light and evasion of measurement backaction, one can enhance the sensitivity and measurement bandwidth of the magnetometer at sensitivity-optimal atom number density~\cite{PhysRevLett.131.133602}. 
%
%

Cold-atom magnetometers generally operate based on the measurement of the collective spin of an ensemble of atoms after a Ramsey interferometry.
The device consists of $N$ two-mode particles (single-particle states) and the two modes are labelled as $\ket{a}$ and $\ket{b}$.
According to the description of the Ramsey interferometry in Sec.~\ref{sec:2}, the magnetic field to be estimated is encoded in the quantum system during the interrogation process.
After the beam splitter, the information of the magnetic field can be obtained via measuring the collective spin.
%
For atomic ensembles, trapped BECs are particularly well suited for sensing applications requiring both high sensitivity and spatial resolution, taking advantage of their small size, high degree of coherence, and the availability of sophisticated techniques for precise controlling of motional degrees of freedom~\cite{PhysRevLett113103004}.
%
%
Recently, Ramsey interferometry using spin-squeezed states, which generated via OAT dynamics with BEC~\cite{PhysRevLett113103004,Gross2010}, has applied to sense magnetic fields.

\begin{figure}[!htp]
 \includegraphics[width=\columnwidth]{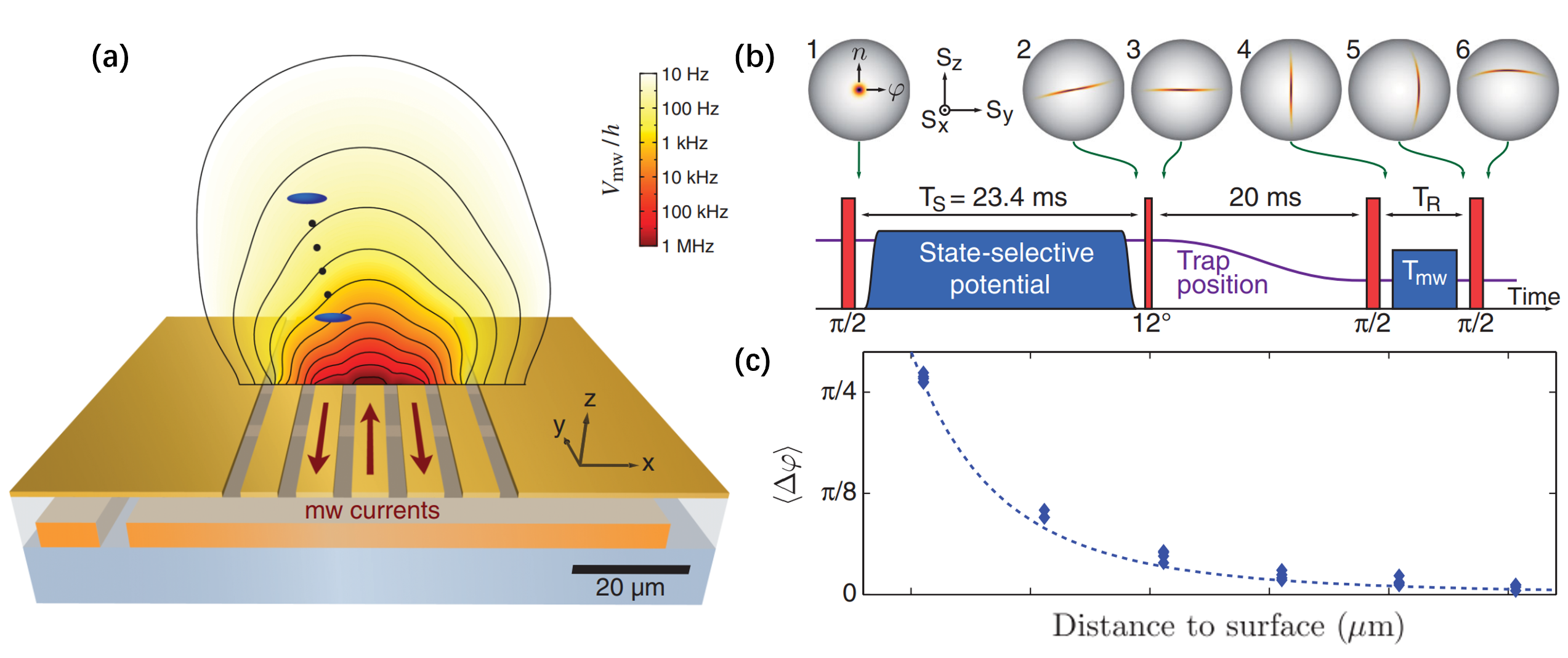}
  \caption{\label{Fig5.2.2}  
  Entanglement-enhanced magnetic field sensing with an atom chip~\cite{Ockeloen2013}.
  (a) Experimental setup. Central region of the atom chip is the atomic probe and the scanning trajectory. The probe is used to measure the magnetic near-field potential generated by an on-chip microwave guide.
  (b) Experimental sequence of the scanning probe interferometer, showing Rabi (red) and on-chip microwave (blue) pulses and the trap position (purple). Spheres show the Wigner function of the collective spin state at different stages of the Ramsey experiment.
  (c) Measured phase shift $\langle \varphi \rangle$ induced by a microwave near-field pulse as a function of the atom-surface distance. Reproduced with permission from Ockeloen \textit{et al.}, Phys. Rev. Lett. \textbf{111}(14), 143001 (2013). Copyright 2013 licensed under a Creative Commons Attribution (CC BY) license.}
\end{figure}

With small BECs on an atom chip, one can overcome the SQL by $4$ dB and yield a microwave magnetic field sensitivity of $77$ pT via using the spin-squeezed state as input~\cite{Ockeloen2013}.
The experimental schematic of the scanning probe interferometer is shown in Fig.~\ref{Fig5.2.2}.
Firstly, a spin coherent state on the equator of the generalized Bloch sphere is created via a $\pi/2$ pulse.
Then, applying OAT Hamiltonian $\hat{H}_\text{OAT}=\chi \hat{J}_{z}^2$ to generate spin squeezed state.
%
Next, applying a suitable rotation around the $J_x$ axis to align the anti-squeezed quadrature with the equator.
%
%
A full Ramsey interferometer sequence consisting of a $\pi/2$ pulse to make it maximally phase sensitive, a free evolution with time $T_R$ for phase accumulation $\phi\propto B T_R$, and a final $\pi/2$ pulse mapping $\phi$ onto the population difference.
The measurements as a function of the atom-surface distance is shown in Fig.~\ref{Fig5.2.2}~(c).
For all positions, the interferometer performs well below the SQL, and the microwave field sensitivity approach to $\Delta B=2.4 \mu T$.

\begin{figure}[!htp]
 \includegraphics[width=\columnwidth]{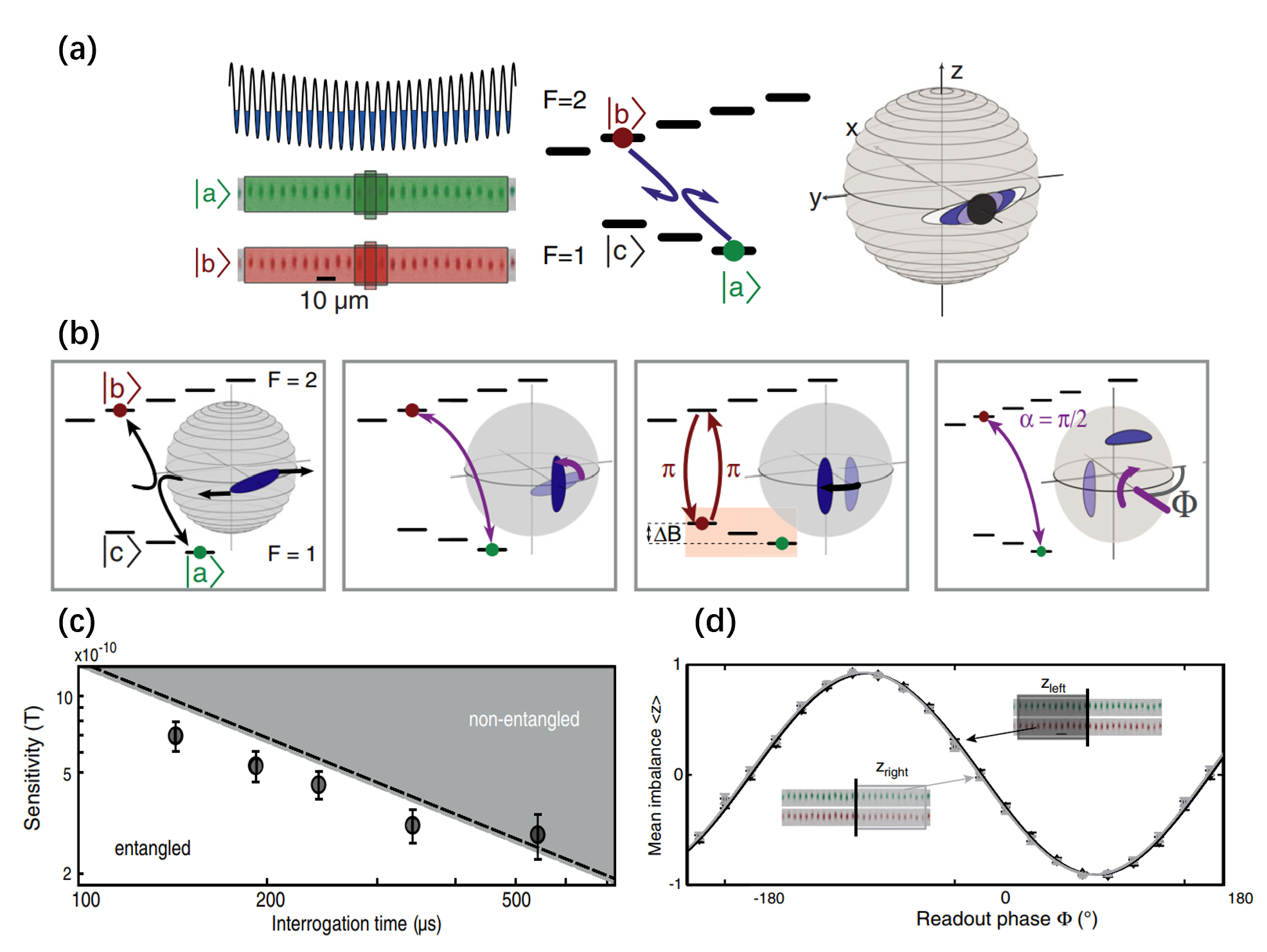}
  \caption{\label{Fig5.2.3}
  Quantum-enhanced magnetometry with BECs in an optical lattice~\cite{PhysRevLett113103004}.
  (a) Independent squeezing of $25$ binary BECs in a 1D lattice.
  %
  %
 (b) The implemented experimental sequence for entanglement-enhanced magnetometer, which are displayed on generalized Bloch sphere.
 (c) Ramsey fringes for left and right parts of the full sample.
 (d) Magnetic field sensitivity of the Ramsey magnetometer versus interrogation time. Reproduced with permission from Muessel \textit{et al.}, Phys. Rev. Lett. \textbf{113}(10), 103004 (2014). Copyright 2014 American Physical Society.}
\end{figure}

With BECs in an optical lattice, one can achieve a quantum-enhanced single-shot sensitivity of $310(47)$ pT for a static magnetic field in a probe volume as small as $90~\mu m^3$~\cite{PhysRevLett113103004}.
The experimental schematic and results of the entanglement-enhanced
magnetometer are shown in Fig.~\ref{Fig5.2.3}.
Firstly, generating the spin squeezed state in the levels $\ket{a}$ and $\ket{b}$ under OAT Hamiltonian $\hat{H}_\text{OAT}=\chi \hat{J}_{z}^2$, the state is rotated to achieve the maximum phase sensitivity.
Subsequently, a microwave $\pi$ pulse transfers the squeezing to a magnetic-field-sensitive hyperfine transition.
After magnetic field-dependent phase evolution with a hold time, swapping the state back to the original level to readout.
During this sequence, the state acquires a phase $\phi\propto B$.
At last, a Ramsey fringe is obtained by a final $\pi/2$ pulse, as shown in Fig.~\ref{Fig5.2.3}~(d).
Magnetic field sensitivity of the Ramsey magnetometer versus interrogation time is shown in Fig.~\ref{Fig5.2.3}~(c), the entanglement-enhanced sensitivity can be maintained up to interrogation time of $342$ ms.
%
%
%
The array configuration is particularly well suited for differential measurements and magnetic-field gradiometry.
\begin{figure}[!htp]
 \includegraphics[width=\columnwidth]{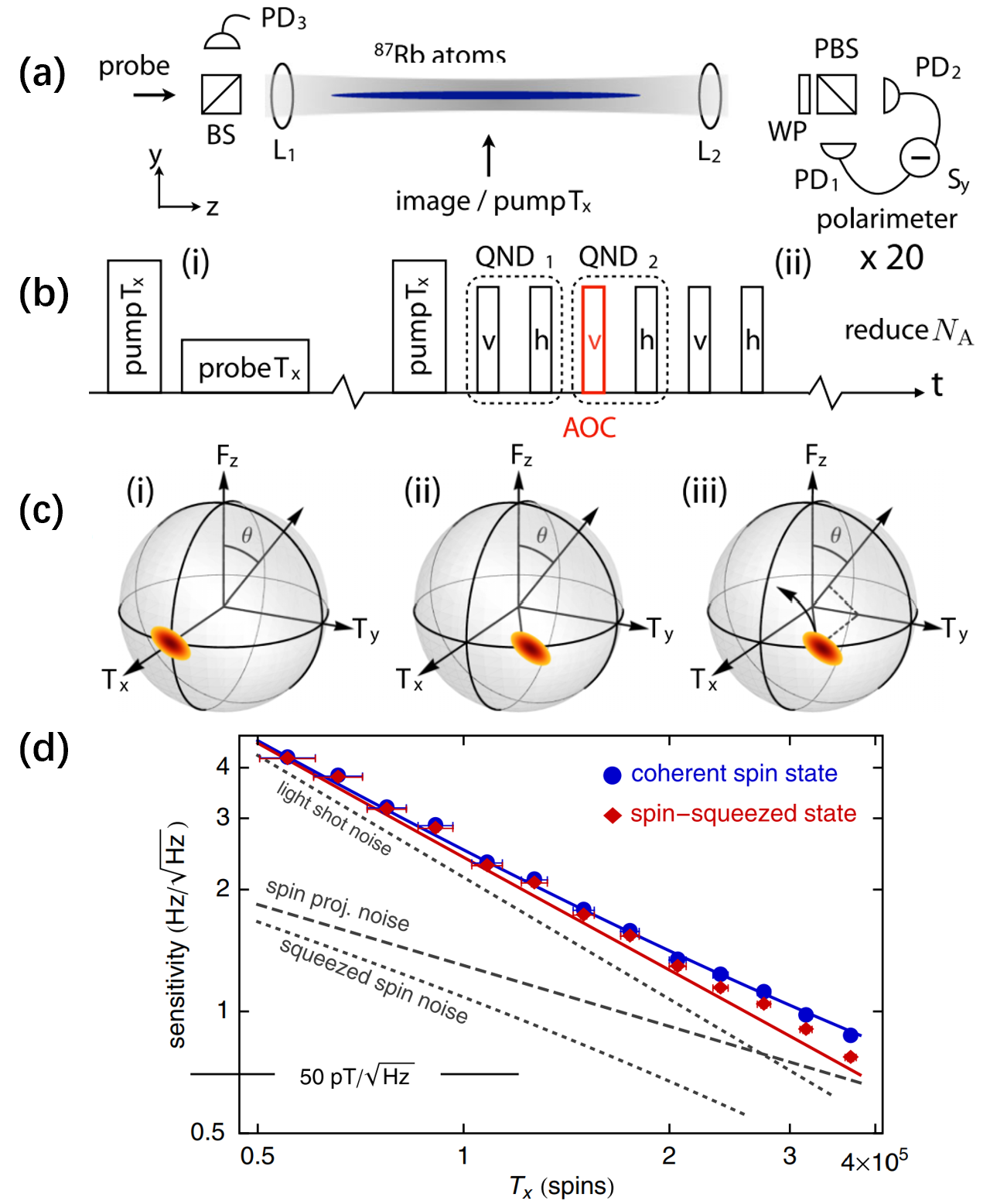}
  \caption{\label{Fig5.2.4}
 Squeezing-enhanced magnetic field sensing with cold atomic ensembles~\cite{PhysRevLett109253605}.
 (a) Experimental geometry. PD: photodiode; L: lens; WP: wave plate; BS: beam splitter; PBS: polarizing beam splitter.
 (b) Measurement sequence for generating spin squeezed state.
 (c) Entanglement-enhanced field measurement.
 (d) Log-log plot of measurement sensitivity with an input spin coherent state (blue circles) and spin squeezed state (red diamonds). Reproduced with permission from Sewell \textit{et al.}, Phys. Rev. Lett. \textbf{109}(25), 253605 (2012). Copyright 2012 American Physical Society.}
\end{figure}

Using synthesized QND measurements in an ensemble of laser-cooled spin-$1$ Rb atoms, the spin squeezed state has been generated and employed to perform high-precision magnetic measurement~\cite{PhysRevLett109253605}.
Based upon a Ramsey sequence, the squeezing-enhanced field measurement in a volume of $V =3.7\times 10^{-6}$ $\textrm{cm}^3$ can achieve a sensitivity of $\Delta B=105fT\sqrt{\textrm{cm}^3 \textrm{Hz}}$ in single-shot measurements.
The experimental geometry is shown in Fig.~\ref{Fig5.2.4}~(a), an ensemble of laser cooled $^{87}$Rb atoms in the ground state.
Then the spin squeezed state is generated via utilizing the pulses sequence, as shown in Fig.~\ref{Fig5.2.4}~(b).
%
%
These pulses are used in pairs to synthesise the QND measurement, or singly for the alignment-to-orientation conversion measurement.
The Ramsey sequence for field measurement is shown in Fig.~\ref{Fig5.2.4}~(c).
Firstly, using synthesized QND measurements to prepares a spin squeezed state.
Then, the state is allowed to evolve for a time due to Zeeman shifts.
Finally, a pulse is used for measurement, giving an integrated signal proportional to the Zeeman shift.
The measurement sensitivity with an input spin coherent state and spin squeezed state is shown in Fig.~\ref{Fig5.2.4}~(d).

In recent, through combining the techniques of optical pumping and continuous QND measurement, a sustained spin squeezed state with $4 \times 10^{10}$ hot atoms has been achieved~\cite{Duan2024}. 
The steady spin squeezing with $\xi^2_R={-3.23 \pm 0.24}$ dB is prepared by applying the prediction and retrodiction QND protocol and can be maintained for about one day. 
The entanglement enhancement due to the steady spin squeezing is utilized by the atomic magnetometer.
The system can be employed to track different types of continuous time-fluctuating magnetic fields, in which deep learning models are constructed to decode the measurement records from the optical signals.  
This work represents an important progress towards using long-lived quantum entanglement resources in realistic sensors.

In principle, utilizing well-developed Ramsey techniques, one can detect DC magnetic fields with ultrahigh sensitivity.
While for measurement of AC magnetic fields, just using Ramsey techniques is insufficient, and various modulations need to be employed.
It has been demonstrated that dynamical decoupling method, one of the well-known quantum control techniques, provides an excellent tool for measuring AC magnetic fields~\cite{science1192739,PhysRevA79062324,PhysRevLett119183603,science1220513,PhysRevA84060302,PhysRevA84042329,science5532,Kotler2011,Shaniv2017,Maze2008,science7009}.
In particular, by combining many-body Ramsey interferometry and rapid periodic $\pi$ pulses, one can estimate DC and AC magnetic fields simultaneously.
Meanwhile, the measurement precisions of DC and AC components can both exhibit Heisenberg scaling simultaneously for entangled particles in GHZ state~\cite{Zhuang2020}.
Moreover, the dynamical decoupling method can be used to realize quantum lock-in measurement for high-precision magnetometers~\cite{science7009,Kotler2011,science7009} and force detector~\cite{Shaniv2017}.
Utilizing many-body entanglement, one can realize many-body quantum lock-in amplifiers, measuring the AC magnetic field with Heisenberg-limited scaling even under a strongly noisy background~\cite{PRXQuantum.2.040317}.

\subsection{\label{sec:5-3}Gravimeters}

Quantum gravimeters make use of quantum systems, such as ultracold atoms or molecules, to probe the gravity and achieve high sensitivity via using quantum resources.
The first-generation of quantum atomic gravimeter~\cite{Kasevich1991, peters1999measurement} was developed in 1990s.
With the unprecedented precision offered by quantum resources, atomic gravimeters are capable of detecting extremely small changes in the gravitational field and provide broad applications in geophysics, geodesy, metrology, and exploration of fundamental physics.
Below we will introduce the progresses in entanglement-enhanced atomic gravimeters.

\begin{figure}[!htp]
 \includegraphics[width=\columnwidth]{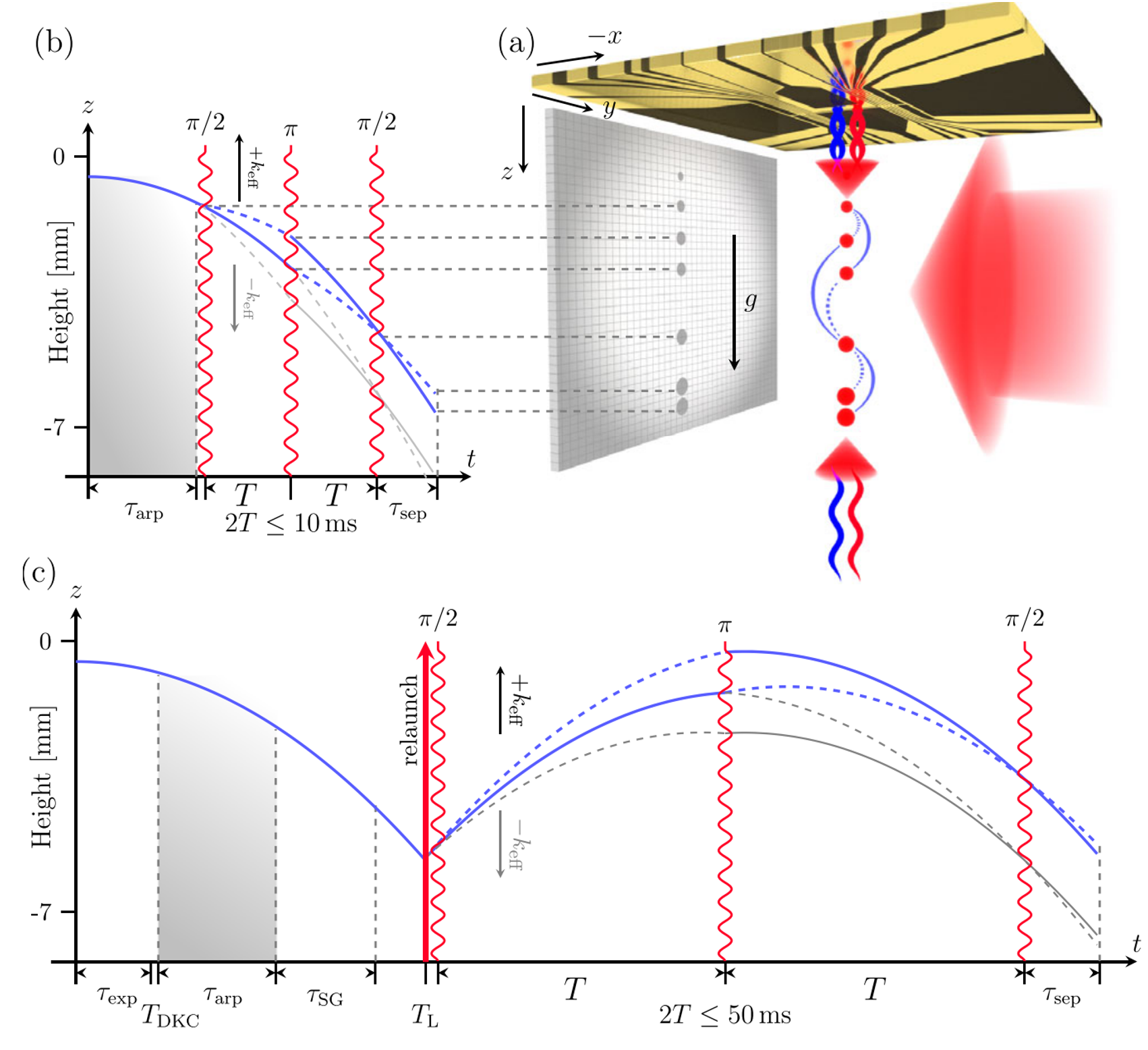}
  \caption{\label{Fig-Sec5-gravity-atomchip} Quantum gravimetry on an atom chip~\cite{PhysRevLett.117.203003}. (a) Schematic diagram of atom interferometry for gravity measurement. (b) Pulse sequence of atom interferometry. (c) Relaunch mechanism using Bloch oscillation and double Bragg mechanism to increase the sensitivity. Reproduced with permission from Abend \textit{et al.}, Phys. Rev. Lett. \textbf{117}(20), 203003 (2016). Copyright 2016 American Physical Society.}
\end{figure}

The basic principle of a quantum gravimeter involves the interferometric measurement of the phase shift induced by gravity on a matter wave~\cite{PhysRevLett.118.183602,PhysRevLett.117.138501}.
As an example, Fig.~\ref{Fig-Sec5-gravity-atomchip} shows a schematic of quantum gravimetry on an atom chip~\cite{PhysRevLett.117.203003}. 
Different from atomic clocks and magnetometers, quantum gravimeters are a type of inertial sensors and they generally require distinct spatial paths.
Generally, quantum gravimeters are implemented via standard Mach-Zehnder interferometers~\cite{peters2001high, PhysRevA.98.023629}, in which beam splitters ($\pi/2$ pulses) and mirrors ($\pi$ pulses) are achieved via Raman transitions. 
The Raman transitions are realized with two counterpropagating lasers of wave vector $k_L$, which coherently couple two internal states.
As a result of momentum conservation, a transition between the two internal states will impart a momentum of $2\hbar k_L$ to an atom, giving the momentum separation needed for gravity measurement. 
The phase of an interferometer is related the gravity acceleration $g$ as $\Phi = k_{eff} \cdot g T^2$ with the effective wave vector $k_{eff}$ and the evolution time $T$.

To improve the precision of quantum gravimeters, in addition to large momentum transfer~\cite{PhysRevLett.118.183602} via Bloch oscillations or double Bragg diffraction, multiparticle entangled momentum states~\cite{Shankar_2019,PhysRevLett.127.140402, Greve2022,PhysRevLett.114.050502} have been introduced.
It has been demonstrated that one can increase the interferometric area via using large momentum transfer~\cite{Leveque2009,Clade2009, Malossi2010,Chiow2011,Altin_2013, Ahlers2016}. 
This allows to improve the sensitivity by increasing the effective separation between the two interfering atomic samples. 
However, there may appear extra phase noises induced by the effective multi-photon process for achieving large momentum transfer. 
Quantum entanglement offers another promising route to improve their measurement precision.
Using spin squeezed states, the sensitivity can be given as 
\begin{equation}
    \Delta g = \frac{\xi_R}{\sqrt{N} k_{eff} T^2}.
\end{equation}
where $N$ is the number of atoms, $\xi_R$ is the squeezing parameter.
In an atom interferometer with entangled momentum states~\cite{PhysRevLett.127.140402}, a spin squeezing of $-3.1$ dB has been achieved in experiments. 
Thus, quantum-enhanced atomic inertial sensors in a more matured form should be promising in the near future~\cite{Greve2022}.

\begin{figure}[!htp]
 \includegraphics[width=\columnwidth]{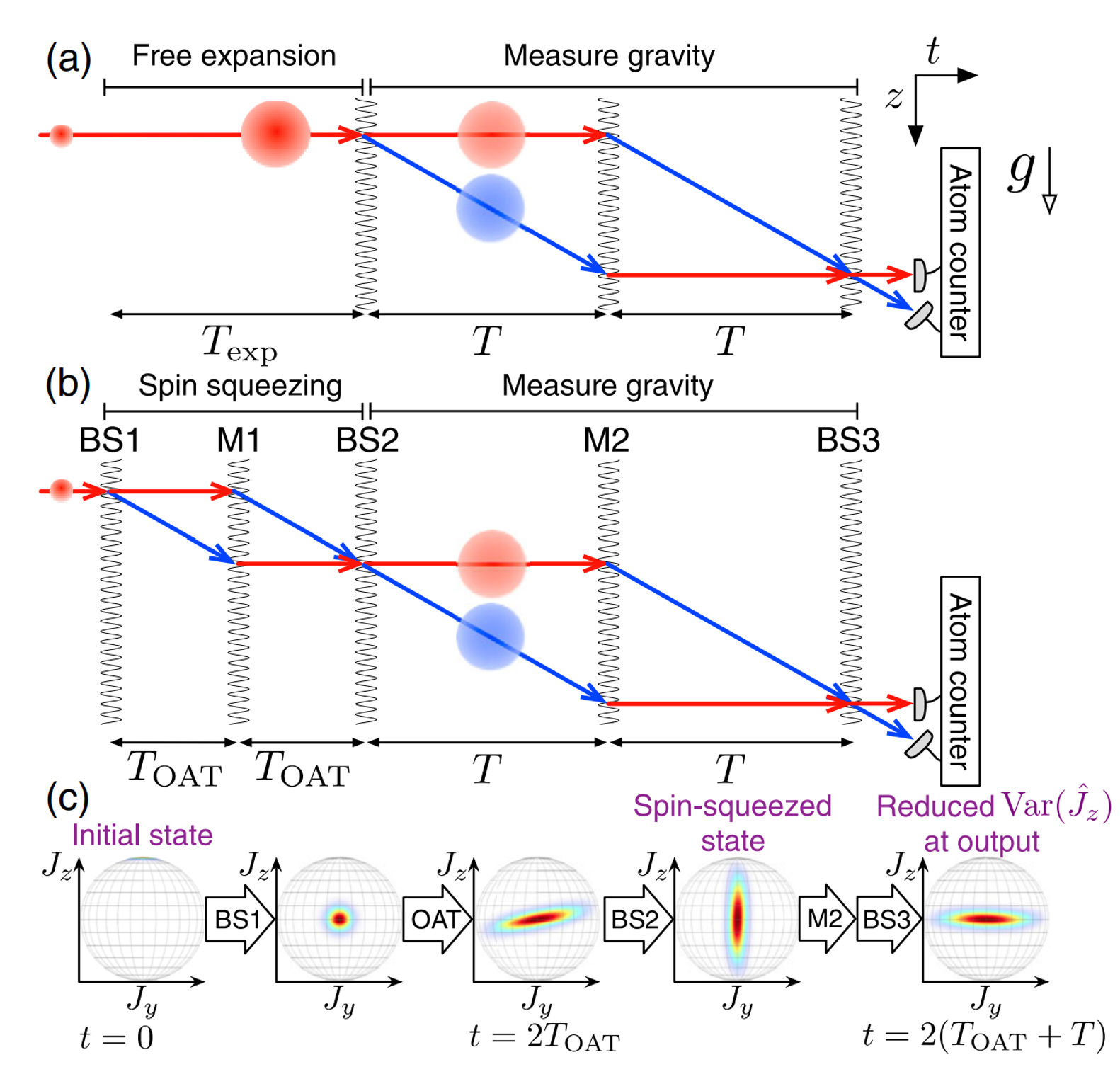}
  \caption{\label{Fig-Sec5-gravity-OAT} Entanglement-enhanced gravity measurement with BEC~\cite{Szigeti2020}. (a) Space-time diagram illustrating gravimetry with a BEC attaining the SQL. Unwanted interatomic interactions are reduced by freely expanding the BEC for duration $T_{\textrm{exp}}$. A $\pi/2 - \pi - \pi/2$ Raman pulse sequence then creates a Mach-Zehnder interferometer of interrogation time $T$. The two interferometer modes correspond to internal states $\ket{1}$ (red) and $\ket{2}$ (blue) with  momentum separation. (b) Quantum-enhanced ultracold-atom gravimetry. During initial expansion duration, the BEC’s interatomic interactions generate spin squeezing via OAT. (c) Bloch sphere representation of state during entanglement-enhanced gravimetry. Reproduced with permission from Szigeti \textit{et al.}, Phys. Rev. Lett. \textbf{125}(10), 100402 (2020). Copyright 2020 American Physical Society.}
\end{figure}

With currently available experimental techniques, a theoretical study~\cite{Szigeti2020} shows that spin squeezed ultracold atoms can improve the sensitivity up to 14 dB beyond the SQL.
As shown in Fig.~\ref{Fig-Sec5-gravity-OAT}~(b), before the standard Mach-Zehnder interference process, an initial state preparation generating spin squeezing via the interatomic interactions during the expansion period is introduced. 
Unlike the trapped schemes, where interatomic collisions cause unwanted multimode dynamics that make it difficult to match the
two modes upon recombination~\cite{PhysRevA.90.023613}, a BEC's spatial mode is almost perfectly preserved under free expansion, even for large atom numbers and collisional energies. 
Furthermore, since the collision energy is converted to kinetic energy during expansion, the interatomic interactions effectively ``switch off'' after $\sim 10$ ms, minimizing their effect during most of the interferometer sequence.
This scheme allows high-precision gravimetry up to a factor of
five below the SQL and is robust to a range of experimental
imperfections.

Due to its compact size, the atomic interferometer via Bloch oscillations~\cite{PhysRevLett.97.060402, PhysRevLett.100.043602, PhysRevLett.106.038501} provides another excellent candidate for building quantum gravimeters. 
In such a setup, cold atoms are held in a vertical optical lattice tilted by the gravity.  
Therefore the cold atoms will undergo Bloch oscillations whose frequency is determined by the gravity and the lattice constant. 
To harness entanglement, a scheme has been developed utilizing an ensemble of ultracold atoms in tilted spin-dependent optical lattices along the direction of gravity~\cite{PhysRevA.98.053826}. 
The fast coherent separation and recombination of atoms can be realized via polarization-synthesized optical lattices. 
The input atomic wave packet is coherently split into two parts by a spin-dependent shift and a subsequent $\pi/2$ pulse. 
Then the two parts are held for accumulating a relative phase related to the gravity. 
Lastly the two parts are recombined for interference by a $\pi/2$ pulse and a subsequent spin-dependent shift. 
The pulses not only preclude the spin-dependent energies in the accumulated phase, but also avoid the error sources, such as dislocation of optical lattices in the holding process. 
In particular, by using entangled states as initial states, the measurement precision of such an interferometer can beat the SQL.

However, even in laboratory-based proof-of-principle apparatus, there is still a considerable gap to overcome in demonstrating entanglement-enhanced gravimetry.   
The main challenge stems from the fact that most methods of generating entangled atomic states do not meet the stringent requirements of precision gravimetry.
Most cold-atom gravimeters necessitate the creation and precise control of distinct and well-isolated atomic matter-wave momentum modes.
To achieve a significant fringe contrast, it is crucial to maintain coherence between these modes for extended periods during the interrogation process, while also ensuring proper alignment of the modes at the output of the interferometer.

\subsection{\label{sec:5-4}Gyroscopes}

High-performance gyroscopes for rotation sensing are crucial for navigation in a wide range of air, ground, marine, and space applications. 
Rotation frequency is commonly measured by using the Sagnac effect, which is the accumulation of a rotation-dependent phase difference by two counter-propagating waves in a revolving loop. 
Gyroscopes operate with either lasers (an ensemble of photons - a type of massless particles) or massive particle such as atoms.

Usually, an optical gyroscope uses two counterpropagating laser beams to sense a rotation. 
If the setup rotates with an angular velocity $\Omega$, the two laser beams will accumulate a relative phase $\Phi = 4 M \Omega A_o /(\lambda c)$ with $M$ the number of rounds, $A_o$ the area enclosed by the gyroscope parallel to the rotation direction, $\lambda$ the wave length, and $c$ the speed of light.
However, the conventional SQL still places a limit on their precision, which is proportional to the surface area contained by the optical path. 
In a recent theoretical proposal, the utilization of distributed quantum sensing is envisioned as a means to enhance the performance of fiber gyroscopes specifically for navigation applications~\cite{PhysRevApplied.14.034065}. 
The idea is to inject quantum-optical squeezed vacuum light, like what was utilized in a RF-sensing experiment~\cite{PhysRevLett.124.150502}, into to an array of fiber gyroscopes to improve the scaling of measurement precisions.
The idea is to inject quantum-optical squeezed vacuum light, like what was utilized in the RF-sensing experiment~\cite{PhysRevLett.124.150502}, into to an array of fiber gyroscopes to improve the scaling of the measurement noise.
Moreover, such a configuration can also be applied to improve the readout signal-to-noise ratio of an array of optomechanical transducers used in inertial sensing~\cite{Hines2020} and atomic force microscopes~\cite{Sugimoto2007}.

Similar to an optical gyroscope, in an atomic gyroscope, two counter-propagating atomic beams accumulate a relative phase $\Phi = 2m \Omega A_a/\hbar$ with $m$ the atomic mass, the angular velocity $\Omega$, and $A_a$ the area enclosed by the gyroscope parallel to the rotation direction.
Theoretically, the sensitivity of an atom interferometer is about ten orders of magnitude larger than the optical counterpart due to $m/\hbar$ is much larger than $1/(\lambda c)$. 
However, the optical interferometers have much larger interferometric area than the atom interferometers, i.e. $4MA_o \gg 2A_a$, practically the sensitivity can be improved by about 3-4 orders of magnitude gain. 
Atomic gyroscopes have made significant advancements since their early demonstrations~\cite{Kasevich1991,Gustavson2000}, achieving a short-term sensitivity of $6 \times 10^{-10}$ (rad/s) $/\sqrt{\textrm{Hz}}$ with a one-second interrogation time~\cite{Gustavson2000}.
Using two counter-propagating atomic beam geometries, one can isolate the phase shifts respectively caused by acceleration and rotation. 
By removing phase shifts caused by non-inertial effects such as magnetic field and ac Stark shift, the long-term sensitivity can be further improved~\cite{PhysRevLett.97.240801}.
To counteract systematic effects and long-term drift, they employed a technique of regularly inverting the propagation vector $k_{eff}$, which resulted in an opposite sign for the inertial phase while preserving the sign of the non-inertial phase. 
By utilizing this approach, the system was able to improve its sensitivity to $3 \times 10^{10}$ (rad/s) $\sqrt{\textrm{Hz}}$ during a 5-hour interrogation duration.

To improve the long-term stability and shorten dead times of atomic gyroscopes, atomic beam source can be replaced by laser-cooled atoms.
By using two cold atomic clouds launched in curved parabolic trajectories in a counter-propagating directions, a cold atomic gyroscope has been demonstrated experimentally~\cite{PhysRevLett.97.010402}, in which the short-term sensitivities of acceleration and rotation were $4.7 \times 10^{-6}$ (rad/s) $/\sqrt{\textrm{Hz}}$ and $2.2 \times 10^{-6}$ (rad/s) $/\sqrt{\textrm{Hz}}$ respectively.
Instead of employing continuously running beams, it utilizes three pulsed Raman beams oriented in three orthogonal directions.
In a similar experiment~\cite{PhysRevA.80.063604}, the long-term stability after an integration time of $1000$ s was shown to be limited by the QPN, $1 \times 10^{-8}$ (rad/s) $/\sqrt{\textrm{Hz}}$.
In recent years, there appear numerous advancements in cold atomic gyroscopes. 
A zero dead-time gyroscope was presented, where continuous operation contributed to enhancing the short-term sensitivity to $1 \times 10^{-7}$ (rad/s) $/\sqrt{\textrm{Hz}}$~\cite{PhysRevLett.116.183003}. 
A Sagnac interferometry with a single-atom clock~\cite{PhysRevLett.115.163001} had been proposed by combining the techniques of state-dependent manipulations and Ramsey pulses.
Based upon an interleaved atom interferometry with a sampling rate of 3.75 Hz and an interrogation time of 801 ms, in which three atomic clouds are simultaneously interrogated, the sensitivity has been improved to $3 \times 10^{-10}$ (rad/s) $/\sqrt{\textrm{Hz}}$, which competes with the best results achieved using fiber-optic gyroscopes~\cite{Savoie2018}.~\cite{Savoie2018}.
In addition, several compact designs utilizing atom chips~\cite{Gersemann2020, Gebbe2021} have shown preliminary rotation sensing capabilities through double Bragg diffraction. 
Furthermore, using the nitrogen vacancy center in diamonds~\cite{PhysRevA.86.052116, PhysRevLett.126.197702}, it is possible to perform rotational measurements, which in turn paves the way for the realization of microchip-size gyroscopes.

Beyond the conventional scheme with individual particles, it is interesting to investigate how to exploit many-body quantum entanglement in measuring rotation frequency. 
Ultracold atomic system provides a feasible experimental platform for achieving this objective~\cite{PhysRevA.93.023616, Luo2017-PRA, PhysRevA.99.052128}.
A spin-squeezed rotation sensor utilizing the Sagnac effect in a spin-1 BEC in a ring trap was proposed~\cite{PhysRevA.93.023616}.
By employing Laguerre-Gauss beams and Raman pulses, the two input states are initially prepared and subsequently amplified through the bosonic enhancement of spin-exchange collisions. 
This process leads to spin-squeezing, which in turn offers quantum enhancement for rotation sensing.
An alternative multi-particle scheme utilizing two-mode Bose condensed atoms was also proposed~\cite{Luo2017-PRA}.
In this scheme, an ensemble of entangled two-state Bose atoms is transported in a ring-shaped path using a state-dependent rotating potential. 
Subsequently, the atoms are recombined for interference by employing Ramsey pulses. 
The duration of phase accumulation is controlled by the state-dependent rotating potential. 
The ultimate sensitivity to rotation can be enhanced to reach the Heisenberg limit by utilizing entangled initial internal degrees of freedom. 
By implementing parity measurement, the measurement precision can be maximized, and the achieved precisions approach the Heisenberg limit.
However, this scheme requires parity measurement based upon single-atom resolution detection. 
The nonlinear detection in Sec.~\ref{sec:4-3} may be helpful for removing the need of single-atom resolution detection.

The journey towards practical entanglement-enhanced quantum sensors may be long and unpredictable. 
A hard and crucial initial step involves surpassing the SQL in measurements to demonstrate the feasibility of a specific inertial quantity. 
Advancements in entanglement-enhanced clocks provide valuable insights that inform the development of design strategies to achieve this goal. 
Furthermore, quantum sensors for acceleration, rotation, gravity, and magnetic fields can be integrated into position, navigation, and timing systems for applications in underwater and underground environments. 
In such applications, quantum sensors surpass their classical counterparts by offering reduced bias, improved precision, and greater stability. 
This may enable navigation with meter-level accuracy without relying on global satellite systems~\cite{Bongs2023}.

\section{\label{sec:6}Summary and outlooks}

In summary, the utilization of multi-particle entanglement offers promising potentials to improve measurement precision beyond conventional shot-noise limit.   
This has sparked an active and dynamic research field focused on characterizing, manipulating, and utilizing entangled multi-particle quantum states~\cite{dai2016generation, PhysRevResearch.3.043122, PhysRevA.108.053327}.
With the rapid progress in creating and detecting multi-particle entangled state, quantum metrology is emerging as a domain capable of achieving quantum advantages over conventional measurements and sensing for practical applications in the near future. 
The field of entanglement-enhanced quantum metrology is poised to shift from theoretical studies~\cite{Zhuang2022} to experimental demonstrations and practical applications, specifically in the realm of precise measurements involving frequency~\cite{PhysRevApplied.16.064056}, electromagnetic fields~\cite{Simons2021, Ji2024}, and forces.
Nevertheless, there remain several challenges in constructing entanglement-enhanced quantum sensors for real-world applications.

Developing a practical entanglement-enhanced quantum sensor poses a significant challenge, not in generating entanglement itself, but in generating entanglement that meets the precise sensing requirements of a particular application. 
These requirements, such as particle number and interrogation time, impose constraints on achievable measurement precision.
One of the difficulties lies in scalability with particle number, as an increase in particles tends to rapidly amplify errors and complicate the experiment.
If only a limited number of particles can achieve a certain degree of entanglement, the improvement over the SQL may not be readily apparent.  
Similarly, if entanglement can only persist for a short interrogation time or is susceptible to weak experimental noises, it may not lead to a significant enhancement in practical sensing tasks.  
Future research should prioritize utilizing entanglement to surpass the SQL and focus on developing entanglement-enhanced quantum metrology protocols specifically tailored for practical sensing scenarios. 

It is worth noting that, while lots of experiments demonstrate improved precision scaling in entanglement-enhanced metrology, their absolute measurement precisions for certain quantities still do not match the highest levels achieved by state-of-the-art measurements performed without entanglement. 
Although various quantum entanglement has been successfully generated in experiments, utilizing quantum entanglement for realistic measurements continues to encounter several challenges, resulting in performance that lags behind state-of-the-art methods without involving entanglement.
Exhilaratingly, a recent experiment of optical lattice clock comparison~\cite{Robinson2024} has demonstrated a stability performance at the level of $10^{-17}$, which is close to the best state-of-the-art clocks~\cite{Bothwell2022}. 
With the growing capabilities of quantum control for engineering many-body entangled states~\cite{PhysRevLett.132.190001}, the prospect of transitioning from demonstration to achieving genuine quantum advantage in the near future is increasingly promising.

Quantum entanglement is highly susceptible to decoherence~\cite{PhysRevA.81.052330,PhysRevA.76.042127,Wan2020,PhysRevLett.109.233601,PhysRevA.106.023703}, leading to a significant reduction in achievable measurement precision~\cite{Jeske2014, Wang2017, Wan2020}.
This fragility is particularly evident when employing long interrogation times to achieve high precision.
Theoretical investigations have demonstrated the detrimental impact of decoherence on surpassing the SQL~\cite{PhysRevA.76.032111, Huelga1997}.
Typically, measurement precision returns to the SQL during an optimal transient interrogation period but diverges in the long-interrogation-time regime due to the effects of decoherence.
To mitigate the negative effects of decoherence on quantum metrology, various strategies have been proposed~\cite{PhysRevLett.115.110401,Berni2015,lu2015robust,PRXQuantum.3.020310,PhysRevLett.130.240803,Kessler2014,PhysRevLett.112.080801,Unden2016,Zhou2018,Nautrup2019,PhysRevA.100.022335,Huo2017,RevModPhys.76.1267,PhysRevLett.130.070803,PhysRevLett.131.050801,PhysRevA.106.023703}.  
These include adaptive~\cite{PhysRevLett.111.090801,PhysRevX.7.041009,PhysRevResearch.5.013138} and quantum nondemolition measurements~\cite{PhysRevLett.125.200505}, correlated decoherence~\cite{Jeske2014}, and dynamical control~\cite{PhysRevA.87.032102,PhysRevA.94.052322,PhysRevX.10.031003}, aiming to restore the Heisenberg limit.
With advancements in strategies such as quantum reservoir engineering~\cite{PhysRevLett.123.040402}, Floquet engineering~\cite{PhysRevLett.131.050801}, quantum error mitigation~\cite{PhysRevX.11.041036, PhysRevLett.129.250503} and other techniques, quantum metrology is on the verge of transitioning from the laboratory to practical use.
Additionally, quantum feedback~\cite{PhysRevLett.115.110401,Berni2015,PRXQuantum.3.020310,PhysRevLett.130.240803}, quantum error correction~\cite{Kessler2014,PhysRevLett.112.080801,Unden2016,Zhou2018,Nautrup2019,Unden2016,Zhou2018,PhysRevLett.112.080801} and other techniques~\cite{PhysRevA.100.022335,Huo2017,PhysRevLett.122.010408,PhysRevA.96.012117,PhysRevLett.131.043602,PhysRevX.7.041009} can be employed to extend coherence time and achieve robust high-precision measurements in experiments.
Moreover, various modulation techniques such as periodic modulation~\cite{PhysRevA84042329,science7009}, quantum heterodyne~\cite{Meinel2021}, quantum mixing~\cite{PhysRevX.12.021061} and quantum lock-in amplification measurement~\cite{Kotler2011,PRXQuantum.2.040317} enable high-precision frequency estimation even in noisy environments. 
The combination of these quantum control techniques with quantum entanglement holds the potential for significant advancements in the field of entanglement-enhanced sensing technologies.

On the other hand, while enhancing measurement precision, it is crucial to improve measurement accuracy by minimizing systematic errors.
In general, systematic errors always introduce inaccuracies that impact the overall performance of measurements~\cite{PRXQuantum.1.010306,Len2022N}.
For instance, in atomic optical clocks, system errors arising from laser probe frequency or intensity instabilities can lead to systematic frequency shifts, compromising clock accuracy.
To mitigate the effects of laser intensity variations and reduce corresponding frequency shifts, composite pulse sequences can be employed~\cite{Willette2018,PhysRevLett.122.113601,PhysRevLett.120.053602}, improving measurement accuracy. 
Other sources of systematic errors, such as collision shifts, Doppler shifts, Stark shifts, and blackbody radiation shifts, can also have adverse effects~\cite{Ludlow2015}. 
Therefore, it is important to assess systematic errors throughout the entire process of quantum parameter estimation and develop methods to suppress specific system errors in many-body sensing systems.

To optimize the performance of a practical quantum metrology, in addition to conventional optimization tools, machine learning has emerged as a powerful tool~\cite{PhysRevLett.121.150503,Cimini2023}.
Machine learning may optimize various aspects of a quantum metrology process~\cite{Huo_2022}, including generation of entangled states, signal accumulation, entangled state readout, and parameter estimation.
In scenarios with limited resources, such as limited coherence time or measurement times, machine learning approaches like reinforcement learning~\cite{Chen2019, Qiu2022, Cimini2023}, quantum algorithms~\cite{PhysRevLett.123.260505,  PhysRevLett.118.150503,Marciniak2022N, PRXQuantum.4.020333}, and Bayesian inference~\cite{Valeri2020,Qiu2022-2, Gebhart2023NRP,PhysRevResearch.6.023201} can guide us in determining the most effective strategies for conducting metrology~\cite{Huang2024}.
The exploration of quantum metrology protocols augmented by machine learning offers an alternative pathway towards achieving practical high-precision measurements.

Besides the typical entangled states discussed in Sec.~\ref{sec:3}, strong types of quantum correlations such as EPR steering and Bell nonlocality are also useful for quantum metrology~\cite{Frerot2023}.
It has been demonstrated that EPR-assisted metrology may precisely estimate a local phase shift and its generating observable~\cite{Yadin2021,PhysRevA.108.012435}.
Bell nonlocality seems to be linked to high QFI, implying a potential connection to metrological advantages~\cite{PhysRevA.99.040101, PhysRevLett.131.070201}.
Moreover, nonclassical states of atomic ensembles can be spatially distributed in different positions~\cite{Fadel2018, Kunkel2018, Lange2018, Jing2019, Vitagliano2023}, which allows for quantum enhanced distributed sensing~\cite{PhysRevA.102.012412, Malia2022,Fadel2023}.

In addition, associated with quantum entanglement, quantum criticality has been recognized as a novel quantum resource for enhancing quantum metrology~\cite{PhysRevLett.121.020402,PhysRevX.8.021022,PhysRevLett.126.200501, PhysRevA.78.042105,PRXQuantum.5.020342}.
On one hand, quantum systems prepared near the critical point exhibit extreme sensitivity to small perturbations~\cite{PhysRevA.78.042106}. 
By leveraging the criticality associated with quantum phase transitions, high-precision measurements of physical quantities can be achieved.
Systems near a quantum phase transition display a divergent susceptibility, indicating the potential for achieving remarkably high precision by utilizing quantum critical systems as probes for parameter estimation~\cite{PhysRevLett.130.240803,liu2021experimental}. 
On the other hand, parameter estimation can be accomplished by measuring the equilibrium state or dynamical properties of the system when it is driven close to the critical point of a quantum phase transition~\cite{PhysRevLett.120.150501,PhysRevA.105.042620}. 
Furthermore, the response of eigenstates in non-Hermitian systems to small perturbations at singularities can also be utilized to detect specific physical quantities~\cite{Chen2017, Lai2019, Wang2020, McDonald2020, Wiersig2020,  PhysRevLett.124.020501, PhysRevLett.123.180501, PhysRevLett.125.180403, PhysRevA.103.042418, Ou2021,Li2023-2}.

At last, it is important to note that multi-parameter  estimation~\cite{Szczykulska2016, ALBARELLI2020126311, PhysRevA.96.042114, PhysRevLett.126.120503, PhysRevLett.119.130504,Liu_2020,PhysRevLett.121.130503,Polino:19} is a crucial problem in both fundamental research and practical applications~\cite{Gessner2020NC,PhysRevLett.126.080502,Pezz2021EntanglementenhancedSN}, alongside single-parameter estimation. 
It has been extensively employed for various purposes, including vector field detection~\cite{wang2015high,Wang2021NanoscaleVA}, imaging and biological sensing~\cite{Taylor2016,Thiel2016}.
Several studies have demonstrated that a multi-parameter  approach can outperform the optimal quantum individual estimation scheme~\cite{ALBARELLI2020126311} and offer advantages in the presence of noise~\cite{Roccia18,Donati2014}.
Although progress has been made in multi-parameter estimation scenarios~\cite{PhysRevA.98.033603,Liu_2016,PhysRevA.103.042615,Liu_2016,PhysRevA.98.012114,PhysRevLett.121.130503,Gessner2020NC,PhysRevA.103.042615,Liu_2020,Donati2014,Roccia18}, including finding the optimal measurement strategies~\cite{PhysRevX.11.011028, Hayashi2023, Hayashi2024}, how to experimentally achieve the optimal multi-parameter measurement in the presence of noise remains a challenge.
Furthermore, the development of entanglement-enhanced multi-parameter  estimation is crucial for practical quantum sensing, such as vector magnetometers~\cite{Wang2021NanoscaleVA,wang2015high} and multimode quantum imaging~\cite{PRXQuantum.4.020333,ALBARELLI2020126311}.

\begin{acknowledgments}
J. Huang and M. Zhuang contribute equally to this work.
The authors acknowledge Matteo Fadel, Kaoru Yamamoto, Geza Toth, Yingkai Ouyang, and Chao-Yang Lu for their valuable suggestions.
The authors thank Jihao Ma, Jungeng Zhou, Sijie Chen, Yi Shen, Yuehua Pang, and Jinye Wei for critical reading and helpful suggestions.
This work is supported by the National Natural Science Foundation of China (12025509, 12305022) and the National Key Research and Development Program of China (2022YFA1404104).
\end{acknowledgments}








%

\end{document}